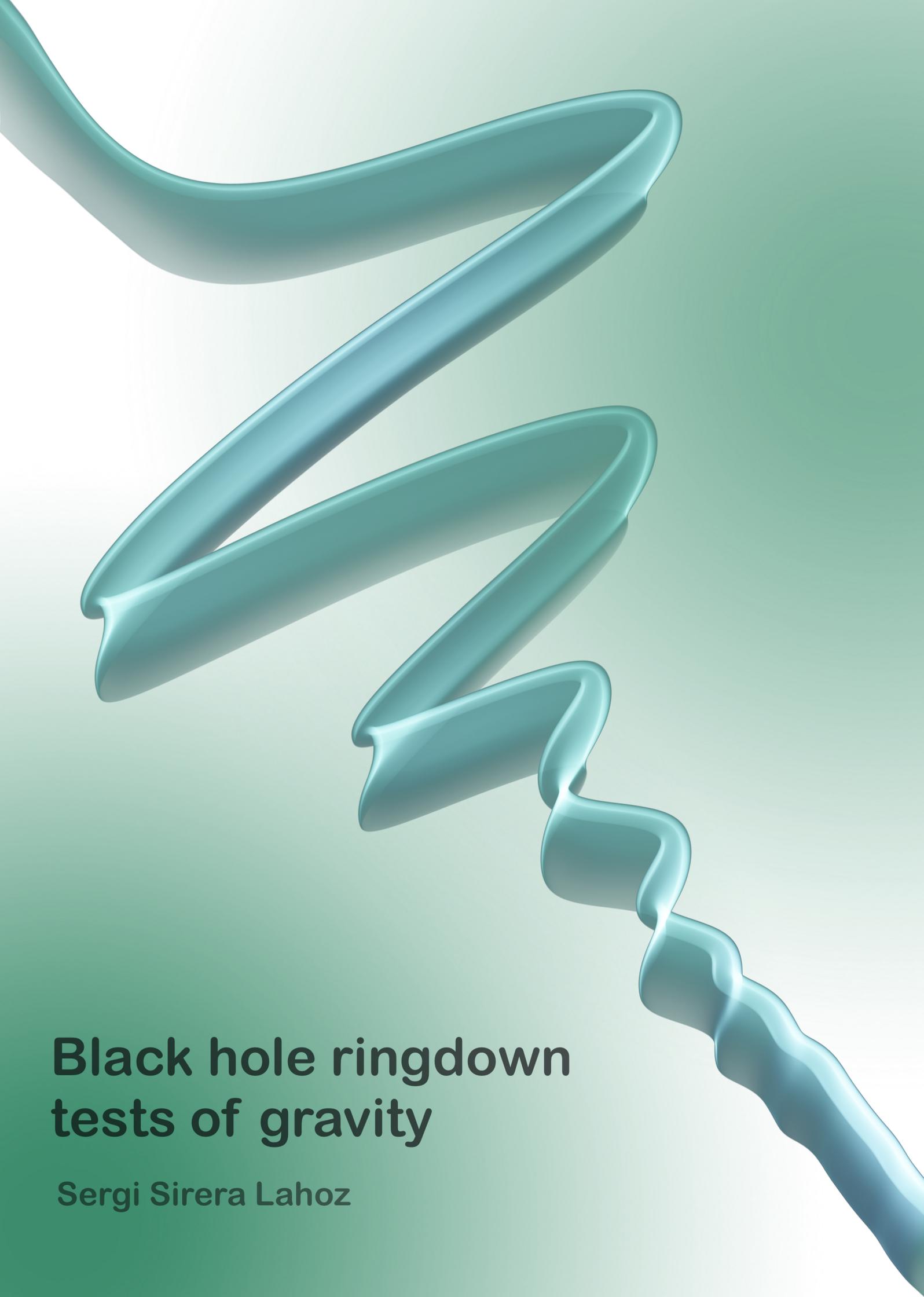

# Black hole ringdown tests of gravity

Sergi Sirera Lahoz

# Black hole ringdown tests of gravity

Sergi Sirera Lahoz

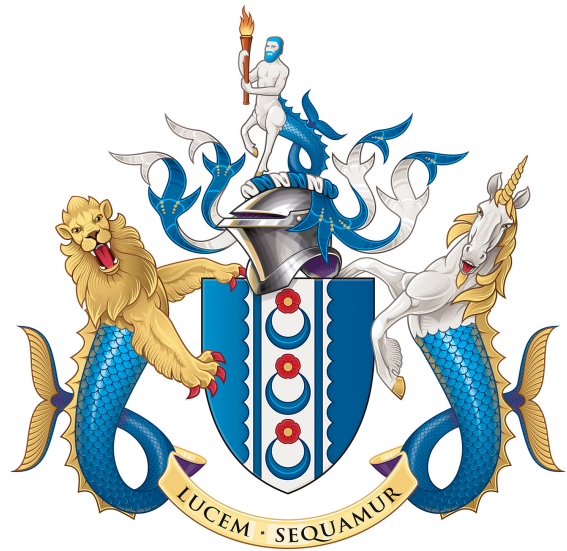

May 2025

*This thesis is submitted in partial fulfillment of the requirements
for the award of the degree of
Doctor of Philosophy of the University of Portsmouth*

Institute of Cosmology and Gravitation
University of Portsmouth

**Declaration**

The work of this PhD thesis was carried out at the Institute of Cosmology and Gravitation, University of Portsmouth, United Kingdom, under the supervision of Professor Johannes Noller.

Whilst registered as a candidate for the above degree, I have not been registered for any other research award. The results and conclusions embodied in this thesis are the work of the named candidate and have not been submitted for any other academic award.

**Word count**

This thesis has an estimated word count of 67,974 as calculated by the LaTeX Workshop extension in VS Code.

**Examination**

The contents of this thesis were examined in a viva voce held on 24th June 2025 by Prof Enrico Trincherini and Prof David Wands, whom I would like to thank for a pleasant and stimulating discussion.

The important thing is not to stop questioning. Curiosity has its own reason for existing. One cannot help but be in awe when he contemplates the mysteries of eternity, of life, of the marvelous structure of reality. It is enough if one tries merely to comprehend a little of this mystery every day.

---

Albert Einstein, *Life magazine, May 2, 1955*

# Acknowledgements

*If I have seen further, it is by standing on the shoulders of giants.*

Newton

I have been immensely fortunate to stand on the shoulders of many giants—both academic and personal—whose support and guidance have sustained me throughout these years. Precisely given how many there have been, I was almost tempted to avoid naming them individually. Thankfully, I have been persuaded otherwise. So here, I would like to extend my heartfelt thanks to each of them.

First and foremost, I am deeply grateful to my supervisor, Johannes. Aside from being nearly a literal giant, your impact on my academic journey has been truly immense. Thank you for placing your trust in me and giving me the chance to pursue my dream of becoming a scientist. Your mentorship has not only guided me through the complexities of research but also inspired me to grow as a thinker and learner. I am especially thankful for the opportunity to explore the fascinating topics we've worked on together. I consider myself incredibly lucky to have had you as a supervisor; it has been a truly rewarding and well-matched partnership that I hope continues for years to come.

I would also like to thank my collaborators in Japan and Thailand. A special thank you to Shinji for hosting me in Kyoto for an unforgettable three months filled with science, travel, food, and adventure. Thank you for your kindness, generosity, and for everything I learnt from you—and for the spare bike, which I promise I almost used. I'm also especially grateful to Kazufumi, who was an absolute pleasure to work with. Thank you for being the kind of collaborator everyone wishes for: approachable, rigorous, knowledgeable, and fully committed. You've made me a better researcher.

For most of my PhD, I've been based at the ICG in Portsmouth. I couldn't have imagined a better place to pursue my doctorate. I'm thankful to the ICG as an institution, and to all its members who make it such a special place. It's been a slightly bittersweet privilege to witness the ICG's evolution over the years—seeing some people leave and others arrive—while it has always maintained the warm, vibrant, and intellectually stimulating atmosphere that defines it. Thank you to the faculty and admin team for helping preserve that. A special thank you to the Theoretical Cosmology and Gravitational Waves groups for welcoming me into such supportive and inspiring research groups.

Over the last couple of years, I've also been fortunate to regularly visit the beautiful UCL campus in London for supervisory meetings. Thank you to the Department of Physics and Astronomy for making that possible and always making me feel welcome.

I don't believe there's a clear divide between academic and personal influences. A perfect example of this is the brilliant people I've met at conferences—scientists from around the world who are now also friends. Too many to name, but you know who you are. Thank you for making academia more human, more inspiring, and a lot more fun.

Speaking of friends, I have been lucky to count on many. This thesis is truly as much yours as it is mine—except you didn't write it and you don't have to defend it. Lucky you.

To my fellow PhD cohort, thank you for sharing this journey from day 1 to the finish line. Fox, I feel very grateful for having shared an office (actually two offices) with you for so long. Thank you for filling up my days with *Star Trek* (or was it *Doctor Who*?) references I never quite got. Molly, you've been an inspiring friend—and together we definitely made the Gravity Meeting the coolest one in the ICG. Susanna, I've loved working on *Physics Chat* with you. Thank you for your life lessons and for keeping me up with the Kardashians.

To my older PhD crew—who I miss so much—thank you for welcoming me into your lives. Rafaela, just thank you. I'm so glad you came into my life. I hope you never leave. You've been the coolest office mate and lifelong friend I could have asked for. Robyn, you've been someone I've looked up to since day one—thank you for that. Chane, thank you for being such a kind and genuine soul. I miss our walks back from the ICG. Bart, I miss our pub chats and swimming sessions, but I'm grateful to have those memories. Arthur, your humour has been a much-needed refreshment throughout this journey. Connor, Jamie, and Lolla—thank you for keeping me grounded whenever I drifted too far into delusion, and for all the laughs, chats, and unforgettable moments at the pub.

To the rest of the PhD community—thank you for making coming into work something to look forward to (beyond the science, of course). Ana, thanks for helping represent the Spanish community and organising the occasional reggaeton night. Tian, your enthusiasm and warmth are infectious. Sophie, thanks for all the tea and never letting me down on our Duolingo friend quests. Sai, thanks for supporting Barça of course. Arthur, thank you for your rare talent for squeezing an absurd amount of humour into a single line. Natalie, thank you for being such a fun and interesting person to sit next to (and sorry for being annoying). I hope you enjoy my desk. Emily, thank you for your fun facts and always bringing good vibes and easy conversation. To everyone else—you may not be listed here, but I am no less grateful.

A very special thank you to Elena, who could easily belong in every single category of these acknowledgements. Thank you for being the extraordinary person you are. I'm grateful for the big things, the small things, your support in the not-so-happy times, and your joy in the good ones. Your presence has made this journey brighter and infinitely more meaningful.

To the postdocs—despite your above-minimum-wage salaries—you still chose to spend time with us, and for that I'm grateful. I always enjoyed your company (and your occasional delusions that we were in the same boat). You're simply great people.

Outside academia, life continues—and I was lucky it included some truly wonderful humans. Within Portsmouth, I am especially gratefull to the Bransbury football gang: thank you for the weekly games and for being such a nice group. Marco, thank you for your invaluable friendship over these years—it's meant a lot. To my Greek brothers George and Aris, thank you for the barbecues, the Champions League nights, and the beers. And to the non-footballing Italians—Luca, Danilo, and Atmadev (yes, you're Italian too)—thank you for the endless board game nights. Finally, Naomi, thank you for always listening—whether I was ranting about something mildly infuriating or rambling about something wildly and unjustifiably interesting. I'm truly grateful to have met you.

Whenever Portsmouth felt too small, I could always escape to London and find comfort in the company of old friends. Irene and Diego, thank you not just for always offering me a place to crash, but more importantly for the many adventures we've shared—I hope there's many more. To Pablo—remember when we were freshers with big dreams of doing research? Look at us now. I'm proud to be your friend. Panos, Nick, and Ethan—our friendship has stood the test of time and distance. That says everything about how much you all mean to me. I hope we're still sahring beers in ten years.

Whenever I wanted to leave the UK (or the UK wanted me to leave), I could always go back to Barcelona and be surrounded by friendships that have shaped who I am. Artur, Emma and Helena, thank you for being a constant part of my life, no matter the distance. Eric, Puchi and Adrià, thank you for always wanting to see me when I was back. Finally, Zizawah—thank you for making my time in Kyoto immeasurably better. I admire your strength, positivity, and generosity, and I'm proud to say that knowing you has made me a better person.

To my family—thank you for always being there, for your unwavering emotional support, encouragement, and genuine interest in this long journey. I'm incredibly lucky to have you in my life. A special thank you goes to my grandparents, Paulino and Dora, who have been, and will always remain, true role models. I am deeply proud that the sacrifices you made helped pave the way for me to reach this point—and I hope you're just as proud of yourselves as I am of you.

Above all, I would like to thank my parents Carles and Merce, and my sister Paula. Thank you for always being my greatest source of strength and support. Without your unconditional love this would not have been possible.

# Abstract

Understanding gravity is at the heart of some of the biggest questions in modern physics. While General Relativity (GR) is a theoretically unique and experimentally well-tested framework, it remains important to question whether it accurately describes gravity at all scales. This motivates the exploration of broader theories of gravity, particularly in regimes where deviations from GR might still arise. Black holes provide ideal natural laboratories for testing gravitational theories in the strong-field regime, and the recent advent of gravitational wave (GW) astronomy has opened a new observational window into these extreme environments. In particular, the final stage of a compact binary merger—the ringdown phase—is of great interest. Here, under the so-called *black hole spectroscopy* program, the study of quasinormal modes (QNMs) offers a powerful tool to probe the fundamental nature of gravity and to extract intrinsic properties of black holes.

This thesis investigates black hole solutions and their QNM spectra within scalar-tensor (ST) theories of gravity—well-motivated extensions of GR that include an additional scalar degree of freedom. In particular, it focuses on 'stealth' black holes, where scalar hair exists without altering the background metric but can modify the QNM spectrum. By analysing perturbations around such spacetimes, we derive forecasted constraints on beyond-GR parameters for current and future GW detectors.

Three main investigations are presented: (i) a novel method to constrain the speed of gravity using ringdown signals alone; (ii) a stability and QNM analysis of black holes with linearly time-dependent scalar hair; and (iii) a general classification of stealth solutions in higher-order scalar-tensor (HOST) theories, including a stability analysis and identification of ringdown observational signatures.

Together, these studies contribute new theoretical tools and observational forecasts that advance our understanding on fundamental gravitational physics in the era of GW astronomy.

# Dissemination

This thesis is based on the following published works:

1. **S. Sirera**, J. Noller
   "Testing the speed of gravity with black hole ringdown"
   Phys. Rev. D 107, 124054 (2023); [2301.10272]
2. **S. Sirera**, J. Noller
   "Stability and quasinormal modes for black holes with time-dependent scalar hair"
   Phys. Rev. D 111, 044067 (2025); [2408.01720]
3. H. Kobayashi, S. Mukohyama, J. Noller, **S. Sirera**, K. Takahasi, V. Yingcharoenrat
   (authors are listed alphabetically)
   "Inverting no-hair theorems: How requiring General Relativity solutions restricts scalar-tensor theories"
   Phys. Rev. D 111, 124022 (2025); [2503.05651]

The code written for this work (mostly in Mathematica) is open source and available at:

▸ **S. Sirera**
   "ringdown-calculations"
   github.com/sergisl/ringdown-calculations

# Contents

# Notation and conventions

The metric signature convention is $(-, +, +, +)$. The following conventions will be used:

| **Spacetime indices** | |
|---|---|
| Greek: $\mu, \nu, \cdots$ | 4 spacetime dimensions, e.g. $x^\mu = (t, r, \theta\phi)$ |
| Lower-case latin: $a, b, \cdots$ | 2 time+radial spacetime dimensions, e.g. $x^\mu = (t, r)$ |
| Upper-case latin: $A, B, \cdots$ | 2 angular spacetime dimensions, e.g. $x^\mu = (\theta, \phi)$ |

| **Metric tensors** | |
|---|---|
| Flat Minkowski metric | $\eta_{\mu\nu}$ |
| Full metric | $g_{\mu\nu}$ |
| Background metric | $\tilde{g}_{\mu\nu}$ |
| Metric perturbations | $h_{\mu\nu}$ |
| Metric determinant, i.e. $\det(g_{\mu\nu})$ | $g$ |
| Trace of metric perturbations, i.e. $h^\mu_{\ \mu}$ | $h$ |

| **Curvature objects** | |
|---|---|
| Partial derivative | $\partial_\mu$ |
| Covariant derivative | $\nabla_\mu$ |
| D'Alembert operator | $\Box \equiv \nabla_\mu \nabla^\mu$ |
| Christoffel Symbol | $\Gamma^\rho_{\mu\nu} \equiv \frac{1}{2} g^{\rho\sigma} \left( \partial_\mu g_{\nu\sigma} + \partial_\nu g_{\mu\sigma} - \partial_\sigma g_{\mu\nu} \right)$ |
| Riemann Curvature Tensor | $R^\rho_{\ \sigma\mu\nu} = \partial_\mu \Gamma^\rho_{\nu\sigma} - \partial_\nu \Gamma^\rho_{\mu\sigma} + \Gamma^\rho_{\mu\lambda} \Gamma^\lambda_{\nu\sigma} - \Gamma^\rho_{\nu\lambda} \Gamma^\lambda_{\mu\sigma}$ |
| Ricci Tensor | $R_{\mu\nu} \equiv R^\lambda_{\ \mu\lambda\nu}$ |
| Ricci Scalar | $R \equiv g^{\mu\nu} R_{\mu\nu}$ |
| Einstein Tensor | $G_{\mu\nu} \equiv R_{\mu\nu} - \frac{1}{2} g_{\mu\nu} R$ |

| **Identities** | |
|---|---|
| Symmetric tensor | $A_{(\mu\nu)} \equiv \frac{1}{2}(A_{\mu\nu} + A_{\nu\mu})$ |
| Antisymmetric tensor | $A_{[\mu\nu]} \equiv \frac{1}{2}(A_{\mu\nu} - A_{\nu\mu})$ |
| Ricci identity | $[\nabla_\mu, \nabla_\nu] A^{\alpha_1 \dots \alpha_r}_{\quad\quad \beta_1 \dots \beta_s} = R^{\alpha_1}_{\ \sigma\mu\nu} A^{\sigma \dots \alpha_r}_{\quad\quad \beta_1 \dots \beta_s} + \cdots + R^{\alpha_r}_{\ \sigma\mu\nu} A^{\alpha_1 \dots \sigma}_{\quad\quad \beta_1 \dots \beta_s}$ $- R^\sigma_{\ \beta_1 \mu\nu} A^{\alpha_1 \dots \alpha_r}_{\quad\quad \sigma \dots \beta_s} - \cdots - R^\sigma_{\ \beta_s \mu\nu} A^{\alpha_1 \dots \alpha_r}_{\quad\quad \beta_1 \dots \sigma}$ |
| First Bianchi identity | $R^\rho_{\ [\mu\nu\sigma]} \equiv R^\rho_{\ \mu\nu\sigma} + R^\rho_{\ \nu\sigma\mu} + R^\rho_{\ \sigma\mu\nu} = 0$ |
| Second Bianchi identity | $\nabla_{[\lambda} R^\rho_{\ \sigma\mu\nu]} \equiv \nabla_\lambda R^\rho_{\ \sigma\mu\nu} + \nabla_\mu R^\rho_{\ \sigma\nu\lambda} + \nabla_\nu R^\rho_{\ \sigma\lambda\mu} = 0$ |

| **Scalar** | |
|---|---|
| Scalar field | $\phi$ |
| Background scalar field | $\tilde{\phi}$ |
| Covariant derivatives on the scalar | $\phi_{\mu_1 \cdots \mu_m}^{\quad\quad \nu_1 \cdots \nu_n} \equiv \nabla^{\nu_n} \cdots \nabla^{\nu_1} \nabla_{\mu_m} \cdots \nabla_{\mu_1} \phi$ |
| Scalar canonical kinetic term | $X \equiv -\frac{1}{2} \phi_\mu \phi^\mu$ |

Note that background quantitites—e.g. $\tilde{g}_{\mu\nu}, \tilde{\phi}, \tilde{R}, \dots$—will oftentimes be written without the overbars in order to avoid cluttering. Alternatively, the background scalar and its canonical kinetic term will sometimes be referred to as $\phi_0$ and $X_0$. This is clarified within the text when relevant.

# Units

Throughout this thesis, unless otherwise stated, we adopt natural and/or geometric units. However, in select cases—especially when interpreting observational data—standard SI units will be used. When such conversions are made, the appropriate values of the universal constants below are reinstated explicitly.

| Quantity | SI Units | Natural Units $(\hbar = c = 1)$ | Geometric Units $(G = c = 1)$ |
|---|---|---|---|
| Speed of light $c$ | $2.998 \times 10^8$ m/s | 1 | 1 |
| Reduced Planck constant $\hbar$ | $1.055 \times 10^{-34}$ J·s | 1 | $2.61 \times 10^{-70}$ m$^2$ |
| Newton's constant $G$ | $6.674 \times 10^{-11}$ m$^3$/kg·s$^2$ | $6.71 \times 10^{-39}$ GeV$^{-2}$ | 1 |
| Planck mass $M_{\text{Pl}}$ | $2.18 \times 10^{-8}$ kg | $1.22 \times 10^{19}$ GeV | 1 |
| Solar mass $M_\odot$ | $1.989 \times 10^{30}$ kg | $1.12 \times 10^{57}$ GeV | 1.477 km |
| Cosmological constant $\Lambda$ | $\sim 1.3 \times 10^{-52}$ m$^{-2}$ | $\sim 5.06 \times 10^{-84}$ GeV$^2$ | $\sim 1.3 \times 10^{-52}$ m$^{-2}$ |
| Mass | 1 kg | $5.62 \times 10^{26}$ GeV | $7.42 \times 10^{-28}$ m |
| Length | 1 m | $5.07 \times 10^{15}$ GeV$^{-1}$ | 1 m |
| Time | 1 s | $1.52 \times 10^{24}$ GeV$^{-1}$ | $2.998 \times 10^8$ m |
| Energy | 1 J | $6.24 \times 10^9$ GeV | $1.6 \times 10^{-10}$ m$^{-1}$ |

Note that the electric constant $\epsilon_o$ and Boltzmann's constant $k_B$ are not included here because the systems and phenomena considered in this thesis do not rely on thermodynamic or electromagnetic interactions in a form that requires those constants explicitly.

# Acronyms

**Gravity theories**

| | |
|---|---|
| SR | Special Relativity |
| GR | General Relativity |
| EFEs | Einstein Field Equations |
| ST | Scalar-tensor |
| (D)HOST | (Degenerate) Higher-order scalar tensor theories |
| EOM(s) | Equation(s) of motion |
| EFT | Effective Field Theory |
| KGB | Kinetic Gravity Braiding |
| GLPV | Gleyzes–Langlois–Piazza–Vernizzi |
| DGP | Dvali-Gabadadze-Porrati |
| dRGT | de Rham–Gabadadze–Tolley |
| EsGB | Einstein-scalar-Gauss-Bonnet |
| EdGB | Einstein dilaton Gauss-Bonnett |
| QFT | Quantum Field Theory |

**Black holes & compact objects**

| | |
|---|---|
| BH | Black hole |
| SMBH | Supermassive black hole |
| NS | Neutron star |
| Schw | Schwarzschild |
| SdS | Schwarzschild-de Sitter |
| RN | Reissner-Nordström |
| KNdS | Kerr-Newmann-de Sitter |
| BBH(s) | Binary black hole(s) |
| BNS(s) | Binary neutron star(s) |
| EMRI(s) | Extreme mass-ratio inspiral(s) |

**Gravitational waves**

| | |
|---|---|
| GW(s) | Gravitational wave(s) |
| QNM(s) | Quasinormal mode(s) |
| SGWB | Stochastic gravitational wave background |
| PTA(s) | Pulsar Timing Array(s) |
| SNR | Signal-to-noise ratio |
| LIGO | Laser Interferometer Gravitational-Wave Observatory |
| LVK | Ligo-Virgo-Kagra |
| ET | Einstein Telescope |
| CE | Cosmic Explorer |
| AEDGE | Atomic Experiment for Dark Matter and Gravity Exploration |
| DECIGO | Deci-Hertz Interferometer Gravitational Wave Observatory |
| LISA | Laser Interferometer Space Antenna |

**Other**

| | |
|---|---|
| dof(s) | degree(s) of freedom |
| NR | Numerical relativity |
| ODE(s) | Ordinary differential equation(s) |
| PDE(s) | Partial differential equation(s) |

# List of Figures

# List of Tables

# Introduction

Einstein's theory of General Relativity (GR) has proven remarkably successful over the past century, accurately describing the natural phenomenon of gravity across different scales and passing all observational tests to date. From its initial predictions of Mercury's orbital motion and light deflection to the recent direct detections of gravitational waves (GWs), GR continues to be a cornerstone of modern physics. However, despite its successes, GR is widely believed to be an incomplete description of gravity. Open questions related to dark energy, dark matter, singularities, and the reconciliation of gravity with quantum theory suggest that new gravitational physics may lie beyond Einstein's framework.

This thesis is devoted to exploring gravitational physics beyond GR, with a particular focus on black holes and their GW observables. Black holes serve as natural laboratories for testing gravity in the strong field regime, where GR still remains largely untested, but now observationally accessible due to the advent of GW science. Among the most promising avenues for probing deviations from GR are black hole perturbations and their associated quasinormal modes (QNMs), which arise during the ringdown phase following a binary merger, as the newly formed black hole settles into equilibrium. QNMs, which have already been measured by GW detectors, carry clean imprints of the black hole spacetime and the underlying gravitational theory governing its dynamics, and so can be used to carry out tests of gravity.

The focus of the work presented here lies in scalar-tensor (ST) theories of gravity, a broad and well-motivated class of extensions to GR that introduce an additional scalar degree of freedom alongside the metric. In particular, we investigate black hole solutions within these theories, which retain background geometries characteristic of GR but are nonetheless supported by non-trivial (radial or time-dependent) scalar field configurations. This scalar *hair* leaves no imprint at the background level but might allow for deviations from GR in the perturbative regime, i.e. in the QNMs, that could be observable by current and upcoming GW detectors.

> Within the context of ST theories of gravity, this thesis will investigate black hole spacetimes with scalar hair by evaluating their stability, calculating their QNM spectrum and deriving explicit forecasted constraints on beyond-GR parameters for different GW detectors.

## Thesis outline

This thesis is structured in two main parts: *Preliminaries* and *Investigations*. As their names suggest, the first provides the necessary theoretical background and contextual framework for the research, while the second presents the original contributions and findings developed in the course of this work. Here we briefly summarise the Chapters in each of these parts.

### Summary of Preliminaries

#### History and philosophy of gravity

This thesis opens with a brief survey of the historical development of humanity's understanding of gravity, highlighting a few instances where concepts, ideas, and attitudes—first conceived or adopted centuries or even millennia ago—have collectively shaped both our modern understanding of gravity as well as the scientific practices we use to explore the natural world.

#### General Relativity, black holes and gravitational waves

This Chapter reviews the foundations of GR, including its observational evidence and unresolved challenges. We summarise black hole solutions within GR, discuss key theorems, and develop the framework of black hole perturbation theory. This includes a detailed discussion of GWs, their generation, propagation, and detection, and how the ringdown phase can be used to test the nature of gravity through the QNM spectrum.

#### Scalar-tensor theories and hairy black holes

Motivated by these prospects, this Chapter shifts focus to ST theories of gravity. We map out the theory space from the simplest ST models to the full Horndeski action and its higher-order extensions (HOST), paying particular attention to their black hole solutions, GW phenomenology and observational constraints. Special attention is given to black holes endowed with scalar hair, but this Chapter also includes a fully-detailed derivation of the ringdown dynamics for a Schwarzschild black hole with a constant (hence not hairy) scalar in Horndeski gravity.

### Summary of Investigations

We then present three original investigations based on three different papers aimed at testing gravity beyond GR using current and future GW observations. Figure 0.1 shows overlapping and non-overlapping topics within these works in terms of the *theories*, *background solutions*, and *perturbation analyses* appearing in them.

#### Speed of gravity—based on [1]

In this Chapter we discuss the possibility of constraining the speed of gravitational waves via the QNM spectrum of hairy black holes. We provide the first ringdown-only forecast constraints on the deviation of the speed of gravity from the speed of light, characterised by the parameter $\alpha_T$ concluding that future space-based detectors such as LISA and TianQin will be able to constrain it at the $\mathcal{O}(10^{-2})$ level from a single supermassive black hole merger. Unlike other constraints in the literature, these do not require an electromagnetic counterpart or multiband observations, and can effectively be applied as soon as LISA detects a single supermassive black hole merger. Deviations from the speed of light, while extremely tight in the LVK band, are still potentially large in the LISA band. In that case, our work can provide one of the fastest most stringent constraints in such band. Our analysis demonstrates the intricate connection between deviations in the speed of gravity and the presence of scalar hair. Hence, this work also contributes to the broad understanding of observable effects of hairy black holes, and constraints obtained through our analysis can also be translated in upper bounds in the amplitude of scalar hair.

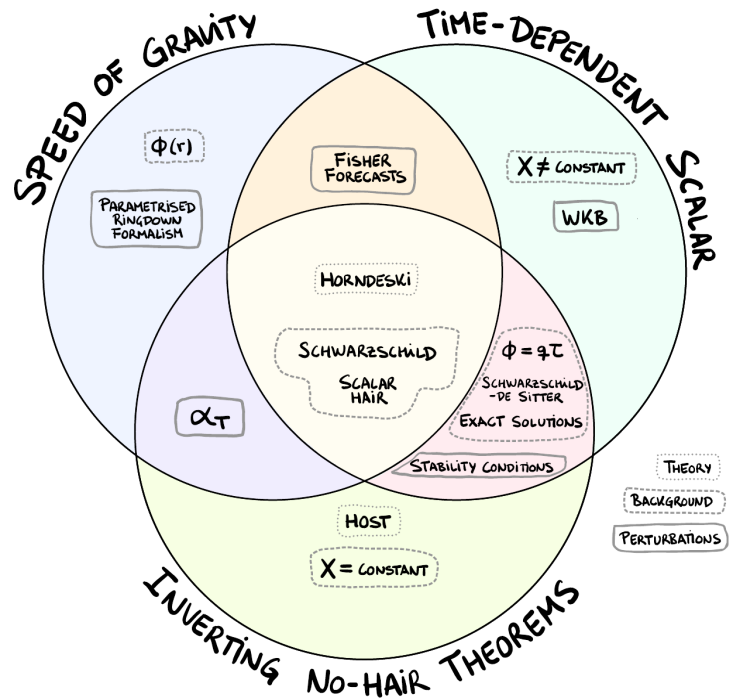

### Time-dependent scalar—based on [2]

In this Chapter we investigate black hole solutions with a scalar endowed with linear time-dependent hair. Known exact solutions of this type exist, and provide well-motivated links with the cosmological regime by embedding the black hole within cosmological asymptotics. However, they are generically plagued by instabilities. We explore a particular solution in which the time-dependence is qualitatively different, hence offering promising stability properties, as compared to previously studied solutions. We analyse the perturbative stability and identify a non-zero surviving parameter space. The QNM spectrum is also extracted and used to establish forecast constraints on the parameter controlling GR-deviations, $\hat{\beta}$. It is shown that a single black hole merger observed by LISA will be able to constrain it at the $\mathcal{O}(10^{-4})$ level, but constraints for other detectors such as LVK are also provided.

### Inverting no-hair theorems—based on [3]

Finally, in this last Chapter we explore the landscape of ST theories that admit general stealth black hole solutions where the metric matches that of GR but supports non-trivial scalar profiles with time-dependent hair. The theories studied here concern the most comprehensive ST frameworks to date, extending Horndeski by including higher orders derivative terms, i.e. by dealing with higher-order scalar-tensor (HOST) theories. We derive the conditions imposed on these theories under which such stealth solutions exist and investigate the associated perturbations in the odd-parity sector. The work presented here includes the most general study of stealth solutions in HOST theories to date. Within the remaining models satisfying the derived existence conditions, we evaluate their stability and identify the allowed deviations from GR. Among these potential deviations, we highlight a specific model where the speed of gravitational waves has a non-trivial behaviour.

This thesis aims to contribute to the theoretical and phenomenological understanding of black holes in modified gravity, providing new tools and predictions for the era of GW astronomy. As the sensitivity of detectors improves, the methods and results presented here will become increasingly relevant for uncovering the true nature of gravity. All calculations presented in the *Investigations* part of this thesis are openly available and reproducible via `Mathematica` notebooks in [4]. The content of this repository is described in Appendix 6.6.

# Preliminaries

# History and philosophy of gravity | 1

Amicus Plato — amicus Aristoteles — magis amica veritas.

Newton

The quest to understand gravity has been at the forefront of science since its inception. As one of the most immediate perceptions of the external world available to humans (i.e. that things have a weight and tend to fall to the ground), gravity has been one of the first physical phenomena humans have attempted to explain in a scientific manner. Perhaps more surprisingly, gravity remains a driving force behind a significant portion of the scientific effort, with its study throughout history consistently uncovering deep non-trivial truths about the world around us. Beyond this, the concept of gravity has transcended disciplinary boundaries, inviting reflection on the nature of scientific progress, the interplay between theory and observation and the limits of human knowledge.

This thesis aims to contribute a small grain of sand to our understanding of what gravity is, what it might be, and what it might not be. In order to place the research presented here in the big picture, we will be taking a brief and incomplete—yet informative—tour on the history of human's understanding of gravity. This is not without further purpose, as one often finds looking to the past to be the best way to forecast the future.

### Ancient Greece

The story of gravity begins, like many other nice things, in the classical world of ancient Greece. Presocratic philosophers were really the first to inquire about the world with a scientific attitude. It is difficult to emphasize how groundbreaking this seemingly small step must have been, as before this, explanations of natural phenomena relied solely on supernatural agents and lacked any true predictive power. These natural philosophers laid down the foundations on which science has been able to progress and, in a way, some of their aspirations and motivations are still present with us, such as the goal of unifying phenomena under one explanation, seeking first principles and guiding thought with an aesthetic sense.

It is difficult to resist not naming a few of them. Anaximander of Miletus (c.610–c.546 BC) has often been referred to as the first scientist [5], as he pioneered in the use of observation and rational thought as mechanisms to reveal laws of nature. Among the ideas proposed by Anaximander that have survived, one finds those related to physics and cosmology to be the most impressive. Before him, other cultures such as the Babylonians and the Egyptians had already looked at the sky with wonder and even registered the movement of stars and planets with outstanding precision. However, what they all lacked was Anaximander's abstraction and desire for *explanation* and *understanding*. Anaximander proposed that the Earth was a finite cylindrical body floating undisturbed and unsupported in infinite empty space. According to Karl Popper, a philosopher of science that we will visit later, this was "*one of the boldest, most revolutionary, and most portentous ideas in the whole*

*history of human thinking*" [6]. Attempting to demystify the origin of the universe, Anaximander proposed that the Earth was at the centre of a series of three concentric hollow wheels corresponding in order to 1) the stars and planets, 2) the Moon and 3) the Sun as the farthest one. Despite getting the order wrong, Anaximander's correct intuition of celestial bodies existing in some sort of *space* with depth and traveling in circles around the Earth would represent a major step in our conception of the universe. These celestial wheels are thought by Anaximander as hollow, filled with fire, and mostly opaque, except for some holes where the fire shines through which correspond to the stars, the Moon and the Sun. At the beggining of the cosmos, Anaximander suggests that a sphere of fire originated from what he called the "*apeiron*", which refers to something without boundaries, a first principle which is infinite, endless, unlimited and boundless and from which everything originates. The *apeiron* would add a new level of abstraction to the toolset available to natural philosophers. The primordial sphere of fire, which might remind one of the primordial hot plasma in the Big Bang theory, would then evolve to be contained in the mentioned celestial wheels around the Earth. With this worldview of the origin of the cosmos, Anaximander was effectively founding cosmology as a science—see e.g. [7][pp. 99-142] for a commentary on Anaximander's surviving fragments.

Another fascinating character in this story is Empedocles (c.494–c.432 BC), who proposed a cyclic cosmology, orchestrated by an endless battle between two cosmic forces: one attractive, which he called Love, and one repulsive, called Strife. These forces alternate their influence in the universe, which undergoes cycles of total Love or Strife domination—see e.g. [8]. According to Empedocles, these forces act upon the fundamental elements, or *roots*, of the universe—earth, air, fire, and water. In this account, we can discern the glimmers of remarkably modern ideas. The notion of attractive and repulsive forces pervading the universe, acting uniformly on all matter and shaping the evolution of the cosmos, evokes parallels with our contemporary understanding of gravity and even dark energy.[1]

After the Presocratic philosophers, the intellectual stage is naturally set for Socrates (c. 470–399 BC). However, Socrates shifts the focus of philosophy to the realm of ethics, emphasizing questions of morality. He is concerned with understanding what is truly right and how we can distinguish genuine moral truths from mere societal or individual claims. Socrates' most distinguished follower, Plato (c. 428/423–348/347 BC), would be deeply inspired by his master—particularly after witnessing his unjust death sentence—and would go on to develop his own philosophy, breaking all frontiers of thought by exploring metaphysics, ethics, politics, epistemology and cosmology in ways that would shape Western intellectual tradition for millennia. Plato's primary focus, one could argue, lies in politics, specifically in determining the most ideal way to organise society. To justify his proposal of philosopher-kings as rulers, however, he developed an expansive worldview that encompasses a detailed account of the origin and nature of the physical world as well as its relationship with and dependence upon a higher, transcendent, non-physical realm.[2] In his account of what constitutes the physical world in a material sense, Plato combines the best aspects of some of his predecessors. He also thinks, like Empedocles, that the four elements constitute all matter. However, he assigns to these elements a more fundamental and mathematical substructure based on two right-angled triangles.[3] This view is reminiscent, on one hand, of Pythagoras' mathematical philosophy, which

---

1: Empedocles' life and philosophical views are brimming with captivating anecdotes and intriguing parallels to modern science. One such story is the myth of his death, which claims he leapt into the flames of Mount Etna in Sicily to prove his immortality. Another is his account of the origin of humans and animals, which bears a striking, albeit peculiar, resemblance to Darwin's theory of evolution.

2: It is worth noting that, unlike the Presocratic philosophers—whose surviving work is limited to fragments and second-hand quotations, if anything at all—we are fortunate to have a relatively complete and accessible account of Plato's and Aristotle's writings.

3: One of the triangles is scalene, with angles 30°, 60° and 90°; and the other one isosceles, therefore with angles 45°, 45° and 90°. These triangles combine in different ways to form the Platonic solids, which correspond to the four elements. The transmutation of elements is therefore elegantly explained by the right-angled triangles rearranging themselves in different ways.

holds that numerical relationships and geometric forms are the fundamental principles underlying the structure and harmony of the cosmos. The Pythagoreans, for instance, famously posited a *musica universalis*, or 'Music of the Spheres', believing that the cosmos was governed by harmonious numerical ratios, a concept that would resonate deeply with later astronomers, most notably with Kepler, seeking mathematical order in the heavens. [4] On the other hand, it also echoes the atomist philosophies proposed by Leucippus and Democritus, where the universe is understood as being composed exclusively of atoms and the void.[5] To all this Plato adds a new innovative concept, that of the Receptacle. This is a substance that provides temporal and spatial location to all material objects. In other words, it is the stage where things happen, come to be and perish. Some aspects of the Receptacle resemble Anaximander's *aperion*, such as its all-encompassing nature. However, some aspects are intrinsically different. In the *Timaeus*, Plato writes about the Receptacle: "*For it is laid down by nature as a molding-stuff for everything, being moved and marked by the entering figures, and because of them it appears different at different times*" (50c) [10]. We therefore see that the Receptacle is a malleable substance, changing in time in relation to the matter that fills it. It is difficult to resist seeing in the Receptacle a proto-version of the modern idea of *spacetime*.

Aristotle (384–322 BC), who was Plato's most famous student, diverged significantly from his teacher's idealism and metaphysics. While Plato emphasised a higher, non-material realm of Forms, Aristotle grounded his philosophy in the observable world, focussing on empirical observation and categorisation. His account of the physical world, however, is also based on the four elements: earth, air, water and fire, to which he assigned different weights. These different weights determined the tendencies of movement for the different elements, with heavier elements being drawn to the centre of the universe, and lighter ones moving away from it. Aristotle also adds an additional element, i.e. a fifth element, or *quintessence*, which he also referred to as the *aether*. The aether is an unchanging substance believed to fill all space in the universe—a concept that would preoccupy and challenge great thinkers for millennia until its experimental refutation in the 19th century by the Michelson–Morley experiment [11]. Interestingly, the term quintessence has re-emerged in academic discourse over the past few decades, this time in the context of explaining the accelerated expansion of the universe.[6] This thesis focuses, in part, on testing this quintessence hypothesis.

Aristotle's physics represents an unparalleled development in the study of the natural world, offering the first comprehensive and systematic attempt to explain motion, change, and the behaviour of physical objects, a framework that would dominate scientific thought for centuries. The main innovation of Aristotle's physics lies in the concept of causation. His framework of the four causes—material (what something is made of), formal (its structure or essence), efficient (the source of change or motion), and final (its purpose or goal)—provided a comprehensive and systematic way to understand why things exist and behave as they do [12]. This teleological perspective, where nature is seen as goal-oriented, was groundbreaking for its time. Aristotle introduced the idea that motion and change are driven by intrinsic tendencies, such as an object's "natural place" in the cosmos. In a sense, Aristotle's view of natural motion can be seen as an early, intuitive precursor to the idea that physical systems tend toward an optimal state or configuration.

4: Aristotle provides an early critique of this concept in *On The Heavens II.9* [9]. For a detailed discussion see also [7][pp. 229-231].

5: Atoms, another word of Greek origin, refer to indivisible, fundamental particles, with the diversity of matter arising from the different arrangements and interactions of these particles.

[10]: Plato (2001), *Timaeus*

[11]: Michelson et al. (1887), "On the relative motion of the Earth and the luminiferous ether"

6: Unlike the classical four elements, the modern counterparts refer to baryons, neutrinos, dark matter, and radiation, which collectively represent the various components shaping the evolution of the cosmos.

[12]: Aristotle (1996), *Physics*

For Aristotle, this meant that objects, like heavy bodies, naturally move toward the Earth's center, while lighter bodies move upward, as if following an optimal path determined by its nature, "minimizing" or "maximizing" certain intrinsic tendencies. This idea parallels, albeit loosely, the principle of minimum action in modern physics. Aristotle also suggested that the speed at which two objects of identical shape fall depends directly on their weights and inversely on the density of the medium they move through. While we know this is not correct, it nonetheless provides a reasonable approximation for objects moving through Earth's gravitational field in air or water, where resistance from the medium plays a significant role, and we would need to wait until Galileo to see it refuted. More notably, it marks an important step from purely qualitative descriptions of motion to a more quantitative relational approach. In the European continent, Aristotle's physics as well as his empiric and teleological philosophy would remain paradigmatic throughout the Middle Ages, as they were extensively adopted and integrated into religious thought. During this time, however, other interesting things were also taking place elsewhere.

### India

[13]: Plofker (2009), *Mathematics in India*

India played a significant role in the development of mathematics and astronomy [13]. Āryabhaṭa (476–550 CE) was a pioneering Indian thinker whose work laid foundational contributions to both fields. In the realm of mathematics, he is thought to have been the first one to introduce the concept of zero and the decimal system, as well as providing major advances in trigonometry. In regards to astronomy, he proposed a heliocentric model with a rotating Earth orbiting the Sun, centuries ahead of its time. His work would also be extremely influential to posterior Indian scholars. Among them, Brahmagupta (598–668 CE) formalised the definition and arithmetic rules of zero while also providing new advances in geometry and other branches of mathematics. Most interestingly to us here, however, are his views on gravity, as he was really the first one to describe it as an attractive force, the *gurutvaakarshanam*, even if only qualitatively. In his own words: "*the earth on all its sides is the same; all people on the earth stand upright, and all heavy things fall down to the earth by a law of nature*" [14]. It would take roughly a millennia for this idea to be famously shaped into a quantitative and predictive universal law by Newton.

[14]: Bīrūnī (2012), *Alberuni's India: An Account of the Religion, Philosophy, Literature, Geography, Chronology, Astronomy, Customs, Laws and Astrology of India about AD 1030*

### Islamic Golden Age

Conveniently located between Greece and India, Baghdad (in modern-day Iraq) emerged as a vibrant center of knowledge and science from the 9th to the 12th century CE. The Abbasid caliphs heavily invested in building libraries, hospitals, and other academic institutions, fostering an environment of intellectual growth. They prioritised translating classical works, which not only preserved these texts, but also led to advancements in the theories described in them. In mathematics, words like *algebra* and *algorithm* are first associated with the Persian scholar Al-Khwarizmi (c.780–c.850 CE)[7] who introduced systematic methods for solving linear and quadratic equations, and contributed to arithmetic, popularising the Hindu-Arabic numeral system, which was later adopted in Europe [15]. Scholars like Ibn Sina

7: In fact, the term *algorithm* comes from a Latinisation of Al-Khwarizmi's name.

[15]: Musa (2013), *The Algebra of Mohammed ben Musa*

(Avicenna) (c. 980–1037) explored motion, force, and dynamics, foreshadowing later ideas of inertia and momentum. Ibn al-Haytham (Alhazen) (c. 965–1040) revolutionised optics, introducing experimental methods and correctly explaining vision as light entering the eye [16]. Advances in mechanics, fluid dynamics, and astronomy also emerged, with thinkers refining Aristotle's ideas and integrating them into new frameworks.

**The Scientific Revolution**

The world we live in is undoubtedly a scientific one. While we now take this for granted, it was not always so. Today, science permeates every aspect of our society and even our language. If this is so, it is largely due to the revolutionary developments that took place in Europe during the Renaissance. The Scientific Revolution is typically agreed to have started with the publishing of Nicolaus Copernicus' (1473–1543 CE) *De Revolutionibus Orbium Coelestium* [17], which he waited until his deathbed to publish, fearing controversy and backlash from religious and academic authorities for challenging the long-held geocentric view of the universe. The idea of a heliocentric universe dates back to ancient Greece, with figures like Aristarchus of Samos (310–230 CE) suggesting that the Earth orbits the Sun [18]. However, Aristarchus's heliocentric model lacked detailed mathematical support and was largely overshadowed by the dominant geocentric model of Ptolemy (c. 100–c. 170 CE) [19]. What made Copernicus revolutionary was his comprehensive mathematical framework and systematic arguments. His work provided a detailed alternative to Ptolemaic cosmology, sparking the Scientific Revolution and influencing subsequent thinkers like Kepler and Galileo.

Johannes Kepler (1571–1630 CE) made groundbreaking contributions to astronomy by formulating the three laws of planetary motion, which described the elliptical orbits of planets around the Sun, their varying speeds, and the relationship between orbital periods and distances from the Sun [20, 21]. His work, based on Tycho Brahe's precise observations [22],[8] refined the Copernican heliocentric model and replaced circular orbits with ellipses. Galileo Galilei (1564–1642 CE) was a pivotal figure in the Scientific Revolution, known for his contributions to physics, astronomy, and scientific methodology. He improved the telescope, enabling observations that supported heliocentrism, such as the moons of Jupiter, the phases of Venus, and the surface features of the Moon. In physics, he studied motion, establishing the principles of inertia and free fall, showing that objects accelerate uniformly under gravity regardless of mass [23]. Galileo supported the use of experimentation and mathematics in science, laying the groundwork for modern empirical methods and challenging traditional Aristotelian views. More broadly, during the Renaissance we in fact witness a rediscovery of Platonic ideals and the underlying Pythagorean emphasis on numerical and geometric harmony as opposed to the long-standing Aristotelian views, which had been refined, reinterpreted and adopted as the canon throughout the Middle Ages.[9]

8: Tycho Brahe (1546–1601 CE) was a Danish astronomer renowned for his extraordinarily precise and systematic observations of the stars and planets, made without a telescope. He established Uraniborg, an unprecedented astronomycal observatory, playing a crucial role in transforming astronomy into a modern, data-driven science.

9: For instance, Kepler believed the orbits of the planets to be governed by pure geometry, and originally attempted to model the Solar System using nested Platonic solids.

Another exemplification of the deep intertwining of philosophical reasoning and mathematical abstraction that defined the Scientific Revolution is offered by René Descartes (1596–1650 CE). Like Plato, Descartes believed that true knowledge could not rest on the senses alone, which he considered unreliable; instead, he sought certainty through reason and deductive logic. His famous dictum "*Cogito, ergo sum*" ("*I think, therefore I am*") [24] reflects his rationalist commitment to grounding knowledge in clear and indubitable ideas.

Descartes applied this methodology to physics and cosmology, aiming to derive the laws of nature from first principles through pure reason. Cartesian physics hence offers a vision of a mechanistic universe—governed by mathematical laws and composed entirely of matter in motion [25]—which, though not always correct, influenced later thinkers, helping to cement a new vision of science rooted in rational abstraction.

Among those thinkers is Isaac Newton (1643–1727 CE), whose work is often seen as the culmination of the Scientific Revolution. Newton formulated the law of universal gravitation, which postulated that every mass attracts every other mass in the universe with a force proportional to the product of their masses and inversely proportional to the square of their separation [26].[10]

10:  Newton's law of universal gravitation can be writen as a formula with

$$F = G\frac{m_1 m_2}{r^2},  \tag{1.1}$$

where $F$ denotes the gravitational force exerted between two bodies of masses $m_1$ $m_2$ separated a distance $r$, with $G$ being the universal garvitational constant.

This law unified celestial and terrestrial mechanics under a single framework, but what it trully singled it out was its predictive power. Equipped with this law, one could predict how many seconds an apple would take to fall to the ground and how many years a planet would take to orbit the Sun, confront the predictions with observations and confirm their validity. Newton's universal gravitation was embedded within a deterministic framework based on differential calculus—developed for the first time by him and Leibniz independently—enabling the precise calculation of trajectories, forces, and accelerations across a wide range of physical systems—what we now refer to as classical mechanics.

The impact of Newtonian mechanics transcended the scientific sphere and profoundly influenced broader intellectual, philosophical, and cultural domains in an unparalleled way. It helped shape the Enlightenment worldview, fostering a belief in the power of reason, order, and natural laws to explain the universe. Philosophers such as Immanuel Kant drew inspiration from Newton's success, seeing in his work a model for how human understanding could uncover the fundamental structure of reality [27].

In sum, the Scientific Revolution was not merely a triumph of observation over dogma, but also a profound philosophical reorientation—one that replaced the qualitative, Earth-centered, and hierarchical cosmos of Aristotle with a mathematically structured, harmonious universe deeply in line with Platonic thought.

**What is science?**

Before arriving at the contemporary understanding of gravity, it is instructive to reflect on foundational questions in the philosophy of science. For instance: What is science? Few questions exist which are so easily posed and seemingly straightforwardly answered, yet prove so difficult to answer with precision. An answer was provided by Karl Popper (1902–1994 CE), who argued that theories must be *falsifiable* in order to be scientific—that is, they must make predictions that can, in principle, be proven wrong by observation or experiment [28]. According to Popper, the hallmark of a scientific theory is not that it can be verified, but that it can be rigorously tested and potentially refuted. This criterion sought to distinguish genuine scientific inquiry from pseudoscience.

Let us now exemplify this distinction through Newton's own interests. According to this definition, his theory of gravity proved to be a truly scientific one, offering uncountable opportunities for its falsification. In fact, it also turned out to be a successful one, passing the majority of such tests and therefore offering a trustable description of gravity within its regime of validity. In his life, Newton also developed a profound interest for alchemy, a tradition based on the transmutation of base metals into noble metals with the aim of achieving immortality. We now know that alchemy is not a scientific practice, but this might not have been so apparent in the 17th century. Alchemy clearly participated in aspects characteristic of the scientific method such as systematic observation and documentation, experimentation and theory-building. However, it was shown over time that through its entanglement with mysticism and esoteric language it lacked a clear commitment to falsifiability or empirical rigor. Its predictions failed to be reproducible or compatible with emerging approaches such as atomic theory or modern chemistry. Does this entail Newton was not being a scientist when engaging with alchemy? Not necessarily. Alchemy could have been considered scientific up to the point where its predictions were shown to be incompatible with the natural world, i.e. falsified. Since this happened gradually over time instead of as a result of a single event or experiment, it results difficult to assess how scientific the attitude of its practitioners was. Alchemy, and similarly other pseudo-sciences, become unscientific when they do not treat their connection with the natural world in an objective manner, either by disregarding or enforcing observations.

The lessons learned by pondering on this question remain important today, where scientific ideas have sometimes been seen to slowly depart from their rigour or abandon attempts to remain connected with observations.

### How does science progress?

Another important question to think about in this context is: How does the progress of science actually take place? It is often conceived that science progresses linearly through the small (although sometimes big) cumulative contributions of individual scientists or groups of scientists. This was however not the view of Thomas Kuhn (1922–1996 CE), who instead argued that science advances via a series of paradigm shifts—i.e. radical changes in the fundamental framework that guide scientific research.

A paradigm, in Kuhn's terminology, is more than just a theory; it encompasses the entire worldview of a scientific community at a given time: its assumptions, methods, values, standards of evidence, and exemplary problems. Kuhn identified two main phases in scientific development. First, *normal science* is a period during which scientists operate under an accepted paradigm, solving puzzles and extending the theory within its established framework. According to Kuhn, this is what mosts scientists ever do. Secondly, we have *scientific revolutions*, disruptive phases that occur when *anomalies* accumulate—empirical results that cannot be explained within the existing paradigm—leading to a *crisis* and ultimately to the adoption of a new paradigm [29].

The transition from Newtonian to Einsteinian gravity is a canonical example of a Kuhnian paradigm shift. For over two centuries, Newton's theory of universal gravitation reigned as the dominant framework for understanding gravity and was the basis of normal science: scientists refined calculations, solved celestial mechanics problems, and applied the theory to new domains, all within the Newtonian worldview. Despite its success, Newtonian gravity began to show cracks by the 19th century. Notably, the precession of

Mercury's perihelion could not be fully explained within Newtonian mechanics. Besides, the concept of instantaneous action at a distance clashed with emerging understandings of field theory and the finite speed of light in electromagnetism. These anomalies accumulated and refused to be explained within Newton's paradigm.

In 1915, Albert Einstein (1879–1955 CE) proposed a fundamentally new theory of gravity, General Relativity (GR) [30]. Rather than viewing gravity as a force acting at a distance, GR described it as the curvature of spacetime caused by the presence of mass and energy, with objects following the straightest possible paths in this curved geometry.

This new paradigm brought with it a redefinition of basic concepts. For instance, space and time became dynamic and interwoven into spacetime. Gravitational interaction became local and geometrical rather than instantaneous and at a distance. Moreover, the mathematical formalism shifted from classical vector calculus to differential geometry and tensor calculus. In Kuhn's view, the Einsteinian paradigm did not simply improve Newtonian theory. Instead, it replaced its foundational assumptions, making it *incommensurable* in some respects, meaning that the lack of a common ground between these paradigms impedes their direct comparison. For example, the concepts of space and time fundamentally differ between these frameworks so one cannot interchangeably employ these terms independently of the paradigm in which they are considered.

In a sense, Newtonian gravity was falsified by the success of GR, resulting from its empirical predictions relating to Mercury's orbit, light bending, and later, gravitational waves among others. However, rather than completely discarding the theory, one should understand Newtonian gravity as a low-energy, weak-field limit of Einstein's theory—valid only under specific conditions.

Kuhn's paradigm shift framework provides a powerful lens through which to view the historical development of gravitational theory. The movement from Newtonian to Einsteinian gravity illustrates how science evolves not just through data collection, but through deep transformations in our conceptual and theoretical frameworks. Understanding this process helps us appreciate the contingent, dynamic, and occasionally disruptive nature of scientific progress. Current persistently unexplained anomalies within GR's framework bear similarities to those once afflicting Newtonian gravity and could indeed be signaling the need for a new paradigm shift. Moreover, according to Popper, as scientists it should be our goal to falsify GR, even if none of the existing reasons to modify it did exist. This motivates the work undertaken in this thesis, which deals with the development of observational probes to test fundamental assumptions within GR's framework.

# General Relativity, black holes and gravitational waves | 2

μηδεὶς ἀγεωμέτρητος εἰσίτω μου τὴν στέγην.

<div align="right">Plato</div>

---

**Chapter summary**

This opening chapter provides a formal introduction to the framework of General Relativity (GR), where we examine its observational confirmations (§2.1.1) and unresolved challenges (§2.1.2). We explore the family of black hole solutions within GR (§2.2.1) and review some theoretical (§2.2.2) and observational (§2.2.3) aspects of these objects. We then present the mathematical formulation of gravitational waves (§3.3.1) and discuss their generation (§3.3.2), propagation (§3.3.3) and detection (§3.3.4). We finally focus on the ringdown stage of gravitational waves within GR, first by formulating the black hole perturbation theory framework (§2.4.1), followed by a discussion of quasinormal modes (§2.4.2), the power of ringdown tests of gravity (§2.4.3) and potential systematics (§2.4.4).

## 2.1  A geometrical theory of gravity

General Relativity remains the cornerstone of our understanding of gravity and exemplifies the highest standards of a scientific theory. Reminiscing of Platonic ideals, GR can be thought to explain the phenomenon of gravity through geometry—more precisely, as the curvature of spacetime—characterised by a metric field $g$ existing in a 4-dimensional manifold $\mathcal{M}$.[1] The action for this theory is the Einstein-Hilbert action

$$S_{EH}[g] = \frac{1}{2\kappa} \int d^4x \sqrt{-g} R, \qquad (2.1)$$

where $R$, the *Ricci scalar*, is the unique non-trivial scalar one can form with the metric and first and second derivatives of the metric.[2] The $\sqrt{-g} \equiv \sqrt{-\det(g_{\mu\nu})}$ factor ensures the measure is Lorentz invariant. Varying $S_{EH}$ with respect to the metric rank-2 tensor $g_{\mu\nu}$ yields the Einstein Field Equations (EFEs) in the absence of matter, i.e. $G_{\mu\nu} = 0$, where $G_{\mu\nu}$ is the *Einstein tensor* defined as $G_{\mu\nu} = R_{\mu\nu} - \frac{1}{2} g_{\mu\nu} R$, and $R_{\mu\nu}$ is the *Ricci tensor*. These tensors encode information about the geometry of the universe (see [31] for a general overview). In Eq. (2.1), $\kappa$ is the *Einstein gravitational constant*. In order to recover Newtonian gravity in the weak field limit for slowly-moving particles, $\kappa$ needs to be set to $\kappa = 8\pi G/c^4$, where $G$ is *Newton's constant of gravitation* and $c$ is the speed of light [32]. In standard units, $\kappa$ has a value of the order of $10^{-43} N^{-1}$, which demonstrates the weakness of the gravitational force as compared to other fundamental forces. For example, the gravitational attraction between two protons is about $10^{36}$ times weaker than their electromagnetic repulsion. Unless otherwise stated, we will be employing *geometrical units* where $\kappa = 8\pi$.

1: $\mathcal{M}$ is a connected 4-dimensional Hausdorff $C^\infty$ manifold and $g$ is a Lorentzian metric.

2: Other scalars in 4 dimensions either are total derivatives or involve higher derivatives of the metric.

[31]: Hawking et al. (1973), *The large scale structure of space-time [by] S. W. Hawking and G. F. R. Ellis*

[32]: Carroll (2019), *Spacetime and Geometry: An Introduction to General Relativity*

To include matter, a standard placeholder action for matter fields is typically used

$$S_M[g, \psi_i] = \int d^4x \sqrt{-g} \mathscr{L}_{\psi_i}[g, \psi_i], \tag{2.2}$$

where $\mathscr{L}_{\psi_i}$ is the Lagrangian density for any matter field of our interest. Varying this action with respect to those fields $\psi_i$ will produce equations of motion for the fields coupled to $g_{\mu\nu}$. Varying with respect to the metric will then give us the matter content appearing in EFEs. Because the variation has to be a scalar, this will be represented by a symmetric rank-2 tensor which we call the *stress-energy tensor* $T_{\mu\nu}$. Hence, the variation

$$\delta_g S_M = \frac{1}{2} \int d^4x \sqrt{-g} T_{\mu\nu} \delta g^{\mu\nu} \tag{2.3}$$

is used to define $T_{\mu\nu}$ as

$$T_{\mu\nu} \equiv \frac{2}{\sqrt{-g}} \frac{\delta_g S_M}{\delta g^{\mu\nu}}. \tag{2.4}$$

Combining the variations with respect to the metric of the Einstein Hilbert action and the matter action gives

$$\delta_g S_{GR} = \int d^4x \sqrt{-g} \left( \frac{1}{16\pi} G_{\mu\nu} - \frac{1}{2} T_{\mu\nu} \right) \delta g^{\mu\nu}, \tag{2.5}$$

and requiring this variation to vanish finally yields the EFEs in the presence of matter

$$G_{\mu\nu} \equiv R_{\mu\nu} - \frac{1}{2} g_{\mu\nu} R = 8\pi T_{\mu\nu}. \tag{2.6}$$

Finally, one can incorporate to the Einstein-Hilbert action a constant $\Lambda$ in the following way:[3]

$$S_{EH+\Lambda}[g] = \frac{1}{16\pi} \int d^4x \sqrt{-g}(R - 2\Lambda). \tag{2.7}$$

Because of the presence of the measure $\sqrt{-g}$, this term still contributes to the field equations, giving the full EFEs

$$G_{\mu\nu} + \Lambda g_{\mu\nu} = 8\pi T_{\mu\nu}. \tag{2.8}$$

The constant $\Lambda$ is referred to as the *cosmological constant* because of the role it can play in the cosmological evolution of the universe. Hence, we see that EFEs symbiotically connect matter with curvature: the presence of matter (or energy) curves the geometry of spacetime and what we experience as gravity are objects moving in straight lines on a curved geometry.

Note that, in differential geometry language, other formulations exist which lead to the same physical predictions, i.e. the same EFEs. In this section we have introduced GR in the metric formulation, where the metric tensor is the only fundamental object. In this formulation, curvature is described with the Levi-Civita connection, which is symmetric in its lower indices, and is given by the metric. The connection is a mathematical object (not a tensor) describing how two points are *connected* in a manifold. In a Lorentzian manifold[4] with an affine metric (i.e. a metric which satisfies $\nabla g = 0$), one can use the Levi-Civita connection to *parallel transport* a vector along different paths and, in the presence of curvature, the vectors will show a misalignment when joined back together. This failure to realign is given by the Riemann tensor.

3: A constant is indeed the only other nontrivial operator besides the Ricci scalar which one can add in 4 dimensions without introducing new degrees of freedom [33, 34].

4: Lorentzian manifolds are a special class of pseudo-Riemannian manifolds, which are smooth differentiable manifolds equipped with a non-degenerate metric. Lorentzian manifolds contain the additional characteristic of having a specific metric signature, which in our convention and dimensionality is $(-, +, +, +)$. Riemannian manifolds are another special case of pseudo-Riemannian manifolds, where the metric is required to be positive definite. Intuitively, Euclidean space is Riemannian, while Minkowski space is Lorentzian.

In other words, the commutator of covariant derivatives acting on a vector is given by $[\nabla_\mu, \nabla_\nu]v^\rho = R_{\mu\nu}{}^\rho{}_\sigma v^\sigma$.

A different perspective comes from the Palatini formulation, where the metric and connection are varied independently in the action, leading to the same EFEs [35]. If one instead assumes the connection to be antisymmetric, one can describe gravity in terms of torsion rather than curvature, which receives the name of teleparallel gravity [36]. A different option is to assume the metric to be non-compatible (i.e. $\nabla g \neq 0$) while keeping torsion and curvature zero. This leads to the formulation of *symmetric teleparallel* gravity, which can also recover the same field equations, see e.g. [37, 38]. Generalisations of the above formulations include Einstein-Cartan (which includes both torsion and curvature but no non-metricity), Einstein-Weyl (which includes non-metricity and curvature but no torsion, see e.g. [39]) and Einstein-Cartan-Weyl (which includes all three of them, see e.g. [40]). Given they give the same equations of motion, these formulations are all (at least classically) equivalent. However, each of them provides unique insights into the nature of gravity and offers alternative ways to explore modifications of GR. In this thesis, however, we only focus on metric formulations of gravity.

### 2.1.1 Observational evidence

General Relativity was put forward by Albert Einstein in 1915 [30] as an extension of his previous work on Special Relativity (SR) in 1905 [41]. SR originated from an attempt to resolve the incompatibility between Maxwell's electromagnetism and Newtonian mechanics. Newtonian mechanics is founded on the Galilean Principle of Relativity, which asserts that the fundamental laws of physics remain the same in all inertial frames—those moving at constant velocity relative to one another. However, Maxwell's equations are not invariant under Galilean transformations, as they predict that the speed of light is the same in all frames. A possible resolution at the time was the existence of a cosmic aether, providing an absolute rest frame where Maxwell's equations would hold. However, this idea was famously disproved by the Michelson-Morley experiment [11]. Einstein elevated the observed constancy of the speed of light to the status of a fundamental principle. He formulated a new principle of relativity, asserting that the speed of light is invariant in all frames of reference. The necessary transformations had been derived earlier by Hendrik Lorentz in 1904 [42], but Einstein reinterpreted them within a radical new framework, deeply challenging long-established views on the nature of space and time. This constituted his theory on Special Relativity.

Einstein spent the next ten years working to incorporate gravitation into his framework of space and time, ultimately succeeding in 1915 [30]. One could argue that the first indications of GR's success were of a theoretical nature, as it provided a description of gravity consistent with the well-tested principles of SR. However, the radical nature of GR's ideas demanded observational confirmation to persuade the scientific community. One of the earliest and most compelling pieces of evidence—sufficient, at least, to convince Einstein—concerned the long-standing problem of Mercury's perihelion precession [43]. This anomaly could not be fully explained within Newtonian mechanics, despite extensive efforts. Scientists had even proposed the existence of an unseen ninth planet, dubbed Vulcan, whose gravitational influence might account for Mercury's orbit, though no such planet was ever observed. Einstein's version of gravity provided corrections to the the Newtonian calculation which accounted precisely for the observed anomaly, something which, in a communication with Adriaan Fokker, Einstein admitted to had given him palpitations of the heart.[5]

Another major piece of evidence in support of GR came a few years later in 1919, not only solidifying its acceptance within the scientific community but also instantly making Einstein world famous. This relates to the bending of light around massive objects, commonly known as *gravitational lensing*. Such a phenomenon already takes place in Newtonian gravity, but Einstein realised that the effect doubles in magnitude if derived with GR when applied to how light would be deflected by the Sun. The first observations of this effect was led by Arthur Eddington during a total Solar eclipse on the 29[th] of May of 1919 [45], providing an agreement with the GR prediction.[6] Such measurements have been refined using radio waves and currently stand compatible with GR's prediction within 0.3 standard deviations, and incompatible with the Newtonian prediction by 600 standard deviations [49]. Similarly, while Newtonian gravity predicts that the deflected signal would arrive before the undeflected one, GR predicts the opposite effect, i.e. that deflected light will be delayed, providing a higher accuracy to distinguish the two theories. This was first done in [50] and then improved in [51], showing data to be within 0.9 standard deviations from Einstein's theory and 130,000 standard deviations away from the Newtonian prediction.

These early successful tests drew the attention of physicists worldwide, who since then contributed to the development of GR both on theoretical and observational fronts. For a recent and more comprehensive review of observational tests passed by GR to date, see [52]. Over time, it became clear that not only did GR appropriately correct the Newtonian results, but also provided unprecedented predictive power, unveiling an entirely new landscape of physical phenomena. The theory was like a vast, uncharted mountain, full of gold and ready to be mined. Among its most profound predictions—particularly relevant to this thesis—are black holes and gravitational waves, which will be examined in detail in the following sections.

### 2.1.2  Unresolved challenges

As of today, we have no strong evidence supporting a better alternative than GR [53]. However, multiple reasons exist, both theoretical and observational in nature, to believe that GR might require some sort of modification to explain gravity in all regimes.

**High-energy completion and singularities:** GR is a low-energy Effective Field Theory (EFT) that breaks down at extremely high energies, such as near singularities inside black holes or at the Big Bang. At these scales, quantum effects are expected to play a crucial role, and could be expected to resolve the existence of singularities. While quantum corrections can be consistently calculated within GR as an EFT at energies below a certain ultraviolet (UV) cutoff, its inherent non-renormalizability prevents a consistent quantum theory at arbitrarily high energies. This breakdown suggests that GR is not the ultimate theory of gravity but rather an effective description valid

5: There are multiple other recollections of Einstein's reaction to his discovery. On a letter to his friend Michelangelo Besso on the 17th of November of 1915, Einstein wrote: "*In these last months I had great success in my work. Generally covariant gravitation equations. Perihelion motions explained quantitatively... you will be astonished*". On January 1916 he wrote to Paul Ehrenfest: "*for a few days I was beside myself with joyous excitement*" [44]. Finally, in his GR paper [30], he concluded: "*These facts must, in my opinion, be taken as a convincing proof of the correctness of the theory*".

[45]: Dyson et al. (1920), "A Determination of the Deflection of Light by the Sun's Gravitational Field, from Observations Made at the Total Eclipse of May 29, 1919"

6: Note that it has been argued, mostly by philosophers and historians of science, that the results obtained by Eddington could have been biased by his own theoretical and political views, and therefore not sufficient to overthrow Newtonian gravity, see e.g. [46, 47]. However, it has also been argued in response that a 1979 reanalysis of the plates used during the 1919 eclipse demonstrates no evidence for such predictor effect taking place, i.e. fitting data to the expected or desired result [48].

[50]: Shapiro et al. (1977), "The Viking Relativity Experiment"

[52]: Clifton et al. (2012), "Modified Gravity and Cosmology"

[53]: Will (2014), "The Confrontation between General Relativity and Experiment"

at low energies. A complete theory of quantum gravity—such as string theory or loop quantum gravity—may resolve singularities and provide a more fundamental understanding of spacetime at the smallest scales. Nonetheless, even in its failure, GR does so elegantly, as it predicts its own demise by showing that singularities are generic within the theory [54–57].

**The ΛCDM cosmological model:** The standard cosmological framework, the ΛCDM model, is built upon GR and stands as arguably the most successful, and definitely the most popular, model of the universe to date (see [58] for a recent review on ΛCDM). However, explaining modern cosmological observations within this model requires invoking several unknown components that play a fundamental role in cosmic evolution: dark energy (Λ), cold dark matter (CDM), and an inflationary field. The nature of these elements remains mysterious and exotic. Hence, the ΛCDM model implies that the Universe today consists of ∼ 95% of unknown 'dark' components, something far from ideal in model building. While they might not necessarily imply that GR is incomplete, multiple strategies aiming to provide more ontologically satisfactory explanations of these phenomena involve modifying the behaviour of gravity at large scales.

**Dark matter:** Observations of rotation curves in spiral galaxies show that the outer stars move at significantly higher speeds than expected based on the distribution of visible matter alone [59]. Dark matter, a hypothetical form of matter whose primary observable interaction is gravitational, is invoked in vast amounts (making up ∼ 27% of the energy content in the Universe) to account for the missing mass within the GR framework [60]. While strong experimental constraints have been placed on the strength of its non-gravitational interactions, for instance with Standard Model particles, these have not been definitively ruled out, and indeed, direct detection experiments are actively designed to probe for such possibilities [61].

**Dark energy and the cosmological constant problem:** In 1998, two independent research teams discovered that the universe is undergoing accelerated expansion [62, 63].[7] This conclusion was mainly drawn from the study of Type Ia supernovae as standard candles. These supernovae appeared dimmer than expected, suggesting that the universe has expanded more than previously thought. While alternative explanations, such as light being absorbed by cosmic obstacles, were considered,[8] they were inconsistent with other observations like the integrated Sachs–Wolfe effect,[9] which shows that photons travelling through potential wells actually gain energy as a result of the accelerated expansion smoothing out the potential [68].

The accelerating expansion is attributed to an unknown component called dark energy, which constitutes about ∼ 68% of the total energy density of the Universe. In GR, this phenomenon is typically modelled by introducing a cosmological constant Λ in the Einstein field equations. However, the origin and nature of Λ remain unknown. In quantum field theory (QFT), vacuum energy contributes a similar term, but calculations of its value famously fail to match the observed value [69].[10] This is known as the *cosmological constant problem*, and constitutes a major unresolved issue in modern physics. As such, Λ remains a phenomenological parameter rather than a fundamental explanation for dark energy.

7: Supernova Cosmology Project [62] and High-Z Supernova Search Team [63].

8: This is often referred to as the 'tired-light' hypothesis, proposed in 1929 by Zwicky [64], but disproved by more detailed brightness observations of supernovae and galaxies [65, 66].

9: Sometimes known as the Rees–Sciama effect [67].

10: It is often stated that the mismatch between the observed value of Λ and the much larger value predicted by QFT for the vacuum energy density is discrepant by up to 120 orders of magnitude. However, this value can be fixed by renormalisation procedures at a specific loop order. The real problem is that such fine tuning is unstable if we go to higher loops, i.e. there is a *radiative instability*. In other words, although one can absorb large quantum contributions into counter-terms and tune the constant by hand, this tuning is highly sensitive to the details of physics at all energy scales. Unlike parameters such as the electron mass–which can be measured and fixed naturally–the cosmological constant requires repeated fine-tuning under radiative corrections, making its small observed value theoretically puzzling and unexplainable.

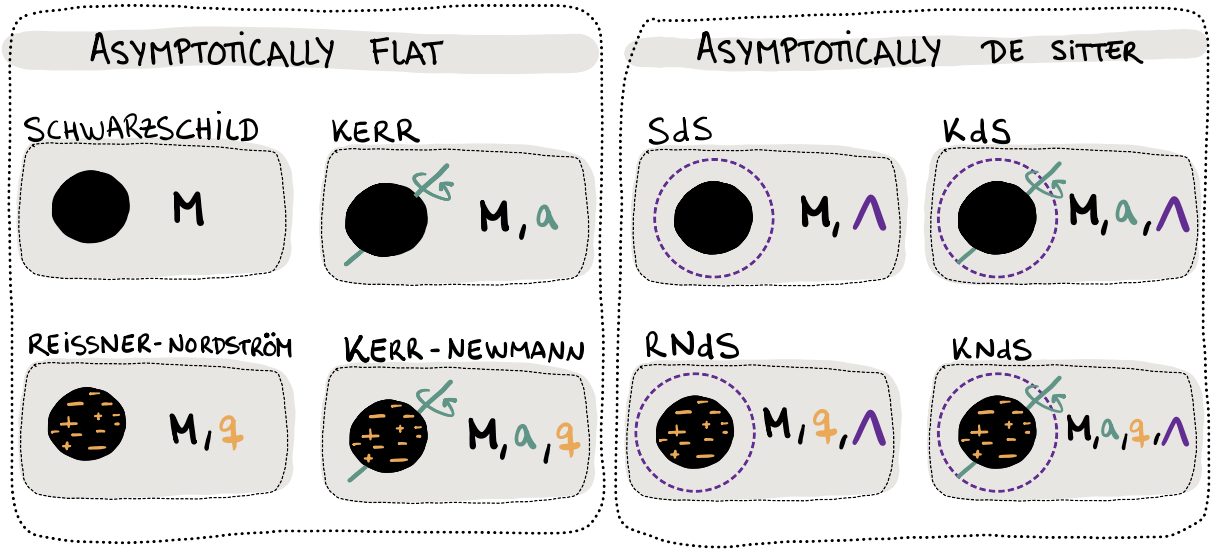

**Figure 2.1:** Schematic representation of black hole solutions in GR.

## 2.2  Black Holes

Black holes are arguably the most fascinating and extreme physical prediction of the last century. They refer to regions of spacetime where gravitational forces are so strong that nothing, not even light, can escape. The defining feature of a black hole is its event horizon, a one-way causal boundary beyond which events cannot influence an outside observer. As incredible as it might seem, black holes exist, and we can routinely observe them today through the gravitational waves emitted when two of them meet and merge. We refer to [70] for an in depth study of black hole solutions and their classification, but here present the most relevant aspects.

[70]: Chruściel et al. (2012), "Stationary Black Holes: Uniqueness and Beyond"

### 2.2.1  The General Relativity black hole family

In this section we briefly review the main exact black hole solutions in GR, schematically organised in Figure 2.1. As will be discussed in Section 2.2.2, these solutions are the subject of uniqueness theorems which demonstrate that, given appropriate assumptions, they constitute the only possible vacuum stationary solutions in GR.

**Schwarzschild black hole**

[71]: Schwarzschild (1916), "On the gravitational field of a mass point according to Einstein's theory"

Roughly a month is all it took Schwarzschild to find the first static and spherically symmetric solution to Einstein's equations after they were first published in 1915. Such solution is given by the metric line element [71]

$$ds^2 = -f(r)dt^2 + f^{-1}(r)dr^2 + r^2 d\Omega_2^2, \qquad f(r) = \left(1 - \frac{r_s}{r}\right), \qquad (2.9)$$

where $d\Omega_2^2$ is the 2-dimensional solid angle and $r_s = 2M$. Physically, this metric describes the spacetime outside a spherical object (e.g. a star, a black hole or a football) of mass $M$. In these coordinates, the metric contains two singularities. One appears at $r \to 0$ and corresponds to a physical singularity, meaning that GR breaks down in a coordinate-independent way. The other one is a coordinate singularity located at the horizon $r_s$.[11] The domain of validity of this solution can be extended to cover the entire manifold by changing to Kruskal-Szekeres coordinates [72, 73]. However, here we will only be interested in the outside region of a black hole where the gravitational waves can be detected, so the Schwarzschild coordinates will suffice. The geometry of this solution is *Ricci-flat* ($R_{\mu\nu} = 0 = R$) and asymptotically flat, i.e. it approaches Minkowski for large $r$. Note that it also smoothly connects to Minkowski flat spacetime in the limit $M \to 0$.

The correspondence between Schwarzschild and Einstein in 1916 quite beautifully describes an educated intellectual debate filled with respect and admiration for each other. Schwarzschild found his solution while serving in the German Army. In his words: "*The war treated me kind enough, in spite of the heavy gunfire, to allow me to get away from it all and take this walk in the land of your ideas.*" At the time, this result was considerably unexpected, as it was thought that EFEs would be challenging to solve due to their non-linear nature. In Einstein's words: "*I would not have expected that the exact solution to the problem could be formulated so simply. The mathematical treatment of the subject appeals to me exceedingly.*" Sadly, Schwarzschild died from disease in 1916 and this laboration could not continue.

**Schwarzschild-de Sitter black hole**

Here we consider an extension to the Schwarzschild metric which incorporates a cosmological constant $\Lambda$ and is given by the metric [74–76]

$$ds^2 = -f(r)dt^2 + f^{-1}(r)dr^2 + r^2 d\Omega_2^2, \quad f(r) = \left(1 - \frac{r_s}{r} - \frac{1}{3}\Lambda r^2\right) \quad (2.10)$$

where now the geometry is no longer Ricci-flat and asymptotes to de Sitter, i.e. we have $R = 12\Lambda$. This solution is therefore named Schwarzschild-de Sitter (SdS). Note that $\Lambda$ is sometimes expressed as a length scale $\Lambda \propto 1/L^2$. This solution can also describe a black hole in AdS (Anti-de Sitter) by changing the sign of $\Lambda$. In that case $L$ would represent the AdS radius. The de Sitter asymptotics at large distances more closely mimic realistic black holes embedded in cosmological spacetimes (in comparison to Schwarzschild solutions with their Minkowski asymptotics). If $\Lambda$ is in the range $0 \leq \Lambda \leq \frac{1}{9M^2}$ this metric possesses two horizons: a black hole horizon $r_b$, which reduces to the Schwarzschild horizon in the absence of $\Lambda$, i.e. $r_b(\Lambda = 0) = r_s$, and a cosmological horizon $r_c$. As shown in Figure 2.2, for cosmological values of $\Lambda$, this is true for all black holes existing in the Universe. As $\Lambda \to \frac{1}{9M^2}$ (or equivalently as we approach $M_{\text{crit}}$ in 2.2) the two horizons become increasingly closer and we obtain the Nariai solution [77, 78]. See also [79] for an alternative interpretation of this saturated point.

**Rotating black holes**

As could be expected, black holes that exist in nature rotate. The spacetime around rotating black holes is described by the Kerr metric, which steeply grows in complexity compared to the non-rotating Schwarzschild case [80].

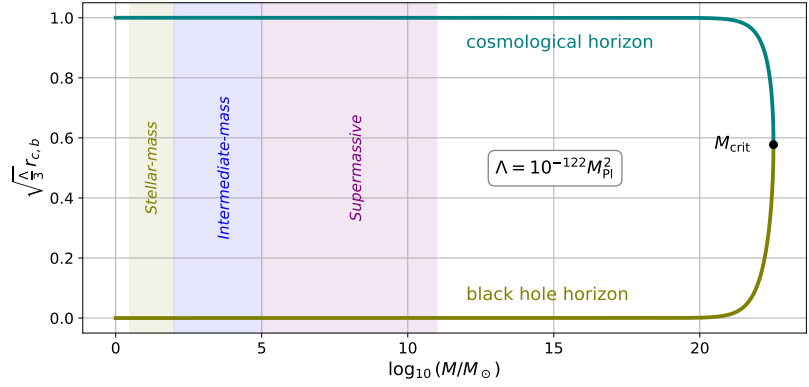

**Figure 2.2:** Cosmological and black hole horizons in Schwarzschild-de Sitter spacetime for a cosmological constant given by $\Lambda = 10^{-122} M_{\text{Pl}}^2$.

The line element is given by

$$ds^2 = -\left(1 - \frac{r_s r}{\Sigma}\right) dt^2 + \frac{\Sigma}{\Delta} dr^2$$
$$+ \Sigma d\theta^2 + \left(r^2 + a^2 + \frac{r_s r a^2}{\Sigma} \sin^2 \theta\right) \sin^2 \theta d\phi^2$$
$$- \frac{2 r_s r a \sin^2 \theta}{\Sigma} dt d\phi, \tag{2.11}$$

where $\Sigma = r^2 + a^2 \cos^2 \theta$ and $\Delta = r^2 - r_s r + a^2$. The structure of the spacetime described by this metric is considerably richer. Note that it took 50 years to generalise the Schwarzschild solution to the rotating case. This solution is also Ricci-flat.

Assuming that the rotation of the black hole is small simplifies the metric greatly. The following line element is obtained by considering the Kerr metric to first order in the rotation parameter $a$, which corresponds physically to a slow rotation

$$ds^2 = -f(r)dt^2 + f^{-1}(r)dr^2 + r^2 d\Omega_2^2 - \frac{4aM}{r} \sin^2 \theta dt d\phi. \tag{2.12}$$

Here $f(r)$ is the same as in Schwarzschild. This geometry is still Ricci-flat to first order in $a$. The Kerr metric is perturbatively smoothly connected to the the slow-rotating solution, and the latter is also perturbatively smoothly connected to the Schwarzschild solution, which can be seen by taking the $a \to 0$ limit. This means that although physical black holes rotate, the Schwarzschild metric constitutes a physically sensible limit to study. One can also introduce the cosmological constant $\Lambda$ to the Kerr metric and therefore construct Kerr-de Sitter black holes.

**Charged black holes**

One can include electromagnetism into the theory by adding the electromagnetic field strength tensor $F_{\mu\nu} = 2\partial_{[\mu} A_{\nu]}$ (where $A_\mu$ is the electromagnetic gauge field) minimally coupled to gravity, resulting in the Einstein-Maxwell action

$$S_{EM}[g] = \frac{1}{2\kappa} \int d^4x \sqrt{-g} \left[\frac{1}{2\kappa} R - \frac{1}{4} F_{\mu\nu} F^{\mu\nu}\right]. \tag{2.13}$$

Solving the equations of motion of this theory for a static, spherically symmetric solution leads to the Reissner-Nordström metric, which takes the

form [81, 82]

$$ds^2 = -f(r)dt^2 + f^{-1}(r)dr^2 + r^2 d\Omega_2^2, \quad f(r) = \left(1 - \frac{r_s}{r} + \frac{Q^2}{r^2}\right), \quad (2.14)$$

where $Q$ is the electric charge endowed to the black hole. This metric contains two horizons at $r_\pm = M \pm \sqrt{M^2 - Q^2}$, where we recall that $r_s = 2M$. In the $Q \to 0$ limit, this metric element is continuously connected to the Schwarzschild one, fully recovering it for $Q = 0$. In that case, $r_+ = r_s$ and $r_- = 0$. Alternatively, if $Q = M$ we say the black hole has reached extremality, and the two horizons coincide $r_+ = r_-$.

**Further generalisations: the Kerr-Newmann-de Sitter family**

We have seen metric solutions for black holes where, in addition to their mass $M$, the geometry is described by an additional parameter, i.e. $\Lambda$ (the cosmological constant), $a$ (rotation) or $Q$ (electric charge). However, as described in Figure 2.1, generalisations of these cases exist which include several or all of these 3 parameters, with the most general one being the Kerr-Newmann-de Sitter (KNdS) metric, which describes a charged rotating black hole embedded in an asymptotically de Sitter spacetime. In Table 2.1 we collect the KNdS black hole family of solutions, all of which are smoothly connected.

| Metric | $M$ | $\Lambda$ | $a$ | $Q$ | Year | Reference |
|---|---|---|---|---|---|---|
| Schwarzschild | ✓ | ✗ | ✗ | ✗ | 1916 | [71] |
| RN | ✓ | ✗ | ✗ | ✓ | 1916 | [81, 82] |
| SdS | ✓ | ✓ | ✗ | ✗ | 1918 | [74−76] |
| RNdS | ✓ | ✓ | ✗ | ✓ | 1918 | [74−76] |
| Kerr | ✓ | ✗ | ✓ | ✗ | 1963 | [80] |
| KN | ✓ | ✗ | ✓ | ✓ | 1965 | [83] |
| KdS | ✓ | ✓ | ✓ | ✗ | 1970-73 | [84−86] |
| KNdS | ✓ | ✓ | ✓ | ✓ | 1970-73 | [84−86] |

In the Petrov classification of GR solutions [87], which categorises spacetimes based on the algebraic structure of the Weyl tensor, the metrics in Table 2.1 correspond to Type D, meaning the Weyl tensor has two distinct repeated principal null directions. For an in-depth discussion on this topic we refer to [88]. Note that other black hole metrics exist in GR with non-zero stress-energy tensors: e.g. the McVittie solutions, which generalises the SdS solution by allowing the cosmological background to evolve dynamically rather than remaining static [89]; the Schwarzschild–Melvin solution, which immerses a Schwarzschild black hole in a background electric or magnetic field [90]; or the Vaidya's radiating Schwarzschild solution, which is used to model the spacetime around a star including its radiation effect with a coordinate-dependent mass function (i.e. $M(t, r)$) [91].

Including more exotic types of matter (i.e. violating some energy conditions) one can construct regular black hole solutions (i.e. singularity-free) such as the Bardeen metric [92]. For a comprehensive review of these solutions as well as other other ones such as accelerating or cylindrical black holes we refer to [88]. With even more exotic stress-energy tensors which often require

12: Note that in some situations, quantum fields can provide the kind of negative Casimir-like energies required to support wormhole constructions, such as for example in [95], where a long traversable wormhole is 'sourced' by a charged massless fermionic field coupled to electromagnetism following a specific trajectory.

severe violations of energy conditions (potentially leading to causality paradoxes), one can also construct metric solutions that might appear as black holes to external observers but actually connect through its interior to other spacelike-separated regions of spacetime, otherwise known as wormholes (see e.g. the Morris-Thorne traversable wormhole solution [93] and [94] for a more general overview of the topic).[12]

**The Kerr hypothesis**

All astrophysical black holes existing in nature are expected to be well described by the Kerr metric, or so does the Kerr hypothesis propose. More specifically, the hypothesis states that the spacetime around a black hole after gravitational collapse is well described by the Kerr metric and therefore contains no hair. This provides a foundation for testing GR in the strong-field regime—any observed deviations from the Kerr geometry could signal a violation of GR. An implicit assumption in this hypothesis is that astrophysical black holes carry negligible electric charge. This expectation is supported by several physical mechanisms that would efficiently neutralise any initial charge (see [96] for a more in-depth evaluation of this assumption).

### 2.2.2 Black hole theorems

The period between 1960 and 1975, often referred to as the Golden Age of General Relativity, saw a revolution in our understanding of black holes, driven by the development of rigorous mathematical theorems. Here we review some of these results.

**Singularity theorems**

Spacetime singularities represent breakdowns in the geometric structure of spacetime, signaling the limits of GR. Their physical significance was initially a matter of debate. Early singularity solutions—such as those in the Schwarzschild and Friedman-Robertson-Walker spacetimes—were thought to be mathematical artifacts arising from the high degree of symmetry in these models. Many physicists believed that a more realistic treatment, incorporating perturbations or more general initial conditions, would eliminate them. However, the development of singularity theorems provided rigorous arguments showing that singularities are not mere curiosities of symmetric solutions but generic predictions of GR under reasonable physical conditions.

Singularities are difficult to pin down in a general and precise mathematical definition. They are usually referred to as the 'place' where curvature becomes infinite, but if by 'place' we mean a location in spacetime, this definition is problematic as the metric is not well-defined at a singularity. Attempts to define singularities as boundaries of the manifold or via the divergence of curvature invariants, while useful in specific cases, fail to fully capture their nature in all scenarios [31].

Quite ironically, the best characterisation of singularities to date is based not on what they are but on what they entail. This is done with *singularity theorems*, which use the notion of *geodesic incompleteness*. The underlying idea is that if singularities represent "holes" in spacetime, then geodesics encountering them should terminate after a finite value of their affine parameter, rendering them inextendible beyond the singularity. A *singular spacetime* is therefore a spacetime which is *geodesically incomplete* and *inextendible*.

The first singularity theorem for a general spacetime was published by Penrose in 1965 [54], a result for which he was awarded a portion of the 2020 Physics Nobel Prize. In the next years new theorems were proved and improved by Hawking and Penrose [55–57].[13] Such theorems require a list of necessary ingredients in order to cook up singularities: 1) a boundary condition, 2) an energy condition, and 3) a causality condition. With different combinations of them one can obtain spacetimes with at least one incomplete causal geodesic. In short, the boundary condition implies that geodesics start focusing, the energy condition ensures that the focusing continues and the causality condition assures that there are no conjugate points along the geodesics. This will inevitably lead to geodesic incompleteness and thus to singularities.

The first singularity theorem for a general spacetime was published by Penrose in 1965 [54], a result for which he was awarded a portion of the 2020 Physics Nobel Prize. In the next years new theorems were proved and improved by Hawking and Penrose [55–57].[14] Such theorems require a list of necessary ingredients in order to cook up singularities: 1) a boundary condition, 2) an energy condition, and 3) a causality condition. With different combinations of them one can obtain spacetimes with at least one incomplete causal geodesic. In short, the boundary condition implies that geodesics start focusing, the energy condition ensures that the focusing continues and the causality condition assures that there are no conjugate points along the geodesics.[15] This will inevitably lead to geodesic incompleteness and thus to singularities.

In this thesis, however, we will not delve any deeper into the topic of singularities, and will remain for most part on the exterior region of black holes. The following theorems address how black holes appear to distant observers.

**Birkhoff's theorem**

Birkhoff's theorem is a fundamental result in GR which states that [99]

**Theorem 2.2.1 (Birkhoff's Theorem)** *Spherically symmetric solutions of the vacuum EFEs must be static and asymptotically flat.*

In other words, this theorem tells us that the Schwarzschild metric with a constant mass is the only solution to EFEs in vacuum and in the absence of $\Lambda$. As a physical corollary of this theorem, we can see how the collapse of a star to a black holes, or equivalently a pulsating star, does not produce gravitational waves as long as the process remains spherically symmetric, because the spacetime around it needs to remain static.

**No hair theorems**

No-hair theorems are a set of results demonstrating that, under certain conditions, black holes are completely described by just three parameters: mass,

13: Previous theorems proving the existence of singularities were restricted to stress-energy tensors for perfect fluids [97, 98].

14: Previous theorems proving the existence of singularities were restricted to stress-energy tensors for perfect fluids [97, 98].

15: Conjugate points (or focal points) along a geodesic are points where nearby, initially parallel geodesics re-converge or intersect. More rigorously, $p$ and $q$ are conjugate points along a geodesic $\gamma$ if there exists a non-trivial Jacobi field (which measures the infinitesimal separation between nearby geodesics) that vanishes at both $p$ and $q$. In the context of singularity theorems, the absence of such points ensures that gravitational focusing continues unimpeded without ever corssing or spreading out again. This unhindered focusing is what inevitably leads to geodesics collapsing to zero cross-sectional area in finite affine time, signifying a singularity [31].

angular momentum, and electric charge. More precisely, they establish that the Kerr-Newman(-de Sitter) solutions are the only asymptotically flat (or de Sitter) and regular black hole solutions to Einstein's field equations in the presence of fundamental fields.

Despite their name, no-hair theorems are not rigorous theorems in the strict mathematical sense, as they lack a fully general proof applicable to all possible scenarios. Instead, they are often referred to as the no-hair conjecture. However, they are widely accepted as "theorems" because they have been shown to hold across a broad range of physically relevant cases. Note as well that such results, in particular in the absence of additional fields, are sometimes referred to as uniqueness theorems.

The first one came in 1967 due to Israel and it established the uniqueness of the Schwarzschild metric as a static asymptotically flat spacetime in vacuum [100], a result which he later extended to incorporate electric charge by establishing that RN black holes are the unique static and asymptotically flat solutions in electro-vacuum (referring to the vacuum in Einstein-Maxwell theory) [101]. Note that Israel's original theorem can be thought of as the converse of Birkhoff's theorem. While the latter tells us that spherically symmetric solutions in vacuum need to be Schwarzschild, Israel's theorem tells us that Schwarzschild is the only static solution in vacuum.

The extension to non-static, rotating solutions was completed in 1975 by Robinson, who proved the uniqueness of the Kerr metric [102]. Later, electric charge was incorporated into the rotating case in [103, 104], establishing that Kerr-Newman black holes are the unique solutions for asymptotically flat, stationary, and axisymmetric spacetimes in electro-vacuum.

Following the establishment of black hole uniqueness results, attention then turned to assessing their robustness under the inclusion of additional fundamental fields. Early investigations suggested that these fields typically do not introduce new independent charges, and no hair theorems were therefore extended for different specific configurations.

In particular, a significant body of no-hair theorems exists for a wide range of scalar-tensor (ST) theories, typically demonstrating that scalar fields must possess a trivial profile around black holes, see [105–118] for the original papers and [119] for a review. The earliest such results include Chase's pioneering work, which established that static, asymptotically flat black hole spacetimes cannot support a regular massless scalar field [105]. This was notably extended by Bekenstein, who demonstrated that even massive scalar fields do not introduce new independent charges, by developing a general method that would then be applied to various configurations [106–108]. Further similar no-hair theorems were subsequently established for conformally-coupled scalars [109], massive complex scalars [110], and other scalar field configurations, typically including the addition of a scalar potential [111–114].

The scope of these theorems was then broadened by Hawking, who showed this also applies to stationary black holes in minimally coupled Brans-Dicke theories [115]. Subsequent research extended these findings to a more general class of ST theories, including those with scalar self-interactions [116], spherically symmetric static black holes in Galilean-invariant theories [117], and slowly rotating black holes in more general shift-symmetric theories

[118]. However, in these beyond-GR models, violations of specific assumptions of such no-hair theorems can be used to construct hairy black hole solutions. This will be reviewed in Section 3.4.

### 2.2.3 Observational evidence

To conclude this section on black holes, let us briefly shift our focus from theoretical considerations to the remarkable body of observational evidence confirming their existence in the universe.

Black holes are by nature difficult to observe. As a result, the first strong pieces of evidence came unsurprisingly from indirect observations, i.e. observations that can only be explained by inferring the presence of black holes with a specific mass and density, despite those not being observable. An example concerns the stars around Sagittarius A* (what we now know to be the supermassive black hole at the centre of our galaxy) whose orbits revolved around a very massive but invisible object [120]. Similarly, gas falling into a black hole from its surroundings forms an accretion disk, where extreme gravitational forces accelerate the gas to high velocities. The resulting friction heats the material to extreme temperatures, causing it to emit X-rays. One of the first confirmed stellar mass black hole candidates, Cygnus X-1, was detected through these emissions (see [121–123] for the original work and [124] for a recent review). Around 50 years ago, Stephen Hawking and Kip Thorne famously bet on whether this system was indeed a black hole or not as, at that time, the community was not yet certain on their existence. Hawking, who bet against it being a black hole, described it as an insurance bet, since all his work on black holes *"would all be wasted if it turned out that black holes do not exist"* [125]. Luckily for us, he lost the bet.

The two most notable pieces of evidence are more direct in nature and have both been recognised with Physics Nobel Prizes. One comes from the Event Horizon Telescope (EHT), which, as its name suggests, enables the direct imaging of the hot plasma surrounding a supermassive black hole, revealing the silhouette of its event horizon. In 2019, the EHT collaboration released the first-ever image of a black hole, capturing the shadow of the M87* black hole, of billions of solar masses [126]. In 2022, they achieved a similar feat for Sagittarius A*, the supermassive black hole at the center of the Milky Way [127]. More recently, advancements in data processing techniques have refined these images, revealing additional details such as polarised light signatures and magnetic field structures [128–131], shown in Figure 2.3.

Secondly, the most compelling evidence for the abundant existence of stellar-origin black holes in our universe is provided by gravitational wave observations, which are discussed in more detail in the next section.

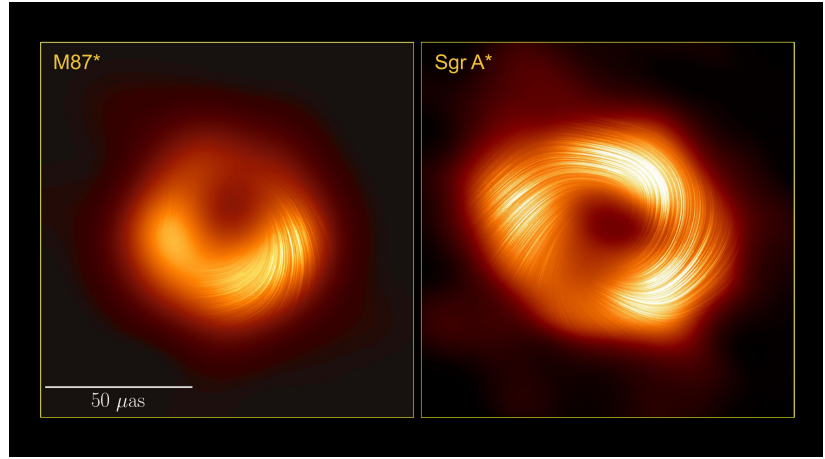

## 2.3 Gravitational waves

If there is one thing about the natural world that we have learnt after centuries of studying it, is that it contains an abounding number of elements that behave as waves. Sound, light, fluids and even the Earth itself are a few examples of the wide range of phenomena that can be accurately modelled by wave equations. According to GR, even spacetime itself can have wave-like behaviour. In this section we review the standard theoretical background for the formulation of gravitational waves (GWs) and the current state of detections as well as forecasted expectations by future missions.

### 2.3.1 Linearised gravity

[132]: Einstein (1916), "Approximative Integration of the Field Equations of Gravitation"

The idea of gravitational waves originated in 1916, less than a year after the formulation of GR, when Albert Einstein demonstrated that, under the weak-field approximation, the equations of GR allow for wave-like solutions [132]. Let us start by considering small perturbations around flat space. At linear order, the spacetime metric can be approximated as:

$$g_{\mu\nu} = \eta_{\mu\nu} + h_{\mu\nu}, \tag{2.15}$$

where $\eta_{\mu\nu} = \mathrm{diag}(-1, 1, 1, 1)$ is the background Minkowski flat metric which, to linear order in perturbations, can be used to raise and lower indices. $h_{\mu\nu}$ describes small perturbations (i.e. $|h_{\mu\nu}| \ll 1$ for all $\mu, \nu$) that will carry the information of gravitational waves, and we use $h \equiv h^\mu_\mu \equiv \eta^{\mu\nu} h_{\mu\nu}$ to denote its trace. The linearised EFEs are obtained by substituting the weak-field approximation (2.15) into (2.8), which gives the following LHS

$$R_{\mu\nu}^{(1)} - \frac{1}{2}\eta_{\mu\nu}R^{(1)} = \frac{1}{2}\Big[ -\eta_{\mu\sigma}\eta_{\nu\rho}\partial_\alpha\partial^\alpha - \eta_{\mu\nu}\partial_\sigma\partial_\rho \\ + \eta_{\nu\sigma}\partial_\mu\partial_\rho + \eta_{\mu\sigma}\partial_\nu\partial_\rho \Big]\overline{h}^{\sigma\rho}, \tag{2.16}$$

where, for convenience, we have employed a trace-reversed version of the metric perturbations defined as[16]

16: The term trace-reverse comes from the fact that $h = -\overline{h}$.

$$\overline{h}_{\mu\nu} = h_{\mu\nu} - \frac{1}{2}\eta_{\mu\nu}h^\sigma_\sigma \tag{2.17}$$

Since $\overline{h}_{\mu\nu}$ is a symmetric 4 by 4 matrix, it requires 10 distinct components to be fully specified but these are not all physical due to the diffeomorphism invariance of GR. It is possible to take advantage of this gauge symmetry to remove some components. To see this explicitly we can perform a general coordinate transformation

$$x'^{\mu} = x^{\mu} + \zeta^{\mu}(x),$$
$$\frac{\partial x'^{\mu}}{\partial x^{\nu}} = \delta^{\mu}_{\nu} + \partial_{\nu}\zeta^{\mu},$$
$$\frac{\partial x^{\mu}}{\partial x'^{\nu}} = \delta^{\mu}_{\nu} - \partial_{\nu}\zeta^{\mu} + \mathcal{O}(|\partial\zeta|^2), \tag{2.18}$$

where the prime denotes the new coordinate system and to write the last equation we assumed $|\partial_{\nu}\zeta^{\mu}| \ll 1$. With this, we can write the metric tensor in these new coordinates as

$$
\begin{aligned}
g'_{\mu\nu}(x') &= \frac{\partial x^{\alpha}}{\partial x'^{\mu}}\frac{\partial x^{\beta}}{\partial x'^{\nu}} g_{\alpha\beta}(x), \\
&= \left(\delta^{\alpha}_{\mu} - \partial_{\mu}\zeta^{\alpha} + \mathcal{O}\left(|\partial\zeta|^2\right)\right)\left(\delta^{\beta}_{\nu} - \partial_{\nu}\zeta^{\beta} + \mathcal{O}\left(|\partial\zeta|^2\right)\right)\left(\eta_{\alpha\beta} + h_{\alpha\beta}\right), \\
&= \eta_{\mu\nu} + h_{\mu\nu} - \partial_{\mu}\zeta_{\nu} - \partial_{\nu}\zeta_{\mu} + \mathcal{O}\left(|\partial\zeta|^2, |\partial\zeta|h\right) \\
&= \eta_{\mu\nu} + h'_{\mu\nu} + \mathcal{O}\left(|\partial\zeta|^2, |\partial\zeta|h\right), \tag{2.19}
\end{aligned}
$$

meaning that another $h'_{\mu\nu}$ in the new coordinates exists such that $|h'_{\mu\nu}| \ll 1$ for all $\mu, \nu$, and is given by

$$h'_{\mu\nu} = h_{\mu\nu} - \partial_{\mu}\zeta_{\nu} - \partial_{\nu}\zeta_{\mu}. \tag{2.20}$$

This is a gauge transformation, and it can be used to modify the components of the perturbed metric. In particular, we can choose to fix the gauge to eliminate non-physical information from the perturbed metric. Going back to our trace-reversed $\overline{h}_{\mu\nu}$, we see that under a change of coordinates it transforms as

$$
\begin{aligned}
\overline{h}'_{\mu\nu} &= h'_{\mu\nu} - \frac{1}{2}\eta_{\mu\nu}h', \\
&= h_{\mu\nu} - \partial_{\mu}\zeta_{\nu} - \partial_{\nu}\zeta_{\mu} - \frac{1}{2}\eta_{\mu\nu}\eta^{\alpha\beta}(h_{\alpha\beta} - \partial_{\alpha}\zeta_{\beta} - \partial_{\beta}\zeta_{\alpha}), \\
&= h_{\mu\nu} - \frac{1}{2}\eta_{\mu\nu}h - \partial_{\mu}\zeta_{\nu} - \partial_{\nu}\zeta_{\mu} + \eta_{\mu\nu}\partial^{\alpha}\zeta_{\alpha}, \\
&= \overline{h}_{\mu\nu} - \partial_{\mu}\zeta_{\nu} - \partial_{\nu}\zeta_{\mu} + \eta_{\mu\nu}\partial^{\alpha}\zeta_{\alpha}. \tag{2.21}
\end{aligned}
$$

We can now choose the so-called de Donder, Lorentz or harmonic gauge, where[17]

$$\partial^{\mu}\overline{h}'_{\mu\nu} = 0, \tag{2.22}$$

implying from (2.21) that $\partial^{\mu}\overline{h}_{\mu\nu} = \partial^{\mu}\partial_{\mu}\zeta_{\nu}$. Note that for smooth $\overline{h}_{\mu\nu}$ functions, this can always be solved for $\zeta_{\nu}$, and actually $\zeta_{\nu} + \varepsilon_{\nu}$ will also be a solution provided that $\partial^{\mu}\partial_{\mu}\varepsilon_{\nu} = 0$, where $\varepsilon_{\nu}$ here encodes another coordinate transformation. In this gauge, the linearised EFEs in vacuum read

$$\Box \overline{h}'_{\mu\nu} \equiv \partial^{\sigma}\partial_{\sigma}\overline{h}'_{\mu\nu} = 0, \tag{2.23}$$

which is nothing else but a wave equation. Writing the equation above in component form and restoring factors of the speed of light in the Minkowski

[17]: Note that the following derivative operators are equivalent at first order in this case, i.e. $\partial_{\mu} = \frac{\partial}{\partial x'^{\mu}} = \frac{\partial}{\partial x^{\mu}}$.

flat metric as $\eta_{\mu\nu} = \text{diag}(-c^2, 1, 1, 1)$ we get

$$\left(-\frac{1}{c^2}\partial_t^2 + \nabla^2\right)\bar{h}'_{\mu\nu} = 0, \tag{2.24}$$

where $\nabla^2 \equiv \partial^i\partial_i$ with $i = 1, 2, 3$, and therefore we see that such waves travel at the speed of light in GR. This wave equation is solved by a standard superposition of plane waves

$$\bar{h}'_{\mu\nu}(t, x^i) = \int \frac{d^3k}{(2\pi)^{3/2}}\left[A_{\mu\nu}(k)e^{ik_\mu x^\mu} + \text{c.c.}\right], \qquad k_\mu k^\mu = 0, \tag{2.25}$$

with $A_{\mu\nu}$ describing the tensorial amplitude and $k$ being the wave vector. This shows that small perturbations of the gravitational field on Minkowski space propagate as waves, i.e. *gravitational waves*. In other words, while GR is a highly nonlinear theory, its linear physics already accepts wave-like equations. Let us look again at the degrees of freedom (dofs) in $\bar{h}'_{\mu\nu}$. Recall that we started with 10 dofs and then we imposed the de Donder gauge, which explicitly gives us 4 relations for $\bar{h}'_{\mu\nu}$, and therefore reduces the number of dofs to 6. Applying the gauge condition (2.22) to the solution (2.25), we can equivalently write

$$k^\mu A_{\mu\nu} = 0, \tag{2.26}$$

meaning that gravitational waves, like electromagnetic ones, propagate in a *transverse* way. For waves traveling in the $z$-direction (i.e. $k^\mu = (-\omega, 0, 0, k_z)$ with $\omega = \pm k_z$), this means $A_{0\nu} = \pm A_{3\nu}$. As was noted below (2.22), one can perform a further coordinate transformation

$$x^{TT\mu} = x'^\mu + \varepsilon^\mu, \tag{2.27}$$

which preserves the condition $\partial^\mu h_{\mu\nu}^{TT} = 0$ as long as $\Box\varepsilon^\mu = 0$, meaning that we still have the freedom to shift $\bar{h}'_{\mu\nu} \to h_{\mu\nu}^{TT} + \partial_\mu\varepsilon_\nu + \partial_\nu\varepsilon_\mu - \eta_{\mu\nu}\partial_\alpha\varepsilon^\alpha$ and therefore remove 4 further components. These extra conditions are usually chosen to make $h_{\mu\nu}^{TT}$ traceless (1 condition: $A_\mu^{TT\mu} = 0$) and set the non-transverse components to zero (3 conditions: $A_{\mu 0}^{TT}$, because we already have $A_{00}^{TT} = A_{30}^{TT}$ from the harmonic gauge).[18] This fully fixes the gauge (known as the traceless-transverse (TT) gauge) and, as a result, we find that we can write the components of the gravitational wave tensor with only 2 independent dofs

$$A_{\mu\nu}^{TT} = \begin{pmatrix} 0 & 0 & 0 & 0 \\ 0 & h_+ & h_\times & 0 \\ 0 & h_\times & -h_+ & 0 \\ 0 & 0 & 0 & 0 \end{pmatrix}, \tag{2.28}$$

where the subscripts $(+, \times)$ are used to denote the two polarisations of gravitational waves. As an example, in Cartesian coordinates, a single plane gravitational wave propagating along the $z$ axis will then have $+$ and $\times$ polarisations such that the associated spacetime line element is

$$ds^2 = -dt^2 + [1 + h_+(t - z)]\,dx^2 + [1 - h_+(t - z)]\,dy^2 + 2h_\times(t - z)dxdy + dz^2. \tag{2.29}$$

Therefore, the $+$ and $\times$ polarisations will distort directions of the space in different ways, as illustrated by Fig. 2.4 which shows how a ring of point particles in the $x - y$ plane is affected by both polarisations.

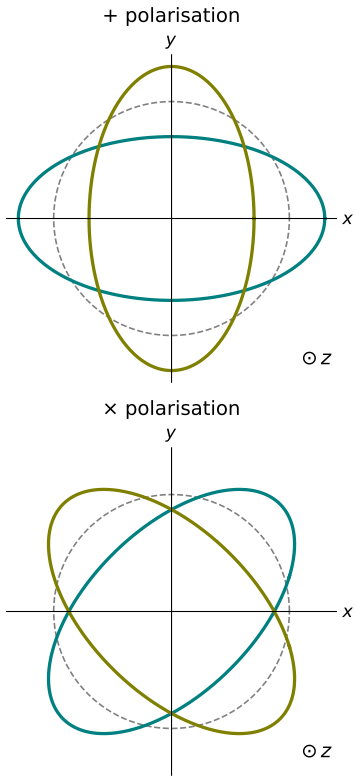

**Figure 2.4:** Illustration of how the polarisations $h_+ = |A|\cos(t - z)$ and $h_\times = |A|\cos(t - z)$ of gravitational waves affect a ring of point particles at different phases. The gray circle is the unperturbed ring (e.g. when the waves are in a node at $t - z = \pi/2$), while the blue and green ellipses show the perturbed ring when $t - z = 0$ and $t - z = \pi$, respectively.

### 2.3.2 Generation

As was shown in 1918 by Einstein, gravitational waves are generated from sources with a varying mass quadrupole moment [133]. These waves, however, are generally weak and can only be observed if they are generated by strongly interacting and/or sufficiently heavy objects. Sources of astrophysical gravitational waves could include: gravitational collapse like supernovae; binary systems of compact objects such as black holes, neutron stars (or alternative compact objects), irregular spinning neutron stars. In the future, it may be possible to also observe cosmological gravitational waves from inflation, or phase transitions.

To date, with the potential exception of the detection of the stochastic gravitational wave background [134–139], whose origin is not yet fully certain,[19] there have only been direct detections of binary black holes and neutron stars, and thus these have become the most important objects of study. The gravitational wave emission of binaries is typically divided into three stages, as depicted in Figure 2.5: inspiral, when the objects are far apart and have low velocities; merger, when the objects touch and form a single heavier object; ringdown, when the final single remnant object settles down into its final stationary state.

19: When convincingly detected, the background will most likely be sourced by a large superposition of binary black hole mergers.

GR is nonlinear so a full detailed GW emission process has to be addressed with numerical relativity simulations. However, perturbative methods allow for analytical approximations to be obtained in the radiation zone (away from the compact sources).

To understand better what can physically generate gravitational waves, let us continue working in the approximation of linear perturbations around Minkowski. In the presence of a matter source, the linearised Einstein equations take the following simple form in the harmonic gauge,

$$\Box \bar{h}_{\mu\nu} = -16\pi \delta T_{\mu\nu} \tag{2.30}$$

where $\delta T_{\mu\nu}$ is the linear matter tensor. The solution to this equation outside the source can be obtained through the use of the retarded Green function[20]

$$\bar{h}_{\mu\nu}(t, x) = 4 \int d^3x \frac{T_{\mu\nu}(t - |x - x'|, x)}{|x - x'|}. \tag{2.31}$$

Let us now assume that the source is moving slowly compared to the speed of light. For example, for a periodic motion we can write

$$T_{\mu\nu}(t, x) = S_{\mu\nu}(x) \cos \omega t, \qquad \omega L \ll 1, \tag{2.32}$$

where $L$ is the characteristic size of the source, and $\omega L$ corresponds to its typical velocity. In Eq. (2.31), $x$ denotes the field point where the gravitational wave is being observed, while $x'$ denotes the source point where it is emitted. Since the energy-momentum tensor $T_{\mu\nu}$ vanishes outside the matter distribution, we have $|x'| \lesssim L$; in other words, contributions to the integral only come from within the source, whose spatial extent is bounded by $L$. Then, because $x'$ is confined to a region of size $L$, and assuming we are evaluating the field far from the source (i.e., $|x| \gg |x'|$), we can expand the distance as

$$\omega |x - x'| \approx \omega |x| + \mathcal{O}(\omega |x'|). \tag{2.33}$$

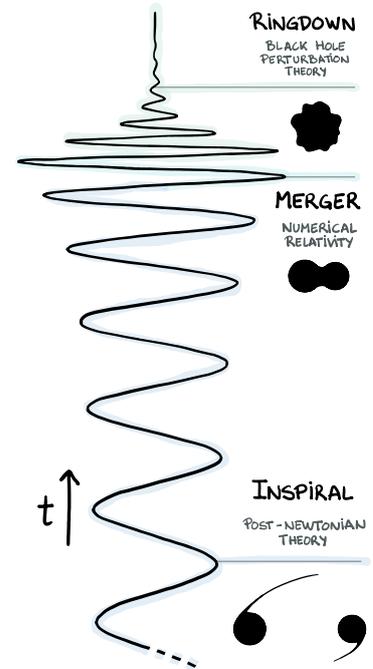

**Figure 2.5:** Illustration of the gravitational wave signal from a compact binary coalescence as a function of time with the three distinct stages and the corresponding mathematical framework in which they are studied.

20: Intuitively, the presence of retarded time in this solution arises from the finite speed of gravitational waves. This means that we observe them as they were in the past rather than in their current state. In other words, by the time we detect the waves from a black hole merger, the merger itself has already occurred long ago.

Furthermore, if we assume the source to be far away from the observer ($|x| \gg L$), we can approximate

$$\frac{1}{|x - x'|} = \frac{1}{|x|} \left( 1 + \mathcal{O}\left(\frac{L}{|x|}\right) \right) \approx \frac{1}{|x|} \equiv \frac{1}{r}, \tag{2.34}$$

and then we can simplify (2.31) to

$$\overline{h}_{\mu\nu}(t, x) = \frac{4}{r} \int d^3x \, T_{\mu\nu}(t - r, r). \tag{2.35}$$

This equation is already in the harmonic gauge, which also ensures conservation of the stress energy tensor in these coordinates, i.e. $\partial^\mu T_{\mu\nu} = 0$. Using this, we can derive the following relation

$$\frac{4}{r} \int d^3x' \, T_{\mu\nu} = \frac{2}{r} \frac{\partial^2}{\partial t^2} \int d^3x' \, T_{00} x_\mu x_\nu \equiv \frac{2}{r} \frac{\partial^2}{\partial t^2} I_{\mu\nu}, \tag{2.36}$$

where $I_{\mu\nu}$ is defined to be the mass quadrupole moment tensor. Finally, we can rewrite the solution for the gravitational field far away from the source (2.35) as [140]

$$h_{\mu\nu}^{TT} = \frac{2}{r} \ddot{I}_{\mu\nu}^{TT}(t - r), \tag{2.37}$$

with overdots denoting time derivatives, and where we have also fixed the TT gauge. This is the *quadrupole formula* and it describes the amplitude of spherical gravitational waves generated by a source in Newtonian motion that decay with the distance $r$ between the source and observer. Substituting (2.28) into (2.37) we can see that the polarisation of GWs emitted in a given direction, say $z$, can then be obtained as:

$$h_+ = \frac{1}{r} \left( \ddot{I}_{11}^{TT} - \ddot{I}_{22}^{TT} \right), \qquad\qquad h_\times = \frac{2}{r} \ddot{I}_{12}^{TT}. \tag{2.38}$$

For slow-moving sources we have that $T_{00} \approx \rho$, with $\rho$ referring to the Newtonian mass density. Hence, $\ddot{I}_{\mu\nu} \neq 0$ requires the distribution of mass in a body to change over time. A rotating spherical body, for instance, will not be able to source gravitational waves. A binary system of gravitationally-bound compact objects, on the other hand, is a natural candidate to emit GWs. Recall that the GW signal generated by binary mergers is characterised by the inspiral, merger and ringdown stages, as depicted in Figure 2.5. We will comprehensively discuss the ringdown stage in Section 2.4, but let us briefly summarise the other two previous stages. Since analytic solutions for the general relativistic two-body problem are not known, one relies on perturbative approximations and/or numerical methods to resolve the dynamics of binary systems and derive the resulting gravitational wave signals.

**Inspiral**

In the inspiral, when the objects are far away from each other moving with velocities $v \ll c$, the quadrupole formula in (2.37) can be used to obtain a leading-order approximation to the GW solution. The emission of GWs causes the system to loose energy throughout this stage, therefore shrinking the orbits and resulting in an accelerating inspiral. For two point masses $m_1$ and $m_2$ in a circular orbit with angular frequency $\omega_{orb}$, the rate of energy loss is given by [141]

$$\frac{dE}{dt} = -\frac{32}{5}\frac{G^{7/3}}{c^5}(\mathcal{M}_c\omega_{orb})^{10/3}, \qquad (2.39)$$

where $\mathcal{M}_c$ is known as the chirp mass and is given by

$$\mathcal{M}_c = \frac{(m_1 m_2)^{3/5}}{(m_1 + m_2)^{1/5}}. \qquad (2.40)$$

According to Kepler's law, the binary velocity is $\omega_{orb}^2 = GM_{tot}/d^3$ for a binary separation $d$ and total mass $M_{tot} = m_1 + m_2$. In the adiabatic regime where $\dot{\omega}_{orb} \ll \omega_{orb}^2$, it is then possible to show that, due to Eq. (2.39), the GW frequency will change slowly in time according to:

$$\dot{f}(t) = \frac{96}{5}(G\mathcal{M}_c)^{5/3}\pi^{8/3}f(t)^{11/3}, \qquad (2.41)$$

which is positive and hence shows how the GW frequency grows in time as the compact objects get closer to the merger, resulting in a characteristic chirping profile for the frequency as a function of time. These analytical expressions only hold during the early inspiral phase. However, the results can be generalised using the *Post-Newtonian* (PN) formalism [142], which is a perturbative approach in which higher order $v/c$ corrections can be obtained. In that case, additional effects coming from the binary mass ratio, object spins and their compositions (e.g. whether they suffer from tidal deformations) become relevant [142]. Nevertheless, as the two compact objects get close to merger, their velocities become relativistic and the PN approximation does not hold.

To model the time evolution of the GW, one can use the following expansion (see e.g. [143, 144]):

$$h_+(t) - ih_\times(t) = \sum_{\ell \geq 2}\sum_{m=-\ell}^{\ell} h_{\ell m}(t; p_i)\,_{-2}Y_{\ell m}(\iota, -\varphi_c), \qquad (2.42)$$

where all angular dependencies appear in the spin-weighted spherical harmonics $_{-2}Y_{\ell m}$ and $p_i$ collects the parameters of the binary (e.g. masses and spins), $\varphi_c$ is the coalescence phase and $\iota$ is the inclination angle.

**Merger**

As the binary nears coalescence, post-Newtonian approximations break down, and full numerical relativity (NR) is required to model the strong-field dynamics.[21] The merger stage is characterised by rapid orbital decay and the eventual collision of the two objects. The gravitational waveform peaks in amplitude just before and during the merger. The NR simulations solve the full Einstein equations in a dynamical, highly nonlinear regime. The resulting waveforms are crucial for detecting and interpreting strong-field gravity effects.

## 2.3.3  Propagation

After being emitted, GWs propagate through the cosmological medium before we can detect them here on Earth. In the cosmological context, the Cosmological Principle states that the universe is homogeneous and isotropic on 22: Note that the FLRW metric in principle also includes a curvature parameter $k = \{0, \pm 1\}$ describing the overall geometry of the spatial dimensions.

$$ds^2 = -d\tau^2 + a^2(\tau)\Big(\frac{dr^2}{1-kr^2} + r^2 d\theta^2 + r^2 \sin^2\theta d\phi^2\Big), \quad (2.43)$$

with $k = 0$ corresponding to a flat universe with Euclidean geometry, $k = +1$ to a closed universe with spherical geometry and $k = -1$ to an open universe with hyperbolic geometry. See Fig. 2.6 for illustrations for these cases. We know from observations that the universe is to a high degree of accuracy spatially flat, and hence in equation (2.44) we have used $k = 0$.

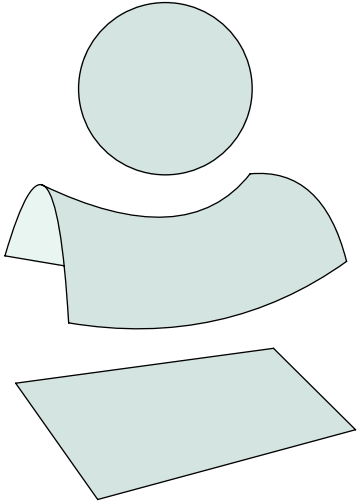

**Figure 2.6:** From top to bottom, illustrations of closed ($k = +1$), open ($k = -1$) and flat ($k = 0$) universes.

23: Scalar-type perturbations account for the matter inhomogeneities in the universe, governing the dynamics of large-scale structures like galaxy clustering. On the other hand, vector-type perturbations are generally disregarded, as they decay with the expansion of the universe and eventually become negligible.

large scales, leading to the Friedmann-Lemaître-Robertson-Walker (FLRW) metric, which describes an expanding (or contracting) universe in GR and is given by the line element[22]

$$ds^2 = -d\tau^2 + a^2(\tau)dx^2. \quad (2.44)$$

Here, $a(\tau)$, known as the scale factor, is a function of cosmic time $\tau$ that describes the expansion of the universe. The spatial coordinates x are called *comoving coordinates*, as the spatial location of gravitationally bound objects remains unchanged despite the expansion of the universe. They relate to *physical coordinates* by the simple rescaling $x_{phys} = a(\tau)x$. It is also often convenient to introduce a new time coordinate known as *conformal time*, denoted by $\eta$. Conformal time is related to cosmic time as $d\tau = a(\tau)d\eta$. In terms of conformal time, the FLRW metric can be rewritten as

$$ds^2 = a^2(\eta)(-d\eta^2 + dx^2). \quad (2.45)$$

This form of the metric is particularly useful because it explicitly factors out the scale factor, making the spacetime locally 'conformally flat'.

At this point, we can define the Hubble parameter $H(\tau)$, and its conformal time version, given by

$$H(\tau) = \frac{\dot{a}}{a}, \qquad \mathscr{H}(\eta) = \frac{a'}{a}, \qquad H = \frac{\mathscr{H}}{a} \quad (2.46)$$

which quantifies the rate of expansion at a given time. Note that we have used the definition $\dot{a} \equiv \frac{da}{d\tau}$ and $a' \equiv \frac{da}{d\eta}$.

The value of the Hubble parameter today, represented as $H_0$, is known as the *Hubble constant* and is the subject of one of the major current unresolved problems in the field of cosmology, i.e. the *Hubble tension*. This issue arises from the discrepancy between local measurements of $H_0$ (using supernovae and Cepheid variables) and early-universe estimates (from the CMB), potentially indicating new physics beyond the standard cosmological model. For reviews of the Hubble tension, see e.g. [149, 150], and for efforts to mitigate it by introducing new physics, such as Early Dark Energy, see e.g. [151–154].

At zeroth perturbative order, using the metric ansatz (2.44) in EFEs (2.8) results in the derivation of Friedmann equations, which are the key feature of the ΛCDM model, as they can be used to describe the dynamical evolution of the different components of the universe, resulting in a set of cosmological parameters that can be matched with observations. Let us, however, focus on gravitational waves, which appear at first order in metric perturbations. Cosmological perturbations divide into scalar, vector and tensor types, whose dynamics decouple at linear order. Here we will solely focus on tensor perturbations, since these are the gravitational waves, but refer to [155] for a comprehensive review on cosmological perturbation theory.[23]

Similarly to the previous discussion on gravitational waves in the transverse-traceless gauge, tensor-type cosmological perturbations have only non-zero spatial components and are both transverse and traceless. These perturbations are then decomposed into two polarisations, with each polarisation satisfying the following linear equation of motion in Fourier space in vacuum

$$h_p'' + 2\mathscr{H}h_p' + k^2 h_p = 0, \quad (2.47)$$

with the prime denoting a conformal time derivative. This equation describes the propagation of gravitational waves in an expanding universe, with the term $2\mathcal{H}h'_p$ accounting for the damping due to the expansion of space. The third term, $k^2$, is the spatial curvature that characterises the wave's propagation.[24] Crucially, we see that all types of matter affect the propagation of GWs only via the Hubble rate parameter $H$, and equally for both polarisations $p = +, \times$. Interestingly, this equation is modified in alternative theories of gravity (such as the ST type that will be the focus of the following chapters, see Eq. 3.38) and therefore observational measurements of how GWs actually propagate through the cosmological medium provides a powerful tool to test gravitational physics beyond GR.

### 2.3.4 Detection

As we have already seen, gravitational waves stretch and squeeze spacetime as they pass through, causing minuscule changes in the distances between objects. These changes are incredibly small—on the order of a fraction of the diameter of a proton—making detection a formidable challenge. The first evidence for gravitational waves came indirectly in 1974, nearly 6 decades after Einstein's original 1916 paper, when R. Hulse and J. Taylor observed the orbital energy loss of the binary pulsar PSR B1913+16, which matched the predictions of GR by the emission of gravitational waves [156].

Nowadays, GWs are detected through the use of highly sensitive interferometers, such as those employed by the LIGO-Virgo-Kagra (LVK) collaboration. In 2015, the LIGO-Virgo-Kagra (LVK) collaboration made the first direct detection of gravitational waves [157]. Since then, LVK has completed 3 observing runs and is currently, at the time of drafting this thesis, undergoing the fourth one, having detected a total of 203 gravitational-wave events [158–170].

The fundamental principle behind interferometric gravitational wave detectors is to split a laser beam into two perpendicular arms of equal length, recombine them, and analyse the resulting interference pattern. When a gravitational wave passes through the detector, it causes tiny, opposite changes in the lengths of the arms, leading to a measurable shift in the interference pattern. Essentially, these interferometers act as extraordinarily precise rulers, capable of measuring distance variations as small as $10^{-18}$ meters.

To better understand the detection process, it is useful to examine the geometry of the problem. Since gravitational waves can originate from any direction in the celestial sphere, their position in the sky is characterised by the usual spherical coordinates ($\theta, \phi$). However, the source of GWs is not spherically symmetric itself and possesses an intrinsic orientation. This orientation is determined by the binary's angular momentum, requiring an additional angle, $\psi$, to describe its inclination relative to the observer's line of sight. These angles are represented in the source, detector and sky frames in Fig. 2.7 [171].

The strain $h$ quantifies the relative change in length of the interferometer's arms due to a passing gravitational wave and is therefore given by how it responds to the two polarisations (2.42)

$$h = F_+(\theta, \phi, \psi)h_+ + F_\times(\theta, \phi, \psi)h_\times. \qquad (2.49)$$

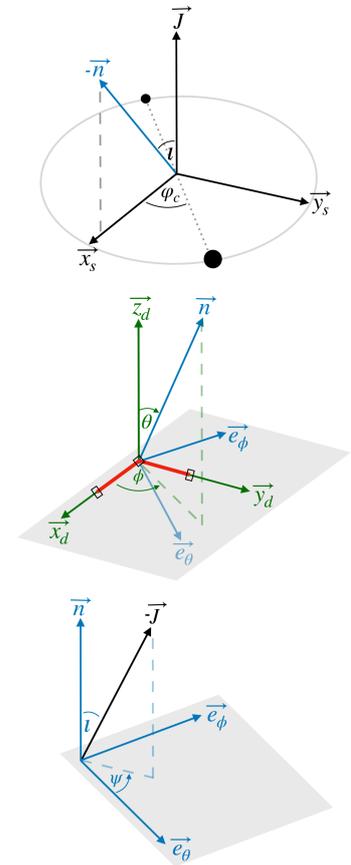

**Figure 2.7:** Angles in the binary system + detector system. From top to bottom different frames are shown: source frame (in black), detector frame (in green), and sky frame (in blue). $\vec{J}$ refers to the total angular momentum, $\iota$ to the orbital plane inclination with respect to the observer, and $\vec{n}$ corresponds to the direction of the line of sight. Figure taken from [171].

Here, $F_{+,\times}$ are the antenna pattern functions specific to each detector and are given by the following combination of angular functions

$$F_+ = \frac{1}{2}\left[1 + \cos^2(\theta)\right]\cos(2\phi)\cos(2\psi) - \cos(\theta)\sin(2\phi)\sin(2\psi), \qquad (2.50)$$

$$F_\times = \frac{1}{2}\left[1 + \cos^2(\theta)\right]\cos(2\phi)\sin(2\psi) + \cos(\theta)\sin(2\phi)\cos(2\psi). \qquad (2.51)$$

Since different detectors are oriented differently, they will have distinct responses to each polarisation mode. In other words, they will have different values of for $F_+$ and $F_\times$. Nonetheless, the characteristic strain functions $h_+$ and $h_\times$, which are intrinsic for the GW, will necessarily be the same. Hence, by combining detections from multiple detectors one can localise the origin direction of GWs with better precision and disentangle the individual polarisation components.

### Current status and future missions

LVK is currently undergoing its fourth observing run (O4) at the time of writing this thesis (see Fig. 2.8) and has detected a total of 203 GW events since the first one in 2015, projecting to detect an event every two to three days [158–170].

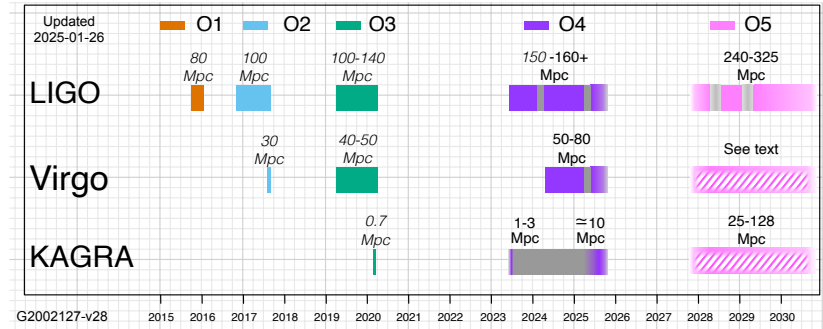

25: These include NANOGrav (North American Nanohertz Observatory for Gravitational Waves) [134], the European Pulsar Timing Array (EPTA) [173], the Parkes Pulsar Timing Array (PPTA, Australia) [139], and the Chinese Pulsar Timing Array (CPTA) [174]. Data from these efforts will be combined under the International Pulsar Timing Array (IPTA) to refine the results; see [175] for a global data release prior to the individual detections reported in June 2023.

Gravitational waves at much lower frequencies (nanohertz) have recently been detected for the first time through precise timing observations of pulsars [134, 139, 173, 174]. Correlated variations in the timing of multiple pulsars—an approach known as Pulsar Timing Arrays (PTAs)—have been independently observed by several collaborations,[25] providing strong evidence for a stochastic gravitational wave background (SGWB). Unlike gravitational waves from individual sources, the SGWB arises from the combined contributions of numerous unresolved sources, forming a pervasive background across the universe.

While the exact origin of this background remains uncertain at the time of writing, the leading explanation is that it results from binaries of supermassive black holes at the centres of merging galaxies. However, alternative origins beyond standard astrophysical sources are also being considered. In particular, a SGWB could arise from exotic early-universe phenomena such as cosmic strings [177], first-order phase transitions [178], or primordial black holes [179] (see [180] for a general review).

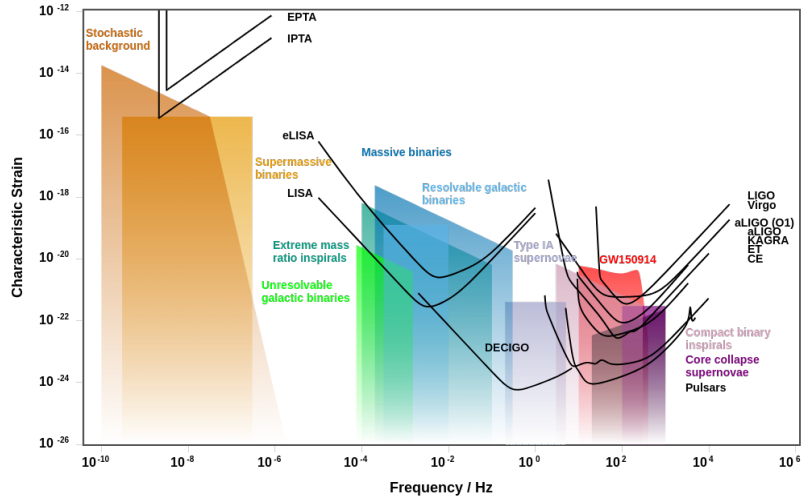

Returning to stellar-origin black holes, the future of GW detections appears exceptionally promising. The LVK collaboration has extended the fourth observing run (O4) until October 2025. Following an engineering break to enhance detector sensitivity and performance, the fifth observing run (O5) is tentatively set to begin in late 2027, though the exact timeline remains uncertain. O5 aims to significantly expand the observable volume of the universe and increase the detection rate of GW events (see e.g. [181] for a visualisation of detection rate forecasts).

Currently, GW observatories are in their second generation, following the limited-sensitivity configurations of first-generation detectors like the early LIGO and Virgo setups. Looking further ahead, the third generation of detectors planned for the 2030s will mark a major leap in sensitivity and range of observable frequencies.

Conceived as a European project, the Einstein Telescope (ET) is planned to be a triangular shaped underground detector with 10km arms with multiple simultaneously operating interferometers [182, 183]. ET will detect gravitational waves from decihertz to several kilohertz with improved sensitivity and sky localisation. With similar scientific objectives, Cosmic Explorer (CE) is a US-proposed detector with a similar L-shaped design as the LVK detectors but with arms 10 times longer (i.e. 40km) [184, 185].

Operating roughly in the millihertz to decihertz range, the upcoming AEDGE (Atomic Experiment for Dark Matter and Gravity Exploration) [186] and DE-CIGO (Deci-Hertz Interferometer Gravitational Wave Observatory) [187, 188] missions will be space-based detectors, extending gravitational wave sensitivity to lower frequencies than ground-based observatories. By accessing this unexplored regime, these detectors will provide complementary avenues for testing gravitational interactions, including the potential to probe specific dark matter models.

Finally, LISA (Laser Interferometer Space Antenna) will be a space-based mission (recently adopted by the European Space Agency) that will detect low-frequency GWs, bridging the gap between ground-based interferometers and PTAs [189]. By employing three spacecraft in a heliocentric orbit, separated by 2.5 million kilometers, LISA will be capable of detecting gravitational waves from sources such as supermassive black hole mergers, extreme mass-ratio inspirals (EMRIs), and potential early-universe relics like primordial gravitational waves or cosmic strings. For LISA, expected event rates are still somewhat uncertain, but most estimates lie in the $O(10 - 100)$ per year range for SMBH mergers – see e.g. [190–197]. In addition, by operating in space—free from the terrestrial noise that ground-based detectors must contend with—LISA is expected to detect gravitational waves with a significantly higher signal-to-noise ratio (SNR). TianQin will be a Chinise-led mission operating with a similar set up and scientific goals as LISA but orbiting the Earth instead of the Sun [198].

## 2.4 Ringdown in General Relativity

After two black holes (or other similar compact objects) merge, the resulting black hole undergoes a rapid phase of relaxation, emitting GWs as it settles into a stable configuration. This stage is known as the ringdown and is charaterised by quasinormal modes (QNMs), a discrete set of damped oscillations whose frequencies are uniquely determined by the remnant's mass, spin, and, in some cases, additional charges or modifications to GR.[26] The corresponding values for quasinormal frequencies are obtained as a result of the application of black hole perturbation theory. Since these are governed solely by the properties of the final object, the ringdown provides a powerful probe of black hole physics, enabling precision tests of GR and potential deviations from the Kerr metric. Studying the ringdown is thus crucial for understanding strong-field gravity, constraining alternative theories, and searching for new physics beyond Einstein's theory. In this section we will discuss in detail the mathematical framework for the study of ringdown signals in GR, its power in testing gravitational physics and the potential systematics one would need to account for in order to fully take advantage of this power.

26: Note that, unlike frequencies, the amplitudes of these oscillations do depend on the initial configuration.

### 2.4.1 Black hole perturbation theory: Regge-Wheeler and Zerilli

In perturbation theory, one exploits the fact that the system under study approximately satisfies the overall characteristics of a much simpler system. By adding small corrections to the *background* solution, one can analytically study the dynamics of *perturbations*.

In gravity territory, there are several ways one can perform a perturbative expansion depending on which 'easier system' we choose to expand around and which small parameter we use to perform the expansion. In our case we work with the black hole perturbation theory framework, as it allows us to define and compute gravitational waves emitted by BBHs in the ringdown.[27]

27: Other examples of perturbative schemes in gravity include: Post-Newtonian theory [199], post-Minkowskian theory [200] and cosmological perturbation theory [155].

The details of black hole perturbation theory in GR were first understood during the 1950s–70s in the seminal papers by Regge and Wheeler (1957) [201] and Zerilli (1970) [202]. Here we will review and rederive these results. Let us start by considering first order perturbations on the metric field

$$g_{\mu\nu} = \bar{g}_{\mu\nu} + h_{\mu\nu}, \qquad (2.52)$$

where $\bar{g}_{\mu\nu}$ is the background metric and $h_{\mu\nu}$ are the first order metric perturbations, which are small in magnitude compared to the background (i.e. $|h_{\mu\nu}| \ll 1$ for all $\mu, \nu$). This is analogous to our earlier discussion in Subsection 2.3.1, where we defined GWs as wave solutions of the evolution equations of metric perturbations in flat spacetime. Here, however, instead of working with a Minkowski spacetime $\eta_{\mu\nu}$ as our background metric, $\bar{g}_{\mu\nu}$ refers to the spacetime metric of a black hole solution (e.g. those discussed in Subsection 2.2.1). As we will see, perturbations of spherically symmetric spacetimes naturally decompose into two distinct sectors, which, at the linear level, evolve independently. These are known as the odd (axial) sector and the even (polar) sector. In each case, the resulting perturbative degree of freedom is governed by a single ordinary differential equation (ODE). For a static and spherically symmetric background given by

$$ds^2 = -f(r)dt^2 + \frac{1}{f(r)}dr^2 + r^2 d\Omega_2^2, \tag{2.53}$$

such as the one corresponding to S(dS) black holes (see (2.9) and (2.10)), the ODEs in vacuum take the form of the following wave equation

$$\left[ \frac{d^2}{dr_*^2} + \left( \omega_{\ell m}^2 - V_{\text{odd/even}}^{\ell m} \right) \right] \psi_{\text{odd/even}}^{\ell m} = 0. \tag{2.54}$$

Given the great importance of this equation throughout this thesis, let us look in detail at its components. Starting from left to right we encounter:

▸ $r_*$, **tortoise coordinate**. The radial coordinate transformation $r_*(r)$ yields a wave equation in a more conveniently solvable (Schrödinger-like) form (2.54) by smoothing out the horizon coordinate singularity. For a metric in the form of (2.53) this transformation looks like

$$\frac{dr}{dr_*} = f(r). \tag{2.55}$$

As an example, in the case of a Schwarzschild metric where $f = 1 - \frac{r_s}{r}$ (2.9), we have the following solution

$$r_* = r + r_s \log \frac{r - r_s}{r_s}, \tag{2.56}$$

which maps the exterior region of the black hole from $r \in [r_s, +\infty]$ into $r_* \in [-\infty, +\infty]$. We therefore see that the precise form of the tortoise coordinate is a background-dependent object.

▸ $\omega_{\ell m}$, **quasinormal frequencies**. Upon employing physically reasonable boundary conditions (as will be exposed in detail in Subsection 2.4.2) the quasinormal frequencies $\omega_{\ell m}$ arise as solutions to the master equation (2.54). These are a discrete set of complex frequencies whose real and imaginary parts respectively describe the oscillation frequency and the damping time of each $(\ell, m)$ mode. They appear in the master equation as a result of taking the master variables to follow a harmonic time dependence, i.e. $\Psi(t, r) = e^{-i\omega t}\psi(r)$, which converts the initial PDE into an ODE with a Schrödinger-like form with $\omega_{\ell m}$ playing the role of eigenvalues and the general solution being constructed as a superposition of the different independent $(\ell, m)$ modes.

▸ $V_{\text{odd/even}}^{\ell m}$, **effective potential**. These potentials govern the evolution of GWs emitted by perturbed black holes and determine the QNM spectrum. The algebraic form for these potentials differs for odd and even perturbations and, for a Schwarzschild metric (2.9) in the action (2.1), these are

$$V_{\text{odd}}^{\ell m} = f\left(\frac{\ell(\ell+1)}{r^2} - \frac{3r_s}{r^3}\right), \tag{2.57}$$

$$V_{\text{even}}^{\ell m} = f\frac{8\lambda^2(\lambda+1)r^3 + 12\lambda^2 r_s r^2 + 18\lambda r_s^2 r + 9r_s^3}{r^3(2\lambda r + 3r_s)^2}, \tag{2.58}$$

with $\lambda \equiv (\ell-1)(\ell+2)/2$. These effective potentials are plotted in Figure 2.10 both in standard Schwarzschild and tortoise coordinates. There, one can see that despite both potentials having considerably differing algebraic forms, their plotted values appear almost indistinguishable, leading to identical QNM spectra.[28] This property is known as *isospectrality* and has been shown to arise from a hidden symmetry between the Regge-Wheeler and Zerilli potentials, which allows them to be mapped onto each other and, more generally, be derived from a single superpotential [203]. This was first found in 1975 by Chandrasekhar and Detweiler in the form of a 1-to-1 mapping between the two potentials [204].[29] This feature applies to Schwarzschild, RN and Kerr black holes, but is generally broken for more general metrics and/or in theories beyond GR (see e.g. [206]).

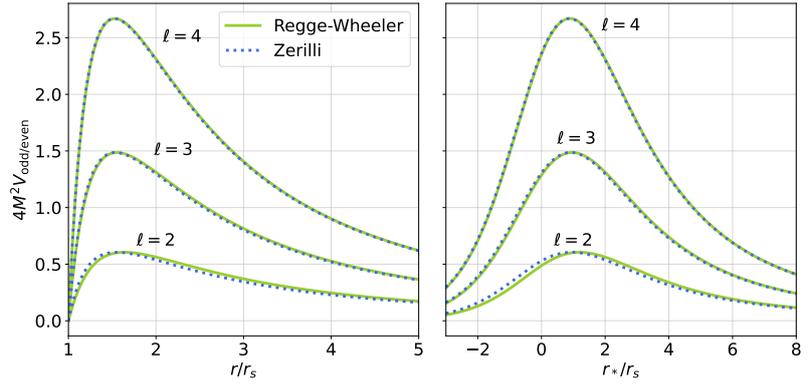

**Figure 2.10:** Regge-Wheeler and Zerilli potentials of a Schwarzschild black hole for different values of $\ell$ as a function of $r/r_s$ (left panel) and the tortoise coordinate $r_*/r_s$ (right panel).

▸ $\psi_{\text{odd/even}}^{\ell m}$, **master variable**. These are scalar functions that encode the gauge-invariant gravitational perturbations of a black hole spacetime, each one carrying one of the 2 degrees of freedom in GR.

In what follows, we will rederive these equations from first principles in two alternative ways. Before doing so, however, we will look at the derivation of the equivalent equation for a test scalar field, which will serve as an easier example to introduce some of the procedures used in the derivations.

### Test scalar on a black hole background

As a warm-up exercise for the more complicated study of tensor perturbations, let us here follow [207] and consider a massless Klein-Gordon scalar in a curved background, whose dynamics is then governed by the equation[30]

$$\Box\phi \equiv \frac{1}{\sqrt{-g}}\partial_\mu\left[\sqrt{-g}\,g^{\mu\nu}\partial_\nu\phi\right] = 0, \tag{2.60}$$

---

30: Note that the d'Alembert operator is defined as $\Box \equiv \nabla^\mu\nabla_\mu$. When acting on a scalar, it can be shown to be equivalent to the form in (2.60) by employing the property $\det(e^A) = e^{\text{Tr}(A)}$, true for any square matrix $A$. This can be used to show that $\frac{\partial_\mu\sqrt{-g}}{\sqrt{-g}} = \frac{1}{2}g^{\alpha\beta}\partial_\mu g_{\alpha\beta}$, and the rest follows straightforwardly:

$$\Box\phi \equiv \partial_\mu\partial^\mu\phi + \Gamma^\mu_{\mu\lambda}\partial^\lambda\phi$$
$$= \partial_\mu\partial^\mu\phi + \frac{1}{2}g^{\mu\nu}\partial_\lambda g_{\mu\nu}\partial^\lambda\phi$$
$$= \partial_\mu(g^{\mu\nu}\partial_\nu\phi) + \frac{\partial_\mu\sqrt{-g}}{\sqrt{-g}}\partial^\mu\phi$$
$$= \frac{1}{\sqrt{-g}}\partial_\mu\left[\sqrt{-g}\,g^{\mu\nu}\partial_\nu\phi\right]. \tag{2.59}$$

where we will take $g_{\mu\nu}$ to be the one corresponding to a Schwarzschild black hole as given by the line element in Eq. (2.9). Here, we intentionally avoid using the overbar to denote background quantities—a common choice to keep the notation uncluttered. Since the background is spherically symmetric, one can exploit this symmetry to separate the $(t, r)$ and $(\theta, \phi)$ dependencies of the test scalar field as[31]

$$\phi(t, r, \theta, \phi) = \frac{1}{r} \sum_{\ell=0}^{\infty} \sum_{m=-\ell}^{\ell} \Phi(t, r) Y_{\ell m}(\theta, \phi). \tag{2.61}$$

Here, $Y_{\ell m}(\theta, \phi)$ are called the *spherical harmonics* and they satisfy the properties described below.

**Spherical harmonics properties**

▶ Orthonormality:

$$\int_0^\pi \int_0^{2\pi} Y_{\ell m}^*(\theta, \phi) Y_{\ell' m'}(\theta, \phi) \sin\theta \, d\theta d\phi = \delta_{\ell\ell'} \delta_{mm'}, \tag{2.62}$$

where $\delta$ correspond to Dirac delta functions.

▶ Completeness:

$$f(\theta, \phi) = \sum_{\ell=0}^{\infty} \sum_{m=-\ell}^{\ell} c_{\ell m} Y_{\ell m}(\theta, \phi), \tag{2.63}$$

where $f$ represents any function on the sphere, and the equation above shows that it can be completely decomposed in the spherical harmonic basis.

▶ Parity transformation:

$$Y_{\ell m}(\pi - \theta, \phi + \pi) = (-1)^\ell Y_{\ell m}(\theta, \phi), \tag{2.64}$$

which describes the transformation $(\theta, \phi) \to (\pi - \theta, \phi + \pi)$, corresponding to an inversion through the origin–i.e. mapping the points $\vec{r} \to -\vec{r}$. This demonstrates that $Y_{\ell m}$ exhibits even parity when $\ell$ is even, and odd parity when $\ell$ is odd.

▶ Solutions to the angular part of the Laplace equation in sphercial coordinates with eigenvalues $\ell(\ell + 1)$:

$$\nabla^2 Y_{\ell m}(\theta, \phi) = \Big[\frac{1}{\sin\theta} \partial_\theta (\sin\theta \partial_\theta)$$
$$+ \frac{1}{\sin^2\theta} \partial_\phi^2 + \ell(\ell + 1)\Big] Y_{\ell m}(\theta, \phi) = 0. \tag{2.65}$$

▶ Written as associated Legendre polynomials:

$$Y_{\ell m}(\theta, \phi) = (-1)^m \sqrt{\frac{(2\ell + 1)}{4\pi} \frac{(\ell - m)!}{(\ell + m)!}} P_{\ell m}(\cos\theta) e^{im\phi}, \tag{2.66}$$

with $P_{\ell m}(\cos\theta)$ being the Legendre polynomials satisfying

$$P_{\ell m}'' = -(\ell(\ell + 1) - m^2 \csc\theta^2) P_{\ell m} - \cot\theta P_{\ell m}', \tag{2.67}$$

where here the prime denotes $\partial_\theta$.

**Table 2.2:** Spherical harmonics $Y_{\ell m}(\theta, \phi)$ for $\ell = 0, 1, 2$ with corresponding visualisations.

| | $m = 0$ | $m = \pm 1$ | $m = \pm 2$ |
|---|---|---|---|
| $\ell = 0$ | $\frac{1}{2}\sqrt{\frac{1}{\pi}}$ 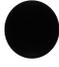 | | |
| $\ell = 1$ | $\frac{1}{2}\sqrt{\frac{3}{\pi}}\cos(\theta)$ 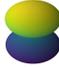 | $\mp\frac{1}{2}\sqrt{\frac{3}{2\pi}}\sin(\theta)e^{\pm i\phi}$ 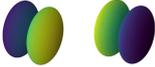 | |
| $\ell = 2$ | $\frac{1}{4}\sqrt{\frac{5}{\pi}}(3\cos^2(\theta) - 1)$ 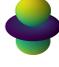 | $\mp\frac{1}{2}\sqrt{\frac{15}{2\pi}}\sin(\theta)\cos(\theta)e^{\pm i\phi}$ 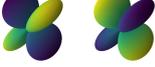 | $\frac{1}{4}\sqrt{\frac{15}{2\pi}}\sin^2(\theta)e^{\pm 2i\phi}$ 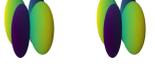 |

Let us now go back to Equation (2.60). Substituting in (2.61) and (2.9) we can write[32]

32: Being $g_{\mu\nu}$ diagonal, its inverse and determinant can be straightforwardly obtained:

$$g_{\mu\nu} = \begin{pmatrix} -f(r) & 0 & 0 & 0 \\ 0 & \frac{1}{f(r)} & 0 & 0 \\ 0 & 0 & r^2 & 0 \\ 0 & 0 & 0 & r^2\sin^2\theta \end{pmatrix},$$

$$g^{\mu\nu} = \begin{pmatrix} -\frac{1}{f(r)} & 0 & 0 & 0 \\ 0 & f(r) & 0 & 0 \\ 0 & 0 & \frac{1}{r^2} & 0 \\ 0 & 0 & 0 & \frac{1}{r^2\sin^2\theta} \end{pmatrix},$$

$$\sqrt{-g} = r^2\sin\theta. \qquad (2.68)$$

$$\begin{aligned}
\Box\phi &= \frac{1}{\sqrt{-g}}\Big[\partial_t\left(\sqrt{-g}\,g^{tt}\partial_t\right) + \partial_r\left(\sqrt{-g}\,g^{rr}\partial_r\right) \\
&\quad + \partial_\theta\left(\sqrt{-g}\,g^{\theta\theta}\partial_\theta\right) + \partial_\phi\left(\sqrt{-g}\,g^{\phi\phi}\partial_\phi\right)\Big]\phi \\
&= \frac{1}{r^2\sin\theta}\Big[-\partial_t\left(r^2\sin\theta\frac{1}{f}\partial_t\right) + \partial_r\left(r^2\sin\theta f\partial_r\right) \\
&\quad + \partial_\theta\left(\sin\theta\partial_\theta\right) + \partial_\phi\left(\frac{1}{\sin\theta}\partial_\phi\right)\Big]\phi \\
&= -\frac{1}{f}\partial_t^2\phi + \frac{1}{r^2}\partial_r(r^2 f\partial_r\phi) + \frac{1}{r^2\sin\theta}\partial_\theta(\sin\theta\partial_\theta\phi) + \frac{1}{r^2\sin\theta}\partial_\phi^2\phi \\
&= \left(-\frac{1}{rf}\partial_t^2\Phi + \frac{1}{r^2}\partial_r\left(r^2 f\partial_r\left(\frac{\Phi}{r}\right)\right)\right)Y_{\ell m} \\
&\quad + \frac{1}{r^3}\left(\frac{1}{\sin\theta}\partial_\theta(\sin\theta\partial_\theta Y_{\ell m}) + \frac{1}{\sin\theta^2}\partial_\phi^2 Y_{\ell m}\right)\Phi \\
&= \frac{1}{rf}\left[-\partial_t^2\Phi + \frac{1}{r}f\left(r\partial_r(f\partial_r\Phi) + f\partial_r\Phi - f\partial_r\Phi - f'\Phi\right) - f\frac{\ell(\ell+1)}{r^2}\Phi\right]Y_{\ell m} \\
&= \frac{1}{rf}\left[\partial_x^2\Phi - \partial_t^2\Phi - f\left(\frac{\ell(\ell+1)}{r^2} + f'\right)\Phi\right]Y_{\ell m} = 0 \qquad (2.69)
\end{aligned}$$

From the last line above, specifying a harmonic time dependence for the master function, i.e. $\Phi(r, r) = e^{-i\omega t}\varphi(r)$,[33] we see that each $(\ell, m)$ mode independently satisfies the master equation

33: Or, in other words, performing a Fourier transform with respect to time

$$\Phi(t, r) = \int_{-\infty}^{\infty}\frac{d\omega}{2\pi}\varphi(\omega, r)e^{-i\omega t}. \qquad (2.70)$$

$$\left[\frac{d^2}{dr_*^2} + \left(\omega_{\ell m}^2 - f\left(\frac{\ell(\ell+1)}{r^2} + \frac{r_s}{r}\right)\right)\right]\varphi = 0, \qquad (2.71)$$

where we have also specified to a Schwarzschild metric.

Writing covariant equations in component form (as done in (2.69)) is sufficiently straightforward for the case of the test scalar to show it here in full detail. When considering tensor perturbations (and even more so in theories beyond GR) it however becomes increasingly more tedious. Hence, when performing the analogous calculations in the works summarised in this thesis [1–3] we employ the tensor algebra `Mathematica` package xAct [208], and make such calculations reproducible and open source at [4]. Some of the key steps in such calculations are:

1. Substitute in component values for tensorial objects,
2. Employ spherical harmonic relations to reduce the problem to $(t, r)$ components,
3. Perform coordinate transformations (e.g. to tortoise coordinates),
4. For tensor perturbations, combine perturbation functions into master variables.

**Tensor perturbations**

Let us now turn our attention to the dynamical study of metric perturbations on a black hole background. Similarly to the case for the scalar, for spherically symmetric backgrounds we can decompose the linear metric perturbations $h_{\mu\nu}$ into *tensorial* spherical harmonics. We here follow the notation of Martel and Poisson [209] (up to a few concessions), where the spacetime coordinates are divided as $x^\mu = (y^a, z^A)$ with $y^a = (t, r)$ and $z^A = (\theta, \phi)$. The metric and Levi-Civita tensor in the spherical coordinates are given by

$$\gamma_{AB} = \begin{pmatrix} 1 & 0 \\ 0 & \sin^2\theta \end{pmatrix}, \qquad \epsilon_{AB} = \begin{pmatrix} 0 & \sin\theta \\ -\sin\theta & 0 \end{pmatrix}. \tag{2.72}$$

This expansion allows us to separate the perturbations into two different sectors depending on how they transform under a parity transformation: $(\theta, \phi) \rightarrow (\pi - \theta, \pi + \phi)$. Even (or polar) components rotate in the same way as the spherical harmonics (2.64), while odd (or axial) components have opposite parity.

$$h_{\mu\nu} = h_{\mu\nu}^{\text{odd}} + h_{\mu\nu}^{\text{even}}, \tag{2.73}$$

which, to first order in perturbations, remain uncoupled and independently-evolving.

In order to decompose tensors into spherical harmonics, we require these harmonics to also incorporate tensorial indices; in other words, we require *tensorial* spherical harmonics. Table 2.3 collects such decomposition basis for up to 2-tensors. Tensorial harmonics also obey a set of orthonormality relations which can be directly derived from the relations shown above—we refer to e.g. [209, 210] for a collection of such relations.

| Rank | Parity | Harmonic |
|---|---|---|
| 1 | Even | $Y_A^{tm} \equiv \nabla_A Y^{tm} = (Y_{,\theta}^{tm}, Y_{,\phi}^{tm})$ |
| | Odd | $X_A^{tm} \equiv -\epsilon_{AC}\gamma^{BC}\nabla_B Y^{tm} = -\epsilon_A{}^B \nabla_B Y^{tm}$ $= \left(-\frac{1}{\sin\theta} Y_{,\phi}^{tm}, \sin\theta Y_{,\theta}^{tm}\right)$ |
| 2 | Even | $Y_{AB}^{tm} \equiv \left(\nabla_A \nabla_B + \frac{1}{2}\ell(\ell+1)\gamma_{AB}\right) \cdot Y^{tm}$ |
| | Odd | $X_{AB}^{tm} \equiv -\frac{1}{2}\left(\epsilon_A{}^C \nabla_B + \epsilon_B{}^C \nabla_A\right) \cdot \nabla_C Y^{tm}$ |

**Table 2.3:** Tensorial spherical harmonics for rank 1 and 2 tensors.

Odd parity perturbations, which ultimately give rise to the Regge-Wheeler equation, can be decomposed as[34]

$$h_{ab}^{\text{odd}} = 0, \tag{2.74}$$

$$h_{aB}^{\text{odd}} = \sum_{\ell m} h_{a,\ell m}(t,r) X_B^{\ell m}, \tag{2.75}$$

$$h_{AB}^{\text{odd}} = \sum_{\ell m} h_{2,\ell m}(t,r) X_{AB}^{\ell m}. \tag{2.76}$$

However, due to diffeomorphism invariance, we know that not all of these components are physical and we can reduce the number of degrees of freedom by fixing the gauge. Under a gauge transformation (2.18), metric perturbations transform as described in Eq. (2.20), now with partial derivatives being promoted to covariant derivatives due to the curved nature of the background. Odd sector functions correspondingly transform in the following way

$$h_a \to h_a' = h_a - \nabla_a \zeta + \frac{2}{r} \frac{\partial r}{\partial x^a}, \tag{2.77}$$

$$h_2 \to h_2' = h_2 - 2\zeta. \tag{2.78}$$

It is convenient to choose the Regge-Wheeler gauge which sets $h_2 = 0$ [201]. This choice leaves $h_a$ invariant under gauge transformations, completely fixing the gauge for $\ell \geq 2$. Hence, this renders the following perturbations in the odd sector, which are controlled by two functions $(h_0, h_1)$

$$h_{\mu\nu}^{\text{odd}} = \begin{pmatrix} 0 & 0 & -\frac{h_0}{\sin\theta}\partial_\phi & \sin\theta h_0 \partial_\theta \\ 0 & 0 & -\frac{h_1}{\sin\theta}\partial_\phi & \sin\theta h_1 \partial_\theta \\ * & * & 0 & 0 \\ * & * & 0 & 0 \end{pmatrix} Y_{\ell m} \tag{2.79}$$

where the $*$ denote symmetric components. Note that $\ell = 0$ and $\ell = 1$, also known as the monopole and dipole modes respectively, require special treatment, as the Regge-Wheeler choices does not fully fix the gauge in these cases. We will however be mostly interested in the quadrupole mode $\ell = 2$ (and occasionally higher multipoles) as it is the lowest contributing mode to the emission of gravitational waves, and so we refer to e.g. [209] for the gauge-fixing details on $\ell = 0$ and $\ell = 1$.[35]

The even parity sector, which gives rise to the Zerilli equation, has perturbations that can be decomposed as

$$h_{ab}^{\text{even}} = \sum_{\ell m} H_{ab}^{\ell m} Y^{\ell m}, \tag{2.80}$$

$$h_{aB}^{\text{even}} = \sum_{\ell m} j_a^{\ell m} Y_B^{\ell m}, \tag{2.81}$$

$$h_{AB}^{\text{even}} = r^2 \sum_{\ell m} (K^{\ell m} \gamma_{AB} Y^{\ell m} + G^{\ell m} Y_{AB}^{\ell m}). \tag{2.82}$$

Under a gauge transformation, these quantities transform as

$$H_{ab} \rightarrow H'_{ab} = H_{ab} - \nabla_a \xi_b - \nabla_b \xi_a. \tag{2.83}$$

$$j_a \rightarrow j'_a = j_a - \xi_a - \nabla_a \xi + \frac{2}{r} \frac{\partial r}{\partial x^a} \xi, \tag{2.84}$$

$$K \rightarrow K' = K + \frac{\ell(\ell+1)}{r^2} \xi - \frac{2}{r} \frac{\partial r}{\partial x_a} \xi_a, \tag{2.85}$$

$$G \rightarrow G' = G - \frac{2}{r^2} \xi. \tag{2.86}$$

In this case the Regge-Wheeler gauge sets $j_a = G = 0$ and leaves the other perturbations invariant. Defining

$$H_{ab} = \begin{pmatrix} H_0 & H_1 \\ H_1 & H_2 \end{pmatrix}, \tag{2.87}$$

we see that the metric perturbations in the even sector are governed by four functions $(H_0, H_1, H_2, K)$ and can be written as

$$h_{ab}^{\text{even}} = \begin{pmatrix} fH_0 & H_1 & 0 & 0 \\ * & \frac{1}{f}H_2 & 0 & 0 \\ 0 & 0 & r^2 K & 0 \\ 0 & 0 & 0 & r^2 \sin^2 \theta K \end{pmatrix} Y_{\ell m}. \tag{2.88}$$

It is important to note that this formalism does not apply to the full Kerr metric, as Kerr spacetime lacks full spherical symmetry and retains only axial symmetry. In that case, one needs to use the Teukolsky equation to perform the equivalent separation [211].

To study the evolution of these perturbations we can follow two equivalent approaches:

1. **EOM approach**: We apply perturbations at the level of the equations of motion (EOM), i.e. EFEs.

$$G_{\mu\nu} = \bar{G}_{\mu\nu} + \delta G_{\mu\nu} = 0. \tag{2.89}$$

Note that the background Einstein tensor for the Schwarzschild metric satisfies $\bar{G}_{\mu\nu} = 0$. Hence, the linearised field equations are fully determined from the first order perturbation, which is given by

$$\delta G_{\mu\nu} = -h_{\mu\nu}\bar{R} + \nabla_\nu \nabla_\mu h - \nabla_\sigma \nabla_\mu h_\nu^\sigma - \nabla_\nu h_\mu^\sigma + \nabla_\sigma \nabla^\sigma h_{\mu\nu}$$
$$+ \bar{g}_{\mu\nu} \left( h^{\sigma\lambda} \bar{R}_{\sigma\lambda} + \frac{1}{2} \bar{g}^{\sigma\lambda} (\nabla_\lambda \nabla_\sigma h - 2\nabla_\rho \nabla_\lambda h_\sigma^\rho + \nabla_\rho \nabla^\rho h_{\sigma\lambda}) \right) = 0, \quad (2.90)$$

where $h = h^\mu_{\ \mu}$.[36] We then substitute in the expressions of the metric perturbations for the odd and even sector and combine different components of the equations to obtain the two differential equations determining their evolution.

2. **Action approach**: We apply perturbations at the level of the action to obtain a quadratic action, substitute the components and then vary with respect to the functions in the perturbations (e.g. in the odd sector either $h_0$ and $h_1$ separately or their redefinition into a master variable). The covariant quadratic Lagrangian for the Einstein-Hilbert action is given by[37]

$$\delta L = \frac{1}{16}\sqrt{-g}\Big[8h_\mu^\sigma h^{\mu\nu}\bar{R}_{\nu\sigma} + h(-4h^{\mu\nu}\bar{R}_{\mu\nu} + h\bar{R}) - 2h_{\mu\nu}h^{\mu\nu}\bar{R}$$
$$+ 2(\nabla_\mu h\nabla^\mu h - 2\nabla^\mu h\nabla_\nu h_\mu^\nu + 2\nabla_\nu h_{\mu\sigma}\nabla^\sigma h^{\mu\nu} - \nabla_\sigma h_{\mu\nu}\nabla^\sigma h^{\mu\nu})\Big]. \quad (2.91)$$

Both methods must of course produce the same results. We now present in explicit detail the methodology to obtain such equations via both methods, as sometimes one is more suited than the other for specific settings.

**EOM approach: Odd sector**

Here we approximately follow the Regge-Wheeler original paper [201] and Maggiore's GW textbook [207]. We begin by substituting the odd-parity components of the metric perturbations (5.36) into Eq. (2.90). While this step is conceptually straightforward, the calculations are lengthy, so they are carried out using `Mathematica`. However, we can leverage the significant symmetries of our system to simplify the computation. First, since the background geometry is Ricci-flat ($\bar{R} = \bar{R}_{\mu\nu} = 0$), all Ricci curvature terms evaluated on the background vanish. Second, in the odd-parity sector, the trace of the perturbation is clearly zero, i.e. $h = 0$. Finally, due to the axial symmetry of the Schwarzschild metric,[38] $\partial_\phi$ is a Killing vector. This implies that all $m$-modes are degenerate in frequency and decouple from each other at linear level, allowing us to take $m$ to a fiducial value without loss of generality, conveniently $m = 0$, giving

$$h_{\mu\nu}^{\text{odd}} = \begin{pmatrix} 0 & 0 & 0 & \tilde{h}_0 \\ 0 & 0 & 0 & \tilde{h}_1 \\ 0 & 0 & 0 & 0 \\ * & * & 0 & 0 \end{pmatrix} \sin\theta \partial_\theta Y_{\ell m} e^{-i\omega t}, \quad (2.92)$$

where note that we have also separated the time-dependence from the perturbation functions as $h_a \equiv \tilde{h}_a e^{-i\omega t}$. We can start by solving for $\tilde{h}_0(\tilde{h}_1)$ the component equation $(\theta, \phi) = 0$ of the equations of motion (2.90), which gives

$$\tilde{h}_0 = \frac{if(\tilde{h}_1 f)'}{\omega}, \quad (2.93)$$

Looking at the $(r, \phi) = 0$ component equation, we substitute (2.93) and solve for $\tilde{h}_1''$, giving

$$\tilde{h}_1'' = \frac{1}{r^2 f^2}\Big(-r^2\omega^2\tilde{h}_1 + lf\tilde{h}_1 + l^2 f\tilde{h}_1 - r^2\tilde{h}_1(f')^2$$
$$+ 2rf^2\tilde{h}_1' - 3r^2 ff'\tilde{h}_1' - r^2 f\tilde{h}_1 f''\Big). \quad (2.94)$$

Defining the Regge-Wheeler function as

$$\psi_{\text{odd}}(r) \equiv \frac{f\tilde{h}_1}{r}, \quad (2.95)$$

we can rewrite equation (2.94) in tortoise coordinates (2.55) as

$$\left[\frac{d^2}{dr_*^2} + (\omega_{\ell m}^2 - V_{\text{odd}}^{\ell m})\right]\psi_{\text{odd}}^{\ell m} = 0, \qquad V_{\text{odd}}^{\ell m} = f\left(\frac{\ell(\ell+1)}{r^2} - \frac{3r_s}{r^3}\right). \quad (2.96)$$

which is the Regge-Wheeler equation in the absence of matter.

38: That is, it contains no $\phi$-dependence.

**EOM approach: Even sector**

Due to its higher number of functions in (2.88), this sector is more laborious to compute. In fact, despite the seminal 1957 Regge-Wheeler paper already setting the basis for solving this sector, it would remain unsolved until 1970 when Zerilli finally completed the analysis of Schwarzschild perturbations in GR [202]. Again, here we approximately follow the Zerilli paper [202] as well as Maggiore's version [207]. As in the odd sector, we also separate the time-dependence from the perturbation functions as $H_{ab} \equiv \tilde{H}_{ab}e^{-i\omega t}$ and $K \equiv \tilde{K}e^{-i\omega t}$. It is also convenient to start with the component equation $(\theta, \phi) = 0$, which gives us $\tilde{H}_0 = \tilde{H}_2$. This is used to eliminate $\tilde{H}_2$ from all the other components. Now the component equations $\{(t,\theta),(t,r),(r,\theta)\} = 0$ give the following equations

$$\tilde{H}_0' = \frac{1}{2}\Big(\frac{-2i\omega\tilde{H}_1 + (-2\tilde{H}_0 + \tilde{K})f'}{f}$$
$$+ \frac{2r\omega\tilde{H}_0 - 2r\omega\tilde{K} + i\tilde{H}_1(\ell^2 + \ell - 2rf' - r^2f'')}{r^2\omega}\Big), \qquad (2.97)$$

$$\tilde{H}_1' = -\frac{i\omega\tilde{H}_0 + i\omega\tilde{K} + \tilde{H}_1 f'}{f}, \qquad (2.98)$$

$$\tilde{K}' = \frac{1}{2}\left(\frac{\tilde{K}f'}{f} + \frac{2r\omega\tilde{H}_0 - 2r\omega\tilde{K} + i\tilde{H}_1(\ell^2 + \ell - 2rf' - r^2f'')}{r^2\omega}\right). \qquad (2.99)$$

These can be used to reduce the order of the derivatives in the component equation $(r,r) = 0$, which then results in an algebraic relation between the three remaining functions

$$4\omega f^2(-\tilde{H}_0 + \tilde{K}) + r^2\omega\tilde{K}(4\omega^2 + f') + f\Big(-2\omega\tilde{K}(\ell^2 + \ell - rf')$$
$$+ 2\omega\tilde{H}_0(\ell^2 + \ell - rf' - r^2f'') - i\tilde{H}_1(4r\omega^2 + 2r(f')^2 - f'(\ell^2 + \ell - r^2f''))\Big) = 0. \qquad (2.100)$$

We can solve this for $\tilde{H}_0$ and substitute it back to the differential equations (2.98) and (2.99). Rewriting $\tilde{H}_1 = \omega G$, we can uncouple $\tilde{K}$ and $G$ from these two equations. The next step is to rewrite these functions in the following way

$$\tilde{K} = f_1\psi_{\text{even}} + f_2 X, \qquad\qquad G = f_3\psi_{\text{even}} + f_4 X, \qquad (2.101)$$

where the $f$ coefficients are functions of $r$ given by

$$f_1 = \frac{1}{2\ell r + 2\ell^2 r - 4rf - 2r^2f' - 2r^3f''}\Big(\ell^2 + 2\ell^3 + \ell^4$$
$$- 2\ell f - 2\ell^2 f - r\ell r f' - 3\ell^2 r f' + 2rff' + 2r^2(f')^2$$
$$- 2\ell r^2 f'' - 2\ell^2 r^2 f'' + 4r^2 ff'' + 3r^3 f'f'' + r^4(f'')^2\Big), \qquad (2.102)$$

$$f_2 = 1, \qquad (2.103)$$

$$f_3 = -\frac{i(-\ell - \ell^2 + 2f + 2rf_1 + rf' + r^2f'')}{2f}, \qquad (2.104)$$

$$f_4 = -\frac{ir}{f}. \qquad (2.105)$$

These coefficients are chosen so that $\partial_*\psi_{\text{even}} - X$ vanishes. To obtain the Zerilli equation, we just need to differentiate $\partial_*\psi_{\text{even}} - X = 0$ with respect to $r_*$, giving

$$\left[\frac{d^2}{dr_*^2} + \left(\omega_{\ell m}^2 - V_{\text{even}}^{\ell m}\right)\right]\psi_{\text{even}}^{\ell m} = 0, \tag{2.106}$$

with

$$V_{\text{even}}^{\ell m} = f\frac{8\lambda^2(\lambda+1)r^3 + 12\lambda^2 r_s r^2 + 18\lambda r_s^2 r + 9r_s^3}{r^3(2\lambda r + 3r_s)^2}, \tag{2.107}$$

where we have redefined $\lambda = (\ell+2)(\ell-1)/2$.

**Action approach: Odd sector**

We turn now to the action approach as an alternative way to obtain the same differential equations. We start by considering the quadratic Lagrangian (2.91) as evaluated on a Schwarzschild background, i.e. with $\bar{R} = 0 = \bar{R}_{\mu\nu}$, $h = 0$ (for odd perturbations), and $m = 0$ in (2.92). After substituting the odd perturbation components (2.92) in (2.91), we also need to integrate the angular components, which involves integrating different combinations of spherical harmonics. In terms of the associated Legendre polynomials the integrals required at this point are

$$\int_0^\pi d\theta\sin\theta P_{\ell m}^2 = \frac{2}{2\ell+1}, \qquad \int_0^\pi d\theta\sin\theta(P'_{\ell m})^2 = \frac{2(\ell(\ell+1))}{2\ell+1}. \tag{2.108}$$

The $\phi$ integral is trivial to compute, as setting $m = 0$ means nothing explicitly depends on $\phi$ and hence just contributes an overall $2\pi$ factor. As a result, we obtain the following expression for the action in components

$$\delta L = \frac{2\pi\ell(\ell+1)}{r^2(2\ell+1)}\Bigg[\left(1 + \frac{(\ell+2)(\ell-1)}{2f}\right)h_0^2 - f\frac{(\ell+2)(\ell-1)}{2}h_1^2$$
$$+ \frac{r^2}{2}\left(\dot{h_1}^2 - 2\dot{h_1}h_0' + (h_0')^2 + \frac{4}{r}\dot{h_1}h_0\right)\Bigg], \tag{2.109}$$

where the dot denotes a time derivative. The step of converting the quadratic Lagrangian from covariant (2.91) to component form is performed in `Mathematica` and invovles performing some integrations by parts [4]. Note that here we do not yet separate the time dependence from the perturbation functions in the same way we did before.[39] To obtain the Regge-Wheeler equation, one can either redefine $h_0$ and $h_1$ into one master variable and hence write a one-variable quadratic action (this will be the procedure used in the following Chapters in this thesis), or vary directly this action with respect to $h_0$ and $h_1$ and combine such equations of motion appropriately to obtain the master equation. Here we will follow this latter route. Varying (2.109) with respect to $h_0$ and $h_1$ we obtain the following expressions respectively

$$\tilde{h}_0'' + i\omega\tilde{h}_1' + 2i\omega\frac{\tilde{h}_1}{r} - \frac{\tilde{h}_0}{r^2}f^{-1}\left(\ell(\ell+1) - \frac{4M}{r}\right) = 0, \tag{2.113}$$

$$f^{-1}\left(2i\omega\frac{\tilde{h}_0}{r} + \omega^2\tilde{h}_1 - i\omega\tilde{h}_0'\right) - \frac{\tilde{h}_1}{r^2}(\ell+2)(\ell-1) = 0, \tag{2.114}$$

39: It is important not to evaluate any of the time derivatives before varying the action, as these operations do not commute. To illustrate this point, let us look at the following toy example:

$$\mathscr{L} = (\dot{h})^2, \qquad \dot{h} = -i\omega h. \tag{2.110}$$

Evaluating the time derivative first and then varying with respect to $h$ gives

$$\mathscr{L} = -\omega^2 h^2, \qquad \frac{\delta\mathscr{L}}{\delta h} = -2\omega^2 h. \tag{2.111}$$

Nonetheless, following the correct route we get

$$\delta\mathscr{L} = 2\dot{h}\delta\dot{h} = -2\ddot{h}\delta h, \qquad \frac{\delta\mathscr{L}}{\delta h} = 2\omega^2 h, \tag{2.112}$$

which results in the opposite sign.

where now we have specified the time dependence. Taking an $r$-derivative of the second equation gives the first one multiplied by an overall factor of $i\omega/f$, which can be simplified to give the same equation as (2.93), obtained in the EOM approach,

$$\tilde{h}_0 = -\frac{f(\tilde{h}_1 f)'}{i\omega}. \tag{2.115}$$

Using the same Regge-Wheeler master variable (2.95) and tortoise coordinates (2.55) we recover the same Regge-Wheeler equation (2.96)

$$\left[\frac{d^2}{dr_*^2} + \left(\omega_{\ell m}^2 - V_{\text{odd}}^{\ell m}\right)\right]\psi_{\text{odd}}^{\ell m} = 0, \qquad V_{\text{odd}}^{\ell m} = f\left(\frac{\ell(\ell+1)}{r^2} - \frac{3r_s}{r^3}\right). \tag{2.116}$$

### Action approach: Even sector

To close our discussion on the derivation of Reggie-Wheeler and Zerilli equations, let us finally derive the Zerilli equation through the action approach. We start by substituting the even metric components (2.88) in the quadratic action (2.91). The resulting expression is long and uninsightful, so we do not show it here and leave it for the interested reader in the `Mathematica` repository [4]. Varying the quadratic action with respect to the four functions $\{H_0, H_1, H_2, K\}$ and integrating the angular dependencies, one obtains the following four respective equations of motion[40]

$$(\ell^2 + \ell + 2)\tilde{H}_2 + \ell^2 + 2 - 2)\tilde{K} - r\tilde{K}' + fr(2\tilde{H}_2' - 5\tilde{K}' - 2r\tilde{K}'') = 0, \tag{2.117}$$

$$\frac{1}{f}(-2i\omega\tilde{K}r + f(2\ell(\ell+1)\tilde{H}_1 - 4i\omega\tilde{H}_2 r + 6i\omega\tilde{K}r + 4i\omega r^2\tilde{K}') = 0, \tag{2.118}$$

$$\frac{1}{f}(2\omega^2\tilde{K}r^2 + f^2r(-2\tilde{H}_0' + \tilde{K}')$$
$$+ f((\ell^2 + \ell)\tilde{H}_0 - 2\tilde{H}_2 - (\ell^2 + \ell - 2)\tilde{K} - 4i\omega\tilde{H}_1 r + r\tilde{K}')) = 0, \tag{2.119}$$

$$\frac{1}{f}(-2i\omega\tilde{H}_1 r + 2\omega^2 r^2(\tilde{H}_2 + \tilde{K})$$
$$+ f((\ell^2 + \ell)(\tilde{H}_0 - \tilde{H}_2) - 2i\omega\tilde{H}_1 r - 3r\tilde{H}_0' - 4i\omega r^2\tilde{H}_1' - r\tilde{H}_2' + 2r\tilde{K}')$$
$$+ f^2r(\tilde{H}_0' - \tilde{H}_2' + 2(\tilde{K}' - r\tilde{H}_0'' + r\tilde{K}'')) = 0, \tag{2.120}$$

where again we have now separated the time dependence, giving rise to factors of $\omega$.

These equations can now be combined in different ways to obtain the Zerilli equation. The intricate details of these manipulations can be found in the same `Mathematica` notebook but we summarise them here in Figure 2.11.

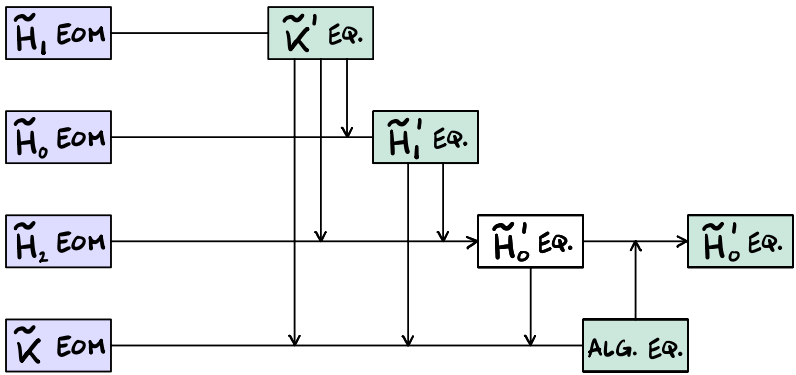

**Figure 2.11:** Equation manipulations for the derivation of the Zerilli equation in the action approach.

After such manipulations, we obtain the same equations as in the EOM approach (2.97), (2.98), (2.99), (2.100).

$$\tilde{K}' + \frac{r - 3M}{r(r - r_s)}\tilde{K} - \frac{1}{r}\tilde{H}_0 - \frac{\ell(\ell + 1)}{2i\omega r^2}\tilde{H}_1 = 0, \tag{2.121}$$

$$\tilde{H}_0' + \frac{r - 3M}{r(r - r_s)}\tilde{K} - \frac{r - 4M}{r(r - r_s)}\tilde{H}_0 + \left(\frac{i\omega r}{r - r_s} + \frac{\ell(\ell + 1)}{2i\omega r^2}\right)\tilde{H}_1 = 0, \tag{2.122}$$

$$\tilde{H}_1' + \frac{i\omega r}{r - r_s}\tilde{K} + \frac{i\omega r}{r - r_s}\tilde{H}_0 + \frac{r_s}{r(r - r_s)}\tilde{H}_1 = 0, \tag{2.123}$$

$$\left(\frac{6M}{r} + (\ell + 2)(\ell - 1)\right)\tilde{H}_0$$
$$- \left((\ell + 2)(\ell - 1) - \frac{2\omega^2 r^3}{r - r_s} + \frac{r_s(r - 3M)}{r(r - r_s)}\right)\tilde{K}$$
$$- \left(2i\omega r + \frac{\ell(\ell + 1)M}{i\omega r^2}\right)\tilde{H}_1 = 0. \tag{2.124}$$

The rest follows in the same way as described in the EOM approach. However, following [212, 213] we present here an equivalent but different way of obtaining the Zerilli equation from the equations above. Let us apply the following field redefinitions

$$\tilde{K} = g_1(r)Z + \left(1 - \frac{r_s}{r}\right)(\psi_{\text{even}}^{\ell m})', \tag{2.125}$$

$$\tilde{H}_1 = -i\omega\left(g_2(r)\psi_{\text{even}}^{\ell m} + r(\psi_{\text{even}}^{\ell m})'\right), \tag{2.126}$$

$$\tilde{H}_0 = \left(\left(1 - \frac{r_s}{r}\right)\left(g_2(r)\psi_{\text{even}}^{\ell m} + r(\psi_{\text{even}}^{\ell m})'\right)\right)' - \tilde{K}, \tag{2.127}$$

where the $g$ functions are given by[41]

$$g_1(r) = \frac{\ell(\ell + 1)r^2 + 3\ell mr + 6M^2}{r^2(\lambda r + 3M)}, \tag{2.128}$$

$$g_2(r) = \frac{\lambda r^2 - 3\lambda Mr - 3M^2}{(r - r_s)(\lambda r + 3M)}. \tag{2.129}$$

Now writing Eq. (2.124) in terms of the Zerilli function and changing to tortoise coordinates we recover the Zerilli equation

$$\left[\frac{d^2}{dr_*^2} + (\omega_{\ell m}^2 - V_{\text{even}}^{\ell m})\right]\psi_{\text{even}}^{\ell m} = 0, \tag{2.130}$$

with

$$V_{\text{even}}^{\ell m} = f\frac{8\lambda^2(\lambda + 1)r^3 + 12\lambda^2 r_s r^2 + 18\lambda r_s^2 r + 9r_s^3}{r^3(2\lambda r + 3r_s)^2}. \tag{2.131}$$

## 2.4.2 Quasinormal modes

Once the Regge-Wheeler and Zerilli equations are obtained, they can be solved to determine the QNM frequencies. Since the two potentials are isospectral in GR, solving just one of these equations is sufficient to obtain the full spectrum.

Unlike normal modes, QNMs are characterised by complex numbers, $\omega_{\ell m n}$,

usually written as

$$\omega_{\ell m} = 2\pi f_{\ell m} + \frac{i}{\tau_{\ell m}}, \qquad (2.132)$$

where the real part represents the physical oscillation frequency of a mode, and the imaginary part determines the damping time—i.e. how quickly the oscillations decay. Normal modes lack an imaginary part, and therefore do not decay over time, i.e. they are not dissipative.

These frequencies depend on the following numbers:

▶ $\ell$, **multipole number**. This is the angular number associated with spherical harmonic decomposition. For GWs, $\ell = 2$ (i.e the quadrupole) is the lowest mode sourcing the generation of GWs (see Subsection 2.3.2). Lower $\ell$ values are prevented to source GWs due to conservation of momentum ($\ell = 1$) or due to their spherical symmetry ($\ell = 0$). Higher $\ell$ values correspond to modes with more angular nodes.

▶ $m$, **azimuthal number**. This describes how the mode behaves under rotations around the symmetry axis. Perturbations on non-rotating black holes—e.g. S(dS)—are invariant under such rotations and are therefore independent of $m$. However, in a rotating Kerr black hole, frame-dragging breaks this degeneracy, and the frequencies depend on $m$, leading to a Zeeman-like splitting.[42] This can be visualised in Fig. 2.13.

▶ $n$, **overtone number**. This labels QNMs by a monotonically increasing $|\mathrm{Im}\{\omega\}|$. The fundamental mode, $n = 0$, has the longest damping time (smallest imaginary part). Higher overtone modes, $n \geq 1$, decay faster, meaning their imaginary parts are larger in magnitude. For each specific overtone $n$, the Regge-Wheeler and Zerilli equations yield two distinct solutions for the frequency: one, typically referred to as the *regular mode*, which possesses a positive real part, and another, known as the *mirror mode*, which has a negative real part. Crucially, both the regular and mirror modes share the identical imaginary component of the frequency.

QNM frequencies are independent of initial perturbations. They only depend on the characteristics of the remnant black hole (mass, angular momentum, charge), meaning they constitute clean tests of the background geometry and/or the underlying theory. However, the amplitude of individual modes does depend on more contingent characteristics of the system [214]. One then may ask: is there an ($\ell m n$) mode that contributes over the others? What is their hierarchy? It turns out that theory and observation both single out the (220) mode as the observationally most relevant one, i.e., as the *dominant mode* containing the majority of power. Actually, it has been shown that only two of the observed GW events to date show evidence for the presence of higher multipoles [215]. More generally, the modes with the largest amplitudes for astrophysical binary compact object mergers are the $\ell = m$ modes [216–221]. Note that, for a non-rotating black hole solution, the equations of motion are independent of $m$. So while $m = 0$ is typically fixed in such setups for simplicity, the results derived apply for any $m$. The relative amplitude of subdominant modes (in particular $\ell = 3$) grows as the mass ratio $q$ and angular momentum $j$ of the remnant black hole increase [217, 219, 222, 223]—also see those references for discussions related to the detectability of such modes. Nonetheless, the $\ell = 2$ mode still generically dominates in all scenarios and higher $\ell$ modes decay more quickly, see Table 2.4. The damping time $\tau$ goes as the inverse of the imaginary component of $\omega$, which increases for higher $\ell$ modes. So in addition to generically possessing a smaller amplitude, these modes also decay faster. Finally, also notice that, for binary systems that have orbited each other for a sufficiently long time for orbits to have approximately circularised, the $\ell = 2$ mode will be additionally enhanced relative to other modes [224–226].

**Table 2.4:** Numerical values for the real and imaginary parts of the QNMs of a Schwarzschild BH in GR for $\ell = 2, 3, 4, 5$ and $n = 0, 1, 2$.

|  | $n = 0$ | | $n = 1$ | | $n = 2$ | |
|---|---|---|---|---|---|---|
|  | $\mathrm{Re}(r_s\omega)$ | $\mathrm{Im}(r_s\omega)$ | $\mathrm{Re}(r_s\omega)$ | $\mathrm{Im}(r_s\omega)$ | $\mathrm{Re}(r_s\omega)$ | $\mathrm{Im}(r_s\omega)$ |
| $\ell = 2$ | 0.7474 | −0.1779 | 0.6934 | −0.5478 | 0.6021 | −0.9566 |
| $\ell = 3$ | 1.1989 | −0.1854 | 1.1653 | −0.5626 | 1.1034 | −0.9582 |
| $\ell = 4$ | 1.6184 | −0.1883 | 1.5933 | −0.5687 | 1.5454 | −0.9598 |
| $\ell = 5$ | 2.0246 | −0.1897 | 2.0044 | −0.5716 | 1.9654 | −0.9607 |

We list QNMs of Schwarzschild black holes in GR in Table 2.4 and plot them in Figure 2.12. Note that different conventions exist when numerically expressing QNM frequencies depending on which of the two dimensionless quantities is chosen: $r_s\omega$ or $M\omega$. Knowing that $r_s \equiv 2M$, one can easily map from one convention to the other simply with a factor of 2. In this thesis we always show QNM values with the quantity $r_s\omega$. Including rotation in the black hole, we show in Figure 2.13 how the fundamental (i.e. $n = 0$) mode $\ell = 2$ changes and splits for different $m$ as the rotation parameter increases. QNM data used in these plots is provided online [227, 228].

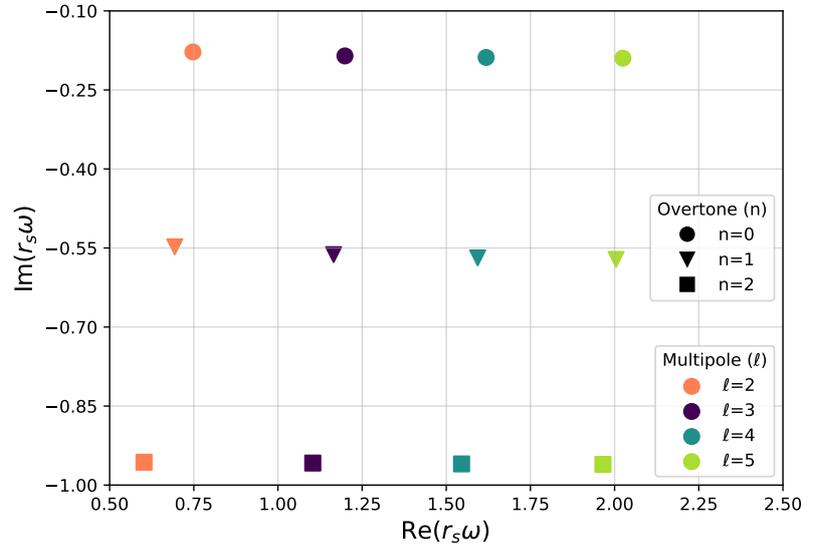

**Figure 2.12:** Quasinormal modes of a Schwarzschild black hole in GR plotted in the complex plane for $\ell = 2, 3, 4, 5$ and overtone numbers $n = 0, 1, 2$.

43: Imposing the (unrealistic) boundary conditions $\psi = 0$ as $r_* \to \pm\infty$ results in solving the Schrödinger equation for bound-state solutions. However, RW and Zerilli potentials cannot support such states [207].

**Definition of QNMs**

Upon imposing the dissipative boundary conditions[43]

$$\psi \sim \begin{cases} e^{-i\omega r_*}, & \text{for } r_* \to -\infty, \\ e^{i\omega r_*}, & \text{for } r_* \to \infty, \end{cases} \qquad (2.133)$$

which correspond to outgoing waves at spatial infinity and ingoing waves at the black hole horizon (corresponding to $r_* \to \pm\infty$) such that on the horizon wavepackets are moving inwards and at infinity wavepackets are moving outwards, discrete values for $\omega$ are selected [229]. To see this, let us first realise that the potential $V(r)$ in the master equation of the form (2.54)

$$\left[\frac{d^2}{dr_*^2} + \left(\omega_{\ell m}^2 - V_{\ell m}\right)\right]\psi_{\ell m} = 0 \tag{2.134}$$

vanishes in the asymptotic limits $r_* \to \pm\infty$, and the equation reduces to one solved by plane waves [207]. Since in this scenario, the black hole itself is the source of the wave emission, we begin by considering an outward-propagating wavepacket, illustrated in red in Fig. 2.14. As it approaches the peak of the potential barrier, a portion of the wave is reflected back toward the black hole (depicted in blue), while the remainder is transmitted toward spatial infinity (shown in green).

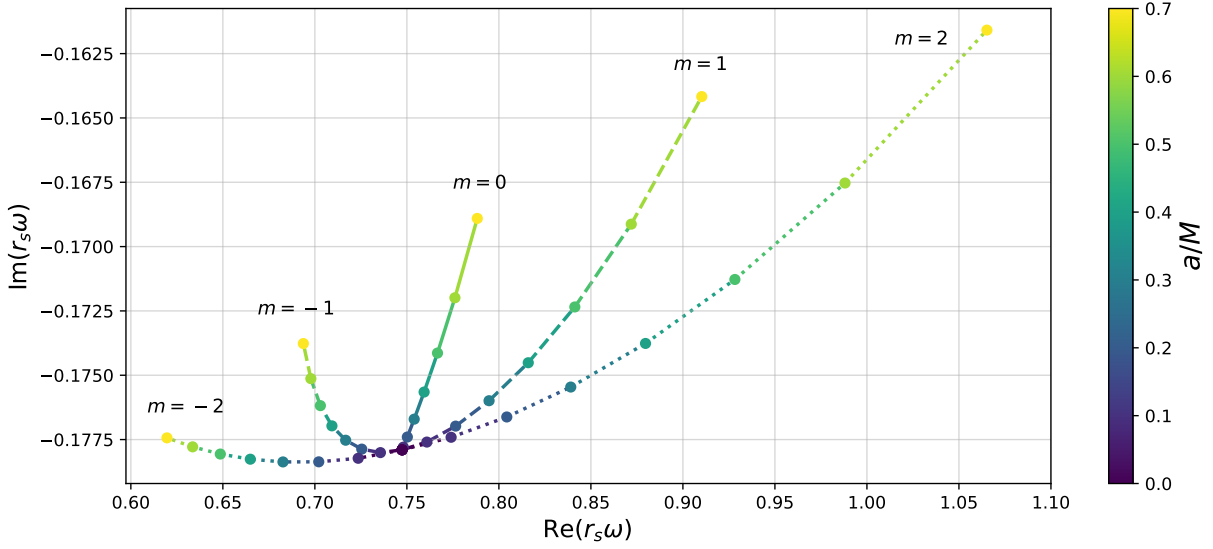

**Figure 2.13:** Quasinormal modes with $\ell = 2$ of a rotating black hole as a function of the dimensionless rotation parameter $a/M$.

Due to conservation of probability, we require that $|A_0(\omega)|^2 = |A_r(\omega)|^2 + |A_t(\omega)|^2$. The imposition of our boundary conditions in (2.133) implies that $A_0(\omega) = 0$, meaning that the scattering amplitude defined as

$$S(\omega) \equiv \frac{A_r(\omega)}{A_0(\omega)} \tag{2.135}$$

picks up some poles, therefore selecting some discrete values for $\omega$ as the resonances of this system. For a more rigorous demonstration of how QNMs are defined we refer to [207], which uses the Laplace transform with respect to time in the Regge-Wheeler or Zerilli equation. This introduces a source term which depends on the initial conditions and Green's functions can be used to solve such sourced equation. The Green's function has an associated Wronskian $W(\omega)$ which vanishes for the values of $\omega$, corresponding to the QNM frequencies. In other words, the 'quantised' values of $\omega$ selected through the imposition of boundary conditions correspond to poles in the Green's function, see Fig. 2.15.

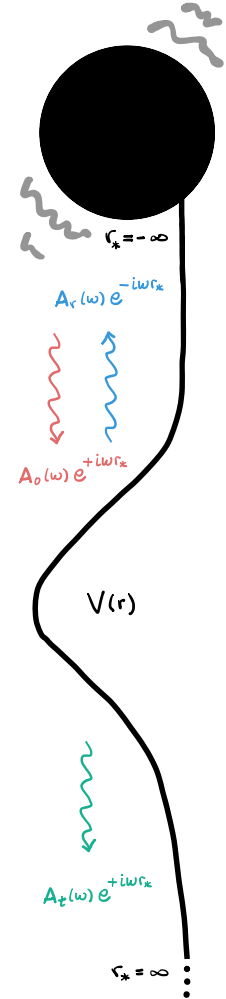

**Figure 2.14:** Inwards and outwards-travelling wavepackets in a black hole spacetime with a Regge-Wheeler or Zerilli-type potential.

### Methods to calculate QNMs

There exist various techniques for computing the numerical values of the QNM frequencies. We refer to [194, 210] for an extensive review of those but here present some key aspects of the most commonly used methods. Importantly, no exact solutions can typically be found, except for some special cases where e.g. the potential is shown to reduce to the Pöschl-Teller potential which offers exact solutions [230, 231].

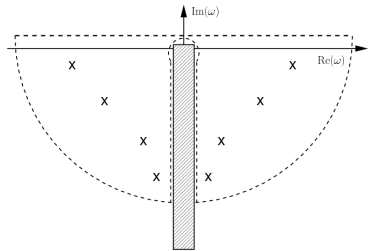

**Figure 2.15:** Integration contour in the complex frequency plane for the Regge-Wheeler or Zerilli function. The shaded region corresponds to a branch cut and crosses correspond to QNM frequencies, i.e. zeros of the Wronskian. Figure taken from [194].

▶ **Leaver's continued fraction method.** Since Leaver's seminal paper in 1985 [232] on the application of the continued fraction method to the computation of black holes QNMs, this method has been further developed [233, 234] and extensively used in the gravitational context for a wide range of black hole solutions (see e.g [235–242]). The method is based on expressing the solutions $\psi$ to the master equation (2.54) as a power series and obtaining, upon the imposition of boundary conditions, a recurrence relation between the coefficients of different powers. The convergence of such power series entails that the coefficients will satisfy an infinite continued fraction equation which can be numerically solved for $\omega$.

▶ **WKB approximation.** The WKB (Wentzel-Kramers-Brillouin) approximation is a semi-analytic method used to compute QNMs which was first applied to black holes in [243] and subsequently extended to higher orders in [244–247].[44] A notable generalisation is the 'uniform approximation method', which extends the WKB approach to provide significantly enhanced accuracy, particularly for large overtone numbers where standard WKB methods can lose precision [251]. The rationale is to match the asymptotic solutions given by (2.133) respectively with a Taylor expansion around the maximum of the potential. We will use this method to compute QNMs for the set up studied in [2] in Section 5.4.2, where we will expand on more technical details. In fact, in such computations we significantly extend an existing `Mathematica` package [252] to allow it to obtain explicit semi-analytic expressions for the QNM frequencies in terms of standard GR or beyond-GR parameters. This is covered in detail in Appendix A, where we exemplify how the extension can be used to quantify the effect of a cosmological constant $\Lambda$ in the QNMs [2].

▶ **Pöschl-Teller potential matching.** In some special cases, the effective potential can be reduced or approximated to the Pöschl-Teller potential, which is analytically solvable. This includes, for instance near extreme SdS, Nariai or asymptotic anti-de Sitter black hole spacetimes [194]. However, this method is usually not available for more realistic black hole backgrounds and so we will not go into further detail here.

▶ **Parametrised ringdown formalism.** More recently, a method has been proposed to compute QNMs which offers a straightforward and easily-applicable procedure to convert terms in the effective potential $V(r)$ to concrete contributions to the QNMs [253]. In practice, one identifies contributions to the potential appearing as different powers of $(r_s/r)^n$, and a numerical basis is provided in [253] which allows one to convert such coefficients to contributions to $\omega$. For this to be applicable, the corresponding master equation must be written in a specific form, specifically with extra terms in the potential appearing as separate inverse powers of the radial coordinate, which is not necessarily true in general. This method, which has been subsequently developed in [254–256], will be used in Section 4.3 when computing QNM corrections corresponding to non-luminal GWs [1].

▶ **Direct integration.** An alternative way to compute QNMs is to employ a direct integration shooting method [210, 257, 258]. This method uses two different ansätze for the two asymptotic regions (horizon and infinity) motivated by the boundary conditions. Then, one integrates the horizon ansatz equation from the horizon to a matching point $r = r_m$ imposing only ingoing wavepackets. Similarly, the infinity ansatz is integrated from infinity to the same matching point imposing only outgoing waves. One can then construct a linear combination of the two solutions which is $C^0$ at the matching point. This is ensured by a specific choice of the coefficients. The QNM frequencies are then found by imposing that the linear combination is $C^1$ so that the derivatives are continuous.

### 2.4.3 Testing gravity with black hole ringdown

Black hole ringdown provides a clean and powerful way to test GR in the strong field regime. We have seen how, as a consequence of no-hair theorems in GR (2.2.2), QNMs are fully specified by the asymptotic parameters of black hole solutions. Since QNMs are complex numbers (therefore with 2 components), assuming the Kerr hypothesis, the detection of 1 QNM (typically the dominant mode) can be used as a measurement for the 2 only intrinsic black hole parameters $(M, a)$. Then, the measurement of an additional mode, whose value will be predicted for a Kerr black hole in GR with specific $(M, a)$ values, can be used as a consistency test of both the Kerr geometry and the underlying theory governing the dynamics of perturbations. This program is often referred to as *black hole spectroscopy*.

### 2.4.4 Ringdown systematics

The use of QNMs in black hole spectroscopy constitutes a powerful way to devise tests of GR. However, in order to fully exploit this power in a consistent and robust way, there are a few questions to be addressed which we discuss here.

**When does the ringdown start?**

GWs are schematically divided into three stages (inspiral-merger-ringdown) as shown in Fig. 2.5. However, this separation is merely artificial, and nature itself does not adhere to such distinctions. In other words, there is no clearly defined boundary separating these stages. In fact, the ambiguity in the definition of ringdown start-time is a known issue in the application and interpretation of ringdown analyses.[45] Ringdown waveforms are modelled as a superposition of exponentially damped sinusoids. Conversely, the merger is a highly non-linear process where the hierarchy of modes present in the ringdown no longer applies. Therefore, starting the ringdown too early, i.e. too close to the merger, can introduce systematic errors in the extraction of QNM frequencies via overfitting the signal [221, 271]. This can be seen in Figure 2.16, which shows estimates for the QNM frequency and damping time as a function of ringdown start-time $t_0$ for the GW150914 event [272]. There,

| Detector(s) | Ringdown SNR |
|---|---|
| LVK | $10$ [183, 259, 260] |
| ET / CE | $10^2$ [182–185] |
| pre-DECIGO | $10^2$ [261] |
| DECIGO / AEDGE | $10^3$ [186, 262][*] |
| LISA | $10^3$ [263–266] |
| TianQin | $10^3$ [265, 266] |
| AMIGO | $10^5$ [267] |

**Table 2.5:** Achievable order-of-magnitude ringdown SNRs for a single observed event for different GW detectors. A star[*] denotes that the quoted forecasted SNR is not ringdown-specific. For ET/CE we have quoted the ringdown-specific ET forecast [183], in the current absence (to our knowledge) of an analogous forecast for CE. For LISA, we quote the maximum SNR estimates from [263–266] going up to ~ $\mathcal{O}(10^3)$, which coincide with the SNRs in the LISA Mock Data Challenge [268, 269], but also note that some more optimistic SNR forecasts (up to ~ $\mathcal{O}(10^3)$) exist for sufficiently nearby and massive events [260, 270]. The same applies for TianQin. This also illustrates that there is still significant variance in the forecasted SNRs relevant for the missions considered here.

45: This ambiguity is due in part to the incomplete and non-orthogonal nature of the QNM 'basis'.

one can see how starting the ringdown too soon after the merger biases the inferred QNM parameters. However, starting the ringdown too late, despite ensuring the applicability of the perturbative prescription, also entails that all testing-GR power (i.e. the SNR) is lost due to the exponentially decaying amplitude. In this case, regions overlap for a starting time of $t_0 = t_M + 3\text{ms}$, with $t_M$ being the time of merger. Expressed in units of the remnant's mass this corresponds to $\sim 10M$ after the merger.[46] A similar choice has been employed in [217, 273], albeit for different interpretations of what $t_M$ constitutes. Other explored alternatives include e.g. selecting the starting time as the one which maximises a measure of the energy contained in the QNMs [218, 274]. A method for assessing different choices of start-time was proposed in [275] which uses a measure of the Kerr nature of the signal to validate the application of a perturbative analysis, finding that a starting time of $t_0 \sim 16.4M$ after merger is preferred and, in particular, that future events with higher expected SNR will require later that $10M$ start times. In [276], the start time for GW150914 was directly inferred to be $\sim 14^{+2}_{-2}M$. In order to perform multiple mode analyses in the future, and hence fully exploit black hole spectroscopy, robust implementations of ringdown starting time will be required.

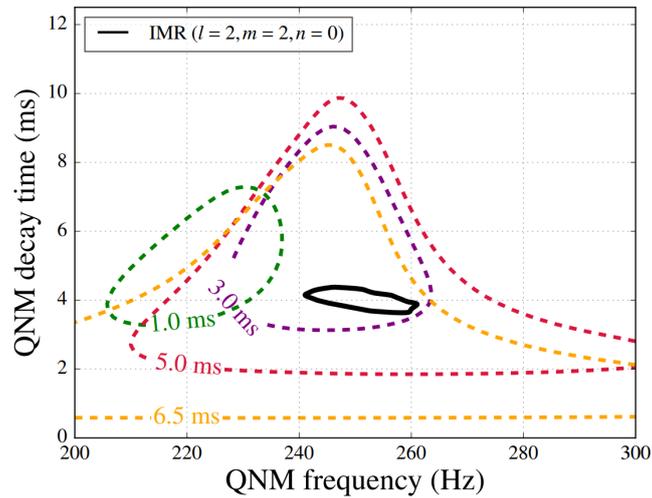

**Figure 2.16:** 90% credible intervals of the posterior probability distributions for the dominant QNM parameters for different ringdown start-times. In solid black is represented the 90% credible interval from the posterior distributions of the remnant mass and spin. Figure taken from [272].

### Non-linear quasinormal modes

Having the longest damping time, the (220) mode dominates the ringdown signal in physically relevant scenarios for binary mergers. As detector sensitivities improve, and in particular in light of next-generation detectors with higher ringdown SNRs (see Figure 2.5) a clear understanding of the hierarchy of subdominant modes will be essential for multimodal analyses. Note that a ringdown SNR of 30 is sufficient to resolve overtone frequencies [221]. In addition to overtones, whose inclusion is required for 1% measurements of the remnant's mass and spin [274], one needs to clearly understand when non-linearities in the ringdown may become relevant.

The impact of non-linearities in black hole ringdown has been recently studied in several works [217, 277–283]. Through the use of numerical simulations, evidence for the presence of modes appearing at second-order perturbative level (in the form of self-interacting first-order modes) has been reported in [279, 280]. For quasi-circular mergers, where (220) dominates at first order, the first second-order contribution comes from the (220)×(220) interaction. Figure 2.17 shows how the amplitude of this mode is comparable, and in fact higher for low mass ratio, to its corresponding (440) linear mode.[47] In [283], a Fisher forecast analysis indeed shows that for some configurations the quadratic mode can be measured with more precision than the linear one. Similarly to (440) (c.f. Table 2.4), the mode (220) × (220) has a frequency twice the one of (220). As shown in Figure 2.17, fits for the amplitude of the second-order mode are consistent with the expected quadratic relationship with the one of (220). These findings suggest that once ringdown SNRs reach sensitivities where $\ell = m = 4$ harmonics are detectable, non-linearities will necessarily need to be included in order to perform precise black hole spectroscopy. In fact, the inclusion of the first quadratic mode will enhance the power of consistency tests of GR for upcoming GW detectors [283].

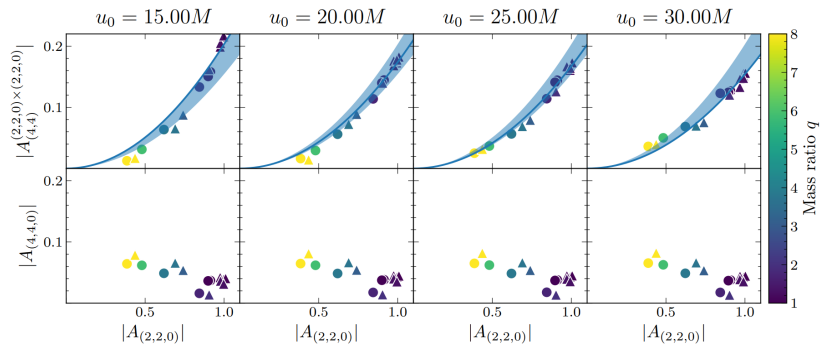

**Figure 2.17:** Amplitudes for the first second-order mode (220) × (220) and the (same harmonic) linear mode (440) as a function of the dominant mode amplitude for different mass ratios and starting times. Figure taken from [280].

### Degeneracies with GR deviations

Incomplete ringdown models can introduce biases in the inferral of black hole parameters, potentially mimicking deviations from GR. As we have seen, an example of this could be the omission of, or overfitting with, non-linear modes, but in reality there is a much larger plethora of potential effects whose impact in the QNMs could be degenerate with standard deviations from GR, see e.g [284].

In [285], a thorough compilation of potential sources of false signals indicating deviations from GR is presented, encompassing experimental systematics, data analysis and processing techniques, as well as missing physics in the theoretical models. Some examples of such effects that might contaminate ringdown tests of GR include:

- Kick-induced effects from the anisotropic emission of GWs, see e.g. [286].
- Large unmodelled eccentricity in the premerger [287].
- Environmental effects from matter or other fields around black holes, see e.g. [288–293].

Before robustly claiming a violation from GR, one should then rule out each of the potential degeneracies discussed in [285].

### Stability under perturbations in the potential

It has been observed that the QNM spectrum of black holes is notably sensitive to small modifications in the effective potential [288]. Such spectral instability threatens the practical realisation of black hole spectroscopy and has therefore motivated further investigations [292–297]. Figure 2.18 displays how such instability takes place for Schwarzschild black holes [296], as the QNM spectrum branches out from the unperturbed values (red circles) once the potential is modified. As can be appreciated, overtones are generally more prone to become unstable.

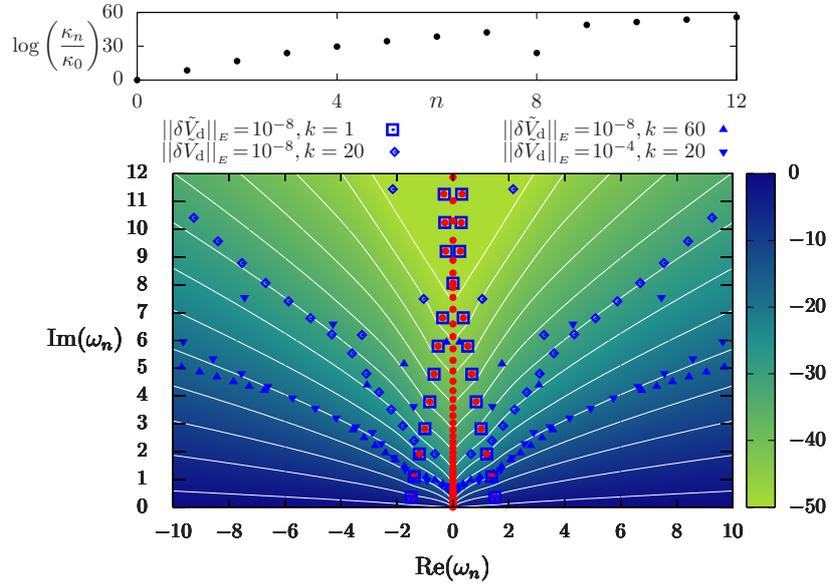

**Figure 2.18:** Spectral (in)stability of a Schwrazschild black hole under small perturbations in the potential. The unperturbed $\ell = 2$ QNM spectrum is shown in red dots. With blue symbols, the spectrum is shown to branch out for different $\delta\tilde{V}_d$, corresponding to small modifications to the effective potential in the form of cos functions with wavenumber $k$. Figure taken from [296].

The stability of the fundamental dominant mode under 'tiny bumps' in the potential was further investigated in [293], where it was shown how such instability depends on the position of the bump. However, the physical origin and significance of such ad hoc potential modifications is in some cases unwarranted and requires exotic types of matter for their realisation [298]. Instabilities potentially affecting the fundamental dominant mode, which, as mentioned in Subsection 2.4.3, has already been measured, can arise from physically reasonable physics but generally appears in the low-amplitude late-time part of the signal [298].

# Scalar tensor theories and hairy black holes | 3

Todas las teorías son legítimas y ninguna tiene importancia. Lo que importa es lo que se hace con ellas.

Jorge L. Borges

**Chapter summary**

After a review of different alternatives for extending GR (§3.1), this chapter will focus on scalar-tensor theories of gravity. We will introduce specific theories in order of complexity building up to Horndeski gravity and its higher-order generalisations (dubbed HOST) (§3.2), therefore comprehensively mapping the theory space of GR plus a scalar field. We will discuss gravitational wave effects (§3.3) as well as hairy black hole solutions within such theories (§3.4) and close this chapter by reviewing basic applications of black hole perturbation theory in the scalar-tensor framework (§3.5).

## 3.1 Extensions to General Relativity

Not only is GR strongly supported by observational evidence but also distinguished by its theoretical elegance and uniqueness. This is encapsulated in Lovelock's theorem [33, 34], which states that GR is the only possible theory of gravity that satisfies a set of fundamental principles. In a schematic form, the theorem states that:

**Theorem 3.1.1** (**Lovelock's Theorem**) *General Relativity (i.e. EFEs) is the only theory satisfying all the following assumptions:*

1. *The theory is local;*
2. *it contains one metric field $g_{\mu\nu}$;*
3. *it lives in 4 dimensions;*
4. *it contains only second-order derivatives of the metric.*

Consequently, any alternative theory of gravity aiming to extend GR must violate at least one of the assumptions outlined above. Below, we briefly examine the implications of breaking each assumption and provide key references in each case. Interestingly, violations of different assumptions are often interconnected and can frequently be recast as the introduction of additional fields—effectively constituting a direct violation of the second assumption. In this thesis, we explore extensions of GR that incorporate an additional scalar degree of freedom and, in some cases, also include higher-derivative terms.[1]

1: It is worth noting that the criterions in Lovelock's theorem are 'classical'. Quantum corrections to gravity, such as those arising in effective field theory approaches or from string theory, generally introduce higher-derivative terms, thereby violating assumption 4.

**Non-locality**

GR is fundamentally a local theory, meaning that the interactions it describes, such as the curvature of spacetime in response to matter and energy, occur at specific points in spacetime and propagate according to local differential equations. This ensures that information cannot travel faster than the speed of light, in accordance with relativity, and that gravitational effects are determined solely by the immediate surroundings rather than by distant influences. Nonetheless, despite seeming a bizarre concept, non-locality is naturally built in our reality, as described by quantum mechanics.[2] One can then attempt to build a non-local theory of gravity by incorporating non-local operators in the action. Examples of such operators include inverse powers of the d'Alembert operator, as in $f(R, \Box^{-1}R)$, see [300–302] for more details.

**Additional fields**

In GR, the mediator of gravity is the metric, a rank-2 symmetric tensor (or spin-2 field). This is not by chance, as it needs to couple to the stress-energy tensor, which is also a rank-2 symmetric tensor. However, this does not entail the absence of additional fields playing a gravitational role. Among the potential list of additional fields, scalar fields represent the simplest and most natural extension. A minimal example is quintessence, where the scalar field is minimally coupled to gravity (as the rest of standard matter components in the universe).[3] Extensions to this model including non-minimal couplings exist, such as in Horndeski, Beyond Horndeski or DHOST, and include some of the best-studied alternatives to GR aimed at explaining late-time cosmic acceleration. These will constitute the main focus of this thesis and will be reviewed in more detail later. Note that one can in principle also add multiple scalar fields as in e.g. multi-galileon theories [303, 304].

Alternatively, one can consider vector-tensor theories, whose generalisation (restricting to second-order equations of motion) is the Generalised Proca theory [305]. Within this family we have for instance the Einstein-Maxwell theory (2.13), Einstein-Aether [306, 307] and others [308, 309]. Such theories often result in a breaking of Lorentz invariance, as they introduce a preferred direction in spacetime. Beyond single-vector extensions, one can also consider frameworks involving multiple interacting vector fields such as multi-Proca theories [310–314], Yang-Mills theories [315–317], and gaugid inflation [318].

Following a logical sequence, one can then consider tensor-tensor theories of gravity, where an additional rank-2 tensor field is introduced alongside the metric. For instance, the massive gravity theory described by the de Rham-Gabadadze-Tolley model [319] introduces a non-dynamical fixed fiducial additional tensor which renders self-interactions of the standard dynamical metric stable, giving rise to a non-zero graviton mass. A natural extension is to then promote the fiducial metric to be dynamical, as is done in bigravity theories [320]. These theories generically contain ghostly degrees of freedom [321], but can also be made stable, see also [322, 323]. Generalisations including multiple interacting spin-2 fields have also been proposed in [324].

There also exist theories including fields of spin higher than two. Those with at least one additional massless field of spin higher than 2 are referred to as Higher-Spin gravity (see e.g. [325–331] for different models and [332] for a more recent review). Such extensions are mostly motivated by the goal of constructing a consistent quantum theory of gravity.

Finally, one can in principle also combine together multiple additional fields of different spin at the same time as is done for instance in scalar-vector-tensor theories [333].

**Extra dimensions**

While physicists generally agree that theories existing in 4 dimensions (3 space + 1 time) are the most sensible ones to study, and that extra dimensions are somewhat exotic, there is in fact no fundamental reason why spacetime should be 4-dimensional. Ideally, the number of dimensions would instead emerge as a theoretical prediction. This is the case, for instance, in string theory, which necessitates a specific minimum number of dimensions to be consistent.[4] Importantly, however, when the number of dimensions is higher than 4, there need to be mechanisms in place to explain why we only experience 4 of them. This is typically done by making extra dimensions either very 'small' (e.g. through compactification) or very 'large'[5] (e.g. branes). Interestingly, it has been shown that extra dimensions can be repackaged as additional fields in an effective 4-dimensional theory. One famous example is that of the Kaluza-Klein theory, which describes gravity in 5 dimensions, but can be reduced to describe Einstein's gravity and Maxwell's electromagnetism in 4 dimensions [336]. More generally, it has been shown that theories of gravity in extra dimensions can often be compactified in such a way that they manifest as an additional scalar degree of freedom in an effective four-dimensional theory, see [337–340] for general overviews. Other interesting models that modify gravity by adding extra dimensions include for instance the Randall-Sundrum models [341, 342], Dvali-Gabadadze-Porrati (DGP) gravity [343] or Einstein dilaton Gauss-Bonnett (EdGB) gravity [344, 345].

**Higher-order derivatives**

GR, like Newtonian mechanics, is governed by second-order equations of motion. This guarantees that the evolution of a system is fully determined by specifying the initial values of a function and its first derivative (e.g. the position and velocity of a particle). As a result, the *initial value problem* is well-posed, meaning that a unique solution exists and depends continuously on the initial conditions, ensuring stable and predictable dynamics. Allowing the field equations to involve higher-order derivatives of the metric, rather than being strictly second order, offers potential extensions to GR, but these often introduce severe instabilities. When these higher derivatives are non-degenerate (i.e. they cannot be eliminated through field redefinitions or constraints), the system introduces extra ghostly degrees of freedom. While these lead to a Hamiltonian that is not bounded from below and thus to negative-energy states that can, in principle, cause runaway behaviour, it is important to stress that such instabilities are only problematic if the mass of the ghost lies below the ultraviolet (UV) cutoff of the theory. If the ghost's mass is above this cutoff, non-degenerate higher-derivative

4: However, note that this does not preclude the possibility that string theory could be embedded in a more fundamental theory with a different or larger number of dimensions, as is the case with M-theory, see [334, 335].

5: And unaccessible by standard model interactions.

interactions do not necessarily render the theory inconsistent, as is often the case in well-defined EFTs. Conversely, if the ghost's mass is below the cutoff, this means the system permits states with arbitrarily negative energy, allowing for uncontrolled energy extraction and rendering the theory physically inconsistent. Nonetheless, not all higher-order modifications are pathological. One example is $f(R)$ gravity, where the Einstein-Hilbert action is generalised to include arbitrary functions of the Ricci scalar, introducing higher-order derivatives in the field equations (see [346] for a review). In this case, instabilities can be avoided as higher order derivatives act on non-propagating modes. More generally, Lovelock gravity in higher dimensions and Horndeski theories in four dimensions introduce higher derivatives at the level of the action while carefully avoiding Ostrogradsky ghosts by ensuring that the equations of motion remain second-order [33, 347]. Even more generally, Degenerate Higher-Order Scalar-Tensor (DHOST) theories introduce higher derivatives but impose degeneracy conditions that prevent the presence of unstable ghost degrees of freedom, maintaining a well-posed initial value problem [348–352]. Another example of a theory which introduces higher-order spatial derivatives is given by Hořava-Lifshitz gravity [353], a theory which abandons Lorentz invariance at high energies by introducing an anisotropy between space and time, allowing it to avoid Ostrogradsky ghosts by keeping time derivatives to second order.

## 3.2  Scalar-tensor gravity: from Branks-Dicke to HOST

Scalar-tensor (ST) theories of gravity originated in the 1960s with the work of Brans and Dicke [354], but did not capture much interest until the beggining of the new century, as potential alternative explanations to the observed late-time accelerated expansion [62, 63]. Since then, the scalar-tensor framework has rapidly evolved, growing in their theoretical construction (therefore providing richer phenomenology) as well as shrinking in viability as a result of their confrontation with observations. Most significantly, Horndeski gravity embeds the most general scalar-tensor theories resulting in second-order equations of motion [355, 356], and the Higher-Order-Scalar-Tensor (HOST) family extends this to include higher-order equations of motion [348–350]. In this overview, we will start by looking at the simplest scalar-tensor extensions to GR and build up to Horndeski and HOST theories.

### 3.2.1  Quintessence

6: This behaviour is analogous to models of inflation.

The simplest scalar-tensor model is quintessence, whose action in vacuum is (see [357–361] for the original studies and [362] for a review)

$$S = \int d^4x \sqrt{-g} \left[ \frac{1}{2\kappa} R + X - V(\phi) \right], \tag{3.1}$$

where recall that $\kappa = 8\pi G/c^4 = 8\pi$ in geometric units (used throughout this Section). Here we have introduced the shorthand $X \equiv -\frac{1}{2}\phi_\mu \phi^\mu$ for the kinetic term of the scalar, where $\phi_\mu \equiv \nabla_\mu \phi$ (and $\phi_{\mu\nu} \equiv \nabla_\nu \nabla_\mu \phi$). In quintessence, the scalar field is minimally coupled to gravity like any other matter field, and dynamically evolves along the potential $V(\phi)$.[6] Such evolution can pro-vide an explanation for the observed late-time acceleration in the universe as an alternative to the cosmological constant. In this case, dark energy is modelled with an evolving equation of state $\omega$ that depends on the shape of the potential, see e.g. [363−367]. Cosmological observations such as the growth rate of matter perturbations are able to constrain quintessence models. It is generally found that cosmic acceleration in accordance with observations requires carefully selected initial conditions and/or fine-tuned parameters. As a result, despite assuming a zero $\Lambda$,[7] quintessence models are not completely free of coincidence and/or fine-tuning problems. In addition, recent DESI DR2 observations suggest that dark energy may evolve and weaken over time [368−372], displaying a specific behaviour referred to as *phantom crossing*, i.e. an equation of state parameter which crosses the $\omega = −1$ divide. In many models, such as quintessence, this behaviour leads to ghost and gradient instabilities, rendering such theories physically inconsistent.

### 3.2.2 $k$-essence

A natural extension to quintessence is to allow the kinetic term to have a non-canonical form. Doing so, one can construct ST theories which offer alternative dark energy explanations without necessarily requiring ad hoc potentials. These models are known as $k$-essence and their action can be written as [373, 374]

$$S = \int d^4x \sqrt{-g} \left[ \frac{1}{2\kappa} R + p(\phi, X) \right], \tag{3.2}$$

with $p$ being a generic function of the scalar field and its kinetic term. In models where $p$ contains non-canonical kinetic terms, the dynamics of the scalar field can be driven by this non-canonical structure rather than a potential and self-tune to the observed accelerated expansion, therefore not relying on fine-tuned initial conditions. Such behaviour is often referred to as a dynamical attractor solution.

However, modifying kinetic terms in the action can also give rise to instabilities within the theory. In particular, the propagation speed of perturbations depends on the functional form of $p$ and can, in some cases, lead to unphysical exponentially growing perturbations. In order to see how instabilities can arise in practice for fields with non-trivial kinetic interactions we will consider closely a toy-example.

**Toy-model: stability of a generic scalar**

Mathematical models are used to describe observed phenomena. In order to provide analytically tractable descriptions, such models often contain idealisations. Nature, on the other hand, is usually far from ideal, making such models good approximations at best rather than exact descriptions. These approximations can only be considered valid if they are unaffected by small changes or, in other words, if they are stable under perturbations. Consider a generic action for a scalar in one spatial dimension with quadratic interactions. Linear perturbations of the scalar

7: In other words, this assumes the cosmolgcial constant problem is solved by another mechanism which effectively sets $\Lambda$ to zero. However, in this case, one still might wonder what protects the form of this potential.

can then be written as [375]

$$S = \int d^4x \frac{1}{2}\Big[a\dot{\varphi}^2 - b\varphi'^2 - c\varphi^2\Big], \qquad (3.3)$$

where here overdots and primes denote time and space derivatives respectively. Assuming that $a$, $b$ and $c$ are constants,[a] the equations of motion are

$$\ddot{\varphi} - \frac{b}{a}\varphi'' + \frac{c}{a}\varphi = 0, \qquad (3.4)$$

which are solved by the ansatz

$$\varphi = Ae^{-i\left(\sqrt{\frac{b}{a}}k\cdot x - \omega t\right)} + A^* e^{i\left(\sqrt{\frac{b}{a}}k\cdot x - \omega t\right)}, \qquad \omega = +\sqrt{\frac{b}{a}|k|^2 + \frac{c}{a}}. \qquad (3.5)$$

Whether this solution is stable or unstable, and in fact which type of instability it displays, depends on the relative signs of the coefficients. To guarante stability, we require $a > 0$, $b > 0$, $c \geq 0$. Otherwise, we encounter the following types of instabilities:

| Tachyonic | Gradient | Ghost |
|---|---|---|
| $a > 0, \ b > 0, \ c < 0$ | $a > 0, \ b < 0,$ or $a < 0, \ b > 0$ | $a < 0, \ b < 0$ |
| $\omega$ purely imaginary, exponentially growing modes | $\omega$ purely imaginary, exponentially growing modes | Unbounded Hamiltonian, negative energies |

Stability issues are more nuanced than outlined here and can depend critically on the characteristic scales of a given theory. In particular, tachyonic instabilities, which arise from negative mass-squared terms, can often be mitigated by making the growth rate very long, which cannot be done for gradient instabilities.

[a] The coefficients $a$, $b$ and $c$ encode information of the background solution and hence can depend on spacetime coordinates. However, their background evolution will typically occur over much longer timescales.

In addition to potential stability issues, which can nonetheless be dealt with, $k$-essence models with late-time attractor solutions do not completely get rid of fine-tuning problems [373].

### 3.2.3 Brans-Dicke

Brans-Dicke theories[8] introduce a non-minimal coupling between the scalar and curvature. Implementing ideas first proposed by Dirac and Jordan questioning the constancy of the gravitational constant [376, 377],[9] Brans and Dicke proposed the action given by [354]

$$S = \int d^4x \sqrt{-g} \frac{1}{16\pi}\Big[\phi R + \frac{2\omega_{\mathrm{BD}}}{\phi}X - V(\phi)\Big], \qquad (3.6)$$

where $\omega_{BD}$ is a free constant. The presence of the non-minimal coupling term $\phi R$ implies that the effective gravitational strength, and thus the effective Planck mass ($M_P \propto \sqrt{\phi}$), becomes spacetime-dependent. Conventionally, the factor $16\pi$ is introduced as an overall multiplicative factor to this scalar-tensor action instead of using $\kappa$ (which is defined to include the gravitational constant $G$). This highlights that the scalar field $\phi$ should be understood as a spacetime-varying substitute for (the inverse of) Newton's gravitational constant $G$. Consequently, matter fields, which are minimally coupled to the Jordan frame metric, effectively experience a gravitational force mediated by a varying 'strength'. One of the key motivations behind Brans-Dicke gravity was to incorporate Mach's principle, which is not fully realised in GR, and states that local inertial properties of bodies are determined by its interactions with all the matter distributed throughout the universe.[10] The principle derived from Mach's attempt, inspired by previous ideas from Berkeley and Leibniz, to question Newtonian substantivalism, i.e. the postulation of space and time as absolute entities whose existence is independent of the matter that exists in them. Instead, Mach suggested that theories of gravity should be relational, where space and time exist only as a system of relations between objects. His perspective strongly influenced Einstein in the development of GR,[11] but GR ultimately failed to fully implement it, as is demonstrated by the existence of inertial frames in vacuum spacetimes such as Schwarzschild, meaning that inertial properties do not arise from their relation with the rest of matter in the universe. Brans-Dicke gravity, by introducing a dynamical scalar field coupled to curvature, takes a step towards a more relational view of gravity. The presence of this additional field means that the gravitational interaction—and consequently the inertial properties of objects—becomes more dependent on the distribution of matter. However, Brans-Dicke theory still retains some non-Machian aspects of GR, such as the existence of vacuum solutions beyond Minkowski spacetime. Thus, while it moves closer to fulfilling Mach's vision, it does not completely eliminate the persistence of absolute structures.

The action in (3.6) is said to be in the Jordan frame, meaning that the scalar is non-minimally coupled to curvature, therefore modifying Einstein field equations. Under a conformal transformation[12] of the metric given as

$$g_{\mu\nu} \to \tilde{g}_{\mu\nu} = \Omega^2(x)g_{\mu\nu}, \qquad \sqrt{-g} \to \sqrt{-\tilde{g}} = \Omega^4(x)\sqrt{-g}, \qquad (3.7)$$

and further specifying $\Omega^2 = \phi = e^{\varphi}$, the Brans-Dicke action can be written as[13]

$$S = \int d^4x \sqrt{-g}\, \frac{1}{16\pi} \left[ \tilde{R} - \left( \omega_{\rm BD} + \frac{3}{2} \right)(\partial\varphi)^2 - 2V(\varphi) \right], \qquad (3.9)$$

where now the scalar is only-minimally coupled to curvature. This is referred to as the Einstein frame, as the field equations take the form of an uncoupled Einstein tensor with the scalar field appearing as a source of the stress-energy tensor. Here, $\tilde{g}_{\mu\nu}$ is the Einstein-frame metric. It is important to note that in writing the actions above we have left out the Lagrangian densities for matter fields $\mathscr{L}_m(\psi)$. These fields minimally couple to the metric $g_{\mu\nu}$, i.e. the metric in the Jordan frame. Performing a conformal transformation introduces non-minimal couplings between the matter fields and the scalar $\varphi$. As a consequence, for instance, the mass of the scalar field can depend on the density of matter fields, which provides a screening mechanism (dubbed chameleon mechanism) to hide fifth forces in solar system scales [380].

Particularly relevant to this thesis is the importance of distinguishing between Jordan and Einstein frames in regards to the interpretation of GWs. Matter, and consequently GW detectors, minimally couple to the Jordan-

14: The Vainshtein radius, i.e. the screening radius, characterises the volume within which the non-linearities become strong enough to suppress deviations from GR. This is given by

$$r_V = \left( \frac{r_S}{\Lambda^3} \right)^{1/3}. \qquad (3.11)$$

For a cosmologically relevant cubic Galileon, i.e. with $\Lambda^3 \sim M_{Pl} H_0^2$, the Sun's Vainshtein radius is $r_V \sim 10^3$ light-years [387], while the Oort cloud, i.e. the edge of the Solar system, is roughly 200 times closer, merely a couple light-years away, meaning the Galileon is well-screened within the Solar system.

[388]: Vainshtein (1972), "To the problem of nonvanishing gravitation mass"

15: Recall that we use the definition $\phi_\mu \equiv \nabla_\mu \phi$.

frame metric. Hence, data obtained from such detectors should be understood in terms of the Jordan-frame metric. However, theoretical calculations are often simplified in the Einstein frame, so it is common to transform back and forth between the two frames, see e.g. [381].

One can further generalise the Brans-Dicke action by allowing the BD coupling to be a function of the scalar field $\omega_{BD}(\phi)$, see e.g. [382–385].

### 3.2.4 Galileon

Another approach to modifying GR while ensuring that deviations remain screened at solar system scales is to introduce non-linear derivative self-couplings of the scalar field. This can be achieved through the following action [355, 386]

$$S = \int d^4x \sqrt{-g} \left[ \frac{1}{2\kappa} R + X \left( 1 - \frac{\alpha}{\Lambda^3} \Box \phi \right) \right], \qquad (3.10)$$

commonly referred to as cubic Galileon. Here, $\alpha$ is a dimensionless constant and $\Lambda$ is a constant with dimensions of mass, determining the scale at which the cubic Galileon interaction enters. The effect of the Galileon is suppressed near massive sources via the Vainshtein mechanism[14] [388], but still provides self-accelerated behaviour at cosmological scales, maintains second-order equations of motion, and avoids ghost instabilities, making it an attractive model to study.

Interestingly, the Galileon emerges as a low-energy effective description of several models, including the 5-dimensional DGP model [343], massive gravity [322], and certain vector theories [305, 309, 389].

The action (3.10) is invariant under the transformation[15]

$$\phi_\mu \to \phi_\mu + b_\mu, \qquad (3.12)$$

called a Galilean transformation. In total, there exist five different Galilean invariant Lagrangians in Minkowski spacetime giving rise to second-order equations of motion [386]

$$\mathcal{L}_1 = \phi, \qquad (3.13)$$

$$\mathcal{L}_2 = \phi_\mu \phi^\mu, \qquad (3.14)$$

$$\mathcal{L}_3 = \Box \phi \phi_\mu \phi^\mu, \qquad (3.15)$$

$$\mathcal{L}_4 = (\Box \phi)^2 \phi_\mu \phi^\mu - 2\Box \phi \phi_\mu \phi^{\mu\nu} \phi_\nu - \phi_{\mu\nu} \phi^{\mu\nu} \phi_\rho \phi^\rho + 2\phi_\mu \phi^{\mu\nu} \phi_{\nu\rho} \phi^\rho, \qquad (3.16)$$

$$\mathcal{L}_5 = (\Box \phi)^3 \phi_\mu \phi^\mu - 3(\Box \phi)^2 \phi_\mu \phi^{\mu\nu} \phi_\nu - 3\Box \phi \phi_{\mu\nu} \phi^{\mu\nu} \phi_\rho \phi^\rho + 6\Box \phi \phi_\mu \phi^{\mu\nu} \phi_{\nu\rho} \phi^\rho$$
$$+ 2\phi^\mu_\nu \phi^\nu_\rho \phi^\rho_\mu \phi^\lambda \phi_\lambda + 3\phi_{\mu\nu} \phi^{\mu\nu} \phi_\rho \phi^{\rho\lambda} \phi_\lambda - 6\phi_\mu \phi^{\mu\nu} \phi_{\nu\rho} \phi^{\rho\lambda} \phi_\lambda. \qquad (3.17)$$

Note that here derivatives are in flat space, i.e. $\nabla_\mu = \partial_\mu$. One can construct a generalised action with a linear combination of all these Lagrangian densities as

$$S = \int d^4x \sqrt{-g} \left[ \frac{1}{2\kappa} R + \sum_{i=1}^{5} c_i \mathcal{L}_i \right], \qquad (3.18)$$

with $c_i$ being free constant coefficients for each term. This construction can be generalised to curved spacetimes while maintaining second-order equa-tions of motion by introducing direct couplings to curvature, though this inevitably breaks Galilean invariance [390]. This led to the rediscovery of Horndeski theories [356], first proposed over 30 years earlier[355].[16]

### 3.2.5 Horndeski

Horndeski gravity is the most general ST theory leading to second-order equations of motion and is governed by the following action [355, 356].

$$S = \int d^4x\sqrt{-g}\Big[G_2 + G_3\Box\phi + G_4R + G_{4X}\left[(\Box\phi)^2 - \phi^{\mu\nu}\phi_{\mu\nu}\right] +$$
$$G_5G_{\mu\nu}\phi^{\mu\nu} - \frac{1}{6}G_{5X}\left[(\Box\phi)^3 - 3\phi^{\mu\nu}\phi_{\mu\nu}\Box\phi + 2\phi_{\mu\nu}\phi^{\mu\sigma}\phi_\sigma^\nu\right]\Big]. \quad (3.19)$$

Here, the $G_i$ are free functions of $\phi$ and $X$ and $G_{iX}$ denotes the partial derivative of $G_i$ with respect to $X$.

Horndeski gravity provides a general consistent construction to explore modifications of GR, motivated in part by the existence of self-accelerating solutions which do not require a cosmological constant as well as evading Solar system constraints. More broadly, they constitute a framework to generically test gravitational interactions. We have already seen how Horndeski gravity includes quintessence, $k$-essence, Brans-Dicke and the cubic Galileon as special members of its family. Other theories within Horndeski deserving a special mention include Einstein-scalar-Gauss-Bonnet (EsGB) [344, 392, 393], Fab Four [394], Kinetic Gravity Braiding (KGB) [395] and $f(R)$ [396–398]. Several aspects of Horndeski gravity will be introduced in the following Sections and Chapters. For a review on Horndeski theories, covering e.g. their cosmological implications, see [399].

### 3.2.6 Beyond Horndeski

One might have initially thought that Horndeski gravity marked the end of the road for general ST theories. However, it turns out that physically viable ST models can exist well beyond the Horndeski class.

A prominent example is the Gleyzes–Langlois–Piazza–Vernizzi (GLPV) theory, sometimes referred to as beyond Horndeski gravity [400]. GLPV extends the Horndeski framework by allowing higher-order field equations, whereas Horndeski is strictly second order. In practice, this means the following Lagrangians are added to the Horndeski action

$$L_4^{\text{bH}} \equiv F_4(\phi, X)\,\epsilon^{\mu\nu\rho\sigma}\epsilon^{\mu'\nu'\rho'}{}_\sigma\,\phi_\mu\phi_{\mu'}\phi_{\nu\nu'}\phi_{\rho\rho'}, \quad (3.20)$$
$$L_5^{\text{bH}} \equiv F_5(\phi, X)\,\epsilon^{\mu\nu\rho\sigma}\epsilon^{\mu'\nu'\rho'\sigma'}\,\phi_\mu\phi_{\mu'}\phi_{\nu\nu'}\phi_{\rho\rho'}\phi_{\sigma\sigma'}. \quad (3.21)$$

Despite this extension, GLPV avoids Ostrogradsky instabilities and remains healthy by satisfying specific degeneracy conditions that eliminate the unwanted ghostly degrees of freedom. As a result, the theory still propagates only three degrees of freedom: two tensor modes and one scalar.

While higher-order equations typically signal the presence of Ostrogradsky ghosts—unphysical instabilities due to an unbounded Hamiltonian—GLPV cleverly circumvents this issue. A simplified demonstration of how these degeneracy conditions work is presented below through a toy example.

**Toy model: dynamics of higher-order particle–Ostrogradsky ghost**

Here we present a toy-model to introduce the main features of the Ostrogradsky instability [401, 402]. Let us consider a field $\varphi(t)$ whose dynamics is given by the action[a]

$$S = \frac{1}{2} \int dt \left( \dot{\varphi}^2 - \omega^2 \varphi^2 + \alpha \ddot{\varphi}^2 \right), \tag{3.22}$$

which is a harmonic oscillator with the addition of a higher-derivative term. Its equation of motion is given by

$$\ddot{\varphi} + \omega^2 \varphi - \alpha \ddddot{\varphi} = 0. \tag{3.23}$$

This theory propagates two degrees of freedom and requires four initial conditions for $\varphi(0)$, $\dot{\varphi}(0)$, $\ddot{\varphi}(0)$ and $\dddot{\varphi}(0)$. The two degree-of-freedom nature of the theory can be made explicit by rewriting the action (3.22) in an equivalent, albeit different form

$$S = \frac{1}{2} \int dt \left[ \dot{\varphi}^2 - \omega^2 \varphi^2 + \alpha(2\dot{\varphi}\dot{\psi} - \psi^2) \right], \tag{3.24}$$

$$= \frac{1}{2} \int dt \left[ (\dot{\varphi} + \alpha\dot{\psi})^2 - \alpha^2 \dot{\psi}^2 - \omega^2 \varphi^2 - \alpha\psi^2 \right]. \tag{3.25}$$

From the first line above, it can easily be seen that solving for $\psi$ and substituting back into the action recovers (3.22). From the second line, one can see that this theory propagates two modes with opposite-sign kinetic terms, meaning the energy of the system is unbounded from above and below, rendering the model unphysical. The additional propagating degree of freedom is referred to as a ghost.

Let us now add another particle dynamically coupled to the original one as

$$S = \frac{1}{2} \int dt \left( \dot{\varphi}^2 - \omega^2 \varphi^2 + \dot{q}^2 - \omega^2 q^2 + 2\alpha\ddot{\varphi}\dot{q} \right), \tag{3.26}$$

with equations of motion

$$\ddot{\varphi} + \omega^2 \varphi - \alpha\ddddot{\varphi} - \alpha\ddot{q} = 0, \qquad \ddot{q} + \omega^2 q + \alpha\dddot{\varphi} = 0. \tag{3.27}$$

In general, these type of models have three propagating degrees of freedom ($\varphi$, $q$ and the ghost). However, the specific form of the dynamical coupling allows for the Ostrogradsky instability to be evaded. This can be seen for our example by using the field redefinition $Q = q + \alpha\dot{\varphi}$, which gives

$$S = \frac{1}{2} \int dt \left( \dot{\varphi}^2 + \dot{Q}^2 - \omega^2 \varphi^2 - \omega^2 Q^2 \right), \tag{3.28}$$

which contains two healthy propagating modes. This property of absorbing higher-order derivatives in the field redefinitions is known as *degeneracy*.

_______________

[a] This example is taken from a lecture by Karim Noui. A more elaborate presentation of the instability can be found in [403, 404].

Another example of a ST theory that lies outside the Horndeski framework is mimetic gravity [405], originally proposed to mimic the behaviour of dark matter. This is achieved by expressing the physical metric as a non-invertible conformal transformation of an auxiliary metric, with the conformal factor determined by the kinetic term of a scalar "mimetic" field. Due to the non-invertibility of this transformation, mimetic gravity falls outside the Horndeski class.

A further example is Hořava-Lifshitz gravity, first introduced in [353]. Motivated by quantum gravity, this theory explicitly breaks Lorentz invariance by introducing higher-order spatial derivatives (hence going beyond Horndeski), while keeping time derivatives strictly second order – thereby evading Ostrogradsky instabilities. The theory was further healthily developed in [406]. In the infrared (low-energy) limit, Hořava-Lifshitz gravity reduces to what is known as khronometric gravity, where a scalar field (the khronon) defines a global time function, thus enforcing a preferred time-foliation of spacetime. Although still Lorentz-violating, the low-energy version remains quadratic in time and space derivatives, and may be extended to include higher derivatives using degeneracy conditions, similar to those discussed previously [407].

An additional example of a theory beyond Horndeski is cuscuton gravity, originally introduced in [408, 409] and later generalised in [410]. The cuscuton is a scalar field with no independent dynamics, i.e. it does not propagate, and the theory only contains two tensorial degrees of freedom. This is accomplished by giving the scalar field a Lagrangian that is linear in first-order time derivatives, e.g., $\mathscr{L} \ni \sqrt{X}$. The theory discussed in Chapter 5 belongs to the extended cuscuton class, and its properties will be explored in more detail there.

### 3.2.7  Higher-Order Scalar-Tensor theories

Theories going beyond Horndeski required a generalised classification, which was provided by the Higher-Order Scalar-Tensor (HOST) framework. HOST theories are the most general ST theories containing up to second-order derivatives in the action. GR and Horndeski gravity also contain second-order derivatives of the metric ($R \sim \partial^2 g_{\mu\nu}$) and the scalar ($\phi_{\mu\nu}$) in the action, but nonetheless lead to second-order Euler-Lagrange equations, as opposed to third-order. This is because the structure of the action is carefully constructed so that higher-order derivatives cancel out in the equations of motion. In HOST theories, the equations of motion are generally allowed to be higher than second-order. This is in principle problematic, as it generically entails the presence of an additional ghostly degree of freedom. However, degeneracies between the different functions appearing in the Lagrangian can be imposed to ensure the absence of such ghost degree of freedom.[17] When such degeneracy conditions are imposed to HOST theories, they become DHOST (Degenerate Higher-Order Scalar-Tensor) theories. DHOST theories were first constructed up to quadratic order in second-derivatives of the scalar (i.e. $\sim \phi_{\mu\nu}^2$) in [348] and further extended in [349]. They were also later constructed to cubic order ($\sim \phi_{\mu\nu}^3$) in [350]. Here, we introduce the

17: Similar constructions also exist for vector-tensor theories [411].

cubic HOST action

$$S = \int d^4x \sqrt{-g} \left[ F_0(\phi, X) + F_1(\phi, X)\Box\phi + F_2(\phi, X)R + \sum_{I=1}^{5} A_I(\phi, X)L_I^{(2)} \right.$$
$$\left. + F_3(\phi, X)G_{\mu\nu}\phi^{\mu\nu} + \sum_{J=1}^{10} B_J(\phi, X)L_J^{(3)} \right] + \int d^4x \sqrt{-g}\, L_m \ , \quad (3.29)$$

where $L_I^{(2)}$ and $L_J^{(3)}$ comprise all possible terms built from $\phi_\mu$ and $\phi_{\mu\nu}$ which are quadratic and cubic in $\phi_{\mu\nu}$ respectively, and are written explicitly as

$$
\begin{aligned}
L_1^{(2)} &= \phi_{\mu\nu}\phi^{\mu\nu}, & L_1^{(3)} &= (\Box\phi)^3, & L_2^{(3)} &= (\Box\phi)\phi_{\mu\nu}\phi^{\mu\nu}, \\
L_2^{(2)} &= (\Box\phi)^2, & L_3^{(3)} &= \phi_{\mu\nu}\phi^{\nu\rho}\phi_\rho^\mu, & L_4^{(3)} &= (\Box\phi)^2\phi_\mu\phi^{\mu\nu}\phi_\nu, \\
L_3^{(2)} &= (\Box\phi)\phi^\mu\phi_{\mu\nu}\phi^\nu, & L_5^{(3)} &= \Box\phi\phi_\mu\phi^{\mu\nu}\phi_{\nu\rho}\phi^\rho, & L_6^{(3)} &= \phi_{\mu\nu}\phi^{\mu\nu}\phi_\rho\phi^{\rho\sigma}\phi_\sigma, \\
L_4^{(2)} &= \phi^\mu\phi_{\mu\rho}\phi^{\rho\nu}\phi_\nu, & L_7^{(3)} &= \phi_\mu\phi^{\mu\nu}\phi_{\nu\rho}\phi^{\rho\sigma}\phi_\sigma, & L_8^{(3)} &= \phi_\mu\phi^{\mu\nu}\phi_{\nu\rho}\phi^\rho\phi_\sigma\phi^{\sigma\lambda}\phi_\lambda, \\
L_5^{(2)} &= (\phi^\mu\phi_{\mu\nu}\phi^\nu)^2, & L_9^{(3)} &= \Box\phi(\phi_\mu\phi^{\mu\nu}\phi_\nu)^2, & L_{10}^{(3)} &= (\phi_\mu\phi^{\mu\nu}\phi_\nu)^3. \quad (3.30)
\end{aligned}
$$

The covariant Euler-Lagrange equations for these broad class of theories were derived in [3], which constitutes the content of Chapter 6. This action encompasses both standard Horndeski [355, 356] and beyond-Horndeski [400, 412, 413], see e.g. [414]. For instance, the Horndeski action in the standard form with the Galileon functions can be recovered with the choices:

$$
\begin{aligned}
F_0 &= G_2 \ , & F_1 &= G_3 \ , & F_2 &= G_4 \ , & F_3 &= G_5 \ , \\
A_1 &= -A_2 = -G_{4X} \ , & 6B_1 &= -2B_2 = 3B_3 = -G_{5X} \ ,
\end{aligned} \quad (3.31)
$$

and $A_I = B_J = 0$ for $I = 3, 4, 5$ and $J = 4, \cdots, 10$. For a review of DHOST theories see [404]. The full classification of quadratic/cubic DHOST theories can be found in [350], and there are a large number of subclasses distinguished by different sets of degeneracy conditions, as shown in Table 3.1.[18]

18: Note that the sum of two degenerate Lagrangians is not necessarily degenerate.

**Table 3.1:** Classification of degenerate DHOST subclasses involving quadratic and/or cubic terms.

| Class of theory | # of degenerate subclasses |
| --- | --- |
| Purely quadratic ($F_3 = 0$, all $B_J = 0$) | 7 total |
| • Non-minimally coupled ($F_2 \neq 0$): $^2$N-I, $^2$N-II, … | 4 |
| • Minimally coupled ($F_2 = 0$): $^2$M-I, $^2$M-II, … | 3 |
| Purely cubic ($F_2 = 0$, all $A_I = 0$) | 9 total |
| • Non-minimally coupled ($F_3 \neq 0$): $^3$N-I, $^3$N-II | 2 |
| • Minimally coupled ($F_3 = 0$): $^3$M-I, $^3$M-II, … | 7 |
| Mixed quadratic + cubic | 25 (out of $7 \times 9 = 63$) |

19: It was shown in [416] that in the context of cosmology the disformal Horndeski class is phenomenologically favoured, unlike other subclasses of DHOST theories.

Among these subclasses, there is one that can be obtained from the Horndeski class via invertible conformal/disformal transformation, which was called "disformal Horndeski" class in [415].[19] (Note that an invertible conformal/disformal transformation preserves the number of physical degrees of freedom [417, 418].) In the terminology of [350], this class corresponds to a sum of the quadratic DHOST of class $^2$N-I and the cubic DHOST of class $^3$N-I. This notation can be better understood by looking at Table 3.1. For the quadratic part (characterised by $F_2$ and $A_I$'s), the degeneracy conditions are given by[20]

$$A_2 = -A_1 \, , \tag{3.32}$$

$$A_4 = \frac{1}{2(F_2 + 2XA_1)^2} \Big[ (3F_2 + 8XA_1 + 2X^2A_3)(A_1 + XA_3 + F_{2X})^2$$
$$- 2A_3(3XA_1 + X^2A_3 + F_2 + XF_{2X})^2 \Big] \, , \tag{3.33}$$

$$A_5 = \frac{(A_1 + XA_3 + F_{2X}) \left[ A_1(A_1 - 3XA_3 + F_{2X}) - 2A_3F_2 \right]}{2(F_2 + 2XA_1)^2} \, , \tag{3.34}$$

where $F_2 (\neq 0)$, $A_1$, and $A_3$ are free functions and the condition $F_2 + 2XA_1 \neq 0$ is assumed. For the cubic HOST characterised by $F_3$ and $B_J$'s, we have

$$-\frac{B_2}{3} = \frac{B_3}{2} = B_1 \, , \quad B_5 = -B_7 = \frac{4XB_4F_{3X} - (6B_1 + F_{3X})^2}{12XB_1} \, ,$$

$$B_8 = \frac{B_5(4XB_4 - 6B_1 - F_{3X})}{12XB_1} \, , \quad B_6 = -B_4 \, ,$$

$$B_9 = \frac{B_4(4XB_4 - 6B_1 - F_{3X})}{6XB_1} \, , \quad B_{10} = \frac{B_4(4XB_4 - 6B_1 - F_{3X})^2}{24X^2B_1^2} \, , \tag{3.35}$$

where $F_3$, $B_1 (\neq 0)$, and $B_4$ are arbitrary functions. Moreover, when both the quadratic and cubic parts are present, one has to impose the following conditions on top of Eqs. (6.6) and (6.7):

$$A_3 = \frac{A_1(4XF_{2X} - 3F_2)}{XF_2} + \frac{XF_{2X} - F_2}{X^2} - \frac{F_{3X}(F_2 + 2XA_1)^2}{6X^2F_2B_1} \, ,$$
$$B_4 = \frac{6B_1(F_2 - XF_{2X}) + F_{3X}(F_2 + XA_1)}{2XF_2} \, . \tag{3.36}$$

### Conformal and disformal transformations in HOST theories

In ST theories, it is often useful to consider transformations of the metric that depend on the scalar field $\phi$ and possibly its derivatives. These are often considered to be of this form [419]

$$\tilde{g}_{\mu\nu} = C(\phi, X) \, g_{\mu\nu} + D(\phi, X) \, \phi_\mu \, \phi_\nu, \tag{3.37}$$

where $C$ is the conformal factor and $D$ is the disformal factor. We have previously seen how a conformal transformation can be used to map back and forth between the Jordan and Einstein frames (see Eq. (3.7)). In the HOST framework, more generally, specific theories can sometimes be mapped into each other by the transformation above. For instance, such transformation can be used to map a subset of DHOST theories to Horndeski [412, 420]. Therefore, disformal transformations can change the appearance of a theory at the Lagrangian level but, as long as the transformation is invertible, preserve the number of degrees of freedom and in fact describe the same physics. These transformations hence help to distinguish physically distinct theories with respect to those which are equivalent up to field redefinitions. Moreover, they have been recently used as black hole solution-generating techniques, by e.g. transforming known GR solutions onto different non-trivial black hole solutions with scalar hair [421, 422]. This often involves ignoring the fact that after the transformation matter couples to a new metric—as otherwise the same physics should be recovered—which somewhat accidentally can lead to new healthy solutions.

## 3.3  Gravitational wave effects of scalar-tensor theories

The beauty and elegance of gravity resides in how it is uncorrupted by the intricate structure of the other fundamental forces. It acts universally on all forms of energy and matter, indifferent to their charge, colour, spin, flavour, or any other quantum property. As a result, when GWs are detected, we are witnessing gravity in its rawest form[21]–disturbances in spacetime itself–allowing us to remarkably precisely measure the masses and spins of black holes billions of light-years away. Moreover, the power of GWs goes beyond astrophysical inference, as it also offers a direct probe into the fundamental nature of gravity. ST theories, in particular, open up a rich landscape of potential deviations from GR in their GW phenomenology. In this Section, we will summarise some of the key GW signatures predicted by such theories, and in some cases discuss how observations have helped constrain them. Schematically, one can divide such effects into emission and propagation categories, based on the context where they arise. Broadly speaking, emission effects probe the strong gravity, high-energy, local regime, while propagation effects probe the weak gravity, low-energy, cosmological regime.

### 3.3.1  Emission

Due to modified gravitational dynamics, the process of a binary merger can in principle differ in ST theories as compared to GR, and ultimately lead to a difference in the emitted signal.

**Inspiral**

One such example is in the inspiral, where the scalar field can alter the trajectories of the compact objects and lead e.g. to a different inspiral rate. Unlike in GR, ST theories allow for dipolar radiation, in which case the energy loss to GWs during the inspiral is enhanced, resulting in a faster orbital decay [423, 424]. In addition, if the effective gravitational constant is allowed to vary in time, the gravitational interaction can be altered, making the inspiral dependent on its environment and hence allowing for constraints on such time evolution [425]. Independently, for scalar dark matter models, an overdensity between binary black holes can arise as a result of scalar accretion leading to a faster inspiral (i.e. a dephasing of GWs) due to the additional loss of energy by friction with the scalar cloud [426–429]. In a similar vein, *spontaneous scalarisation* is a process that may occur in which compact objects (black holes or neutron stars differently) acquire a non-trivial scalar configuration [430], leading to deviations from GR in the strong gravity regime, but maintaining agreement with GR predictions elsewhere. Scalarisation is ignited by a tachyonic instability settled by non-linear self-interactions of the scalar. This process typically requires a certain property of the compact object such as its compactness or spin to be above a specific threshold, see [431] for a general review. Scalarisation is most typically relevant in the inspiral stage, where it can get activated and leave imprints on the GW.

Effects on the inspiral rate have been accurately measured and constrained with binary pulsars. In Section 2.3.4 we have already explained how the discovery of the Hulse-Taylor binary pulsar constituted the first indirect evidence of GWs [156]. After several decades of continuous observations, this system has been able to place constraints on modified gravity theories, see e.g. [432] where the orbital decay rate after 30 years of observation was found to be given by $0.997 \pm 0.002$ times the one expected for GR. Another interesting binary is PSR J1738+0333, composed of a pulsar and a white dwarf, and shown to also agree with the GR orbital decay prediction, resulting in an upper bound on the dipolar emission effectively translated to a constraint on the parameter coupling matter with the scalar [433]. For Brans-Dicke gravity, a 10-year observation of this binary set a constraint of $\omega_{BD} > 25000$ [433],[22] slightly weaker than the constraints set by the Solar system test from the Cassini spacecraft [51]. In terms of scalarisation, the original model proposed in [430] has been ruled out by binary pulsar observations [434–436]. However, different scalarisation sources involving scalar couplings to curvature, as opposed to non-linear matter sources, have been proposed in [437–441], which have in some part also been constrained by binary pulsar data [442]. An interesting prospect for probing scalarisation with GWs are extreme mass-ratio inspirals (EMRIs), which will be observed with high SNR by future detectors. In these binaries, one of the compact objects is much smaller than the other, and hence the former orbits the latter for a long time. Since scalarisation turns on e.g. for high compactness, a reasonable scenario is where the smaller (and more compact) object is scalarised while the supermassive black hole it orbits around is not, leaving an imprint in the GW in form of a dephasing, which should be observable by LISA [443].

Additional inspiral observables signalling deviations from GR are tidal Love numbers (TLNs), which measure how much an object deforms under an external tidal field, with black holes in GR famously having vanishing TLNs [444]. Analyses of TLNs have been completed in a number of GR extensions, see e.g. [445–452].

### Ringdown

In Section 2.2.2, we have seen how special black holes are in GR, while in Section 2.4.2 we have shown how this simplicity can be translated to ringdown observables, in particular to QNMs. In the following section, §3.4, we will summarise different ways in which black holes can grow in complexity in ST theories. As a result, QNMs of these ST black holes will in principle differ from the GR ones, rendering ringdown observations as a powerful tool to test and constrain such models, as argued in Section 2.4.3.

The use of QNMs to test modified gravity theories (scalar-tensor or otherwise) is known as the *black hole spectroscopy* program and has received significant attention in the recent years. Agnostically parametrised deviations have been introduced in the data analysis pipelines, where current GW QNM data seems to provide a small hint of a GR deviation, as shown in Figure **??** [453]. However, in order to gain understanding of the microphysics responsible for QNM deviations, ringdown calculations need to be repeated for different combinations of modified gravity theories and black hole backgrounds. In Section 3.5, we will introduce some key aspects of how such calculations take place in ST theories, but here we will summarise some key literature on the topic. First, let us highlight that in theories of gravity that include additional fields, perturbations of such fields on black hole backgrounds will also lead to QNM emission. As such, in ST theories one also obtains scalar QNMs. As shown in Section 2.4.1, these are much easier to compute than tensor perturbations and, as a result, a significant portion of new literature on QNMs beyond GR relates to scalar modes. However, distortions in spacetime as observed by GW detectors come in the form of tensorial perturbations, making them more interesting objects of study, despite them being significantly more challenging to compute. Moreover, scalar and tensorial modes can couple to one another, so computing the whole spectra is highly preferred over computing each in isolation.

QNMs for Schwarzschild black holes with a constant scalar field in Horndeski gravity were studied in [213]. There, it is shown that odd parity modes evolve just like in GR, i.e. they obey the same Regge-Wheeler equation 2.96, while even modes are mixed with the scalar by the function $G_{4\phi}$ (i.e. the $\phi$ derivative of $G_4$ evaluated on the background). In this scenario, which will be shown in more detail in Section 3.5, the only Horndeski fingerprint is the introduction of an effective mass term for scalar perturbations, which is forecasted to be constrained by LISA to the order $\sim 10^{-17} eVc^{-2}$ [454]. More rich QNM phenomenology can be achieved if black holes acquire hair, i.e. are different from the GR ones. Small parametrised hairy deviations from the background solution studied in [213] (i.e. a Schwarzschild black hole with a constant scalar) were introduced in [455], where the odd sector was studied and a modified Regge-Wheeler equation containing corrections from the parametrised hair was derived. In Chapter 4, we will employ a similar construction to study the effect of non-luminal GWs on QNMs [1]. QNM calculations have also been completed for more concrete theory choices, such as Einstein-dilaton-Gauss-Bonnet gravity [456–458], Chern-Simons gravity [459–461], Einstein-Aether [462, 463] and (curvature-only) higher-derivative gravity [446, 464, 465].

Black holes, like any other astrophysical object, rotate [466]. Therefore, in order to fully exploit black hole spectroscopy, it is crucial to study QNMs of rotating black holes. However, the inclusion of rotation comes with increased computational difficulties, and QNM calculations beyond GR have only been computed in some special cases, mostly relying on slow-spin approximations. For instance, QNMs for slowly rotating black holes in Einstein-dilaton Gauss-Bonnet gravity have been calculated to first [467] and second order in rotation [468], where it is found that QNM shifts are significantly enhanced by rotation. In addition, slowly-rotating QNMs have been calculated in dynamical Chern-Simons [469–472] and in higher-derivative gravity [473, 474]. More recently, QNMs calculations have in some cases been extended to rapidly spinning black holes. In particular, [475–477] calculated QNMs for rotating black holes in higher-derivative gravity to 18$^{th}$ order in rotation. The QNM shifts calculated in these works were then implemented into `pyRing`, where GW data from GWTC-3 was used to set an upper bound of $\ell \lesssim 35$ km on the length scale of new physics [478]. Such QNM computations were made possible by the development of the Modified Teukolsky Equation (or modified Teukolski formalism) [475, 479, 480]. In addition, a framework named *METRICS* based on a spectral decomposition of metric perturbations has been developed [481, 482] and employed to calculate QNMs for rapidly rotating black holes in Einstein-dilaton-Gauss-Bonnet gravity [483–485] and dynamical Chern-Simons [486].

Besides QNMs, greybody factors exist as alternative ringdown observables, quantifying how much of a wave gets transmitted versus reflected by the compact object's effective potential. These frequency-dependent factors can serve as powerful diagnostic tools for probing the nature of compact objects. For example, horizonless compact objects—as predicted in various quantum gravity models, such as fuzzballs [487], gravastars [488, 489], or boson stars [490–492]—can possess a physical surface or structure that partially reflects incoming gravitational waves. This reflection gives rise to delayed secondary signals or *echoes* in the ringdown phase, see e.g. [493]. The presence, structure, and frequency-dependence of these echoes are given by the modified greybody factors.[23] In [497, 498], greybody factors have been proposed as ringdown waveform modelling alternatives to QNMs, in particular motivated by their robustness against ringdown start-time choices, which, as explained in Section 2.4.4 can constitute a problem for QNM interpretation.

**Polarisations**

GWs in modified theories of gravity can carry additional polarisations beyond those predicted by GR [499]. In particular, in ST theories the scalar field can manifest itself as a breathing mode—an isotropic expansion and contraction in the plane transverse to the direction of propagation—and, when the scalar is massive, a longitudinal mode, where the displacement of test particles occurs along the direction of propagation [500], as depicted in Fig. 3.1. Detecting or constraining such non-tensorial modes would provide a powerful observational test of GR. In the present case, since the LIGO Hanford and Livingston detectors are closely aligned, they can not by themselves detect additional polarisations, but would require instead a third detector, such as Virgo to be simultaneously active [501]. This was the case for GW170814, where simplified analyses showed that data prefer pure tensor modes over pure scalar or vector ones, without including mode-mixing [502, 503].

### 3.3.2 Propagation

After being emitted, GWs will propagate through cosmological distances before reaching our detectors, and the specific way that occurs will generically differ in ST theories. We can discuss some of these effects by looking at how the propagation equation is modified in beyond GR theories[24][505]

$$h_p'' + (2 + \nu)\mathcal{H}h_p' + (c_{\mathrm{GW}}^2 k^2 + a^2\mu^2)h_p = S(h_p, t_p), \quad (3.38)$$

where here the prime denotes derivatives with respect to conformal time and $\mathcal{H}$, $\nu$, $c_{\mathrm{GW}}$, $\mu$ and $S$ are potential modifications to the GW propagation equation in GR (2.47), which is recovered in the limit of $\mathcal{H}$ being the same Hubble parameter as in GR, $c_{\mathrm{GW}} = c = 1$ and $\nu = \mu = S = 0$.

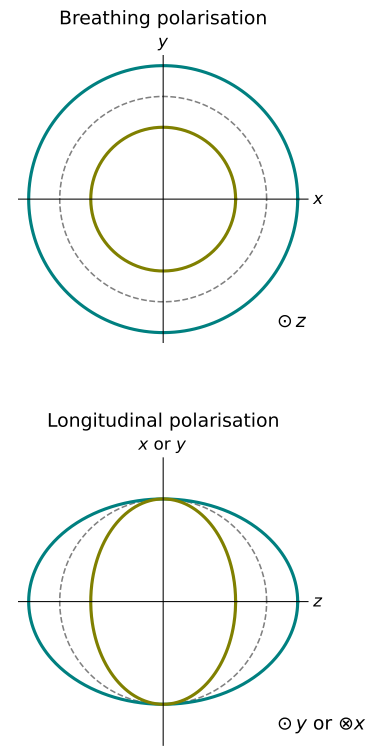

**Figure 3.1:** Illustration of scalar mode polarisations on a ring of particles at different phases (c.f. to the tensor modes in 2.4).

25: This includes theories than can be disformally related to Horndeski.

26: We will later see how on black hole backgrounds one can also define effectively the same parameters albeit with a spatial dependence instead.

27: In particular, the luminosity distance relation is given by

$$d_L^{gw} = (1+z) \int_0^z \frac{c}{H(z)} dz, \qquad (3.40)$$

which is the same as the electromagnetic luminosity distance, $d_L^{gw} = d_L^{em}$.

28: The terminology being inspired by *standard candles* (supernovae observations) and *standard rulers* (baryon acoustic oscillation (BAO) measurements).

29: Current neutron star merger rates are inferred to be between $10 - 1700$ Gpc$^{-1}$yr$^{-1}$[513].

In the context of ST theories, the Bellini & Sawicki parametrisation compresses the functions appearing in the full covariant Lagrangian into their effective combination appearing in the equations of motion on a cosmological background [506, 507]. This way, for Horndeski theories we have $\alpha_K$ (*kineticity*), $\alpha_B$ (*braiding*), $\alpha_M$ (*Planck-mass run rate*), and $\alpha_T$ (*tensor speed excess*), with the addition of $\alpha_H$ in beyond Horndeski theories.[25] Note that these 'parameters' are actually time-dependent functions on a cosmological background.[26] In particular, $\alpha_K$ and $\alpha_B$ affect the evolution of scalar perturbations and are therefore probed by data from the cosmological microwave background (CMB) or the large-scale structure (LSS), while $\alpha_M$ and $\alpha_T$ affect tensorial (as well as scalar) perturbations in the form of Eq. (3.38) with

$$\nu = \alpha_M, \qquad c_{GW}^2 - 1 = \alpha_T, \qquad \mu = 0, \qquad S = 0. \qquad (3.39)$$

The main type of effects or tests in GW propagation appear as modifications to the following (see [501] for a comprehensive review):

▶ **Expansion history with standard sirens**. In astronomy, the *luminosity distance* $d_L$ is a quantity that connects an observed signal to how far it is in physical space. In the context of GWs, the amplitude of the signal is inversely proportional to their $d_L^{gw}$. The expansion of the universe introduces a friction term in (3.38) which stretches waves as they propagate, affecting their observed redshift and therefore implying that GWs are sensitive to the expansion history. In the absence of the additional modified gravity terms, in particular the absence of the extra friction term $\nu = 0$ but also $c_{GW} = 1$ and $\mu = S = 0$, luminosity distance $d_L^{gw}$, redshift $z$ and the Hubble parameter $H$ are interconnected.[27] As a result, because $d_L^{gw}$ is known from the GW amplitude, if the redshift of the source is known, the Hubble parameter H can be inferred, therefore potentially providing an independent observational channel for the resolution of the Hubble tension. Because of this, GW signals with an identified redshift have been referred to as *standard sirens*[28] [508]. However, GW redshifts are not easy to come by. In the special case of neutron star mergers, such as GW170817, their electromagnetic counterpart can be used to obtain the redshift, and thus these class of events are referred to as *bright* sirens. This particular event, GW170817, was used to set the first broad bounds on the Hubble parameter from GWs: $H_0 = 70^{+12}_{-8}$ km s$^{-1}$ Mpc$^{-1}$ [509]. Competitive 1% measurements of $H_0$ would require $\mathcal{O}(100)$ bright sirens [510–512]. However, due to the scarcity of neutron star mergers observed,[29] alternative techniques to identify redshifts for the largely more abundant number of black hole mergers have been devised. In particular, galaxy catalogues can be used in combination with GW sky localisation to obtain redshifts from candidate host galaxies while accounting for the uncertainty in host identification. These are instead referred to as *dark* sirens, due to the lack of an intrinsic electromagnetic counterpart, and their constraining power resides in the stacking of multiple observations. Employing a fully Bayesian method developed in [514], an LVK analysis including GW events up to O2 constrained $H_0 = 68.7^{+17}_{-7.8}$ km s$^{-1}$ Mpc$^{-1}$ [515]. This method can also be implemented through the construction of cross-correlations between GWs and galaxy catalogues, see e.g. [516–522]. Another method to identify the redshift of dark sirens is to model how the mass spectrum scales with luminosity distance. Being the observed masses redshifted, spe-cific known features in the mass distribution of compact objects (e.g. the neutron star–black hole mass gap) can be used to infer the redshift [523–525]. In addition, both methods can be used in conjunction as in [526–528] which, including data from GWTC-3, resulted in a shrinking of the error bars in the $H_0$ measurement, e.g. $H_0 = 69^{+12}_{-7}$ km s$^{-1}$ Mpc$^{-1}$ in [527].

This way, in the absence of specific modified GR parameters, GW propagation depends on the nature of dark energy indirectly through $H$, and standard sirens can therefore probe the dark energy equation of state, see e.g. [529–531].

▶ **GW damping**. As we have seen, the GW propagation equation in GR contains an friction term given by the Hubble parameter. In addition, a non-zero $\nu$ in (3.38) will modify the GW damping rate and will lead to a GW luminosity distance different to the electromagnetic one.[30] Therefore, one can use the same standard siren methods described above to study the GW damping rate as a probe of physics beyond GR [532]. Modified gravity theories which modify the GW luminosity distance include higher-dimensional gravity [532], non-local gravity [533] and ST theories with a non-zero $\nu$ which, in Horndeski, is written as a time-varying effective Planck mass

$$\nu = \alpha_M \equiv \frac{d \ln M_*^2}{d \ln a}, \tag{3.42}$$

with $M_*^2 \equiv 2(G_4 - 2XG_{4X} + XG_{5\phi} - \dot{\phi}HXG_{5X})$ in Horndeski language, with the dot denoting a $t$-derivative (c.f. Eq. (2.44)). Constraints on $\nu$ (or alternatively $\alpha_M$) have already been placed by GW observations, but remain somewhat weak [534–536]. Looking forward, we can expect competitive constraints on the GW damping rate from dark sirens, provided a sufficiently abundant number of them. In terms of bright sirens, despite their enhanced constraining power, competitive results can only be expected for third-generation detectors, assuming they can achieve a significantly higher neutron star merger rate [537].

▶ **GW speed**. In GR, GWs are predicted to propagate at the speed of light, a result easily broken in theories which modify the dispersion relation, i.e. with $c_{GW} \neq c = 1$ (tensor speed) and/or $\mu \neq 0$ (effective graviton mass) in (3.38). In Horndeski gravity, non-luminal speed is sourced by a non-zero $\alpha_T$ (3.39), which on a cosmological background is given by

$$\alpha_T = \frac{1}{M_*^2} \left[ 2G_{4X} - 2G_{5\phi} - (\ddot{\phi} - \dot{\phi}H) G_{5X} \right]. \tag{3.43}$$

This quantity was famously strongly constrained by GW170817 to [538][31]

$$-3 \times 10^{-15} \leq \frac{\alpha_T}{2} \leq +7 \times 10^{-16} \tag{3.45}$$

thanks to an electromagnetic counterpart in the form of a short duration gamma-ray burst (SGRB) observed 1.74±0.05s after the GW signal, and similarly named GRB 170817A [539, 540]. Here, the bounds are set assuming a constant $\alpha_T$. More concretely, the lower bound is set by assuming $\alpha_T < 0$ and a SGRB emitted with a 10s delay after the slower GW.[32] Conversely, the upper bound is set by assuming $\alpha_T > 0$ while attributing the entire $1.74 \pm 0.05$s lag due to the faster propagation of the GW signal.

This observational milestone quickly motivated theorists to evaluate

30: For theories with a luminal GW propagation we have [501]

$$\frac{d_L^{GW}(z)}{d_L^{em}(z)} = \exp \left[ \frac{1}{2} \int_0^z \frac{\nu}{1 + z'} dz' \right]. \tag{3.41}$$

31: Note that the factor of 1/2 comes from the fact that constraints are applied on the quantity (reinstating now standard units) $\delta c/c$ with $\delta c \equiv c_{GW} - c$, which is equivalent to $\alpha_T$ up to a factor of 2 for $\delta c \ll 1$ as shown below

$$\alpha_T \equiv \frac{c_{GW}^2 - c^2}{c^2} = 2\frac{\delta c}{c} + \frac{\delta c^2}{c} \approx 2\frac{\delta c}{c}. \tag{3.44}$$

32: It is not certain whether the SGRB emission might be relatively retained compared to the GW one due to local physical processes, therefore introducing some uncertainty in emission time delays.

33: The cutoff is the largest possible energy/frequency scale, where the high energy completion can take over, but this can already take place at significantly lower energies/frequencies. Theoretically predicting the precise scale would require detailed knowledge about such a fiducial (currently unknown) high energy completion.

its implications for beyond GR models, with the initial conclusions being that Horndeski theories required $\alpha_T = 0$ to remain viable. Moreover, it was argued that such vanishing would more naturally be achieved by having $G_{4X} \approx 0$ and $G_5 \approx constant$ independently, as opposed to a non-trivial cancellation of non-zero terms in (3.43) [541–543]. Note how the Horndeski functions which affect $\alpha_T$ are particularly those which introduce second derivatives of the scalar in the action (3.19), and thus according to these conclusions the ST landscape would be restricted to their simplest forms. In addition, the most general luminal DHOST theory was identified in [544].

However, it was later realised that interpreting such GW observation as a constraining factor for modified gravity theories was more subtle than initially imagined, and relied on concrete assumptions on the range of energy scales in which the theory in question is applicable [545]. Interestingly, it turns out that requiring higher-order operators to be cosmologically relevant, as should be the case for dark energy models, leads to a cutoff[33] scale given by

$$\Lambda_{\text{Horndeski}} \sim (M_{\text{Pl}} H_0^2)^{1/3} \sim 260\text{Hz} \qquad (3.46)$$

expressed in units of frequency, which coincides with the frequency range at which LVK operates, as displayed in Figure 3.2. This entails that, for such theories, an (unknown) high energy completion of the fiducial new dark energy physics ought to take over as one approaches this cutoff, i.e. close to or somewhat below the LVK band, i.e. $\sim 20 - 2000$Hz. This high energy completion will naturally enforce $c_{\text{GW}} = c$ at high energies if it permits Lorentz invariant solutions, so LVK measurements such as GW170817 may simply confirm that feature of the high energy completion instead of probing the original (low energy) dark energy physics itself. In other words, in theories that do affect $c_{\text{GW}}$ at cosmological scales, one therefore naturally expects a frequency-dependent transition back to $c_{\text{GW}} = c$ upon approaching the LVK band. With frequencies in the LISA band $\sim 10^{-4}$–$10^{-1}$Hz being significantly lower, upcoming [189] and TianQin [198] measurements therefore provide a much cleaner probe of $c_{\text{GW}}$ in such dark energy related theories.

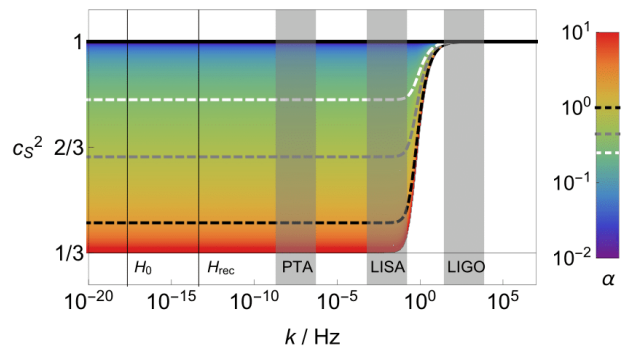

**Figure 3.2:** Frequency-dependent sound speed for scalar perturbations in a toy model incorporating corrections from UV physics, restoring luminality around the cutoff. Taken from [545].

We will describe in more detail different probes of $\alpha_T$ in Chapter 4, where we will introduce the first ringdown-only forecasted constraints on $\alpha_T$ [1].

▸ **Polarisations**. The different + and × polarisations evolve equally and independently of each other in Eq. (3.38). However, theories of gravity that break parity-symmetry can introduce couplings between polarisations. These are usually called chiral models, with Chern-Simons gravity being a standard example. In these cases, transforming the polarisation basis from $\{+, \times\}$ to $\{R, L\}$, i.e. right and left-handed,[34] the metric propagation equations can be decoupled and therefore studied independently, but will generically evolve differently. In particular, effects on the morphology of GWs due to modified polarisation evolution include amplitude birefringence (modifying the relative amplitude of different polarisations) and velocity birefringence (modifying the relative phase of different polarisations) [546, 547]. The amount of amplitude birefringence present in GWTC-3 data has been shown to be consistent with the GR prediction, i.e. no birefringence, providing bounds on the opacity parameter $\kappa$, quantifying the amount of birefringence, and translated to constraints on Chern-Simons gravity [548]. Note that while these constraints already experienced a significant enhancement compared to pervious results using GWTC-2 data [549], they are expected to further improve with third-generation detectors [550]. Different sources and probes of amplitude birefringence have been discussed in the literature [551–554]. Concerning velocity birefringence, it has been found to be highly degenerate with the orientation angle of the binary's angular momentum $\psi$, hindering its capability of being constrained by current data [555, 556], and meaning that such effect requires multimessenger GW events to be constrained, where the orientation angle can be obtained from the electromagnetic counterpart [557].

Note that there exist other types of GW modified phenomenology not captured by the above points, specially for non-ST theories. For instance, the source term $S(h_p, t_p)$ captures the potential interaction between standard metric perturbations with other tensor-like fields $t_p$. These fields naturally originate in bimetric theories, e.g. dRGT massive bigravity, but can also be sourced by tensor-like arrangements of vector fields in multi-proca theories, and even in ST theories, provided the background is inhomogeneous [504]. These 'GW interactions' can induce exchanges of energy and momentum between the tensor fields, and result in copies of the original GW signal (potentially arriving at different times, i.e. GW echoes), which can manifest time-dependent oscillations in their observed amplitude, phase and non-zero birefringence factors [555].

Looking forward, it will be important to model and probe these effects simultaneously due to different degeneracies between them.

## 3.4 Hairy black holes

In GR, no-hair theorems ensure that black holes are fully described by their mass, spin and charge, as we have seen in Section 2.2.2. However, in some ST theories such theorems can be evaded, resulting in black holes with extra structure, extra hair. Hairy black hole solutions in ST theories can broadly be divided in two categories, depending on whether the scalar is static or contains a linear time-dependence, each violating no-hair theorems in different ways. In the context of shift-symmetric theories ($\phi \rightarrow \phi + c$), [558] proved that black holes cannot support non-trivial scalar profiles if the following assumptions apply [35]

1. The spacetime is spherically symmetric, static and asymptotically flat,
2. The scalar field is static $\phi(r)$ and has a vanishing derivative $\phi'$ at infinity,
3. The norm of the current associated with the shift-symmetry is finite down to the horizon,
4. The action contains a canonical kinetic term $X \subset G_2$,
5. All $G_i(X)$ functions are analytical at $X = 0$.

In Chapter 4, where we describe the work in [1], we investigate black holes with parametrised radial hairy deviations. In that case, since the family of theories studied includes all Horndeski (including its non-shift-symmetric branch), hair does not arise as a result of a direct violation of a specific assumption. Conversely, in Chapters 5 and 6, respectively covering the work in [2] and [3], the studied hairy black hole solutions directly violate assumption 2 by dealing with time-dependent scalars. In addition, the specific solution studied in 5 (i.e. [2]) also violates assumptions 4 and 5.

### 3.4.1 Static scalar

We begin by considering solutions where the metric and the scalar are both static and spherically symmetric

$$ds^2 = -B(r)dt^2 + \frac{1}{B(r)}dr^2 + d\Omega^2,$$
$$\bar{\phi} = \phi(r). \tag{3.48}$$

Arguably the most common and minimal setup to achieve such kind of radial scalar hair is via a linear coupling to the Gauss-Bonnet invariant [118, 561], given by the choices

$$G_2 = \eta X, \qquad G_4 = \zeta, \qquad G_5 = \alpha \ln |X|, \tag{3.49}$$

where $\eta$, $\zeta$ and $\alpha$ are constants and the other $G_i$'s are set to zero. The $G_5$ term is indeed equivalent to the linear coupling $\mathscr{L} \supseteq \alpha \phi \mathscr{G}$, where $\mathscr{G}$ is the Gauss-Bonnet invariant [391]. This configuration in particular violates assumptions 3 and 4 above.

There exist, however, more general ways of breaking assumptions in no-hair theorems and thus constructing hairy solutions of this kind. [562] showed that including at least one of the following

$$G_2 \supseteq \sqrt{X}, \qquad G_3 \supseteq \ln |X|, \qquad G_4 \supseteq \sqrt{X}, \qquad G_5 \supseteq \ln |X|, \tag{3.50}$$

to an overall shift-symmetric action is sufficient to potentially enact a radial-dependent scalar profile. These additional operators are in principle of notable interest as, unlike in the scalar Gauss-Bonnet case, they result in a current with a finite norm at the horizon [562]. They do, however, come with potential issues, as it was shown that operators of the type (3.50) other than the Gauss-Bonnet coupling lack a Lorentz invariant ($X = 0$) solution in Minkowski spacetime [563]. If such solutions are desired, this requirement can therefore be used to eliminate all solutions in (3.50), except for the scalar-Gauss-Bonnet solution. Having said this, note that the Lorentz-invariant $X = 0$ vacuum is also not a stable solution in Galileon cosmologies consistent with CMB constraints [564], so the absence of a healthy $X = 0$ solution need not imply the absence of relevant solutions in settings where Lorentz invariance is (already) spontaneously broken.

## 3.4.2 Time-dependent scalar

Let us now turn our attention to linearly time-dependent scalars, which directly violate assumption 2 and therefore require in principle no additional violations to generate hair. These solutions are given by

$$ds^2 = -B(r)dt^2 + \frac{1}{B(r)}dr^2 + d\Omega^2,$$
$$\bar{\phi} = qt + \psi(r), \tag{3.51}$$

where $q$ is a (dimensionful) constant. Table 5.1 in Chapter 5 provides a comprehensive summary of the discussion presented here.

### Existence of solutions

This class of solutions were firstly found in [565] for the theory given by

$$G_2 = -2\Lambda + 2\eta X, \qquad G_4 = \zeta + \beta X, \tag{3.52}$$

with $\eta$, $\zeta$ and $\beta$ being constants. The $\beta$-term introduces a non-minimal coupling of the scalar field to the Einstein tensor (i.e. $\beta G^{\mu\nu}\phi_\mu\phi_\nu$)[36] which has been referred to as the 'John term' in [394] and plays an essential role in driving expansion. This result was then generalised in [566] to the shift-symmetric ($\phi \rightarrow \phi + c$) and reflection-symmetric ($\phi \rightarrow -\phi$) Horndeski subset, of which (3.52) is an example.[37] In particular, it is the shift symmetry of the scalar which allows it to posses a linear time-dependence while keeping the background metric static, as the background equations of motion will only depend on derivatives of $\phi$ [567].

As a specific case of (3.50), [562] considered the theory given by

$$G_2 = \eta X, \qquad G_4 = \zeta + \beta\sqrt{X} \tag{3.54}$$

and extended the construction of hairy black holes to the time-dependent background. The metric solution in this case was found to be non-stealth.

It was also investigated in [568, 569] whether these time-dependent backgrounds could also be solutions of Horndeski theories incorporating $G_3$ such as the Cubic galileon, given by

$$G_2 = -2\Lambda\zeta + 2\eta X, \qquad G_3 = -2\gamma X \qquad G_4 = \zeta. \tag{3.55}$$

This choice breaks the reflection symmetry of the previously studied theories but keeps the shift symmetry. However, no exact stealth black holes solutions of the form of (3.51) were found. Instead, some approximated analytical expressions were given for different asymptotics, which we will not discuss here.

A breaking of shift-symmetry was also studied in [570], finding that solutions of the form (3.51) cannot exist there. This is no longer true, however, in beyond Horndeski theories, where such background configurations were found

36: To check this, note that the $\beta$-term contributes to $G_{4X} = \beta$. One can then use the Ricci identity of commuting covariant derivatives alongside integration by parts to show that

$$G_{4X}\left[(\Box\phi)^2 - \phi^{\mu\nu}\phi_{\mu\nu}\right] = \beta R^{\mu\nu}\phi_\mu\phi_\nu. \tag{3.53}$$

Combining this with $G_4 R \ni -\beta\frac{1}{2}g^{\mu\nu}R\phi_\mu\phi_\nu$ gives us the coupling to the Einstein tensor.

37: Other non-stealth black hole solutions were also found for this Horndeski subclass.

to be solutions of shift-symmetric [571, 572] and shift-symmetry breaking quadratic DHOST [573].

**Stability**

Despite being exact solutions to a considerably large class of scalar-tensor theories, backgrounds of the form of (3.51) were quickly shown to be prone to instability issues. Odd parity perturbations were initially argued in [574] to possess either ghost or gradient instabilities close to the black hole horizon. It was later pointed out in [575] that this statement, made on the unboundedness of the Hamiltonian density, was coordinate-dependent and therefore not a good stability criterion, which should be independent of coordinates. In fact, it was shown there that stability could be attained in some time slicings for odd parity modes and even parity modes with $\ell = 0$. The assumption of reflection symmetry was dropped from the stability analysis in [576], thus including also the non-stealth solutions found in [569], and it was found that odd parity perturbations could be stable in some subclasses of shift-symmetric Horndeski. This analysis was then extended to the shift- and reflection-symmetric subclass of DHOST theories in [577], showing that odd parity perturbations there are stable and propagate with the same effective potential as in GR, i.e. the Regge-Wheeler potential.

However, higher-$\ell$ even modes for the same stealth solutions in DHOST were then investigated in [578], finding that they were unfortunately plagued with instability or strong coupling issues. In particular, it was found that even modes coming from scalar perturbations were always everywhere unstable (c.f. to just close to the horizon in [574]). This result was later confirmed in [579], which also found even gravitational modes to suffer from instabilities or be strongly coupled. This then leaves us with no known well-behaved exact stealth black hole solutions with a linearly time-dependent scalar. Being $X = const$ a key requirement for the results of [578] to hold, the possibility of stable non-constant $X$ solutions was investigated in [2], where odd parity perturbations were shown to be stable for some region of parameter space, and deviations in the QNM spectrum were computed.

In the context of non-exact stealth black hole solutions, [580] looked at static solutions with a non-constant $X$ and their stability against odd parity perturbations, finding that some configurations can be stable. When including time-dependence, however, no stable configuration was found.

## 3.5  Black hole perturbation theory in scalar-tensor gravity

Before entering the *Investigations* part of this thesis, it is instructive to study a simple example of black hole perturbations in ST theories. This will enable us to introduce some key features of such computations. With this aim, let us consider the following perturbations on the metric and scalar fields

$$g_{\mu\nu} = \bar{g}_{\mu\nu} + h_{\mu\nu}, \qquad\qquad \phi = \phi_0 + \delta\phi. \qquad (3.56)$$

Here, we separate the quantities into the background $\{\bar{g}_{\mu\nu}, \phi_0\}$ and perturbations $\{h_{\mu\nu}, \delta\phi\}$ as shown in Table 3.2, where it is also highlighted that scalar perturbations are purely of even parity.

| Field | Background | Perturbations | |
|---|---|---|---|
| | | Odd | Even |
| Metric | $\bar{g}_{\mu\nu}$ | $h_{\mu\nu}^{\mathrm{odd}}$ | $h_{\mu\nu}^{\mathrm{even}}$ |
| Scalar | $\phi_0$ | ✗ | $\delta\phi$ |

Henceforth, by *background solution* we will mean the set $\{\bar{g}_{\mu\nu}, \phi_0\}$. ST theories thereby expand the range of possibilities for studying black hole perturbations: one must specify not only a background metric, such as those listed in Table 2.1, but also a scalar background configuration. Moreover, the background metric in ST theories can itself differ from the GR solutions shown in Table 2.1. As for the scalar background, the range of alternatives is somewhat more constrained. The simplest option–which we will explore in this section–is a constant scalar field, $\phi_0 = constant$. As we will see, this choice significantly simplifies the perturbation analysis, to the extent that the dynamics of metric perturbations closely resemble those in GR. As discussed in Section 3.4, more general scalar profiles, such as spacelike $X_0 < 0$ or timelike $X_0 > 0$, can lead to much richer phenomenology, and will be studied in the following chapters. However, as previously mentioned, these more general cases can sometimes lead to overly unconstrained dynamics, resulting in fully unstable and hence unphysical solutions.

### 3.5.1 Schwarzschild black hole with a constant scalar field in Horndeski

Let us here investigate the black hole perturbations in Horndeski gravity– with the action (3.19)–where the background solution is given by a stealth Schwarzschild GR metric with a constant scalar field, $\phi_0 = constant$, meaning that the kinetic term evaluated on the background is $X_0 = 0$.[38] Within this simplified set up, the metric and scalar equations of motion become

$$G_4 G_{\mu\nu} - \frac{1}{2} g_{\mu\nu} G_2 = 0, \qquad G_{2\phi} + R G_{4\phi} = 0, \qquad (3.57)$$

where the $G$-functions are evaluated in the background.[39] We see that in this setting, only $G_2$ and $G_4$ contribute to the equations of motion and, in the GR limit, this clearly recovers EFEs. Choosing to work with a stealth Ricci-flat metric (such as Schwarzschild), where $R_{\mu\nu} = R = 0$ on the background, the equations of motion give us the following constraint equations

$$G_2 = G_{2\phi} = 0. \qquad (3.59)$$

Perturbing the action to quadratic order in perturbations, i.e. up to two powers of the perturbative fields $\{h_{\mu\nu}, \delta\phi\}$ appearing together, we initially find a quadratic Lagrangian given by

$$\frac{8\mathscr{L}^{(2)}}{\sqrt{-g}} = G_2\left(h^2 - 2h_{\mu\nu}h^{\mu\nu}\right)$$

$$+ 2G_4\left[2R_{\mu\nu}\left(2h^{\mu\sigma}h_\sigma^\nu - hh^{\mu\nu}\right) - R\left(h_{\mu\nu}h^{\mu\nu} - \frac{1}{2}h^2\right) + \nabla_\mu h(\nabla^\mu h - 2\nabla_\nu h^{\mu\nu}) + \nabla_\sigma h_{\mu\nu}(2\nabla^\nu h^{\sigma\mu} - \nabla^\sigma h^{\mu\nu})\right]$$

$$+ 4(G_{2\phi\phi} + RG_{4\phi\phi})\delta\phi^2 - 4G_{2X}\nabla_\mu\delta\phi\nabla^\mu\delta\phi - 8G_{3\phi}\delta\phi\Box\delta\phi + 8G_{5\phi}G^{\mu\nu}\delta\phi\nabla_\mu\nabla_\nu\delta\phi$$

$$+ 4G_{4X}\left[2(\Box\delta\phi)^2 - 2\nabla_\mu\nabla_\nu\delta\phi\nabla^\mu\nabla^\nu\delta\phi - R\nabla_\mu\delta\phi\nabla^\mu\delta\phi\right]$$

$$+ 4G_{2\phi}h\delta\phi - 4G_3\left[(\nabla_\mu h - 2\nabla_\nu h_\mu^\nu)\nabla^\mu\delta\phi - 2h^{\mu\nu}\nabla_\mu\nabla_\nu\delta\phi + h\Box\delta\phi\right] + 8G_{4\phi}(\nabla_\mu\nabla_\nu h^{\mu\nu} - \Box h - G_{\mu\nu}h^{\mu\nu})\delta\phi$$

$$+ 4G_5\left[R_{\mu\nu}\left(h\nabla^\mu\nabla^\nu\delta\phi + h^{\mu\nu}\Box\delta\phi\right) + R\left(h^{\mu\nu}\nabla_\mu\nabla_\nu\delta\phi - \frac{1}{2}h\Box\delta\phi\right) + G_{\mu\nu}\left((\nabla^\sigma h^{\mu\nu} - 2\nabla^\nu h^{\mu\sigma})\nabla_\sigma\delta\phi - 4h^{\mu\sigma}\nabla_\sigma\nabla^\nu\delta\phi\right)\right.$$

$$\left. + \left(\Box h - \nabla_\mu\nabla_\nu h^{\mu\nu}\right)\Box\delta\phi + \left(2\nabla_\sigma\nabla_\mu h_\nu^\sigma - \nabla_\mu\nabla_\nu h - \Box h\right)\nabla^\mu\nabla^\nu\delta\phi\right].$$

$$(3.60)$$

However, we expect many of these terms to cancel out under closer inspection, as we have seen that only $G_2$ and $G_4$ appear in the equations of motion. The perturbed field equations to linear order should be indistinguishable independently of how they are derived. As we have seen exemplified in Section 2.4.1, one can perturb the background equations of motion to first order, or equivalently obtain directly the perturbed equations of motion from the quadratic action. The equivalence between the two methods tells us that we should expect many of the terms in (3.60) to vanish. In fact, by restricting to Ricci-flat metrics, we can employ the constraints (3.59) on the action above, thus eliminating the first line and the first term in the fifth line.[40] In a similar vein, restricting at this point to a Ricci-flat background means we can substitute $G_{\mu\nu} = R_{\mu\nu} = R = 0$ back in the quadratic Lagrangian, considerably reducing the terms appearing in it.

Let us now look at the $G_3$ term, which can be shown to vanish up to integration by parts.

$$G_3\left[(\nabla_\mu h - 2\nabla_\nu h_\mu^\nu)\nabla^\mu\delta\phi - 2h^{\mu\nu}\nabla_\mu\nabla_\nu\delta\phi + h\Box\delta\phi\right]$$

$$= G_3\left[-\cancel{h\Box\delta\phi} + \cancel{2h^{\mu\nu}\nabla_\mu\nabla_\nu\delta\phi} - \cancel{2h^{\mu\nu}\nabla_\mu\nabla_\nu\delta\phi} + \cancel{h\Box\delta\phi}\right] = 0, \qquad (3.61)$$

40: Note that $G_{2\phi} = 0$ does not imply $G_{2\phi\phi} = 0$, as is exemplified by the function $G_2 = \phi\phi_0 - \frac{1}{2}\phi^2$.

**Integration by parts**

In order to simplify a quadratic action, integration by parts is widely used as a technique to move derivatives around between fields to rewrite different terms in the same form and eliminate total derivative terms. Integration by parts directly follows from the product rule (or Leibniz rule) for derivatives. I covariant form, for any two generic fields $A$ and $B$ one can express this rule as

$$\nabla_\mu(AB) = \nabla_\mu A \cdot B + A \cdot \nabla_\mu B. \qquad (3.62)$$

Let us see how integration by parts is actually implemented for one of the terms in the quadratic action (3.60)

$$-\int d^4x \sqrt{-g}\, G_{3\phi}\delta\phi \Box \delta\phi = -\int d^4x \sqrt{-g}\, G_{3\phi}\delta\phi \nabla_\mu \nabla^\mu \delta\phi$$

$$= \int d^4x \sqrt{-g}\Big[ G_{3\phi}\nabla_\mu \delta\phi \nabla^\mu \delta\phi + \underline{\nabla_\mu G_{3\phi}\delta\phi \nabla^\mu \delta\phi} - \overbrace{\nabla_\mu\big(G_{3\phi}\delta\phi\nabla^\mu\delta\phi\big)} \Big]$$

$$= \int d^4x \sqrt{-g}\, G_{3\phi}\nabla_\mu \delta\phi \nabla^\mu \delta\phi. \tag{3.63}$$

The term $\nabla_\mu G_{3\phi}$ cancels due to the constant field nature of our background, i.e.

$$\nabla_\mu G_{3\phi} = G_{3\phi\phi}\phi_\mu + G_{3\phi X}\phi^\nu \phi_{\nu\mu} = 0. \tag{3.64}$$

This applies to all $G$-functions in the quadratic Lagrangian. The last term $\nabla_\mu\big(G_{3\phi}\delta\phi\nabla^\mu\delta\phi\big)$ cancels because it is a total derivative, which can be rewritten as a boundary integral using Gauss's theorem (also known as the divergence theorem). Employing the principle of least action, we vary the fields and can discard boundary terms assuming that the fields and their variations vanish at the boundary.

Looking now at the $G_{4X}$ term, we see that using a combination of integration by parts and the Ricci identity, we can rewrite one of its terms as

$$-2G_{4X}\nabla_\mu\nabla_\nu \delta\phi \nabla^\mu\nabla^\nu\delta\phi = 2G_{4X}\nabla^\mu\nabla_\mu\nabla_\nu\delta\phi\nabla^\nu\delta\phi$$

$$= 2G_{4X}\nabla_\mu\nabla_\nu\nabla^\mu\delta\phi\nabla^\nu\delta\phi$$

$$= 2G_{4X}\Big[\nabla_\nu\nabla_\mu\nabla^\mu\delta\phi\nabla^\nu\delta\phi + R_{\mu\nu}\nabla^\mu\delta\phi\nabla^\nu\delta\phi\Big]$$

$$= G_{4X}\Big[ -2\nabla_\mu\nabla^\mu\delta\phi\nabla_\nu\nabla^\nu\delta\phi + 2G_{\mu\nu}\nabla^\mu\delta\phi\nabla^\nu\delta\phi + R\nabla_\mu\delta\phi\nabla^\mu\delta\phi\Big], \tag{3.65}$$

meaning the whole $G_{4X}$ term can be rewritten as

$$G_{4X}\Big[2(\Box\delta\phi)^2 - 2\nabla_\mu\nabla_\nu\delta\phi\nabla^\mu\nabla^\nu\delta\phi - R\nabla_\mu\delta\phi\nabla^\mu\delta\phi\Big] = 2G_{4X}G_{\mu\nu}\nabla^\mu\delta\phi\nabla^\nu\delta\phi, \tag{3.66}$$

and therefore it is shown to vanish for Ricci-flat backgrounds.

**Ricci and Bianchi identities**

The Ricci identities express how covariant derivatives fail to commute due to the curvature of spacetime. Starting with a generic scalar field, covariant derivatives are known to commute as

$$[\nabla_\mu, \nabla_\nu]A \equiv \nabla_\mu\nabla_\nu A - \nabla_\nu\nabla_\mu A = 0. \tag{3.67}$$

In our context, this is equivalent to the symmetry $\phi_{\mu\nu} = \phi_{\nu\mu}$, or equivalently $\phi_{[\mu\nu]} = 0$. For a vector (and co-vector) field, we instead have the Ricci identity

$$[\nabla_\mu, \nabla_\nu]A^\alpha = R^\alpha_{\sigma\mu\nu}A^\sigma, \qquad [\nabla_\mu, \nabla_\nu]A_\beta = -R^\sigma_{\beta\mu\nu}A_\sigma. \tag{3.68}$$

More generally, the Ricci identity can be written for any tensor as

$$[\nabla_\mu, \nabla_\nu] A^{\alpha_1 \ldots \alpha_r}{}_{\beta_1 \ldots \beta_s} = R^{\alpha_1}{}_{\sigma \mu \nu} A^{\sigma \ldots \alpha_r}{}_{\beta_1 \ldots \beta_s} + \cdots + R^{\alpha_r}{}_{\sigma \mu \nu} A^{\alpha_1 \ldots \sigma}{}_{\beta_1 \ldots \beta_s}$$
$$- R^{\sigma}{}_{\beta_1 \mu \nu} A^{\alpha_1 \ldots \alpha_r}{}_{\sigma \ldots \beta_s} - \cdots - R^{\sigma}{}_{\beta_s \mu \nu} A^{\alpha_1 \ldots \alpha_r}{}_{\beta_1 \ldots \sigma}. \tag{3.69}$$

Bianchi identities express symmetries of the Riemann tensor. The first identity is an algebraic relation relating permutations in the indices

$$R^\rho{}_{[\mu \nu \sigma]} \equiv R^\rho{}_{\mu \nu \sigma} + R^\rho{}_{\nu \sigma \mu} + R^\rho{}_{\sigma \mu \nu} = 0. \tag{3.70}$$

The second Bianchi identity also relates different index permutations, now in a differential way

$$\nabla_{[\lambda} R^\rho{}_{|\sigma| \mu \nu]} \equiv \nabla_\lambda R^\rho{}_{\sigma \mu \nu} + \nabla_\mu R^\rho{}_{\sigma \nu \lambda} + \nabla_\nu R^\rho{}_{\sigma \lambda \mu} = 0. \tag{3.71}$$

Contracting the second identity leads to the conservation of the Einstein tensor $\nabla^\mu G_{\mu \nu} = 0$.

Let us now move to the more cumbersome $G_5$ term. After restricting to Ricci-flat backgrounds we simply need to inspect the last line in (3.60). Again, integrating by parts multiple times we can move all the derivatives away from the scalar perturbation, and implementing the Ricci and Bianchi identities we get (now skipping more intermediate steps)

$$G_5 \Big[ \left( \Box h - \nabla_\mu \nabla_\nu h^{\mu \nu} \right) \Box \delta \phi + \left( 2 \nabla_\sigma \nabla_\mu h^\sigma_\nu - \nabla_\mu \nabla_\nu h - \Box h \right) \nabla^\mu \nabla^\nu \delta \phi \Big]$$
$$= G_5 \nabla^\mu \delta \phi \Big[ \nabla_\sigma \nabla_\nu \nabla_\mu h^{\nu \sigma} - \nabla_\sigma \nabla^\sigma \nabla_\nu h^\nu_\mu - 2 \nabla^\rho R_{\mu \nu \sigma \rho} h^{\nu \sigma} - R_{\mu \nu \sigma \rho} \nabla^\rho h^{\nu \sigma} \Big]$$
$$= G_5 \delta \phi \Big[ - \nabla^\mu \nabla_\sigma \nabla_\nu \nabla_\mu h^{\nu \sigma} + \nabla^\mu \nabla_\sigma \nabla^\sigma \nabla_\nu h^\nu_\mu + 2 \nabla^\mu (\nabla^\rho R_{\mu \nu \sigma \rho} h^{\nu \sigma}) + \nabla^\mu (R_{\mu \nu \sigma \rho} \nabla^\rho h^{\nu \sigma}) \Big]$$
$$= G_5 \delta \phi \Big[ - \nabla^\mu \nabla_\sigma \nabla_\nu \nabla_\mu h^{\nu \sigma} + \nabla^\mu \nabla_\sigma \nabla^\sigma \nabla_\nu h^\nu_\mu + 2 \nabla^\mu ((\nabla_\mu R_{\sigma \nu} - \nabla_\nu R_{\sigma \mu}) h^{\nu \sigma})$$
$$+ (\nabla_\sigma R_{\mu \nu} - \nabla_\nu R_{\mu \sigma}) \nabla^\sigma h^{\mu \nu} + R_{\mu \nu \sigma \rho} \nabla^\mu \nabla^\rho h^{\nu \sigma} \Big]. \tag{3.72}$$

Now we can employ Ricci and Bianchi identities to rewrite the second term in the last line as

$$\nabla^\mu \nabla_\sigma \nabla^\sigma \nabla_\nu h^\nu_\mu = \nabla^\mu \nabla_\sigma \nabla_\nu \nabla_\mu h^{\nu \sigma}$$
$$+ \Box R_{\rho \nu} h^{\rho \nu} - 2 \nabla^\sigma \nabla^\nu R_{\rho \sigma} h^\rho_\nu + 2 \nabla_\sigma R_{\rho \nu} \nabla^\sigma h^{\rho \nu} - 3 \nabla_\nu R_{\rho \sigma} \nabla^\sigma h^{\rho \nu}$$
$$+ R_{\mu \rho \sigma \nu} \nabla^\sigma \nabla^\mu h^{\rho \nu} - \nabla^\sigma R_{\rho \sigma} \nabla^\mu h^\rho_\mu - R_{\rho \sigma} \nabla^\sigma \nabla^\mu h^\rho_\mu. \tag{3.73}$$

Hence, we see that the two first terms in the last equality line in (3.72) can be made to cancel out up to some Ricci-flat vanishing terms. Putting together everything, after a few more manipulations one can show that

$$G_5 \Big[ \left( \Box h - \nabla_\mu \nabla_\nu h^{\mu \nu} \right) \Box \delta \phi + \left( 2 \nabla_\sigma \nabla_\mu h^\sigma_\nu - \nabla_\mu \nabla_\nu h - \Box h \right) \nabla^\mu \nabla^\nu \delta \phi \Big]$$
$$= -G_5 \Big[ \Box R_{\mu \nu} h^{\mu \nu} + \nabla_\mu G_{\nu \sigma} \nabla^\sigma h^{\mu \nu} + R_{\mu \nu} \nabla^\nu \nabla^\sigma h^\mu_\sigma \Big], \tag{3.74}$$

which vanishes for Ricci-flat backgrounds.

Finally, looking now at the $G_{5\phi}$ term, it cannot be made to vanish by any of the techniques used above. However, it can instead be absorbed into the other $G$-functions, due to the fact that the quadratic Lagrangian contains no $G_{5X}$-explicit terms. In order to show this, let us begin by assuming $G_5$ has no $X$ dependence (i.e. $G_5 = G_5(\phi)$). Then,[41]

$$
\begin{aligned}
\mathscr{L}_5 &\equiv G_5(\phi)G_{\mu\nu}\phi^{\mu\nu} = -\nabla^\mu G_5(\phi)G_{\mu\nu}\phi^\nu = -G_{5\phi}(\phi)G_{\mu\nu}\phi^\mu\phi^\nu \\
&= -XG_{5\phi}(\phi)R - G_{5\phi}(\phi)R_{\mu\nu}\phi^\mu\phi^\nu \\
&= -XG_{5\phi}(\phi)R - G_{5\phi}\phi^\mu(\phi_{\mu\nu}{}^\nu - \phi^\nu{}_{\nu\mu}) \\
&= -XG_{5\phi}(\phi)R - G_{5\phi}\left[(\Box\phi)^2 - \phi^{\mu\nu}\phi_{\mu\nu}\right] + 2XG_{5\phi\phi}\Box\phi + G_{5\phi\phi}\phi_\nu\phi^\mu\phi^\nu{}_\mu \\
&= -XG_{5\phi}(\phi)R - G_{5\phi}\left[(\Box\phi)^2 - \phi^{\mu\nu}\phi_{\mu\nu}\right] + 3XG_{5\phi\phi}\Box\phi - 2X^2 G_{5\phi\phi\phi},
\end{aligned}
$$
(3.75)

where we have used the conservation of the Einstein tensor, i.e. $\nabla^\mu G_{\mu\nu} = 0$ and the Ricci identity. Now, redefining the original $G_i$ functions as

$$
\tilde{G}_2 \equiv G_2 - 2X^2 G_{5\phi\phi\phi}, \qquad \tilde{G}_3 \equiv G_3 + 3XG_{5\phi\phi}, \qquad \tilde{G}_4 \equiv G_4 - XG_{5\phi}, \quad (3.76)
$$

the full $\mathscr{L}_5$ Lagrangian with $G_5 = G_5(\phi)$ is reabsorbed into $\mathscr{L}_2$, $\mathscr{L}_3$ and $\mathscr{L}_4$.[42] In practice, this means that the $G_{5\phi}$ term in (3.60) is redundant, and hence can be reabsorbed into the other terms.

This all leaves us with the covariant quadratic Lagrangian

$$
\begin{aligned}
\frac{8\mathscr{L}^{(2)}}{\sqrt{-g}} =& G_4\left[\nabla_\mu h(\nabla^\mu h - 2\nabla_\nu h^{\mu\nu}) + \nabla_\sigma h_{\mu\nu}(2\nabla^\nu h^{\sigma\mu} - \nabla^\sigma h^{\mu\nu})\right] \\
&+ 2G_{2\phi\phi}\delta\phi^2 + 4\left(2G_{3\phi} - G_{2X}\right)\nabla_\mu\delta\phi\nabla^\mu\delta\phi \\
&+ 4G_{4\phi}(\nabla_\mu\nabla_\nu h^{\mu\nu} - \Box h)\delta\phi.
\end{aligned}
$$
(3.77)

This reproduces the result in Equation (9) in [213], with the exception of a non-vanishing $G_{4X}$ term there, which was later acknowledged by tha authors to be vanishing [454]. We see that the quadratic Lagrangians for the metric and scalar perturbations look quite straightforward, with the exception of the kinetic mixing between them given by the $G_{4\phi}$ term. By applying a transformation to the metric perturbation, it can be uncoupled from the scalar perturbation. In practice, we transform

$$
h^{\mu\nu} \to h^{\mu\nu} - \frac{G_{4\phi}}{G_4}g^{\mu\nu}\delta\phi,
$$
(3.78)

which results in the quadratic Lagrangian

$$
\begin{aligned}
\frac{8\mathscr{L}^{(2)}}{\sqrt{-g}} =& G_4\left[\nabla_\mu h(\nabla^\mu h - 2\nabla_\nu h^{\mu\nu}) + \nabla_\sigma h_{\mu\nu}(2\nabla^\nu h^{\sigma\mu} - \nabla^\sigma h^{\mu\nu})\right] \\
&+ C\delta\phi(\Box - \mu^2)\delta\phi,
\end{aligned}
$$
(3.79)

where[43]

$$
\mu^2 = -\frac{G_{2\phi\phi}}{C}, \qquad\qquad C = G_{2X} - 2G_{3\phi} + 3\frac{G_{4\phi}^2}{G_4}. \quad (3.82)
$$

Now the equation of motion for the scalar results in a Klein-Gordon equation with a mass term, which, as shown in detail in Subsection 27, can be written in component form as

$$
\left[\frac{d^2}{dr_*^2} + (\omega_{\ell m}^2 - V_s)\right]\varphi = 0, \qquad V_s = f\left(\frac{\ell(\ell+1)}{r^2} + \frac{2M}{r^3} + \mu^2\right), \quad (3.83)
$$

where $\varphi$ is defined in (2.61).

Hence, we see that for this simplified setting with a constant background scalar field in a Schwarzschild black hole, Horndeski gravity boils down to an additional parameter governing black hole perturbations, namely the effective mass term for the scalar $\mu^2$. This was first studied in [213, 454], where QNMs were calculated and forecasted constraints derived on $\mu$. However, as previously mentioned, it was initially thought that another parameter, named $\Gamma$ in [213], sourced by the $G_{4X}$ term, also played a role. Here we have provided a detailed exposition of why that is not the case, and shown $\mu$ to be the only non-trivial beyond-GR parameter.

# Investigations

# Speed of gravity | 4

Ka-chow!

——————————————

Lightning McQueen

**Chapter summary**

In this Chapter we investigate how the speed of gravitational waves, $c_{GW}$, can be tested by upcoming black hole ringdown observations. We do so in the context of hairy black hole solutions, where the hair is associated with a new scalar degree of freedom, forecasting that LISA and Tian-Qin will be able to constrain deviations of $c_{GW}$ from the speed of light at the $O(10^{-2})$ level from a single supermassive black hole merger. We collect and discuss the relevant results from black hole perturbation theory in Section (§4.2, both for 'bald' and 'hairy' black hole solutions. We extract the observable quasinormal spectrum from the relevant solutions in Section (§4.3), discussing issues related to the parametrisation of $\alpha_T$ in the process. Parametrised constraints are then presented in Section (§4.4), where we analytically compute the precision with which upcoming ringdown observations will be able to probe $\alpha_T$ for a generic observation. We discuss correlations between different constrainable parameters and how constraints depend on the underlying interactions. Forecasted constraints for a range of specific missions and experiments are then discussed in Section (§4.5). Finally, we will summarise conclusions in Section (§6.6).

This Chapter is based on the findings in [1], and all calculations shown here are reproducible in the companion `Mathematica` notebook [4].

## 4.1 Introduction

Measuring the speed of gravitational waves, $c_{GW}$, places strong constraints on the 'medium' gravitational waves are propagating through and hence on the particle content of the Universe. In the strong gravity regime, binary compact object mergers – e.g. binary black hole (BBH) or binary neutron star (BNS) mergers – are one of the cleanest probes of this particle content. Here interactions associated with novel particles can leave an imprint in the inspiral, merger and ringdown phases. These systems can therefore act as a particle detector, identifying or constraining the new physics that would be a consequence of such particles and associated 'fifth forces'.

One of the smoking gun signals for the presence of such new physics is a $c_{GW}$ different from the speed of light, and indeed it has been shown that binary compact object mergers can place powerful constraints on $c_{GW}$ [538, 581–584] and hence on the presence and potential dynamics of new degrees of freedom – see e.g. [395, 541–543, 585–598] and references therein. These previous constraints on $c_{GW}$ from binary compact object mergers have mostly focused on propagation effects (see e.g. [541–543, 585] and references therein)

1: The same is true for bounds from (the absence of) gravitational Cherenkov radiation [599], which place a lower bound on $c_{GW}$ at energy scales of order $\sim 10^{10}$ GeV, i.e. far above the energy scales probed by gravitational wave detectors.

or emission effects during the inspiral phase of such systems (see e.g. [595]).

In this Chapter we instead investigate what bounds can be derived on $c_{GW}$ from the ringdown phase alone. This phase is particularly amenable to being understood perturbatively and hence promises an especially clean analytic understanding. While strong constraints on $c_{GW}$ exist, most notably from the binary neutron star merger GW170817, it is important to keep in mind that these are for frequencies in the LIGO-Virgo-KAGRA (LVK) band, i.e. $\sim 20 - 2000$ Hz [538, 581–584]. Expressed as an energy scale, this corresponds to $\sim 10^{-14} - 10^{-12}$ eV. This range of values is important, because dark energy-related physics is one of the primary targets that can be constrained with measurements of $c_{GW}$ and dark energy theories that do affect $c_{GW}$ generically come with a cutoff around $O(10^2)$ Hz [545]. This means that, for such theories, as one approaches the cutoff (close to the LVK band), a high energy completion of the theory ought to take over. This has been described in more detail in Figure 3.2. In short, one can expect luminality to be restored, i.e. $c_{GW} = c$, at high energies if the high energy completion permits Lorentz invariant solutions. In that case, GW170817 would be understood as a confirmation of such aspect of the high energy completion.[1] Hence, it is natural to expect a frequency-dependent transition back to $c_{GW} = c$ upon approaching the LVK band for theories that modify $c_{GW}$ at cosmological scales. As such, probing $c_{GW}$ at lower frequencies where it is allowed to deviate from luminality (both theoretically and observationally) constitutes a much cleaner probe of $c_{GW}$ in such dark energy related theories. This will be realised by third-generation detectors such as LISA [189] and TianQin [198], with frequencies in the $\sim 10^{-4} - 10^{-1}$ Hz band (and the corresponding energies $\sim 10^{-19} - 10^{-16}$ eV).

**Existing and upcoming constraints on** $c_{GW}$: Existing constraints on $c_{GW}$ from lower frequencies, i.e. frequencies below the LVK band, are comparatively weak, so it is particularly interesting to forecast constraints from and for LISA. Here, it is worth emphasising that a frequency-dependent $c_{GW}$, so e.g. different speeds in the LVK and LISA bands, is a generic consequence of the afore-mentioned dark energy theories – see [545, 600, 601] for more detailed discussions on this point. The existing relevant constraints closest to (in fact, below) the LISA band are from binary pulsars, in particular from the Hulse-Taylor binary, and place a bound of $|\alpha_T| \lesssim O(10^{-2})$ for frequencies $f \sim 10^{-5}$ Hz [595]. Here we have conveniently expressed bounds on $c_{GW}$ in terms of the dimensionless $\alpha_T$ parameter

$$\alpha_T \equiv (c_{GW}^2 - c^2)/c^2, \qquad (4.1)$$

which we will use throughout this Chapter. Bounds from even lower frequencies $f \sim 10^{-18} - 10^{-14}$ Hz come from the cosmic microwave background and large scale structure measurements (see [534, 602–626] and references therein) and require $|\alpha_T| \lesssim O(1)$. Finally there are already a number of $c_{GW}$-related forecasts for upcoming measurements in the LISA band:

1) [627] forecasted that a multi-messenger observation in the LISA band using observations of an eclipsing white-dwarf binary will be able to constrain $|\alpha_T| \lesssim 10^{-12}$ (in the event of a non-detection of any $\alpha_T$-related effect).

2) For the case when there is a significant frequency-dependence for $c_{GW}$ already within the LISA band, [600] used redshift-induced frequency dependence imprinted on waveforms to be observed in the LISA band (i.e. without the need for an optical counterpart) to forecast a constraint of $|\alpha_T| \lesssim 10^{-4}$.

3) Also for frequency-dependent $c_{GW}$ within the LISA band, [601] forecasted that a bound $|\alpha_T| \lesssim 10^{-17}$ can be placed by using the fact that waveforms to be observed by LISA will be squeezed/stretched/scrambled due to the different speeds with which different frequencies will propagate (for frequency-dependent $c_{GW}$ and again without the need for an optical counterpart).

4) Finally, if there is no detectable frequency-dependence in both the LISA or LVK bands individually, but a transition in between, [601, 628] showed that multiband observations using systems such as GW150914 – that are first observable in the LISA band and later enter the LVK band [629] – will constrain $|\alpha_T| \lesssim 10^{-15}$ (again in the event of a non-detection).[2]

Looking forward to upcoming LISA observations, this leaves us with the following situation when looking for the strongest possible upcoming bounds. If there is any significant frequency-dependence in the LISA band, a strong $|\alpha_T| \lesssim 10^{-17}$ bound will very quickly be established once a single sufficiently loud SMBH (super massive black hole) merger has been observed. No optical counterpart or multi-band observation is required for this. If no such frequency-dependence is present, multi-messenger events and multi-band observations will eventually place bounds at the $10^{-12}$ level and $10^{-15}$ level, respectively.[3] Here we will show that additional bounds at the $10^{-2}$ level can be derived from the ringdown phase of an observed SMBH merger. These bounds are model-dependent (we will detail how below), but will effectively be obtainable as soon as LISA goes online (given an expected $O(10 - 100)$ observable SMBH mergers per year [190−196] ). In the LISA context, these bounds are therefore most relevant in the event that no significant frequency-dependence is detectable within the LISA band itself, e.g. when $c_{GW}$ quickly asymptotes to a constant value for high and low frequencies and its frequency-dependence and transition between those asymptotes is effectively localised to a narrow band between the frequencies accessible by LISA and LVK. We will further discuss this setup – as pointed out above, this is the same basic setup as explored in [601, 628] in the context of multi-band observations – below, as well as how our analysis is affected when frequency-dependence leaks into the frequency band under investigation. As we will show, the ringdown bounds on $c_{GW}$ discussed here are also eventually expected to tighten by up to two orders of magnitude when stacking observations of multiple events. We will also highlight that such bounds are not just a complementary and independent constraint on $c_{GW}$, but the fact that they are derived for a different background space-time compared with constraints from gravitational wave propagation (black hole vs. cosmological space-times) also allows us to extract novel insights about the underlying physics.

**Scalar-tensor theories**: We will focus on theories where the fiducial new physics is minimal in the sense that it is described by a single scalar degree of freedom $\phi$, so that we are dealing with a scalar-tensor theory. The most general such theory which results in second-order equations of motion is commonly known as Horndeski scalar-tensor theory [355, 356]. This has been introduced within the broad context of scalar-tensor theories in Section 3.2.5. Its action is first stated in (3.19), but we repeat it here for con-

2: Several of the forecasted constraints listed here were computed considering single waveforms. For upcoming future detectors signal overlap will likely be a regular occurrence, so understanding to what extent this impacts the above bounds will be an interesting issue to explore going forward.

3: Note that the galactic eclipsing white dwarf binary considered in [627] is a known system which is expected to be clearly observable in LISA, whereas detection rates for multi-messenger events more akin to GW170817 (i.e. compact object mergers with a clearly identifiable optical counterpart that pinpoints the merger itself) are highly uncertain [630]. Multiband observations as discussed in [601, 628, 629] will take several years to constrain $c_{GW}$, given the signal has to 'migrate' from the LISA to the LVK frequency band for the constraint to arise.

venience:

$$S = \int d^4x \sqrt{-g} \Big[ G_2 + G_3 \Box\phi + G_4 R + G_{4X} \big[ (\Box\phi)^2 - \phi^{\mu\nu}\phi_{\mu\nu} \big] +$$
$$G_5 G_{\mu\nu} \phi^{\mu\nu} - \frac{1}{6} G_{5X} \big[ (\Box\phi)^3 - 3\phi^{\mu\nu}\phi_{\mu\nu}\Box\phi + 2\phi_{\mu\nu}\phi^{\mu\sigma}\phi_\sigma^\nu \big] \Big]. \qquad (4.2)$$

Recall that we have employed the shorthands $\phi_\mu \equiv \nabla_\mu \phi$ and $\phi_{\mu\nu} \equiv \nabla_\nu \nabla_\mu \phi$, and the $G_i$ are free functions of $\phi$ and $X$, where $X \equiv -\frac{1}{2}\phi_\mu \phi^\mu$. $G_{iX}$ denotes the partial derivative of $G_i$ with respect to $X$.

Most relevant for our purposes will be the ($X$-dependent parts of the) $G_4$ interactions and the $G_5$ interactions, since (as we will discuss below) these are the only interactions affecting $\alpha_T$. Also note that, for simplicity, we will often focus on the case where $G_4$ is $\phi$-independent, so that $G_{4\phi} = 0$ (where $G_{i\phi}$ denotes the partial derivative of $G_i$ with respect to $\phi$) – we will discuss in more detail what this assumption entails.

In a consistent effective field theory (EFT), the functions $G_i$ typically incorporate specific energy scales—often denoted $\Lambda_i^n$—which are associated with irrelevant operators and, in particular, with the number of derivatives acting on the scalar field or metric. These scales implicitly suppress the higher-dimensional operators in the action. For instance, $\Lambda_2$ is typically associated with the $G_2$ function, which governs terms involving no derivatives of $\phi$ (such as a scalar potential) or up to two derivatives (like a canonical kinetic term). Similarly, $\Lambda_3$ is related to the $G_3$ function, suppressing terms that are effectively quadratic in derivatives of $\phi$ through the $\Box\phi$ term. The lowest of these energy scales fundamentally signifies the intrinsic energy threshold where the current EFT ceases to be a complete description of physics, i.e. the cutoff. Below this scale, the theory provides a reliable description, but at or above it, the full underlying theory would be required to accurately describe phenomena. This is precisely analogous to the way in which GR is at most a valid description of gravitational phenomena up to the Planck scale. In the action (4.2), these fundamental energy scales are implicitly absorbed into the definitions of the $G_i$ functions themselves, and are generally assumed to be sufficiently high for the theory to remain valid within the frequency range of interest for black hole ringdown. For the purposes of this Chapter, particularly when (4.2) is considered as a fiducial dark energy theory influencing $c_{GW}$ on cosmological scales, the relevant scales become $\Lambda_2 \sim (M_{Pl}H_0)^{1/2}$ and $\Lambda_3 \sim (M_{Pl}H_0)^{1/3}$. These values are required for the dark energy contributions to the Friedmann equations to appear at the same order as the standard $\Lambda$CDM terms. The lowest of these is $\Lambda_3$, which then sets the cutoff of the EFT. Converting its value to frequency units, the cutoff is expected to be at or below $O(10^2)$ Hz. This energy scale marks the regime where new, unknown physics would become relevant and the full theory would need to be considered.

These observations have an important practical implication when computing BBH merger observables as we do here: $c_{GW}$ and hence $\alpha_T$ derived from (4.2) are in fact frequency-independent as a consequence of the structure of (4.2) imposed by the requirement of 2nd order equations of motion. The frequency-dependence of $c_{GW}$ alluded to above only enters as a consequence of the unknown UV (high energy) completion of (4.2), in other words once we are about to leave the regime of validity of (4.2). Throughout most of this Chapter we will compute and analyse ringdown predictions derived from (4.2), so we are implicitly assuming that we are operating within a frequency-window where I) (4.2) is firmly within its regime of validity, and hence II) $c_{GW}$ is effectively frequency-independent *within* this window. Rigorously computing analogous predictions in frequency-windows where there is significant frequency-dependence for $c_{GW}$ would require incorporating at least some of the effects of the UV completion and hence supplementing/replacing (4.2) with the relevant interactions. We will point out the implications of this assumption in more detail below as well as when one can extrapolate to more general scenarios.

## 4.2 Black hole perturbation theory

Since the ringdown phase of BBH mergers can be well-described perturbatively, we first ought to discuss the relevant setup in black hole perturbation theory. This has been introduced in detail in Section 2.4.1 but here we repeat its main aspects relevant for this analysis. We will consider static and spherically symmetric background solutions that are Ricci-flat ($R_{\mu\nu} = 0 = R$) here, in particular Schwarzschild spacetimes. We therefore write the background metric $\overline{g}_{\mu\nu}$ as

$$ds^2 = \overline{g}_{\mu\nu}dx^\mu dx^\nu = -A(r)dt^2 + \frac{1}{B(r)}dr^2 + C(r)d\Omega^2, \tag{4.3}$$

where $A, B, C$ are general functions of the radial coordinate $r$ and $d\Omega^2$ is the line element of the standard 2-sphere. We now consider metric perturbations $h$ around this background, where

$$g_{\mu\nu} = \overline{g}_{\mu\nu} + h_{\mu\nu}. \tag{4.4}$$

Around the static and spherically symmetric backgrounds considered here such perturbations can be decomposed into odd and even parity perturbations (under rotations), which decouple from one another at linear order (i.e. they evolve independently from one another and can therefore be treated separately). We will work to leading (linear) order in this Chapter, but note that this decoupling does not hold at higher orders – see [278–280, 631–633] for details on the behaviour of higher order modes. In this Chapter we will exclusively focus on odd perturbations, which can be written as

$$h_{\mu\nu}^{\text{odd}} = \begin{pmatrix} 0 & 0 & 0 & h_0 \\ 0 & 0 & 0 & h_1 \\ 0 & 0 & 0 & 0 \\ h_0 & h_1 & 0 & 0 \end{pmatrix} \sin\theta \partial_\theta Y_{\ell m} \tag{4.5}$$

where we have used the Regge-Wheeler gauge [201] and, since we assume a static background metric, we have set $m = 0$ without loss of generality. $h_0$ and $h_1$ are functions of $(r, t)$, where the $t$-dependence will be taken to be of the form $e^{-i\omega t}$. Since perturbations of the scalar $\phi$ are even under parity transformations and we focus on parity odd modes, we will therefore only be concerned with metric perturbations. These perturbations are however affected by the background solution they are propagating on, so odd metric perturbations will nevertheless be sensitive to the new physics encoded by the (background solution of the) fiducial scalar degree of freedom we are probing here.

### 4.2.1  Schwarzschild black holes without hair

4: For scalar-tensor theories, it was first shown that stationary black holes in minimally coupled Brans-Dicke theories contain no hair [115], a result which was extended to a more general class of scalar-tensor theories including self-interactions of the scalar [116], to spherically symmetric static black holes in Galilean-invariant theories [117], and for slowly rotating black holes in more general shift-symmetric theories [118].

No-hair theorems guaranteeing a trivial scalar field profile exist for a wide range of scalar-tensor theories [115–118]. These theorems have been broadly exposed in Section 15.[4]  A natural starting point are therefore Schwarzschild spacetimes with a constant scalar field background profile $\overline{\phi}$ as e.g. investigated by [213, 634, 635] and described in detail in Section 3.5

$$ds^2 = -\left(1 - \frac{2M}{r}\right)dt^2 + \frac{1}{\left(1 - \frac{2M}{r}\right)}dr^2 + d\Omega^2,$$

$$\overline{\phi} = \text{constant}. \tag{4.6}$$

Note that, when we mention the 'background' or 'background solution' going forward, we refer to *both* the metric *and* scalar background solutions, as e.g. provided in (4.6). Around the background (4.6) odd metric perturbations trivially behave just as in GR, since they are unaffected by the even sector (where scalar perturbations do induce non-trivial effects) and also do not feel any effects from the scalar background solution (since this is trivial in the present no-hair setup). So in order to explore potentially observable effects induced by the scalar, one ought to either investigate different background solutions or consider even perturbations. For detailed discussions of the second option we refer to [212, 213, 634, 636–643] for work in the context of Horndeski gravity, and to [212, 578, 644–648] for work in the context of other theories (scalar-tensor or otherwise). However, here we will proceed along the first route, considering the dynamics of odd perturbations around different background solutions. We leave an investigation of how the speed of gravity impacts quasi-normal modes in the even sector in the presence of a non-trivial background solution (i.e. combining the two options discussed above) for future work.

### 4.2.2  Hairy black holes: Background

If $\phi$ acquires a non-trivial background profile, this will provide a medium for gravitational waves (i.e. here in particular $h_{\mu\nu}^{\text{odd}}$) to travel through and hence can affect $c_{\text{GW}}$. Probing $c_{\text{GW}}$ therefore constitutes a powerful test for departures from GR in such cases, as neatly illustrated in the aforementioned cosmological context. Hairy black holes in scalar-tensor theories have been discussed in detail in Section 3.4. For the black hole solutions we focus on here, a well-known scalar-tensor theory example that can have scalar hair are scalar-Gauss-Bonnet (sGB) theories [118, 561]. In the context of Horndeski theories, these are described by an action that (in addition to a standard kinetic terms) contains a $G_5$ interaction where $G_5 \sim \ln|X|$ [391]. However, instead of focusing on a specific hairy solution, we will here follow the approach of [455] and parametrise the scalar-induced hair in a perturbative fashion, but otherwise remain agnostic about the precise nature of the hair. More specifically, we will consider a no-hair Schwarzschild black hole solution at lowest order and introduce small hairy deviations away from this. These can in principle manifest themselves both in the background solution for the metric as well as in the scalar profile, so [455] proposed the following parametrised ansatz

$$A(r) = B(r) = 1 - \frac{2M}{r} + \epsilon \delta A_1 + \epsilon^2 \delta A_2 + O(\epsilon^3),$$
$$C(r) = (1 + \epsilon \delta C_1 + \epsilon^2 \delta C_2)r^2 + O(\epsilon^3)$$
$$\overline{\phi} = \hat{\phi} + \epsilon \delta \phi_1 + \epsilon^2 \delta \phi_2 + O(\epsilon^3). \tag{4.7}$$

Here $\delta A_i, \delta C_i, \delta \phi_i$ are functions of $r$ and $\epsilon$ is simply a useful order parameter, since we will work perturbatively up to quadratic order in the (small) hair $\delta \phi$ – $\epsilon$ has no physical meaning beyond this.[5] Note that we will denote quantities which are evaluated on the background (so $h_{\mu\nu}$ is set to zero, recall odd scalar perturbations vanish identically) with a bar, so $\overline{\phi}$ denotes the scalar field as evaluated on the background. Quantities where in addition the small (background) scalar hair $\delta \phi_i$ is set to zero are denoted by a hat, so e.g. $\hat{\phi}$ denotes the scalar field as evaluated on the background in the absence of any non-trivial scalar hair. As a consequence we have $\hat{X} = 0$, while $\overline{X}$ here acquires non-zero contributions via the $\delta \phi_i$. While (4.7) is a very general parametrisation, for our purposes we will be able to work with a highly simplified subset. We are interested in probing the effect of $c_{GW}$ (or equivalently $\alpha_T$) on the ringdown phase. Since deviations from $c_{GW} = c$ (or equivalently $\alpha_T = 0$) arise due to a non-trivial scalar field profile acting as a medium for gravitational waves passing through, it is unsurprising that at lowest order in $\epsilon$ any $\alpha_T$-dependent contribution only depends on $\delta \phi_1$ in (4.7) and not on $\delta A_i, \delta C_i$, or $\delta \phi_2$. We collect results showing this explicitly in Equations (4.21) and (4.22), but here we will therefore proceed by working with the much simpler parametrised ansatz

$$ds^2 = -\left(1 - \frac{2M}{r}\right)dt^2 + \frac{1}{\left(1 - \frac{2M}{r}\right)}dr^2 + r^2 d\Omega^2,$$
$$\overline{\phi} = \hat{\phi} + \epsilon \delta \phi. \tag{4.8}$$

Recall that $\delta \phi$ is a small deviation in the background solution $\overline{\phi}$. This will allow us to identify the leading order contributions imprinted by a non-zero $\alpha_T$, so is ideally suited for our purposes. We will later discuss to what extent the constraints we will derive on $\alpha_T$ may be contaminated/weakened in the presence of non-zero $\delta A_i, \delta C_i$, but for now proceed with (4.8) as a proof of principle. However, do note that our simplified ansatz is (partially) motivated by sGB-like hair. There, when working perturbatively in a small sGB coupling, at leading order only the scalar background acquires a non-trivial contribution, while the metric remains Schwarzschild [118, 561],[6] i.e. we are working with a so-called 'stealth' solution for the metric.

### 4.2.3 Hairy black holes: Quadratic action

In order to extract the ringdown signal we need to compute the behaviour of (odd parity) perturbations on top of the background (4.8). Working out the quadratically perturbed action, substituting the components of (5.36) as well as our background solution (4.8), integrating over the angular coordinates and performing several integrations by parts, we recover the action [649, 650]

$$S^{(2)} = \int dt dr \left[ \bar{a}_1 h_0^2 + \bar{a}_2 h_1^2 + \bar{a}_3 \left( \dot{h}_1^2 + h_0'^2 - 2\dot{h}_1 h_0' + \frac{4}{r} \dot{h}_1 h_0 \right) \right], \qquad (4.9)$$

where a dot and a prime denote derivatives with respect to $t$ and $r$, respectively, and we have dropped an overall multiplicative factor of $2\pi/(2\ell+1)$ coming from angular integration. The expressions for the $\bar{a}_i$ agree with those found by [649, 650] and satisfy

$$\bar{a}_1 = \frac{\ell(\ell+1)}{4C} \left[ \left( C' \sqrt{\frac{B}{A}} \mathscr{H} \right)' + \frac{(\ell-1)(\ell+2)}{\sqrt{AB}} \mathscr{F} + \frac{2C}{\sqrt{AB}} \epsilon_A \right],$$

$$\bar{a}_2 = -\frac{\ell(\ell+1)}{2} \sqrt{AB} \left[ \frac{(\ell-1)(\ell+2)}{2C} \mathscr{G} + \epsilon_B \right],$$

$$\bar{a}_3 = \frac{\ell(\ell+1)}{4} \sqrt{\frac{B}{A}} \mathscr{H}, \qquad (4.10)$$

where the $\bar{a}_i$ are to be evaluated on the background (4.8) (to avoid clutter, bars are implied, but not written explicitly, for all expressions on the right hand side). $\epsilon_{A,B}$ are contributions that vanish on-shell,[7] and

$$\mathscr{F} = 2 \left( G_4 + \frac{1}{2} B \phi' X' G_{5X} - X G_{5\phi} \right),$$

$$\mathscr{G} = 2 \left[ G_4 - 2X G_{4X} + X \left( \frac{B'}{2} \phi' G_{5X} + G_{5\phi} \right) \right],$$

$$\mathscr{H} = 2 \left[ G_4 - 2X G_{4X} + X \left( \frac{B}{r} \phi' G_{5X} + G_{5\phi} \right) \right]. \qquad (4.11)$$

Note that, in our case of interest, i.e. a stealth Schwarzschild black hole, the expressions above can be simplified by substituting in $A = B$ and $C = r^2$. The quadratic action (5.41) contains two fields ($h_0, h_1$), but describes only one dynamical degree of freedom. [649, 650] show how the action can be rewritten to make this manifest. To do so an auxiliary field $q$ is defined and then re-defined into a field $Q$, satisfying[8]

$$h_0 = -\frac{(r^2 \bar{a}_3 q)'}{r^2 \bar{a}_1 - 2(r\bar{a}_3)'}, \qquad h_1 = \frac{\bar{a}_3}{\bar{a}_2} \dot{q}, \qquad q = \frac{\sqrt{\mathscr{F}}}{r \mathscr{H}} Q. \qquad (4.12)$$

Re-writing the quadratic action in terms of $Q$ in tortoise coordinates $r_*$ (defined as $dr = B dr_*$), one then finds

$$S^{(2)} = \frac{\ell(\ell+1)}{4(\ell-1)(\ell+2)} \int dt dr_* \left[ \frac{\mathscr{F}}{\mathscr{G}} \dot{Q}^2 - \left( \frac{dQ}{dr_*} \right)^2 - V(r) Q^2 \right], \qquad (4.13)$$

where the potential is given by

$$V = \ell(\ell+1) \frac{A}{C} \frac{\mathscr{F}}{\mathscr{H}} - \frac{C'^2}{4C''} \left( \frac{ABC'^2}{C^3} \right)'$$

$$- \frac{C^2 \mathscr{F}^2}{4\mathscr{F}'} \left( \frac{AB\mathscr{F}'^2}{C^2 \mathscr{F}^3} \right)' - \frac{2A\mathscr{F}}{C\mathscr{H}}. \qquad (4.14)$$

Using our background ansatz (4.8) ($A = B, C = r^2$) to simplify this expression, we can write the potential as

$$V = (\ell+2)(\ell-1) \frac{B}{r^2} \frac{\mathscr{F}}{\mathscr{H}} - \frac{r^3}{2} \left( \frac{B^2}{r^4} \right)' - \frac{r^4 \mathscr{F}^2}{4\mathscr{F}'} \left( \frac{B^2 \mathscr{F}'^2}{r^4 \mathscr{F}^3} \right)'. \qquad (4.15)$$

### 4.2.4 Modified Regge-Wheeler equation

In order to obtain the analogue of the Regge-Wheeler equation, we now vary the action with respect to the field $Q$ and find

$$\frac{\partial^2 Q}{\partial r_*^2} - \frac{\mathscr{F}}{\mathscr{G}} \frac{\partial^2 Q}{\partial t^2} - VQ = 0. \tag{4.16}$$

We assume that the time dependence of Q is given by $e^{-i\omega t}$, substitute $\mathscr{F}$, $\mathscr{G}$, $\mathscr{H}$ and $V$ for our background (4.8), and finally obtain the modified Regge-Wheeler equation

$$\frac{d^2 Q}{dr_*^2} + \left[\omega^2(1 + \epsilon^2 \alpha_T) - B(V_{RW} + \epsilon^2 \delta V)\right] Q = 0, \tag{4.17}$$

where $\alpha_T$ satisfies

$$\alpha_T = -B \frac{G_{4X} - G_{5\phi}}{G_4} \delta\phi'^2. \tag{4.18}$$

Note that, given the background we are considering, $\alpha_T$ naturally is a function of $r$ as well as of the Schwarzschild mass $M$. In general there are further contributions to $\alpha_T$ depending on $G_{5X}$, but these only enter at cubic order in $\epsilon$ as can be deduced from (5.43), so do not contribute here.[9] $V_{RW}$ is the well-known Regge-Wheeler potential in GR first introduced in Equation (2.58)[10]

$$V_{RW} = \frac{\ell(\ell+1)}{r^2} - \frac{6M}{r^3}, \tag{4.19}$$

and $\delta V$ is given by

$$\delta V = \alpha_T \left[ \frac{M(2r - 5M)}{r^3(r - 2M)} + \frac{(\ell+2)(\ell-1)}{r^2} \right.$$
$$\left. - \frac{r - 2M}{2r}\left(\left(\frac{\delta\phi''}{\delta\phi'}\right)^2 - \frac{\delta\phi'''}{\delta\phi'}\right) + \frac{r - 5M}{r^2}\frac{\delta\phi''}{\delta\phi'} \right]. \tag{4.20}$$

While it has been re-arranged into a more concise form here, this as well as the above expressions in this subsection agree with the corresponding results given in [455], when specialised to our ansatz (4.8). Note that we have implicitly assumed that $G_{4\phi} = 0 = G_{4\phi\phi}$ here, as would e.g. be the case in shift-symmetric theories. We do this to isolate the effect of $\alpha_T$ on the ringdown spectrum. Considering the full parametrised hair ansatz (4.7) [455], the modified Regge-Wheeler equation becomes

$$\left[\frac{d^2}{dr_*^2} + \omega^2\left(1 + \epsilon^2 \alpha_T(r)\right) - A(r)\left(\frac{\ell(\ell+1)}{r^2} - \frac{6M}{r^3} + \epsilon\delta V_1 + \epsilon^2 \delta V_2\right)\right] Q = 0, \tag{4.21}$$

where the potential perturbations are given by

9: This also implies that (4.18) can be simplified further by integrating the contributing $G_5$ terms by parts in the original theory, but as we will work with a general $\alpha_T$ here, this does not affect our subsequent expressions.

10: Note that in (2.58), the metric function $B$ (named $f$ there) is also included in the definition, while here this is not the case.

$$\delta V_1 = \frac{1}{2r^2}\Big[4\delta A_1 - 2r\delta A_1' - 2(\ell+2)(\ell-1)\delta C_1 + 2(r-3M)\delta C_1' - r(r-2M)\delta C_1''$$
$$- \frac{G_{4\phi}}{G_4}\big(r(r-2M)\,\delta\phi_1'' - 2(r-3M)\delta\phi_1'\big)\Big] \tag{4.22}$$

$$\delta V_2 = \frac{1}{4r^2}\Big[8\delta A_2 - 4r\delta A_2' + 4(\ell+2)(\ell-1)\big(\delta C_1^2 - \delta C_2\big) + 3r(r-2M)\delta C_1'^2 + 4(r-3M)\delta C_2' - 2r(r-2M)\delta C_2''$$
$$+ 4r\delta A_1\delta C_1' - 2r^2\big(\delta A_1'\delta C_1' + \delta A_1\delta C_1''\big) - 4(r-3M)\delta C_1\delta C_1'' + 2r(r-2M)\delta C_1\delta C_1''\Big]$$
$$- \frac{1}{2r^2}\frac{G_{4\phi}}{G_4}\Big[-2(r-3M)\delta\phi_2' + r\big(r\delta A_1'\delta\phi_1' - \delta A_1\big(2\delta\phi_1' - r\delta\phi_1''\big) + (r-2M)\big(\delta\phi_2'' - \delta C_1'\delta\phi_1'\big)\big)\Big]$$
$$+ \frac{1}{4r^2}\left(\frac{G_{4\phi}}{G_4}\right)^2\Big[3r(r-2M)\delta\phi_1'^2 + 2\delta\phi_1\big(r(r-2M)\delta\phi_1'' - 2(r-3M)\delta\phi_1'\big)\Big]$$
$$- \frac{1}{2r^2}\frac{G_{4\phi\phi}}{G_4}\Big[r(r-2M)\delta\phi_1'^2 + \delta\phi_1\big(r(r-2M)\delta\phi_1'' - 2(r-3M)\delta\phi_1'\big)\Big]$$
$$- \frac{\alpha_T(r)}{2r^3}\Big[-5M + Mr(r-2M)^{-1} - 2r(\ell+2)(\ell-1) + r^2(r-2M)\left(\frac{\delta\phi_1''}{\delta\phi_1'}\right)^2$$
$$+ r\left(r(r-2M)\frac{\delta\phi_1'''}{\delta\phi_1'} - 2(r-5M)\frac{\delta\phi_1''}{\delta\phi_1'}\right)\Big]. \tag{4.23}$$

There are several observations we can make from these expressions. Let us first point out that the modified Regge-Wheeler equation in (4.17) with (4.20) can be recovered by setting $\delta A_{1,2} = \delta C_{1,2} = G_{4\phi} = G_{4\phi\phi} = 0$. However, we see that $\delta A_1, \delta C_1$ would contribute at lower order (as well as $\delta\phi_1$, if $G_{4\phi} \neq 0$). If present, such terms can therefore significantly contribute to the ringdown signal. Indeed, if the fiducial scalar hair is highly suppressed, i.e. when $\epsilon$ is very small, such lower-order-in-$\epsilon$ contributions would be expected to dominate over any $\alpha_T$-induced contributions. The motivation behind our simple setup then is not to fully explore all parametric effects and degeneracies in a comprehensive parameter space, but rather to isolate and investigate observable signatures of $\alpha_T$. Having said this, it has been shown that for known scalar-tensor theories that have hairy black holes, namely scalar-Gauss-Bonnet theory, the metric at leading perturbative order remains Schwarzschild (i.e. $\delta A_1 = \delta C_1 = 0$) [561], suggesting that several aspects of our simple setup are concretely realised in relevant theories.

In addition, one could consider cases where $G_{4\phi} \neq 0 \neq G_{4\phi\phi}$. It is interesting to point out that interactions contributing to $G_{4\phi}$ in our background can be removed with a conformal transformation of the metric. This is because we still have $\hat{X} = 0$,[11] meaning that all terms in $G_{4\phi}$ are $X$-independent or, in other words, those interactions are of the form $f(\phi)R$ in Jordan frame, which is well known to be convertible to the usual Einstein-Hilbert term by a conformal transformation. In the resulting Einstein frame representation, one then naturally finds $G_{4\phi} = 0 = G_{4\phi\phi}$. This would indeed make our assumptions more general and, because the Horndeski group is closed under conformal transformations and we are not including matter fields, our calculations would follow exactly in the same way. The price to pay for working with the metric in the Einstein frame is that observations in GW detectors are coupled to matter and therefore measure the metric in Jordan frame. Hence, forecasts should ultimately be made for gravitational waves as observed in the detector/Jordan frame. We therefore abstain from removing any interactions via a conformal transformation here and explicitly high-

11: Recall that $\hat{X}$ refers to the kinetic term $X$ evaluated on the 'reference background' of $\hat{\phi}$ (i.e. $\delta\phi = 0$).

light setting $G_{4\phi} = 0 = G_{4\phi\phi}$ as an additional simplifying assumption. Note that this assumption is trivially satisfied in theories where the scalar $\phi$ is endowed with a shift symmetry $\phi \to \phi + c$.

In the case where many of $G_{4\phi}, G_{4\phi\phi}, \delta A_2, \delta C_2$ are non-vanishing, an analogous analysis to the one performed in this Chapter can still be carried out, where the functional form of any non-vanishing such functions would need to be specified as above. The additional parameters introduced in this way mean a higher parameter space would then have to be constrained, presumably degrading constraints on individual parameters. Breaking degeneracies in such a higher-dimensional parameter space would likely require the measurements of multiple QNMs. We leave such a more comprehensive exploration to future work.

## 4.3 Quasinormal modes

Having derived and collected the relevant results from black hole perturbation theory in the previous Section, we are now in a position to extract the key observable in the black hole ringdown context: the quasinormal modes (QNMs), see Section 2.4.2 for an in-depth introduction to QNMs. As before, we will be focusing on the perturbations of odd modes and the modified Regge-Wheeler equation (4.17) governing them. This equation can now be solved to obtain the frequencies of the associated quasinormal modes $\omega$. Recall that, unlike normal modes, these frequencies are complex numbers, where the real part represents the physical oscillation frequency and the imaginary part represents the exponential damping due to dissipation in the system. The QNM spectrum only depends on the properties of the final black hole (mass, angular momentum, charge) as well as on the structure of the underlying theory. Detecting and measuring this spectrum is hence a powerful way to constrain the presence of novel degrees of freedom and interactions, as well as to generally test the Kerr hypothesis [651],[12] a scheme that has received the name of black hole spectroscopy. Note that, while the QNM spectrum does not depend on initial conditions, the amplitude of individual modes does and this will be relevant for us in Section 4.4.

12: See Section 12 for a more detailed exposition of the hypothesis.

### 4.3.1 Parametrised ringdown

There are a number of techniques one can use to obtain the QNM themselves (see e.g. [231, 232, 243, 652–655]) but we refer to Section 43 for a review of those. In this Chapter, we will make use of the parametrised ringdown formalism [253], the relevant key aspects of which we will now summarise. In order to apply this formalism, our modified Regge-Wheeler equation has to be re-cast into the following form

[253]: Cardoso et al. (2019), "Parametrized black hole quasinormal ringdown: Decoupled equations for nonrotating black holes"

$$B\frac{d}{dr}\left(B\frac{dQ}{dr}\right) + \left[\omega^2 - B(V_{RW} + \delta\tilde{V})\right]Q = 0. \tag{4.24}$$

This can easily be achieved by absorbing the $\epsilon^2\omega^2\alpha_T$ term into $\delta V$ in (4.17). Because this term is a small correction to $\delta\tilde{V}$, we can take $\omega$ to be the unperturbed frequencies $\omega_0$ of the unmodified Regge-Wheeler equation characteristic of GR, around which we will compute the leading order $\delta\omega$ corrections below.[13] Doing so, we obtain a modified Regge-Wheeler equation in the form of (4.24) with

$$\delta \tilde{V} = \epsilon^2 \left( \delta V - \frac{1}{B} \omega_0^2 \alpha_T \right), \tag{4.25}$$

with $\delta V$ being given by (4.20). It is then instructive to express the modification to the potential $\delta \tilde{V}$ as an expansion in powers of $(2M/r)$

$$\delta \tilde{V} = \frac{1}{(2M)^2} \sum_{j=0}^{\infty} a_j \left( \frac{2M}{r} \right)^j. \tag{4.26}$$

Once expressed in this form, [253] show that the quasinormal frequencies are determined by the same $a_j$ coefficients as follows

$$\omega = \omega_0 + \delta\omega \equiv \omega_0 + \sum_{j=0}^{\infty} a_j e_j, \tag{4.27}$$

given that a smallness criterion on the coefficients $|a_j| \ll (1 + 1/j)^j (j+1)$ is satisfied. The $e_j$ are a complex 'basis' and we summarise the low-order $e_j$ most relevant here in Table 4.1 – for more details and an explicit computation of this basis see [253].

|        | Re($2Me_j$) | Im($2Me_j$) |
|--------|-------------|-------------|
| $j = 4$ | 0.03668    | –0.00044    |
| $j = 5$ | 0.02404    | 0.00273     |
| $j = 6$ | 0.01634    | 0.00484     |
| $j = 7$ | 0.01136    | 0.00601     |
| $j = 8$ | 0.00795    | 0.00654     |

**Table 4.1:** Real and imaginary components of the $e_j$ 'basis' functions for $\ell = 2$, taken from [253]. Note that we start with $j = 4$ as this is our lowest-order non-zero $a_j$. For the full collection up to $j = 50$ for each $\ell$ up to $\ell = 10$, together with the 'basis' for even-gravitational and even-scalar perturbations, see [253].

At this point we can already appreciate an important subtlety from the structure of equations (4.26) and (4.27). Each coefficient $a_j$ contributing to $\delta \tilde{V}$ enters with different (increasing) powers of $\sim 1/r$ (4.26). While this does mean that those contributions to the potential are suppressed in the far distance limit, i.e. far away from the horizon, it does not entail that these contributions are providing a sub-dominant contribution to the frequency spectrum for the QNMs. Indeed, from (4.27) we explicitly see that this $\sim 1/r^j$ suppression does not play a role in determining the QNM frequencies. The $j$-th correction enters as $a_j e_j$ and, while the $e_j$'s tend to slowly decrease in size as $j$ increases, there is no parametric suppression of higher $j$ contributions. Also note that situations where (some) higher $j$ contributions dominate over lower $j$ contributions do arise rather generically – we will see explicit examples below. Finally, note that the smallness criterion mentioned above guarantees that the $j$-th contribution to the QNM frequencies is a small correction to $\omega_0$, but this does not entail that the sum of all corrections has to be parametrically suppressed.

### 4.3.2 Parametrizing scalar hair and $\alpha_T$

Before proceeding with the QNM computation and applying the above formalism to our (4.25), we require more information about the functional form of $\delta\phi$ and $\alpha_T$. In close analogy to the above discussion, it is natural to to view these functions as an expansion in powers of $(2M/r)$ as well. Starting with the scalar hair function $\delta\phi$, in the main text we will follow [455] and focus on a scalar hair profile parametrised as

$$\delta\phi = \varphi_c \left( \frac{2M}{r} \right), \tag{4.28}$$

where $\varphi_c$ is a constant.[14] In Section 4.4.4 we discuss the more general parametrisation $\delta\phi = \varphi_c \left( \frac{2M}{r} \right)^n$ (where one remains agnostic of the leading order in $1/r$ at which the hair enters) as well as superpositions of different $r$-dependencies in the scalar hair profile. We leave an investigation of even more general (non-power-law) parametrisations for the future. Note that, while in this Section we will focus on the $n = 1$ scalar hair profile (4.28), we will also discuss how different profiles affect eventual constraints on $\alpha_T$ in Section 4.4.4. Having parametrised $\delta\phi$, we turn our attention to the remaining function of $r$ affecting $\delta\tilde{V}$, namely $\alpha_T$. To this end it will be useful to separate out the dependence on the scalar hair background profile and other geometric factors from $\alpha_T$ in (4.18) as follows

$$\alpha_T = -B(2M)^2 G_T \delta\phi'^2, \qquad G_T \equiv \frac{1}{(2M)^2} \frac{G_{4X} - G_{5\phi}}{G_4}. \qquad (4.29)$$

Here the dimensionless $G_T$ parameter has been defined to isolate the dependence of $\alpha_T$ on the Lagrangian $G_i$ functions, as opposed to the $r$-dependence following directly from the scalar profile or via the dependence on the function $B(r)$. We will find this separation especially useful later on when investigating what constraints on $\alpha_T$ can tell us about scalar hair and vice versa.[15] Because we have a non-trivial scalar profile, all the $G_i$ (and hence also $G_T$) are functions of $r$ and so to fully specify the $r$-dependence of $\alpha_T$ we finally also expand $G_T$ in powers of $r$ along the same lines as discussed for $\delta\phi$ above. Doing so we can write

$$G_T = \sum_i G_{Ti} \left(\frac{2M}{r}\right)^i, \qquad (4.30)$$

where each of the $G_{Ti}$ are constant coefficients. Putting everything together, i.e. substituting $\delta\phi$ (4.28) and $G_T$ (4.30) into the expression for $\alpha_T$ (4.29), we can finally write

$$\alpha_T = -\sum_{i=0}^{\infty} G_{Ti}\varphi_c^2 \left(1 - \frac{2M}{r}\right)\left(\frac{2M}{r}\right)^{i+4},$$
$$= -\sum_{i=0}^{\infty} A_{Ti} \left(1 - \frac{2M}{r}\right)\left(\frac{2M}{r}\right)^{i+4}. \qquad (4.31)$$

In the last line, we have implicitly defined a final shorthand as part of our notational setup. The dimensionless amplitude parameters $A_{Ti}$ neatly encapsulate the coefficients controlling $\alpha_T$ and satisfy

$$A_{Ti} \equiv G_{Ti}\varphi_c^2. \qquad (4.32)$$

As one may expect, these are also the effective constant parameters that, as we will find below, QNM observations will constrain observationally. To provide some intuition on the relationship between $A_{Ti}$ and $\alpha_T$, we illustrate the dependence of $\alpha_T$ on $2M/r$ for various choices of the amplitude coefficients $A_{Ti}$ in figure 4.1.

### 4.3.3 Parametrised QNMs

Having parametrised all the functional freedom encoded within $\delta\tilde{V}$ above, it is now straightforward to combine the above expressions. Doing so we can

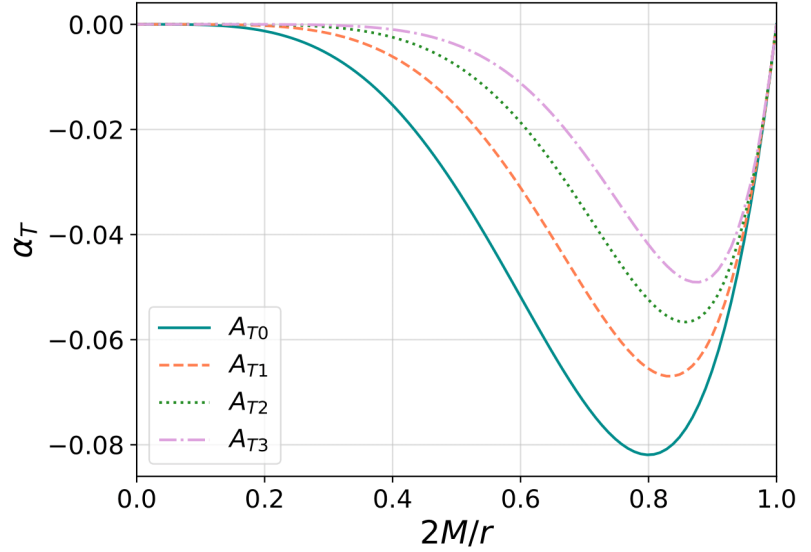

**Figure 4.1:** Here we plot $\alpha_T$, the deviation of the speed of gravitational waves from that of light defined via $\alpha_T \equiv (c_{GW}^2 - c^2)/c^2$, as a function of $2M/r$. We show $\alpha_T$ for different example choices of $i$ in (4.31), where only the amplitude $A_{Ti}$ corresponding to this specific $i$ is non-zero (and fixed to a fiducial value of 1) for each plotted curve. Note that $2M/r = 0$ therefore corresponds to spatial infinity whereas $2M/r = 1$ corresponds to the Schwarzschild radius, so we are interested in this range of values/distances. One can clearly see that $\alpha_T(r) = 0$ at spatial infinity and at the horizon, but displays non-trivial behaviour in-between with the overall amplitude and sign determined by $A_{Ti}$ while the radial dependence depends on the choice of $i$ considered.

express $\delta \tilde{V}$ as

$$\delta \tilde{V} = \frac{1}{(2M)^2} \sum_{i=0}^{\infty} A_{Ti} \left[ \left( \frac{2M}{r} \right)^{4+i} (2M\omega_0)^2 \right.$$
$$+ \left( \frac{2M}{r} \right)^{6+i} (-\ell(\ell+1) + 9)$$
$$+ \left( \frac{2M}{r} \right)^{7+i} (\ell(\ell+1) - 20)$$
$$\left. + \left( \frac{2M}{r} \right)^{8+i} \frac{45}{4} \right], \tag{4.33}$$

Note that we have dropped the order parameter $\epsilon^2$ at this point. From this we can read off the $a$-coefficients defined in (4.26)[16] and from (4.27) we also obtain the following expression for the quasinormal frequencies

$$\delta \omega = \sum_{i=0}^{\infty} A_{Ti} \cdot E_i^1,$$
$$= \sum_{i=0}^{\infty} A_{Ti} \left[ (2M\omega_0)^2 e_{4+i} - (\ell(\ell+1) - 9) e_{6+i} \right.$$
$$\left. + (\ell(\ell+1) - 20) e_{7+i} + \frac{45}{4} e_{8+i} \right], \tag{4.35}$$

where $E_i^1$ has been defined for convenience, and its subscript 1 refers to $n = 1$. A set of $E_i^n$ 'basis' functions for general $n$ are provided in Section 4.4.4.

Anticipating some of our later discussion, we can already see that, for small $\ell$ and $2M\omega_0 \sim O(1)$, the higher $e_j$ terms are enhanced relative to smaller $j$ terms (for a given $A_{Ti}$). For additional details on how different contributions enter into $\delta\omega$, see Section 4.4.4, and especially Figure 4.4. There we also explicitly discuss how this picture changes for different scalar field profiles. Also note that the consistency criterion we alluded to above, $|a_j| \ll (1 + 1/j)^j (j + 1)$, places an implicit bound on the amplitudes $A_{Ti}$ for the per-

16: For a given $i$ the contributions to these coefficients are

$$a_{4+i} \ni A_{Ti}(2M\omega_0)^2,$$
$$a_{6+i} \ni A_{Ti}(-\ell(\ell+1) + 9),$$
$$a_{7+i} \ni A_{Ti}(\ell(\ell+1) - 20),$$
$$a_{8+i} \ni A_{Ti} \frac{45}{4}. \tag{4.34}$$

By using the 'element sign' $\in$ we stress that a given $a_j$ can be built from contributions from different $i$'s. For instance, $a_6$ obtains contributions from $i = 0$ and $i = 2$.

turbative treatment we have outlined to be valid. As an example, for the case where only $i = 0$ terms contribute this bound requires that $A_{T0} \ll 1.5$.[17] Including higher order $i$'s will generate similar joint constraints on different $A_{Ti}$. As we will see, observational bounds will constrain the $A_{Ti}$ at the $10^{-1}$ level or stronger. Given we are measuring deviations away from the standard GR expectation $A_{Ti} = 0$, we therefore expect these consistency bounds to be satisfied in all relevant scenarios here.

## 4.4 Parametrised constraints

In the previous Section we derived an analytic expression for the QNM frequencies, assuming these to be close to the corresponding GR frequencies for a Schwarzschild black hole with the small perturbations encoding information about interactions in the underlying scalar-tensor theory, in particular about $\alpha_T$. We would now like to use this to forecast how well future GW experiments will be able to constrain $\alpha_T$ using ringdown. More specifically, we perform a Fisher forecast to estimate the error in the $A_{Ti}$ parameters (4.32). In this Section we therefore derive general expressions for the resulting constraints and discuss their overall features, following this up in the next Section by forecasting and discussing constraints for specific upcoming experiments. Note that, throughout this Section, we ubiquitously use the techniques developed in [216] for our analysis.

### 4.4.1 Fisher forecast setup

Exhaustive details of the Fisher forecast construction are collected in Appendix 3. Here, it will suffice to say that the waveform is modelled as [216]

$$h = h_+ F_+ + h_\times F_\times, \tag{4.36}$$

where $h_{+,\times}$ represent the strain in the two polarisations of the gravitational wave and $F_{+,\times}$ are the angular antenna pattern functions (specific to each detector) given by Equation (2.51). The strain functions for the ringdown are functions of the parameters $\{\omega_{\ell m}, f_{\ell m}, \tau_{\ell m}, Q_{\ell m}, S_{\ell m}, A^+_{\ell m}, A^\times_{\ell m}, \phi^+_{\ell m}, \phi^\times_{\ell m}\}$. $A$ and $\phi$ are the amplitudes and phases for the two polarisations, and $S_{\ell m}$ are the spheroidal harmonics as in Table 2.3. Finally, $f_{\ell m}$ and $\tau_{\ell m}$ characterise the real and imaginary parts of $\omega_{\ell m}$ in the following way

$$\omega_{\ell m} = 2\pi f_{\ell m} + \frac{i}{\tau_{\ell m}}, \qquad\qquad Q_{\ell m} = \pi f_{\ell m} \tau_{\ell m}, \tag{4.37}$$

where $Q_{\ell m}$ is the quality factor.

With the strain functions, one can compute the signal-to-noise-ratio (SNR) $\rho^2$ (C.4) and the Fisher Information Matrix $\Gamma_{ab}$ (C.7).[18] Such matrix can be inverted (giving the covariance matrix) to find the error for a given parameter $a$.

As an initial estimate we will here study the simplified case where all the usual parameters of the waveform are known ($A, \phi^+, ...$) and our only free parameters are the $A_{Ti}$'s (4.32). We leave forecasting full joint constraints to future work. This simplified setup means the estimates which we will compute below effectively are upper bounds on the precision one can expect. For a setup as considered here, where the only waveform parameters we want to constrain are those appearing inside the quasinormal frequencies $\omega$ (i.e. inside $f$ and $Q$), general expressions for the errors can be analytically derived. These only depend on the number of parameters one wants to constrain. In this Chapter we will constrain up to two $A_{Ti}$ together so we provide here the expression for single-parameter constraints

$$\sigma_{A_{Ti}}^2 \rho^2 = \frac{1}{2}\left(\frac{f}{Qf'}\right)^2, \qquad (4.38)$$

where the prime denotes a derivative with respect to $A_{Ti}$, and for double-parameter constraints

$$\sigma_{A_{Ti}}^2 \rho^2 = \frac{\dot{f}^2}{2}\frac{(2Q)^2 + \left(1 - \frac{f\dot{Q}}{\dot{f}Q}\right)^2}{(\dot{Q}f' - \dot{f}Q')^2},$$

$$\sigma_{A_{Tj}}^2 \rho^2 = \frac{f'^2}{2}\frac{(2Q)^2 + \left(1 - \frac{fQ'}{f'Q}\right)^2}{(\dot{Q}f' - \dot{f}Q')^2}, \qquad (4.39)$$

where again a prime denotes a derivative with respect to $A_{Ti}$, and a dot represents a derivative with respect to $A_{Tj}$. For more details on the derivation of these equations, see Appendix 3, and the repository [4].

Before deriving error estimates on different parameter combinations, let us briefly return to the question of which $(\ell, m)$ modes are of most interest.[19] As discussed above, while the QNM spectrum does not depend on initial conditions, the amplitude of individual modes does. The dominant observable contributions, i.e. the modes with the largest amplitudes for astrophysical binary compact object mergers, generically are the $\ell = m$ modes, more specifically the $(2, 2)$ mode [216–221].[20] Note that, for a non-rotating black hole solution as we are focusing on here, the equations of motion are independent of $m$ [212]. So while $m = 0$ is typically fixed in such setups for simplicity, as we have done here, the results derived apply for any $m$. The relative amplitude of subdominant modes (in particular $\ell = 3$) grows as the mass ratio $q$ and angular momentum $j$ of the remnant black hole increases [217, 219, 222, 223] – also see those references for discussions related to the detectability of such modes. Nonetheless, the $\ell = 2$ mode still generically dominates in all scenarios and higher $\ell$ modes decay more quickly, see Table 4.2. Note that the damping time $\tau$ goes as the inverse of the imaginary component of $\omega$, which increases for higher $\ell$ modes. So in addition to generically possessing a smaller amplitude, these modes also decay faster. Finally, also notice that, for binary systems that have orbited each other for a sufficiently long time for orbits to have approximately circularised, the $\ell = 2$ mode will be additionally enhanced relative to other modes [224–226]. While it is straightforward to repeat the analysis for other higher $\ell$ modes, the above rationale truly singles out the $\ell = 2$ mode as the observationally most relevant. We will therefore focus on this mode in what follows. Having said this, a multiple mode analysis will of course be a powerful tool to probe higher dimensional parameter spaces using ringdown alone tests in the future. While, as we have seen, quasinormal modes are independent of $m$ for static black holes, all astrophysical black holes do in fact rotate. For those, $(2, 2)$ is truly the dominant mode, and hence we will focus on this

|  | $\mathrm{Re}(2M\omega_0)$ | $\mathrm{Im}(2M\omega_0)$ |
|---|---|---|
| $\ell = 2$ | 0.7474 | -0.1779 |
| $\ell = 3$ | 1.1989 | -0.1854 |
| $\ell = 4$ | 1.6184 | -0.1883 |
| $\ell = 5$ | 2.0246 | -0.1897 |

**Table 4.2:** Real and imaginary components of the quasinormal frequencies $\omega_0$ of a Schwarzschild BH in GR for $\ell = 2, 3, 4, 5$. Quasinormal data is provided online [227, 228].

19: As exposed in Section 2.4.2, there is a third index characterising the QNM spectrum, the overtone number $n$. Here we only focus on the 'fundamental mode' $n = 0$. Modes with higher $n$'s (i.e. overtones) are more suppressed by virtue of having increasing values of $|\mathrm{Im}(\omega)|$.

20: Note that the discussion in Section 4.2 only applies for $\ell \geq 1$ modes and the dipole perturbation $\ell = 1$ requires special treatment, as the Regge-Wheeler gauge used in Section 4.2 does not fully fix all gauge degrees of freedom for this mode, see [649] for details. However, the contribution from the $\ell = 1$ mode can be shown to be negligible for the background solutions we are probing [649], so this is of no concern here.

one to perform the Fisher forecast and leave a more detailed study of constraints in the presence of several detected modes for the future. Extending the quasinormal mode calculations in Sections 4.2 and 4.3 to rotating black holes is an interesting way forward for which some machinery already exists, at least for slowly-rotating black holes (see e.g. [656–660]). However, such metrics and the Schwarzschild metric are smoothly connected (i.e. taking the limit of zero rotation $j \to 0$ recovers the Schwarzschild line element) so one expects that the non-rotating scenario still captures the leading order information in the quasinormal frequencies for sufficiently slow rotation.

### 4.4.2  Constraining $A_{T0}$

We begin by considering a minimal setup, where there is only a single relevant $A_{Ti}$ parameter, namely $A_{T0}$. From (4.35) we then find the QNM shift to be given by

$$\delta\omega = A_{T0} \cdot E_0^1. \tag{4.40}$$

where $E_0^1$ is shown in (4.35) and we quote it here for reference

$$E_0^1 = \left[ (2M\omega_0)^2 e_4 - (\ell(\ell+1) - 9)e_6 + (\ell(\ell+1) - 20)e_7 + \frac{45}{4}e_8 \right]. \tag{4.41}$$

Substituting in the numerical values for the $e_j$ from Table 4.1, we obtain

$$M\delta\omega = -[0.00070 + 0.00306i] \cdot A_{T0}. \tag{4.42}$$

In evaluating this, we have also used the $\ell = 2$ mode in Table 4.2. This now allows us to obtain parametric expressions for the $\alpha_T$-induced deviations in the QNM spectrum. From (4.42) and Table 4.2 we find the following percentage differences for the real and imaginary parts, respectively

$$\frac{\delta\omega_R}{\omega_{0R}} \approx -0.19 \cdot A_{T0}\%, \qquad \frac{\delta\omega_I}{\omega_{0I}} \approx 3.44 \cdot A_{T0}\%. \tag{4.43}$$

Finally, we are also in a position to extract an expression for the accuracy with which an experiment with ringdown SNR $\rho$ will be able to measure $A_{T0}$. Reading off $f$ and $Q$ from (4.42), as defined in (4.37), and substituting them into the single-parameter error expression (4.38), we obtain an estimate on its detectability in the same fashion as [455].[21] This gives us[22]

$$\sigma_{A_{T0}}\rho \approx 181. \tag{4.44}$$

The numerical value of this error calculation, as well as the analogous ones which will follow in this Section, will be translated into $\alpha_T$ constraints for specific detectors in Section 4.5. There, we will compare our results with other existing and forecasted constraints.

### 4.4.3  Constraining multiple $A_{Ti}$

Having considered the single-parameter case above, a natural next step is to consider a more complex functional form for $\alpha_T$ and hence for the $A_{Ti}$. Here we consider the case where $\alpha_T$ is controlled by two parameters, $A_{T0}$ as

21: Note that, in evaluating the final expression, we set $A_{T0}$ to zero. This should simply be understood as capturing the leading order contributions to the error – depending on the actual value of $A_{T0}$ the precise error can differ by $\leq O(10\%)$.

22: More precise results are provided in [4]. Ultimately, we will only be interested in the robust order-of-magnitude constraints here, so e.g. in Table 4.4 we will approximate $\sigma_{A_{T0}}\rho \approx O(10^2)$.

before and a second parameter $A_{T1}$. Proceeding as before, we then have

$$\delta\omega = A_{T0} \cdot E_0^1 + A_{T1} \cdot E_1^1. \quad (4.45)$$

Reading off expressions for $f$ and $Q$ from equation (4.45) as before, we find the following error estimates from (4.39)

$$\sigma_{A_{T0}}\rho \approx 302, \qquad\qquad \sigma_{A_{T1}}\rho \approx 465. \quad (4.46)$$

One may wonder how we can constrain two parameters with the measurement of a single mode. To this end note that the measurement of a single mode carries information about the oscillation frequency of that mode as well as for the associated damping time and these independent pieces of information allow constraining two parameters here. Once future observations are capable of measuring multiple modes [194, 220, 267], this will of course allow constraining a correspondingly larger parameter space.

Equation (4.46) shows that $A_{T0}$ and $A_{T1}$ can be constrained to a similar order of precision. This re-iterates that terms at higher order in a $1/r$ expansion are not parametrically suppressed in their contribution to the QNM frequency spectrum, so a $1/r$ expansion is not an ideal basis in terms of observational constraints. Indeed, upon closer inspection, we find that constraints on $A_{T0}$ and $A_{T1}$ are strongly correlated, as can be seen from the off-diagonal elements of the covariance matrix for these two parameters

$$\Sigma_{ab} \sim 10^4 \begin{pmatrix} 9 & -16 \\ -16 & 22 \end{pmatrix}, \quad (4.47)$$

We can therefore diagonalize the covariance matrix to obtain the eigenmodes that will be constrained by the data, i.e. a more optimal basis from a detectability point of view. Under standard matrix diagonalization procedures we obtain

$$\tilde{\Sigma}_{ab} = S^{-1}\Sigma_{ab}S \sim 10^3 \begin{pmatrix} 5 & 0 \\ 0 & 303 \end{pmatrix}. \quad (4.48)$$

This transformation amounts to identifying the combinations of $A_{T0}$ and $A_{T1}$ that yield uncorrelated parameters $A_{TA}$ and $A_{TB}$, i.e. we have performed the parameter transformation $(A_{T0}, A_{T1}) \rightarrow (A_{TA}, A_{TB})$ such that the covariance matrix of the latter is the one given by equation (4.48). More explicitly, the relevant eigenmodes here are $A_{TA} = -0.84A_{T0} - 0.54A_{T1}$ and $A_{TB} = -0.54A_{T0} + 0.84A_{T1}$.[23] Finally, the errors for the new parameters are

$$\sigma_{A_{TA}}\rho \approx 68, \qquad\qquad \sigma_{A_{TB}}\rho \approx 550, \quad (4.50)$$

where we indeed see that $A_{TA}$ can be constrained more tightly than any parameter in the previous basis.

### 4.4.4 Dependence on scalar hair profile

Above we have derived expressions for the precision with which a generic future experiment with SNR $\rho$ will be able to constrain the relevant parameter combinations affecting QNM frequencies, namely the $A_{Ti}$. Here we would like to investigate to what extent the specific form of the scalar hair profile affects this. As we will argue, in certain cases this argument can then also be inverted to place constraints on the scalar hair itself. Recall that we are parametrising the scalar hair profile as

$$\delta\phi = \varphi_c \left(\frac{2M}{r}\right)^n.\tag{4.51}$$

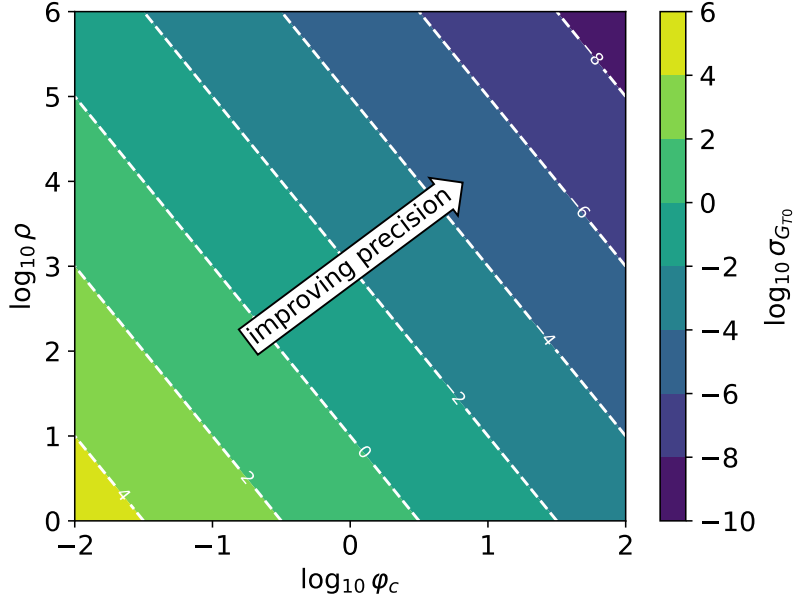

Here $\varphi_c$ effectively captures the scalar field amplitude, while $n$ carries information about the radial dependence of this profile. Until now we have set $n = 1$, but here we will investigate the case with generic $n$.

### Amplitude

The QNM frequencies derived above are functions of the $A_{Ti}$, which we recall depend both on the scalar amplitude $\varphi_c$ as well as on the $G_{Ti}$ (i.e. the interactions in the underlying theory) via (4.32). This has an immediate important consequence, namely that a detection of the specific QNM shifts discussed here implies *both* a detection of scalar hair and of non-trivial $G_4$ and/or $G_5$ interactions contributing to the $G_{Ti}$ – cf. (4.29). The scalar amplitude $\varphi_c$, analogously to the amplitudes of QNMs, will depend on the 'initial conditions' for the ringdown phase. It is worth emphasising that, at present, it is not yet well understood how the non-linear merger stage affects this amplitude in scalar-tensor theories of interest, so we will leave $\varphi_c$ as a free parameter.[24] It is interesting, then, to disentangle the effect of $\varphi_c$ and of $G_{Ti}$ on the constrained $A_{Ti}$ parameter(s). This is shown in Figure 4.2. As can clearly be seen, and indeed as expected from (4.32), in the presence of a larger scalar hair amplitude $\varphi_c$ the constraint on the $G_{Ti}$ becomes stronger. More explicitly

$$\sigma_{G_{Ti}} = \sigma_{A_{Ti}} \varphi_c^{-2}.\tag{4.52}$$

Interestingly, this implies that one can I) infer a constraint on the scalar hair amplitude from measurements of the QNMs, *given* another non-trivial

24: $\varphi_c$ may be significantly enhanced or suppressed during the non-linear merger stage, so in the absence of comprehensive numerical (merger) simulations for the theories in question, even an order of magnitude estimate appears premature. Note that, in cases where the scalar hair does affect the black hole geometry (so unlike the 'stealth' solutions (4.8) we consider here), this effect on the geometry can be used to place additional constraints on the nature and amplitude of the hair e.g. along the lines presented in [661–665].

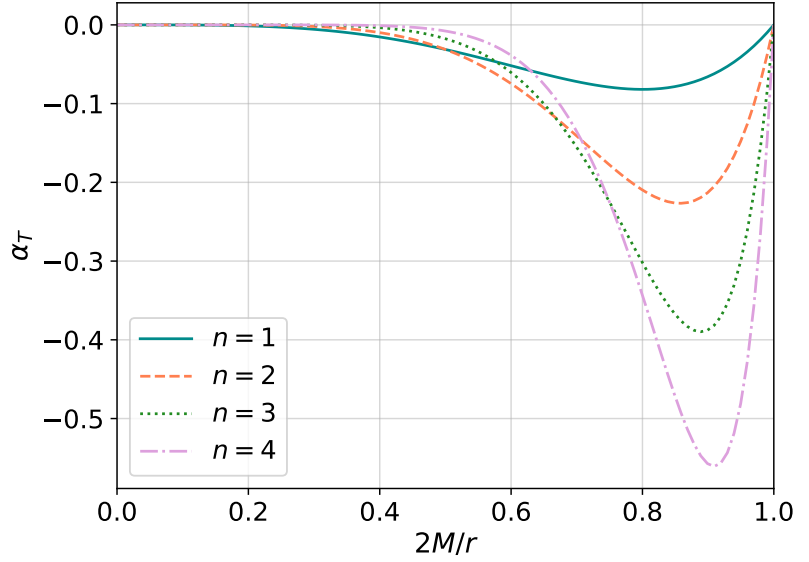

**Figure 4.3:** Here we show $\alpha_T$ (4.53) for different choices of $n$ in (4.51) as a function of $2M/r$. $i = 0$ has been set such that $A_{T0}$ is the only non-zero parameter (and fixed to a fiducial value of 1). We see that $\alpha_T(r) = 0$ at spatial infinity and at the horizon, but again observe non-trivial behaviour in the intermediate region. The size of $\alpha_T$ (partially controlled by $A_{T0}$) is considerably enhanced by increasing $n$. This is mainly due to the factor $n^2$ accompanying all $\delta\omega$ (as can be seen from (4.57) and (4.58)).

bound on the $G_{Ti}$ e.g. from non-ringdown related constraints on $c_{GW}$ (in other words: in the event of a future detection of a $c_{GW} \neq c$ from another probe), and II) infer a constraint of the $G_{Ti}$, *given* other independent information about the amplitude $\varphi_c$ (in other words: in the event of a complementary detection of scalar hair).

### Radial dependence

Having considered the effect of the scalar hair amplitude above, we now investigate how our analysis is affected when the functional form of the scalar hair, i.e. its $r$-dependence and hence $n$ in (4.51), changes. We have considered $n = 1$ above and here we repeat the derivation of quasinormal corrections for the case with generic $n$, and repeat the Fisher analysis for $n = 2$ as a complementary example. In this case, $\alpha_T$ is now given by

$$\alpha_T = -\sum_{i=0}^{\infty} A_{Ti} \cdot n^2 \left(1 - \frac{2M}{r}\right) \left(\frac{2M}{r}\right)^{2n+i+2}, \tag{4.53}$$

where the new $n$-dependence is plotted in Figure 4.3. Substituting this back into $\delta\tilde{V}$ (4.25) we get

$$\begin{aligned}
\delta\tilde{V} = \left(\frac{n}{2M}\right)^2 \sum_{i=0}^{\infty} A_{Ti} \Bigg[ &\left(\frac{2M}{r}\right)^{2n+i+2} (2M\omega_0)^2 \\
&+ \left(\frac{2M}{r}\right)^{2n+i+4} \left(-\ell(\ell+1) + \frac{9}{2} + \frac{7}{2}n + n^2\right) \\
&+ \left(\frac{2M}{r}\right)^{2n+i+5} \left(\ell(\ell+1) - \left(\frac{19}{2} + \frac{17}{2}n + 2n^2\right)\right) \\
&+ \left(\frac{2M}{r}\right)^{2n+i+6} \left(\frac{21}{4} + 5n + n^2\right) \Bigg].
\end{aligned} \tag{4.54}$$

Note that setting $n = 1$ recovers the expression (4.33). Again, this can be written as

$$\tilde{\delta V} = \frac{1}{(2M)^2} \sum_{j=0}^{\infty} a_j \left(\frac{2M}{r}\right)^j \tag{4.55}$$

with the only non-zero a-parameters for a given $i$ contributing as

$$a_{2n+i+2} \ni A_{Ti} \cdot n^2 (2M\omega_0)^2$$

$$a_{2n+i+4} \ni A_{Ti} \cdot n^2 \left(-\ell(\ell+1) + \frac{9}{2} + \frac{7}{2}n + n^2\right)$$

$$a_{2n+i+5} \ni A_{Ti} \cdot n^2 \left(\ell(\ell+1) - \left(\frac{19}{2} + \frac{17}{2}n + 2n^2\right)\right)$$

$$a_{2n+i+6} \ni A_{Ti} \cdot n^2 \left(\frac{21}{4} + 5n + n^2\right). \tag{4.56}$$

From these we find the following quasinormal frequency corrections

$$\delta\omega = \sum_{j=0}^{\infty} a_j e_j = \sum_{i=0}^{\infty} A_{Ti} E_i^n, \tag{4.57}$$

where we have defined the following new basis for convenience

$$E_i^n = n^2 \Bigg[ (2M\omega_0)^2 e_{2n+i+2}$$

$$+ \left(-\ell(\ell+1) + \frac{9}{2} + \frac{7}{2}n + n^2\right) e_{2n+i+4}$$

$$+ \left(\ell(\ell+1) - \left(\frac{19}{2} + \frac{17}{2}n + 2n^2\right)\right) e_{2n+i+5}$$

$$+ \left(\frac{21}{4} + 5n + n^2\right) e_{2n+i+6} \Bigg]. \tag{4.58}$$

The basis for $n = 1$ (4.35) can be straightforwardly recovered from this. Note that here, because we have chosen to remain agnostic about $n$, we have ended up with two indices, i.e. $n$ and $i$ that need to be chosen in order to obtain numerical results. This is shown explicitly in the super and subscripts of the newly defined basis $E_i^n$. In diagram 4.4 we display the corrections coming from different choices of $(i, n)$, and make some observations about their structure.

The step from these analytically calculated corrections to the quasinormal modes to the errors on different $A_{Ti}$ parameters (and hence corrections on $\alpha_T$) is straightforwardly repeated in the same fashion as shown in Section 4.4. We show in Table 4.3 the results for a few more illustrative cases, but stress that such an analysis can easily be repeated for any combination and superposition of $(i, n)$ by adapting the companion notebook provided in [4]. The general trend we find that can already be appreciated in Table 4.3 is that errors decrease for increasing $i$ and $n$ (at least for the single-parameter cases).

|  | $i = 0$ | $i = 1$ | $i = \{0, 1\}$ |
|---|---|---|---|
| $n = 1$ | 181 | 131 | (68, 550) |
| $n = 2$ | 132 | 72 | (27, 250) |
| $n = \{1, 2\}$ | 153 | 93 | (40, 344) |

**Table 4.3:** Values for $\sigma_{A_{Ti}}\rho$ for a scalar profile $\delta\phi = \varphi_c \left(\frac{2M}{r}\right)^n$. The last column ($i : 0, 1$) displays the errors for the two parameters $(A_{TA}, A_{TB})$, obtained from $(A_{T0}, A_{T1})$ via the same diagonalization procedure as shown in Section 4.4.3.

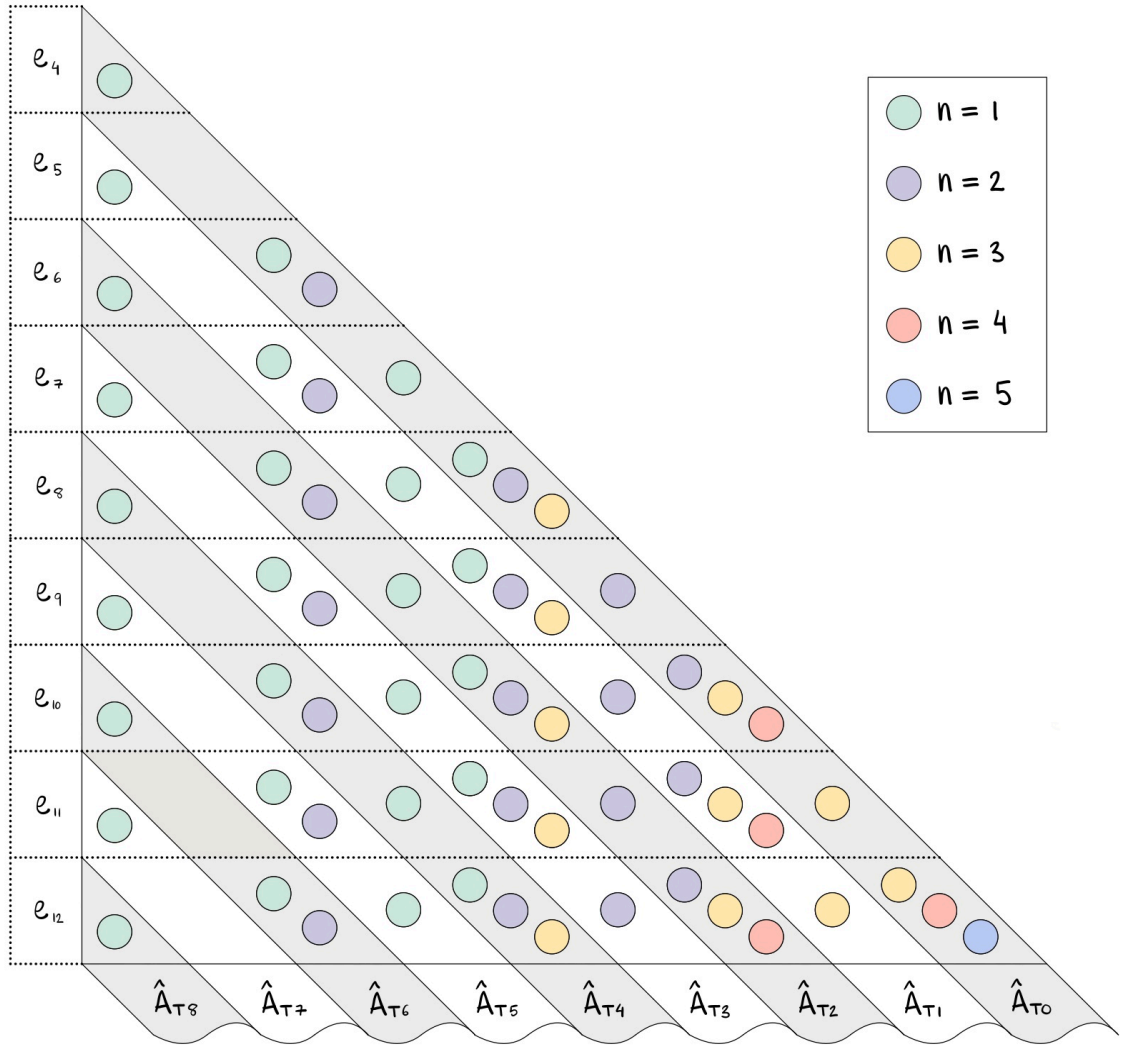

**Figure 4.4:** In this diagram we display the contributions to $\alpha_T$ (4.53) coming from different $n$ values. These are ordered such that all contributions to a specific $A_{Ti}$ parameter appear in one diagonal, which in principle is extendible *ad infinitum*. Contributions from different $n$ also follow a pattern, which can be appreciated here by focusing on the colour of the balls. The vertical axis $e_j$ corresponds to the parametrised basis from [253], which tells us the power of $(1/r)^j$ at which each contribution appears in the potential.

## 4.5 Forecasting observational constraints

In the previous Section we derived parametric expressions for the precision with which the parameters controlling the behaviour of $\alpha_T$ and hence $c_{GW}$ can be measured for probes with a general ringdown SNR $\rho$. In this Section we now summarise and briefly discuss what this concretely implies for a range of current and future missions, spanning the frequency range from $10^{-4}$ Hz to $10^3$ Hz. The main results are collected in Table 4.4.[25]

Before discussing forecasted constraints in detail for the respective missions and frequency bands, this is a good point to recall our introductory discussion in Section 4.1 about how and where the frequency-dependence in $c_{GW}$ may be localised and how this is tied to the regime (i.e. frequency range) where the underlying theoretical framework is valid. Rather obviously, any prediction derived from our starting point – the Horndeski scalar-tensor action (4.2) – is only trustworthy when (4.2) is a faithful description of the relevant physics. Since (4.2) gives rise to a frequency-independent $c_{GW}$, we are therefore implicitly assuming that at the very least in the frequency-window spanning the ringdown frequencies in question, $c_{GW}$ is constant as a function of frequency to high accuracy. A natural scenario to consider would therefore be the one alluded to in the introduction: $c_{GW}$ effectively becomes a constant $c_{GW}^{(0)} \neq c$ at low frequencies where (4.2) applies and may indeed be intimately linked to dark energy phenomenology on cosmological scales. Now we consider the ringdown following a SMBH merger observable in the LISA band and effectively have $c_{GW} = c_{GW}^{(0)}$ there. We can therefore straightforwardly use (4.2) to compute this ringdown signal. In this scenario we also assume (4.2) stops being an accurate description of the relevant gravitational physics between the LISA and LVK bands and its unknown UV (high energy) completion takes over there, resulting in a transition back to $c_{GW} = c$ at high frequencies due to the Lorentz invariant nature of the UV completion. The frequency-dependence in $c_{GW}$, induced by the UV completion is sharply localised in frequency-space between the LISA and LVK bands and so fully consistent with existing bounds on $c_{GW}$ from the LVK band. Now this scenario – as explored in detail in the context of forecasting upcoming multi-band constraints in [601, 628] – is only illustrative and the frequency-dependence of $c_{GW}$ and the regime of validity of (4.2) can easily be altered depending on the UV completion and if the connection to dark energy is loosened or severed completely. We refer to [545, 600, 601, 628] and [667] for more detailed dis-

25: It is worth pointing out that the ringdown SNRs quoted implicitly depend on when precisely the transition from merger to ringdown phase is assumed to take place. While discussing this in detail is beyond the scope of this Chapter, we point the interested reader to [217, 218, 272–275, 666] for discussions on this.

| Detector(s) | Ringdown SNR ($\rho$) | Error on $\alpha_T$ |
|---|---|---|
| LVK | $10$ [183, 259, 260] | $1$ |
| ET / CE | $10^2$ [182–185] | $10^{-1}$ |
| pre-DECIGO | $10^2$ [261] | $10^{-1}$ |
| DECIGO / AEDGE | $10^3$ [186, 262]* | $10^{-2}$ |
| LISA | $10^3$ [263–266] | $10^{-2}$ |
| TianQin | $10^3$ [265, 266] | $10^{-2}$ |
| AMIGO | $10^5$ [267] | $10^{-4}$ |

**Table 4.4:** Order-of-magnitude errors on $\alpha_T$ according to the achievable order-of-magnitude ringdown SNRs for a single observed event for different GW detectors. More details on the quoted forecasted SNR values can be found in the caption of Figure 2.5. Errors in this Table are computed assuming $A_{T0}$ is the only amplitude parameter contributing to $\alpha_T$ in (4.31), as an example. The error on $\alpha_T$, $\sigma_{\alpha_T}$, is quoted as one order-of-magnitude less than the corresponding error on $A_{T0}$, as observed in Figure 4.1. We stress that the precise mapping of underlying amplitude parameters to $\alpha_T$ mildly depends on the precise functional form of the scalar hair and the underlying interactions, but note that errors on other $A_{Ti}$ parameters and hence $\alpha_T$ are qualitatively similar – see e.g. Table 4.3.

cussions of those two points, respectively, and note that in this Chapter this is especially relevant in the context of forecasts for frequency-bands above (i.e. at higher frequencies than) the LISA band. We will come back to this point below.

What would it take to extrapolate/extend the results from the above Sections to cases where $c_{GW}$ is frequency-dependent in the frequency-window associated with ringdown signals of interest? On the theoretical side, we already pointed out that this would involve supplementing/replacing (4.2) with the interactions inducing the frequency-dependence of $c_{GW}$, which requires knowledge of (or assumptions about) the UV completion of (4.2). The resulting action could then be used to repeat the analysis for this frequency-window. It is worth highlighting that the results of Sections 4.3 and 4.4 only know about the Horndeski scalar-tensor action by assuming the corresponding modified form of the Regge-Wheeler equation (4.17). So any UV completion that does not modify this form other than inducing a frequency-dependent $c_{GW}$ and hence $\alpha_T$ is covered by the analysis in Sections 4.3 and 4.4. We leave an exploration of how UV completions might otherwise affect the modified Regge-Wheeler equation and how this affects the subsequent analysis for future work. On the observational side, a frequency-dependent $c_{GW}$ would introduce another challenge in the ringdown analysis. Since different parts (i.e. frequencies) of the waveform then travel at different speeds, the received signal at the detector will be stretched/squeezed/scrambled with respect to the signal emitted at the source, see [601, 668] – also see formally related discussions in [669, 670]. Specifically in the ringdown context, this can make identifying the correct frequencies more challenging and this therefore requires a dedicated analysis [668]. In practice, this means that the strain functions (C.2) accurately describe the signal at emission but will be altered via non-trivial dispersion effects by the time they reach the detector, so this needs to be taken into account to correctly forecast constraints when a frequency-dependent $c_{GW}$ affects the frequency-window associated with the signal under investigation. We will leave such a dedicated analysis to future work and (as also motivated by the theoretical considerations above) in this Section forecast constraints for different frequency bands, assuming an effectively frequency-independent $c_{GW}$ within the band under investigation (i.e. the LISA forecasts assume a frequency-independent $c_{GW}$ in the LISA band and so on).

### 4.5.1 LISA band forecasts

As motivated above and in the introduction, the LISA band is particularly promising in terms of testing for deviations of $c_{GW}$ from the speed of light, given that a frequency-dependent transition of $c_{GW}$ just or somewhat below the LVK band is a natural prediction in a range of candidate dark energy models. In terms of the amplitude parameters $A_{Ti}$, we see from Table 4.4 that one expects the leading order such parameter to be constrained at the $10^{-1}$ level with future LISA/TianQin observations that are forecasted to yield a ringdown SNR of $\sim O(10^3)$ [263–266, 270]. Mapping this back to $\alpha_T$ itself (4.31), this implies one will be able to detect deviations down to the $\alpha_T \sim O(10^{-2})$ level from LISA band ringdown alone in the context of the models we consider,[26] i.e.

26: Note that, $\alpha_T$ is generically about one order of magnitude smaller than the dominant $A_{Ti}$ – see Figure 4.1. Also, as should be obvious from (4.31), this mapping is mildly dependent on the $i$ coefficient.

$$\sigma_{\alpha_T}^{\text{LISA/TianQin}} \sim 10^{-2}. \tag{4.59}$$

Note that present forecasts for far-future missions such as AMIGO predict higher order-of-magnitude ringdown SNR, so this would enhance the constraints on $c_{GW}$ in comparison with those expected from LISA/TianQin for a single event to $\alpha_T \sim O(10^{-4})$.

It is worth emphasising that the main bounds discussed here are forecasted for a single ringdown observation with the SNR achievable by the relevant detector. It is reasonable to expect that qualitatively improved constraints will be obtained when combining multiple observations. Indeed, for sufficiently large $N$ (where $N$ is the number of detected events) the measurement precision for QNMs is expected to improve as $N^{-1/2}$ [671, 672], assuming $N$ identical events. For the LISA band, expected event rates are somewhat uncertain, but most estimates lie in the $O(10 - 100)$ per year range for SMBH mergers – see e.g. [190–197]. An improvement of up to two orders of magnitude on the above constraints therefore seems achievable after several years of operation, so that one may hope to ultimately reach a precision of close to $\sigma_{\alpha_T}^{LISA/TianQin} \sim 10^{-4}$. We again emphasise that the $N^{-1/2}$ scaling discussed here assumes an idealised case with $N$ identical events with the large SNRs considered here and so the above should be taken as an optimistic bound (e.g. many events to be detected will be at higher redshifts and have correspondingly reduced SNRs), also depending on the precise SMBH merger and detection rates as discussed above.

### 4.5.2 LVK band forecasts

Having summarised results for the LISA band above, let us consider the LVK band. The situation here is qualitatively different, given that there already are tight constraints on $c_{GW}$ specific to this frequency band. From measuring the coincidence of the GW170817 signal in GW and optical counterpart observations, one finds that $\alpha_T \lesssim 10^{-15}$ [538, 581–584]. When even a very mild frequency-dependence of $c_{GW}$ is present in the LVK band, this bound can be strengthened $\alpha_T \lesssim 10^{-17}$ [601]. Contrast this with bounds from ringdown observations alone, where the $A_{Ti}$ can be constrained at the $\sim O(10)$ level with LVK observations with an improvement by approximately an additional order of magnitude to be expected from the future Einstein Telescope(ET)/Cosmic Explorer(CE) missions – cf. Table 4.4 and see [182, 183] and [184, 185] for ET and CE, respectively. Again mapping this to constraints on $\alpha_T$ itself, we therefore ultimately expect

$$\sigma_{\alpha_T}^{ET/CE} \sim 10^{-1}. \tag{4.60}$$

once ET/CE are collecting data in the future.

The bound (4.60) given above is again for a single event with the SNR achievable by ET/CE. Taking into account the $N^{-1/2}$ improvement of the measurement precision for $N$ detected events discussed above, we can again extrapolate how this precision might be improved over time. For ET $O(10^4 - 10^5)$ events with a ringdown SNR of $O(10)$ are expected per year [651]. One may therefore reasonably expect that constraints can eventually be improved by about two orders of magnitude to $\sigma_{\alpha_T}^{ET/CE} \sim 10^{-3}$. In the LVK context it is also interesting to point out that existing (non-ringdown-specific) constraints on $c_{GW}$ from GW waveforms in the LVK band have already seen similar improvements by stacking events. More specifically, when comparing I) constraints on $c_{GW}$ from GW170817 data alone (i.e. without using an optical counterpart)

27: This improvement, while still partially driven by the larger number $N$ of observations included, is also partially due to other events having higher individual SNRs than GW170817.

[587] with II) constraints obtained using a LVK catalog of 43 confident binary black hole mergers (used to obtain bounds on the graviton mass in [453], but straightforwardly re-interpretable to place bounds on $c_{GW}$), this improves these bounds on $c_{GW}$ by around two orders of magnitude.[27]

At first sight (4.60) as well as the improved $\sigma_{\alpha_T}^{ET/CE} \sim 10^{-3}$ bound reachable by stacking events are rather weak, albeit complementary, constraints on $c_{GW}$ when compared with the existing GW propagation bounds from GW170817 discussed above. Also note that, for the purposes of this subsection and as discussed in detail above, we are assuming that (4.2) is a valid description of the underlying physics in (at least part of) the LVK band. As discussed, in dark energy-related theories within (4.2) where $c_{GW}$ receives order one corrections on cosmological scales one would not expect this to be the case. One can remedy this (i.e. 'return' the LVK band to within the regime of validity of (4.2)) in two different ways. First, by severing the connection to cosmology/dark energy and looking at the constraints derived here in their own right. Or second, by suppressing the cosmological $\alpha_T$ from the beginning while not precluding a more sizeable $\alpha_T$ around black hole space-times. A specific scenario related to the second case below will be discussed in Section 6.4.2, where such phenomenology will be shown to naturally stem in the context of higher-order scalar-tensor (HOST) theories.

However, a more general related point is the following: The fact that the constraints derived in this Chapter are computed for a different background solution than cosmological background GW-propagation constraints derived e.g. from GW170817 means that they nevertheless contain some interesting new information on the scalar hair profile and the underlying interactions encoded in $G_T$ along the lines discussed in Section 4.4.4 – we show this in Figure 4.2. More specifically, reinstating units of $M_{Pl}$, from (4.52) and in the event of a scalar hair amplitude $\varphi_c \sim O(10^8)M_{Pl}$, the constraint on the underlying interactions will be as strong as constraints on the same interactions from GW170817—and even stronger for a larger amplitude $\varphi_c$.

Reversing the argument, if future observations were to identify a small but non-zero cosmological $\alpha_T$, this would allow placing a bound on the scalar hair amplitude from the ringdown constraints investigated here. For concreteness, consider the following setup: The higher derivative scalar interactions in the $G_{3,4,5}$ terms in (4.2) come with an implicit mass scale $\Lambda$. In cosmology this scale is typically chosen to be $\Lambda = \Lambda_3 \equiv (M_{Pl}H_0^2)^{1/3}$, where this choice ensures those interactions give $O(1)$ contributions to cosmology. However, if a different $\Lambda$ is chosen, the cosmological $\alpha_T$ (and hence $G_T$) scales as $\alpha_T \sim (\Lambda_3/\Lambda)^6$. So raising the interaction scale $\Lambda$ by just three orders of magnitude suppresses the cosmological $\alpha_T$ down to a level of $O(10^{-18})$, comfortably consistent with bounds from GW170817. This setup also allows the full LVK band to be within the regime of validity of the physics described by (4.2) – see [667] for further details on this scenario. Now suppose that a future constraint indeed establishes $G_T \sim O(10^{-18})$, while future ringdown constraints from ET/CE along the lines investigated here do not yield evidence for a non-zero $\alpha_T$. From (4.32) this would allow us to derive a bound on the scalar hair $\varphi_c \lesssim O(10^9)M_{Pl}$ for frequencies in the LVK band, where again we have momentarily abandonded geometric units. Note that other complementary bounds on $\varphi_c$ may be obtainable e.g. from considering even perturbations or going beyond linear theory.

### 4.5.3 Intermediate band forecasts

With LISA and LVK forecasts discussed above, the intermediate frequency band stands out as a third region of interest. Here the upcoming AEDGE [186] and DECIGO [187, 188] experiments will detect and investigate GWs in the future. In the introduction, we motivated probing $c_{GW}$ in the LISA band by pointing out that a frequency-dependent transition from a nearly constant $c_{GW} = c$ at LVK frequencies to a different low-frequency $c_{GW}$ naturally occurs just or somewhat below the LVK band in large classes of dark energy theories. This motivation of course equally applies to the frequencies probed by AEDGE/DECIGO. Candidate transitions in this intermediate band may 'leak out' into the LISA and/or LVK bands, in which case the considerations outlined above for those bands already promise tight constraints. However, another interesting class of transitions are those investigated by [601, 628], where the transition is effectively completely contained within the intermediate frequency band and no detectable frequency-dependence leaks out into the LISA and/or LVK bands. In such a case multiband observations using systems such as GW150914 that are first observable in the LISA band and eventually enter the LVK band can be used to obtain an integrated constraint on any features residing at intermediate frequencies and indeed will be able to constrain $\alpha_T$ down to a level of $O(10^{-15})$ [601, 628]. In addition, once AEDGE/DECIGO observations are available, direct constraints on $c_{GW}$ from this band will be obtainable in analogy to the LVK/LISA analyses discussed above. Whenever there is significant frequency-dependence for $c_{GW}$ in band, a complementary ringdown-specific analysis faces similar theoretical challenges as discussed for the LVK band above, as well as the observational modelling challenges mentioned earlier in this Section. So, as before, the bounds forecasted in this subsection will be for the case where (4.2) applies within (at least part of) the AEDGE/DECIGO band and hence $c_{GW}$ is frequency-independent in this band to high accuracy. With these assumptions and from ringdown alone, we find that AEDGE/DECIGO will be able to constrain the $A_{Ti}$ at the $\sim O(10^{-1})$ level, c.f. Table 4.4. Mapping this to constraints on $\alpha_T$ itself, as before, this implies

$$\sigma_{\alpha_T}^{\text{AEDGE/DECIGO}} \sim 10^{-2}. \tag{4.61}$$

While weaker in magnitude than the integrated multiband constraints discussed above, these bounds are complementary in the same sense as discussed in the LVK Section above. Note that one may again expect this bound to be improved significantly when stacking multiple observed events: Several dozen intermediate mass black hole (IMBH) mergers with an SNR $O(10^3)$ should be observable with AEDGE per year [186], so optimistically an improvement up to $\sigma_{\alpha_T}^{\text{AEDGE/DECIGO}} \sim 10^{-4}$ appears feasible eventually.

## 4.6 Conclusions

In this Chapter, which collects the results obtained in [1], we have investigated how the speed of gravitational waves $c_{GW}$ can be probed using black hole ringdown observations. Focusing on scalar-tensor theories of the Horndeski type and on odd parity quasinormal modes (QNMs), our key findings are as follows:

▶ In the context of non-rotating black holes where the metric background solution is given by Schwarzschild, we find that deviations of $c_{GW}$ from the speed of light only affect the QNMs in the presence of a non-trivial scalar hair profile $\phi = \phi(r)$ in agreement with the results of [455]. Any deviations from $c_{GW} = c$ are then proportional to the square of the amplitude of the scalar hair.

▶ For a single event, ringdown observations from LISA, TianQin and AEDGE/DECIGO will be able to constrain $c_{GW}$ at the $O(10^{-2})$ level. When stacking observations over several years, both constraints may be improved by up to two orders of magnitude, depending on precise event rates. While those constraints are weaker than existing constraints on $c_{GW}$, e.g. from GW170817, they importantly probe different frequency ranges. This is particularly relevant in the context of testing $c_{GW}$, given large classes of dark energy models naturally give rise to a frequency dependent transition in $c_{GW}$ below the LVK band.

▶ With ringdown constraints we are testing the effect of deviations from $c_{GW} = c$ on a different background solution than that relevant for GW propagation constraints on $c_{GW}$ (black hole vs. cosmological spacetimes). The precise dependence of $c_{GW}$ on interactions in the underlying theory is different for these two backgrounds. We have highlighted examples where, in the presence of a sufficiently large scalar hair profile, ringdown observations can provide novel constraints on those interactions. Likewise, given complementary information on those underlying interactions, we have shown how ringdown observations can constrain the nature of scalar hair. We stress that therefore even LVK band ringdown measurements, where we find that $O(10^{-1})$ level will be obtainable from the Einstein Telecope/Cosmic Explorer for a single event, can yield valuable information complementary to existing constraints on $c_{GW}$.

Overall we have therefore derived forecasts for the precision with which ringdown observations will be able to constrain the speed of gravitational waves $c_{GW}$ for various detectors throughout the $O(10^{-4}) - O(10^3)$ Hz frequency range. Our study has been idealised in the sense that we have assumed I) the 'usual' binary black hole merger parameters (masses, amplitudes, phases) to be known and focused on the effect of novel parameters associated with $c_{GW} \neq c$, II) focused on a specific background solution for the black hole geometry and scalar hair profile, and III) by working with Horndeski scalar tensor theories, we have implicitly assumed that $c_{GW}$ is approximately constant in specific frequency windows/bands when forecasting constraints for those respective bands. A more comprehensive analysis, extending the present work, investigating degeneracies and constraints in higher-dimensional parameter spaces as well as a wider range of hairy black hole solutions and underlying theoretical setups, will therefore be an interesting next step. As has been mentioned, another promising route to make our setup more physically realistic is to extend the quasinormal mode calculations to rotating black hole solutions [656–660]. Another step in this direction would be to include in the analysis surrounding matter fields that dynamically interact with the black hole in a way that also affects the emitted quasinormal modes – see e.g. [289–291, 428, 673–676]. It is also worth emphasising that there are several complementary probes of $c_{GW}$ in addition to the gravitational wave probes discussed throughout this Chapter and corresponding to energy/frequency scales outside of the range considered here. These include constraints from cosmological large scale structure, currently at the $O(1)$ level

– see e.g. [602, 603] and references therein – which are expected to improve to $O(10^{-1})$ in the near future [609]. While we have not mandated a specific sign for any potential deviation of $c_{GW}$ away from $c$, theoretical bounds from requiring causality, locality and unitarity at high energies can further yield information on these deviations at the (comparatively) low energies probed by gravitational waves and cosmology, noticeably mandating $c_{GW} \geq c$ for large classes of models [677, 678]. We close by re-emphasising that we have mostly focused on investigating how well $c_{GW}$ can be tested by ringdown observations, for a single detected odd-parity quasinormal mode. As more sources and modes are detected in the future and the theoretical machinery to analyse them is further developed, we fully expect further tightened constraints to become obtainable.

L'originalitat consisteix a tornar a l'origen.

Antoni Gaudí

### Chapter summary

In this Chapter we investigate black hole solutions with time-dependent (scalar) hair in scalar-tensor (ST) theories. Known exact solutions exist for such theories at the background level, where the metric takes on a standard GR form (e.g. Schwarzschild-de Sitter), but these solutions are generically plagued by instabilities, as recapped in Section (§5.2). Recently, a new such solution was identified in [679], in which the time-dependent scalar background profile is qualitatively different from previous known exact solutions – specifically, the canonical kinetic term for the background scalar $X$ is not constant in this solution. In Section (§5.3) we examine the theory in question at the background and perturbative level. We investigate the stability of this new solution by analysing odd parity perturbations, identifying a bound placed by stability and the resulting surviving parameter space. In Section (§5.4) we derive the corresponding modified Regge-Wheeler equation and employ the WKB method to extract the quasinormal mode spectrum predicted by the theory, finding a concrete shift of quasinormal mode frequencies and damping times compared to GR. Finally, in Section (§5.5) we forecast constraints on these shifts (and the single effective parameter $\hat{\beta}$ controlling them) from current and future gravitational wave experiments, finding constraints at up to the $\mathcal{O}(10^{-2})$ and $\mathcal{O}(10^{-4})$ level for LVK and LISA/-TianQin, respectively. Finally, we will summarise conclusions in Section (§5.6).

This Chapter is based on the findings in [2], and all calculations shown here are reproducible in the companion `Mathematica` notebook [4].

[679]: Bakoupoulos et al. (2023), "Black holes with primary scalar hair"

[2]: Sirera et al. (2025), "Stability and quasinormal modes for black holes with time-dependent scalar hair"

## 5.1 Introduction

**Ringdown tests of gravity**: The rise of gravitational wave science offers a new way to probe gravity in the strong field regime. In particular, the final stage of gravitational wave signals emitted by binary compact object mergers, known as the ringdown, can be notably sensitive to new gravitational physics. The ringdown phase is well modelled by linear perturbations on a black hole background, whose evolution is described by a superposition of complex decaying frequencies, or quasinormal modes (QNMs) [194].[1] General Relativity (GR) predicts that the full set of QNMs is fixed by the black hole's mass and angular momentum (and charge, if present). This is a consequence of no-hair theorems in GR[2] [680], and is therefore generically violated in extensions of GR which allow for hairy solutions.[3] In such cases, QNMs can depend on extra parameters associated with the hair, and so look-

1: See Section 2.4.2 for more details.

2: See Sections 2.2.2 and 3.4 for more details.

[680]: Mazur (2000), "Black hole uniqueness theorems"

3: Note, however, that the presence of hair is not a necessary requirement to violate the GR spectrum. In particular, the equations governing perturbations (hence QNMs) can differ from those of GR, even in cases where the background solutions are those of GR without any additional hair [681]. For example, even parity QNMs can be modified in Horndeski ST theories in the absence of non-trivial hair [213].

ing for deviations from the GR-predicted QNM spectrum provides a powerful null test of GR. More quantitatively, QNM measurements can therefore be used to place constraints on (additional) fundamental gravitational degrees of freedom generically associated with extensions of GR that leave an imprint on the QNMs [647]. This program, sometimes referred to as testing the *Kerr hypothesis*[4], or *black hole spectroscopy* [683], has been employed to probe the strong gravity regime in a diverse number of ways, see e.g. [1, 2, 213, 276, 465, 467, 469, 634, 672, 684–691].

**Black hole-scalar solutions**: In this Chapter we will work in the context of ST theories, more specifically Horndeski gravity [355, 356], which is the most general theory built with a metric tensor and a scalar field yielding second-order equations of motion.[5] Black hole solutions to such theories broadly divide into two categories. On one hand, we have *stealth* black hole solutions, where the background metric takes the same form as known GR solution such as Schwarzschild or Schwarzschild-de Sitter (SdS).[6] Any potential 'hair' associated to such stealth solutions is then associated only with the profile of the background scalar. On the other hand, due to the increased complexity of scalar tensor actions with respect to the Einstein-Hilbert action, more general black hole solutions can also exist in such theories. Thus, the other category of black hole solutions involve new background solutions for the metric itself (as well as for the scalar) – see e.g. [422, 692]. In this Chapter, we focus on the former, i.e. stealth black hole solutions with non-trivial scalar profiles. In particular, we will focus on Schwarzschild and SdS black hole solutions. Especially SdS solutions have been a focus of attention in previous studies [562, 565–569, 571–573, 579, 679, 693], since the de Sitter asymptotics at large distances more closely mimic realistic black holes embedded in cosmological space-times (in comparison to Schwarzschild solutions with their Minkowski asymptotics). In Section 5.2 we will recap S(dS) solutions found in ST theories, where the scalar has a time-dependent profile. But looking ahead, known exact solutions 1) generically suffer from instabilities, and 2) have been investigated for scalar profiles where $X \equiv -\frac{1}{2}\phi_\mu\phi^\mu$, the kinetic term for the scalar $\phi$, is constant.

In this Chapter we therefore focus on a novel black hole solution that was recently found within Horndeski ST theories [679] and is a promising candidate going beyond the constant $X$ assumption and possibly providing stable dynamics as well. This theory is given by the following Lagrangian

$$\mathscr{L} = 2\eta\sqrt{X} - 2\Lambda + R(1 + \lambda\sqrt{X}) + \frac{\lambda}{2\sqrt{X}}\left[(\Box\phi)^2 - \phi^{\mu\nu}\phi_{\mu\nu}\right], \qquad (5.1)$$

where one can see that standard GR plus a decoupled scalar (albeit with non-standard kinetic term) is recovered in the $\lambda \to 0$ limit. This theory possesses an exact background solution of the form [679]

$$ds^2 = -B(r)dt^2 + \frac{1}{B(r)}dr^2 + d\Omega^2,$$
$$\bar{\phi} = qt + \psi(r), \qquad (5.2)$$

with

$$B(r) = 1 - \frac{2M}{r} - \frac{1}{3}\Lambda r^2. \qquad (5.3)$$

The background solution for the metric is therefore manifestly of SdS form and $q$ is a (dimensionful) constant characterising the linear time-dependence of the scalar field, with $\psi$ encoding its radial dependence. As we are dealing with a shift symmetric theory, the derivatives of the scalar field encode its most important features. In particular, the radial derivative of the scalar field and the kinetic term $X$ are given by

$$\psi'^2 = \frac{q^2}{B^2}\left(1 - \frac{\lambda B}{\lambda + \eta r^2}\right), \qquad\qquad X = \frac{1}{2}\frac{q^2\lambda}{\lambda + \eta r^2}, \qquad (5.4)$$

where it is immediately apparent that this is a scalar profile with non-constant $X$ and we illustrate the dependence of $X$ on $r$ in Figure 5.1. Our focus in this Chapter will therefore be to investigate this novel solution further, specifically its stability and QNM spectrum.

## 5.2 Scalar-tensor theories and hairy black holes

In this Section we briefly review the current state of hairy black hole solutions and their known stability properties. Our focus is on theories where the new fundamental physics is characterised by a single scalar degree of freedom $\phi$, thus constituting a ST theory. As previously noted, Horndeski gravity is the most general such theory resulting in second-order equations of motion [355, 356][7], and is governed by the following action

$$S = \int d^4x\sqrt{-g}\Big[G_2 + G_3\Box\phi + G_4R + G_{4X}\left[(\Box\phi)^2 - \phi^{\mu\nu}\phi_{\mu\nu}\right] +$$
$$G_5G_{\mu\nu}\phi^{\mu\nu} - \frac{1}{6}G_{5X}\left[(\Box\phi)^3 - 3\phi^{\mu\nu}\phi_{\mu\nu}\Box\phi + 2\phi_{\mu\nu}\phi^{\mu\sigma}\phi^\nu_\sigma\right]\Big]. \qquad (5.5)$$

Recall that we have employed the shorthands $\phi_\mu \equiv \nabla_\mu\phi$ and $\phi_{\mu\nu} \equiv \nabla_\nu\nabla_\mu\phi$, and the $G_i$ are free functions of $\phi$ and $X$, where $X \equiv -\frac{1}{2}\phi_\mu\phi^\mu$. $G_{iX}$ denotes the partial derivative of $G_i$ with respect to $X$. When discussing previous work on black hole solutions in ST theories, we will occasionally also refer to solutions derived in extensions of Horndeski theories, in particular within DHOST [348–350, 694, 695].

Horndeski gravity admits a richer variety of black hole solutions compared to GR. Yet, akin to GR, there exist a number of no-hair theorems for a wide range of ST theories which enforce the scalar to possess a trivial profile [115–118]. These theorems have been broadly exposed in Section 2.2.2.[8] Stealth black holes with a constant scalar field background profile $\bar\phi$ have therefore been investigated in [213, 634, 635]. Around this background odd metric perturbations trivially behave just as in GR, since they are unaffected by the even sector (where scalar perturbations do induce non-trivial effects) and also do not feel any effects from the scalar background solution (since this is trivial in the present no-hair setup).[9] Consequently, in order to explore potentially observable effects induced by the scalar, one ought to either investigate different background solutions (and hence consider Horndeski theories that evade no-hair theorems) or consider even perturbations. For detailed discussions of the second option we refer to [212, 213, 634, 636–643] for work in the context of Horndeski gravity, and to [212, 476, 578, 644–647] for work in the context of other theories (ST or otherwise). However, like in the previous Chapter, here we will proceed along the first route, considering the dynamics of odd perturbations around background solutions which evade no-hair theorems. There exist a number of loopholes around no-hair theorems and one can broadly classify the constructed hairy solutions depend-

7: For the equivalence between the formulations of [355] and [356], see [391].

8: For scalar-tensor theories, it was first shown that stationary black holes in minimally coupled Brans-Dicke theories contain no hair [115], a result which was extended to a more general class of scalar-tensor theories including self-interactions of the scalar [116], to spherically symmetric static black holes in Galilean-invariant theories [117], and for slowly rotating black holes in more general shift-symmetric theories [118].

9: This set up has been portrayed in full detail in Section 3.5.

**Table 5.1:** Stealth black hole solutions with a linearly time-dependent scalar (5.2). S(dS) corresponds to Schwarzschild(-de Sitter), while (K)RN(dS) refers to (Kerr-)Reissner-Nordsröm(-de Sitter). The cosmological constant in dS can either come from a bare cosmological constant in the action as in GR, or from an effective combination of beyond-GR parameters (i.e. self-tuned). We denote cases which can fall in the latter category with ∗. The symbol ∄ is used to indicate the non-existence of stealth black hole solutions – in particular such solutions were shown to be absent in large classes of shift symmetry breaking Horndeski theories in [570]. Further details related to all the solutions shown in this Table are discussed in Section 3.4.2. Let us only point out here that even modes have been shown to suffer from instabilities in all known hairy solutions for which $X$ = const [578]. This Table builds on a pre-existing one [571].

| $\mathcal{L}$ | Background solution | | Stability | |
|---|---|---|---|---|
| | $g_{\mu\nu}$ | $X$ | odd | even |
| (Shift + refl)-sym Horndeski [565, 566] | S(dS)∗ | $\frac{q^2}{2}$ = const | ✓ | ✗ |
| Cubic Galileon [568, 569] | S(dS)∗ (non-exact) | non-const (non-exact) | ? | ? |
| Shift-sym breaking Horndeski [570] | ∄ (for large subclasses) | ∄ (for large subclasses) | – | – |
| $G_2 = \eta X, G_4 = \zeta + \beta\sqrt{X}$ [562] | S(dS)∗ + RN(dS)∗ (non-exact) | non-const | ? | ? |
| Shift-sym beyond Horndeski [693] | SdS∗ | $\frac{q^2}{2}$ = const | ✓ | ✗ |
| Shift-sym breaking quadratic DHOST [573] | S(dS)∗ | $\frac{q^2}{2}$ = const | ✓ | ✗ |
| Shift-sym quadratic DHOST [571, 696] | S(dS)∗ + K∗ | const | ✓ | ✗ |
| Quadratic DHOST [572] | S(dS)∗ + (K)RN(dS)∗ | const | ✓ | ✗ |
| $G_2 = -2\Lambda + 2\eta\sqrt{X}, G_4 = 1 + \lambda\sqrt{X}$ [679] | S(dS) | $\frac{1}{2}\frac{q^2\lambda}{\lambda+\eta r^2}$ | ✓ | ? |

ing on whether the background scalar is static or contains a (typically linear) time-dependence. We have summarised in Section 3.4 the existence of static and time-dependent scalar hair solutions. Here, we focus here exclusively on subsets of Horndeski which admit stealth black hole metrics with a time-dependent scalar as a background solution as in (5.2) of Schwarzschild or Schwarzschild-de Sitter (SdS) form

$$B = 1 - \frac{2M}{r}, \qquad\qquad \text{Schwarzschild},$$

$$B = 1 - \frac{2M}{r} - \frac{1}{3}\Lambda r^2, \qquad \text{SdS}. \qquad (5.6)$$

Here $M$ is the mass of a spherically symmetric compact object (e.g. a black hole) and $\Lambda$ is a cosmological constant. Note that since we are focusing on stealth solutions where $B(r)$ is given by a GR metric, the scalar hair introduced by $q$ is of secondary nature.[10]

10: Non-stealth solutions with primary hair have also been found in [679], but we will not be considering here.

Solutions of the form of (5.2) have been found to be of special cosmological interest, where the time-dependent scalar can affect cosmological dynamics and e.g. play the role of dark energy. More generally speaking, when embedding black hole solutions in a cosmological spacetime time-diffeomorphisms are naturally broken in the long-distance limit (unlike for Schwarzschild solutions with Minkowski asymptotics) and hence a a time-dependent solution for the scalar is a natural occurrence in such settings. Calculating and measuring physical effects arising from such a time-dependence on the gravitational waves emitted by black holes therefore promises to provide informative constraints on models such as (5.1). In this work we will focus on the imprints left on the quasinormal modes in the ringdown signal of binary black hole mergers by the non-trivial nature of the scalar field. However, a key requirement that needs to be satisfied prior to carrying out a ringdown study is for the solution to be stable, i.e. to avoid an unphysical (exponential) growth of perturbations. And indeed, while SdS solutions (i.e. approximants to "cosmological black hole" solutions) have been found for several Sections of the Horndeski family, they have also been generically found to suffer from insta-bilities when the scalar has a time-dependent profile. Table 5.1 summarises the existence and stability of these types of solutions, and a more in-depth and historical examination of the results can be found in Section 3.4.2. Here, it suffices to say that, while such exact SdS and scalar profile solutions have been found for large sub-classes of Horndeski theories, higher-$\ell$ even modes around backgrounds of the form of (5.2) have been shown to generically suffer from instability or strong coupling issues [578].

This then leaves us with no known well-behaved stealth black hole solutions with a linearly time-dependent scalar. However, as pointed out above, $X = const$ is a key requirement for the results of [578] to hold, so an obvious question is whether solutions with different scalar profiles exist and, if so, whether perturbations can be stable on such other backgrounds where $X \neq$ const.[11] While previous solutions have frequently been constructed by first imposing $X = const$ for simplicity (see e.g. a related discussion in [573]) and then finding the form of $\psi$ satisfying this condition, the theory (5.1) was recently identified as possessing a stealth SdS solution for which $X \neq const$ [679]. In terms of Horndeski functions, this theory is given by

$$G_2 = -2\Lambda + 2\eta\sqrt{X}, \qquad G_4 = 1 + \lambda\sqrt{X}, \qquad G_3 = 0 = G_5. \qquad (5.7)$$

As pointed out above, unlike previous solutions, $X$ then adopts a non-trivial radial profile, cf. equations (5.2) and (5.4). In the next Section we will therefore examine this theory at the background and perturbative level, showing that the solution can be stable under odd parity perturbations, and showing the effects on its (likewise odd parity) quasinormal mode spectrum.

## 5.3 Background stability and perturbations

Re-expressing (5.7) as a full action for the theory, we have

$$S = \int d^4x \sqrt{-g} \Big[ R\left(1 + \lambda\sqrt{X}\right) - 2\Lambda + 2\eta\sqrt{X}$$
$$+ \frac{\lambda}{2\sqrt{X}} \left((\Box\phi)^2 - (\phi_{\mu\nu})^2\right) \Big], \qquad (5.8)$$

where the second line comes from the $G_{4X}$ term. It is worth making explicit that, just as for the metric determinant, the square root of $X$ is taken to be the principal (positive) square root. Upon a redefinition of the scalar field one can absorb into the scalar field either of the two parameters $\eta$ or $\lambda$ and hence redefine the theory without loss of generality in terms of only one parameter. This will be important when investigating and constraining physical deviations from GR in Sections 5.4 and 5.5. In particular, we will redefine the scalar field as $\phi \to \frac{\phi}{\eta}$ and use $\beta^2 \equiv \frac{\lambda}{2\eta}$ as the single parameter controlling (small) departures from GR. In this Section, in order to facilitate comparison with [679], we will however keep both $\eta$ and $\lambda$ as bookkeeping parameters. We show in the next Section expressions for the covariant equations of motion, the current associated with the theory's shift-symmetry (demonstrating its regularity), and show a relation between the scalar and metric equations.

### 5.3.1 Equations of motion and conserved current

Writing the Lagrangian (5.8) as

$$\mathcal{L} = \mathcal{L}_{GR} + \mathcal{L}_\eta + \mathcal{L}_\lambda, \tag{5.9}$$

with

$$\begin{aligned}
\mathcal{L}_{GR} &= R - 2\Lambda, \\
\mathcal{L}_\eta &= 2\eta\sqrt{X}, \\
\mathcal{L}_\lambda &= \lambda\left(R\sqrt{X} + \frac{1}{2\sqrt{X}}\left((\Box\phi)^2 - (\phi_{\mu\nu})^2\right)\right),
\end{aligned} \tag{5.10}$$

the equations of motion from for the metric tensor and the scalar can respectively be written as

$$\mathcal{E}_{\mu\nu} = \mathcal{E}_{\mu\nu}^{GR} + \mathcal{E}_{\mu\nu}^{\eta} + \mathcal{E}_{\mu\nu}^{\lambda} = 0, \tag{5.11}$$

$$\mathcal{E}_\phi = \mathcal{E}_\phi^{\eta} + \mathcal{E}_\phi^{\lambda} = 0, \tag{5.12}$$

where we introduce the notation $\mathcal{E}_{\mu\nu}^i \equiv \frac{\delta S_i}{\delta g^{\mu\nu}}$, $\mathcal{E}_\phi^i \equiv \frac{\delta S_i}{\delta \phi}$, and we have used the fact that $\mathcal{E}_\phi^{GR} = 0$ since it is $\phi$-independent. Full expressions for these terms are given by

$$\mathcal{E}_{\mu\nu}^{GR} = R_{\mu\nu} - \frac{1}{2}g_{\mu\nu}(R - 2\Lambda), \tag{5.13}$$

$$\mathcal{E}_{\mu\nu}^{\eta} = -\frac{\eta}{\sqrt{X}}\left(Xg_{\mu\nu} + \frac{1}{2}\phi_\mu\phi_\nu\right), \tag{5.14}$$

$$\begin{aligned}
\mathcal{E}_{\mu\nu}^{\lambda} = \frac{\lambda}{2\sqrt{X}}\Bigg[ &-2X\left(R_{\mu\nu} - \frac{1}{2}g_{\mu\nu}R\right) + \frac{1}{2}\phi_\mu\phi_\nu\left(R - \frac{1}{2}(\Box\phi)^2\right) + \frac{1}{2}g_{\mu\nu}\left(2R_{\sigma\rho}\phi^\sigma\phi^\rho + \phi_{\sigma\rho}\phi^{\sigma\rho} - (\Box\phi)^2\right) + \phi_{\mu\nu}\Box\phi \\
&- 2\phi^\sigma\left(\phi_{(\mu}R_{\nu)\sigma} + \phi_{(\mu}\phi_{\nu)\sigma}\right) - \phi_{\nu\sigma}\phi_\mu^\sigma - R_{\mu\sigma\nu\rho}\phi^\sigma\phi^\rho \\
&+ \frac{1}{2X}\left(2\phi^\sigma\phi_{\sigma\rho}\phi_{(\mu}\phi_{\nu)}^\rho - \phi_{\mu\nu}\phi_\sigma\phi_{\rho\sigma} - \frac{1}{2}\phi_\mu\phi_\nu\phi_{\sigma\rho}\phi^{\sigma\rho} + \phi_{\mu\sigma}\phi^\sigma\phi_{\nu\rho}\phi^\rho \right. \\
&\left.+ g_{\mu\nu}\phi^\sigma\phi^\rho\left(\Box\phi\phi_{\sigma\rho} - \phi_{\rho\gamma} - \phi_{\rho\gamma}\phi_\sigma^\gamma\right)\right)\Bigg],
\end{aligned} \tag{5.15}$$

$$\mathcal{E}_\phi^{GR} = 0, \tag{5.16}$$

$$\mathcal{E}_\phi^{\eta} = \frac{\eta}{\sqrt{X}}\left(\Box\phi + \frac{1}{2X}\phi^\mu\phi^\nu\phi_{\mu\nu}\right), \tag{5.17}$$

$$\begin{aligned}
\mathcal{E}_\phi^{\lambda} = \frac{\lambda}{\sqrt{X}}\Bigg[ &\frac{1}{2}R\Box\phi - 2\phi_{\mu\nu}\phi^{\mu\nu} + \Box^2\phi - X\Box\phi - R_{\mu\nu}\phi^\mu\phi^\nu - \frac{3}{2}\phi_{\mu\nu}\phi^\mu\phi^\nu \\
&+ \frac{1}{2X}\left(-3\phi^\mu\phi^\nu\phi_{\mu\sigma}\phi_\nu^\sigma + \phi_{\sigma\nu}\left(-\phi^{\mu\nu}\phi_\mu^\sigma + \frac{1}{2}\Box\phi\phi^{\sigma\nu}\right) + \phi^\mu\left(\frac{1}{2}R\phi^\nu\phi_{\mu\nu} - 2\phi_\mu^\nu\phi^\rho R_{\rho\nu} - \phi^\rho\phi_\nu^\sigma R_{\rho\sigma\mu}^\nu\right)\right) \\
&+ \frac{3}{(2X)^2}\phi^\mu\phi^\nu\phi_{\sigma\rho}\left(-\phi_\mu^\sigma\phi_\nu^\rho + \frac{1}{2}\phi_{\mu\nu}\phi^{\sigma\rho}\right)\Bigg],
\end{aligned} \tag{5.18}$$

12: Symmetric and antisymmetric tensors are defined as

$$\begin{aligned}
A_{(\mu\nu)} &\equiv \frac{1}{2}(A_{\mu\nu} + A_{\nu\mu}), \\
A_{[\mu\nu]} &\equiv \frac{1}{2}(A_{\mu\nu} - A_{\nu\mu}).
\end{aligned} \tag{5.19}$$

where we have used again the standard definition of symmetric and antisymmetric tensors (5.19).[12] We note that the above equations of motion satisfy an interesting relation given by

$$\nabla^\mu\mathcal{E}_{\mu\nu} = -\phi_\nu\mathcal{E}_\phi. \tag{5.20}$$

This is in fact satisfied independently by the three different lagrangians in (5.9). One can easily check from the expressions (5.14) and (5.17) that $\nabla^\mu \mathscr{E}^\eta_{\mu\nu} = -\phi_\nu \mathscr{E}^\eta_\phi$. On the other hand, showing that $\nabla^\mu \mathscr{E}^\lambda_{\mu\nu} = -\phi_\nu \mathscr{E}^\lambda_\phi$ is more cumbersome and is therefore shown in [4] for the interested reader. Then, one is left with $\nabla^\mu \mathscr{E}^{GR}_{\mu\nu} = -\phi_\nu \mathscr{E}^{GR}_\phi$. As was argued before, $\mathscr{E}^{GR}_\phi = 0$ due to the lack of $\phi$-dependence, which in turn reassures that Bianchi identities are satisfied, i.e. $\nabla^\mu \mathscr{E}^{GR}_{\mu\nu} = 0$.

In this Chapter we are focusing on so-called 'stealth' solutions, where despite of dynamical modifications induced by the scalar field, the metric background solution is as it would be in standard GR (in our case: a standard SdS solution) and hence $\mathscr{E}^{GR}_{\mu\nu} = 0$. In that case, equations (5.11) and (5.12) imply that the scalar background profile satisfies both

$$\mathscr{E}^\eta_{\mu\nu} = -\mathscr{E}^\lambda_{\mu\nu}, \tag{5.21}$$

$$\mathscr{E}^\eta_\phi = -\mathscr{E}^\lambda_\phi. \tag{5.22}$$

As a consequence of the shift-symmetry of the theory considered here, the scalar equation of motion can be rewritten as a conservation equation

$$\mathscr{E}_\phi = -\nabla_\mu J^\mu = 0, \tag{5.23}$$

where the current $J^\mu$ can similarly be written as

$$J^\mu = J^\mu_\eta + J^\mu_\lambda \tag{5.24}$$

with [561]

$$J^\mu_\eta = -\frac{\eta}{\sqrt{X}} \phi^\mu, \tag{5.25}$$

$$J^\mu_\lambda = \frac{\lambda}{2\sqrt{X}} \left[ G^{\mu\nu} \phi_\nu - \frac{1}{2X} \left( \phi^\mu \left( (\phi_{\nu\sigma})^2 - (\Box\phi)^2 \right) \right. \right. $$
$$\left. \left. + 2\phi^{\mu\nu} \left( \Box\phi\phi_\nu - \phi_{\nu\sigma}\phi^\sigma \right) \right) \right]. \tag{5.26}$$

Note that $\mathscr{E}^\eta_\phi = -\nabla_\mu J^\mu_\eta$ and $\mathscr{E}^\lambda_\phi = -\nabla_\mu J^\mu_\lambda$. The norm of the current is given by

$$J^2 \equiv J_\mu J^\mu = \frac{1}{6r(\lambda + \eta r^2)} \left[ 384M\eta^2 - r \left( 27\lambda^3\Lambda^2 + 108\eta^3 r^2 \right. \right.$$
$$\left. \left. + 9\eta\lambda^2\Lambda(20 + 3\Lambda r^2) + 4\eta^2\lambda(75 + 29\Lambda r^2) \right) \right]. \tag{5.27}$$

Hence, the norm of the current is found to be regular for all $r$ down to the black hole horizon.

## 5.3.2 Cosmological background

We will focus on black hole solutions and related perturbations in this Chapter and hence primarily investigate (5.8) as a fiducial effective description of physics on the corresponding scales. Nevertheless, given the Schwarzschild-de Sitter stealth solution for the metric, it is interesting to briefly discuss the long-distance, 'cosmological' de Sitter limit of the dynamics encoded in (5.8). Focusing on the background evolution in the cosmological $r_s/r \ll 1$ limit, one can map the de Sitter metric in spherically symmetric coordinates – the long distance limit of (5.6) – to standard cosmological coordinates by using the transformations [568]

$$t = \tau - \frac{1}{2H} \ln \left[ 1 - \left( He^{H\tau}\rho \right)^2 \right]$$
$$r = e^{H\tau}\rho,$$
(5.28)

where $\Lambda = 3H^2$. This transformation then yields the metric

$$ds^2 = -d\tau^2 + e^{2H\tau}(d\rho^2 + \rho^2 d\Omega^2),$$
(5.29)

up to corrections $\mathcal{O}(r_s/r)$, i.e. terms strongly suppressed in the cosmological long distance limit. The Hubble parameter $H$ is a constant in the de Sitter limit we are considering here. In these new coordinates, the canonical scalar kinetic term $X$ is given by

$$X = \frac{1}{2} \frac{q^2\lambda}{\lambda + e^{2\sqrt{\frac{\Lambda}{3}}\tau}\eta\rho^2} = \frac{e^{-2\sqrt{\frac{\Lambda}{3}}\tau}q^2\lambda}{2\eta\rho^2} + \mathcal{O}\left(\frac{1}{\rho^4}\right),$$
(5.30)

where we again see that $X$ is asymptotically suppressed at large distances. The fact that the effective cosmological constant is $\Lambda$ (i.e. is not redressed by contributions from the scalar) as well as (5.30) serve to highlight two key points: 1) In the solution we are considering, the scalar field $\phi$ does not affect the cosmological background solution and e.g. is therefore not playing the role of dark energy. While, as discussed above, time-dependent scalar hair is often motivated by the way in which this time-dependence can be linked to cosmological background dynamics, this is therefore not a primary motivation for the specific solution we investigate here. 2) The form of $X$ (5.30) and the fact that it is asymptotically suppressed at large distances means that this is not a solution where one obtains *both* $\phi \propto \tau$ in cosmological coordinates *and* a metric solution of the simple cosmological form (5.29). Instead, a significant (unsuppressed) $\rho$-dependence remains present in the $\phi$ profile.

The behaviour of this solution, particularly the kinetic term $X$, warrants careful consideration. Perturbations can become problematic when $X \to 0$ on a Minkowski background, as non-analytic terms in the action involving $1/\sqrt{X}$ would diverge, potentially leading to pathologies or ill-defined equations of motion for perturbations [703]. However, it is crucial to distinguish between the asymptotic flat (Minkowski) background, where $X \to 0$ at spatial infinity might signal a genuine strong coupling problem, and the de Sitter background considered here. In a de Sitter spacetime, the asymptotic region is fundamentally altered by the positive cosmological constant $\Lambda$, leading to a cosmological horizon. This implies that physical processes are confined to regions within this horizon. The behaviour of perturbations near the de Sitter horizon and the existence of this horizon can prevent the system from truly probing the $X \to 0$ regime in a pathological manner relevant for observable physics, effectively shielding such strong coupling issues from being physically problematic. In order to fully determine the strong coupling scale of this theory, one would be required to go beyond the linear theory. Moreover, in order to fully assess weather strong coupling issues as described in [578] persist for such solutions with a non-constant $X$, a full analysis of the even sector would be required. This questions lie beyond the scope of this Chapter and therefore constitute potential directions for future research.

Having briefly considered the background evolution, especially in the context of perturbative dynamics it is interesting to note that the theory we are investigating (5.7) belongs to the class of 'extended cuscuton' theories [410] – also see [408, 409] for the original cuscuton theory. These theories satisfy the following condition

$$\mathscr{G}_T \mathscr{K} + 6\mathscr{M}^2 = 0, \tag{5.31}$$

where $\mathscr{G}_T, \mathscr{K}, \mathscr{M}$ are defined for general Horndeski theories in [410], but specialised to our setup are given by

$$\mathscr{G}_T = 2(G_4 - 2G_{4X}), \qquad \mathscr{M} = 2H\partial_\tau\phi(G_{4X} + 2XG_{4XX}),$$
$$\mathscr{K} = G_{2X} + 2XG_{2XX} + 6H^2(G_{4X} + 8XG_{4XX} + 4X^2G_{4XXX}). \tag{5.32}$$

It is straightforward to check that (5.7) satisfies $\mathscr{K} = 0$ and $\mathscr{M} = 0$, hence trivially meeting the condition (5.31). For such theories, [410] show that the scalar manifestly does not propagate when expressed in coordinates such that $\phi \propto \tau$. While we do not work in a coordinate system where $\phi \propto \tau$ here, the timelike nature of the derivative of $\phi$ means a transformation to new coordinates $\tilde{x}^\mu$ such that $\phi \propto \tilde{\tau}$ is expected to exist. Note that the metric is likewise expected to take on a form different from (5.29) in this new coordinate system. While we will not investigate the cosmological limit in different coordinate systems or further detail here, it will be interesting to investigate the link to cosmological cuscuton analyses further in the future. In the same vein it will also be interesting to in the future go beyond the odd parity perturbations around black hole backgrounds we focus on here. Investigating the even parity sector of (5.7) will in particular allow an assessment of whether scalar perturbations (which are even parity and hence do not show up in the odd sector) propagate in the black hole solutions considered here and precisely how this connects to the cosmological limit at large distances.[13]

### 5.3.3 Black hole background

Having briefly discussed cosmological limits, let us now move to the black hole backgrounds which are the main focus of this Chapter. We recall from equations (5.3) and (5.4) that it was shown in [679] that the theory (5.1) has a solution of the form (5.2) with [14]

$$B(r) = 1 - \frac{2M}{r} - \frac{\Lambda r^2}{3},$$
$$\psi'^2 = \frac{q^2}{B^2}\left(1 - \frac{\lambda B}{\lambda + \eta r^2}\right), \tag{5.33}$$

where the scalar kinetic term is consequently given by

$$X = \frac{q^2}{2B} - \frac{1}{2}B\psi'^2 = \frac{1}{2}\frac{q^2\lambda}{\lambda + \eta r^2}. \tag{5.34}$$

13: For a related study of black hole perturbations on a cuscuton-like model see [704], where it is found that in their set up odd parity modes evolve in the same way as in GR. Related to cosmological dynamics, while the model investigated here does not alter the cosmological expansion and dark energy is just given by a cosmological constant, we note that other cuscuton-related models can affect dark energy dynamics [705].

14: We refer to our reproducible **Mathematica** notebook for the details [4].

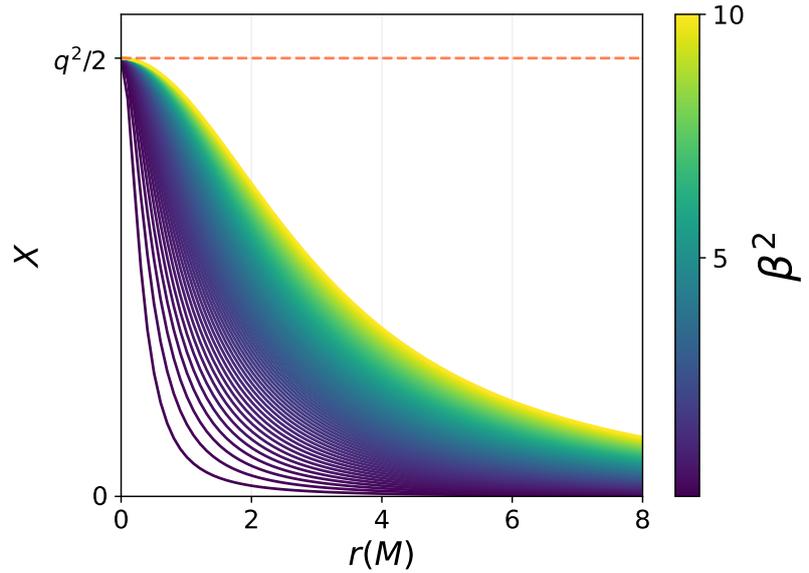

**Figure 5.1:** Here we plot $X$ (5.34), the standard kinetic term for the background scalar as a function of $r$. We show different choices for $\beta$, defined as $\beta^2 = \frac{\lambda}{2\eta}$. Smooth continuous lines exist only for $\beta^2 > 0$, i.e. $\eta$ and $\lambda$ having the same sign. In fact, it will be shown in Section 5.3.5 that $\beta^2 > 0$ is necessary to guarantee stability for all $r$. The horizontal orange dashed line at $X = q^2/2$ corresponds to the limit $\eta = 0$.

As mentioned above, unlike for other known time-dependent solutions, the kinetic term $X$ here is not a constant but rather contains a specific r-dependence (see Figure 5.1). Note that the presence of *both* $\eta$ and $\lambda$ is required in order to have $X \neq const$. The form of $\psi$ in (5.33) ensures that both (5.21) and (5.22) are satisfied and thereupon guarantees the existence of stealth black hole solutions such as SdS.

### 5.3.4 Perturbations

15: We will work to leading (linear) order in this Chapter, but note that this decoupling does not hold at higher orders – see [278–280, 631–633] for details on the behaviour of higher order modes. This linear order decoupling enables us to study the odd sector in isolation.

In order to investigate the linear stability of the solution (5.33) in the theory (5.8) and extract the quasinormal spectrum we need to study the evolution of linear perturbations. As presented in detail in Section 2.4.1, recall that around the static and spherically symmetric backgrounds considered here such perturbations can be decomposed into odd and even parity perturbations (under rotations), which decouple from one another at linear order (i.e. they evolve independently from one another and can therefore be treated separately). [15] Perturbations on the scalar are purely of even parity and therefore will not be considered here. Odd perturbations are however affected by the background solution they are propagating on, so they will nevertheless be sensitive to the new physics encoded by the (background solution of the) scalar field $\phi$. We split the full metric as

$$g_{\mu\nu} = \bar{g}_{\mu\nu} + h_{\mu\nu}, \qquad (5.35)$$

where $\bar{g}_{\mu\nu}$ is given by (5.2) and (5.33), and $h_{\mu\nu}$ are small perturbations on top of it. As we are considering the odd sector and hence no scalar perturbations will be present in our analysis, we will (in an abuse of notation) use the same symbol for the scalar field $\phi$ and its background value. In the Regge-Wheeler gauge [201], these look like

$$h_{\mu\nu}^{\text{odd}} = \begin{pmatrix} 0 & 0 & 0 & h_0 \\ 0 & 0 & 0 & h_1 \\ 0 & 0 & 0 & 0 \\ h_0 & h_1 & 0 & 0 \end{pmatrix} \sin\theta \partial_\theta Y_{\ell m}, \tag{5.36}$$

where we have set $m = 0$ without loss of generality as a consequence of the background metric being static. $h_0$ and $h_1$ are functions of $(r, t)$, where the $t$-dependence will be taken to be of the form $e^{-i\omega t}$. To obtain the evolution of such linear perturbations, we need to work at quadratic order in the perturbed action, which we define as

$$S^{(2)} = \frac{1}{2} \int d^4 x \sqrt{-g} \Big[ \mathcal{L}_{GR}^{(2)} + \mathcal{L}_\eta^{(2)} + \mathcal{L}_\lambda^{(2)} \Big], \tag{5.37}$$

where the expressions for the corresponding quadratic Lagrangians are

$$\mathcal{L}_{GR}^{(2)} = \frac{1}{2} \left( \nabla^\sigma h^{\mu\nu} (2\nabla_\nu h_{\mu\sigma} - \nabla_\sigma h_{\mu\nu}) + 2\Lambda h_{\mu\nu} h^{\mu\nu} \right), \tag{5.38}$$

$$\mathcal{L}_\eta^{(2)} = \frac{-\eta}{\sqrt{X}} \left( X h_{\mu\nu} h^{\mu\nu} + \phi^\mu \phi^\nu h_\mu^\sigma h_{\nu\sigma} \right), \tag{5.39}$$

$$\begin{aligned}
\mathcal{L}_\lambda^{(2)} = \frac{-\lambda}{2\sqrt{X}} \Big[ & \nabla_\sigma h_{\mu\nu} \left( X (\nabla^\sigma h^{\mu\nu} - 2\nabla^\nu h^{\mu\sigma}) + \phi^\sigma \phi^\rho \left( \frac{1}{2}\nabla_\rho h^{(\mu\nu)} - 2\nabla^\nu h_\rho^\mu \right) + \phi^\mu \phi^\nu \nabla_\rho h^{\sigma\rho} + 2\phi^\mu \phi^\nu \nabla^{(\nu} h_\rho^{\sigma)} \right) \\
& + h_{\mu\nu} \Big( \frac{1}{2} h^{\mu\nu} ((\Box\phi)^2 - \phi_{\rho\gamma} \phi^{\rho\gamma}) + 4 h_\sigma^\nu (\Lambda\phi^\mu \phi^\sigma - \phi^{\sigma\mu}\Box\phi + \phi_\rho^\mu \phi^{\rho\sigma}) + 4 h_{\sigma\rho} \phi^{\mu[\sigma} \phi^{\nu]\rho} \\
& \quad + \frac{1}{2X} \left( h_\sigma^\nu \phi^\mu \phi^\sigma (\phi_{\rho\gamma} \phi^{\rho\gamma} - (\Box\phi)^2) + 2\phi^{\sigma\rho} (\phi^\mu \phi^\nu \phi_{(\sigma}^\lambda h_{\rho)\lambda} - h_{\sigma\rho} \phi^\mu \phi^\nu \Box\phi) \right) \\
& \quad + 4\phi^\rho \phi^{\mu\sigma} (2\nabla_{[\sigma} h_{\rho]}^\nu + \nabla^\nu h_{\rho\sigma}) + 2\phi^\mu \phi^{\sigma\rho} (2\nabla_\sigma h_\rho^\nu - \nabla^\nu h_{\rho\sigma}) + 2\phi^\mu \phi^{\nu\sigma} \nabla_\rho h_\sigma^\rho \\
& \quad + 2\phi^\sigma \phi_\sigma^\mu \nabla_\rho h^{\nu\rho} + 2\phi^\sigma \phi_{\sigma\rho} \nabla^{[\nu} h^{\rho]\mu} + 2\Box\phi (\phi^\sigma \nabla_\sigma h^{\mu\nu} - 2\phi^\mu \nabla_\sigma h^{\nu\sigma} - 2\phi^\sigma \nabla^\nu h_\sigma^\mu) \\
& \quad + \frac{1}{2X} \phi^\mu \phi^\nu \phi^\sigma \left( \phi_\sigma^\rho \nabla_\gamma h_\rho^\gamma + \phi^{\rho\gamma} (2\nabla_{(\rho} h_{\gamma)\sigma} - \nabla_\sigma h_{\rho\gamma}) \right) \Big) \Big],
\end{aligned} \tag{5.40}$$

Here, we have used the fact that some terms vanish for odd perturbations. e.g. the trace $h \equiv h_\mu^\mu = 0$ as can be seen from (5.36).[16] We have used the metric equations of motion for this background, i.e. $R_{\mu\nu} = \Lambda g_{\mu\nu}$ and $R = 4\Lambda$, and the notation for symmetric and antisymmetric tensors. From the form of the quadratic action we can already make the following observations: (5.38) describes how odd parity modes propagate in GR, (5.39) provides modification to the effective potential only while (5.40) also provides modifications to the kinetic term.

Substituting the components (5.36) and the solution for the background scalar (5.2) and (5.33) into the quadratic action (5.37), integrating over the angular coordinates and performing several integrations by parts, we can write the action in the following form:

$$S^{(2)} = \int dt dr \Big[ a_1 h_0^2 + a_2 h_1^2 + a_3 \left( \dot{h}_1^2 + h_0'^2 - 2\dot{h}_1 h_0' + \frac{4}{r} \dot{h}_1 h_0 \right) + a_4 h_0 h_1 \Big], \tag{5.41}$$

where a dot and a prime denote derivatives with respect to $t$ and $r$, respectively, and we have dropped an overall multiplicative factor of $2\pi/(2\ell + 1)$

16: Note that this is not true for even parity perturbations - we refer to [4] for full details and recall that we are working in Regge-Wheeler gauge here.

coming from angular integration. The $a$-coefficients are given by

$$a_1 = \frac{\ell(\ell+1)}{r^2}\left[(r\mathscr{H})' + \frac{(\ell-1)(\ell+2)\mathscr{F}}{2B}\right],$$

$$a_2 = -\frac{\ell(\ell+1)(\ell-1)(\ell+2)}{2}\frac{B}{r^2}\mathscr{G},$$

$$a_3 = \frac{\ell(\ell+1)}{2}\mathscr{H},$$

$$a_4 = \frac{\ell(\ell+1)(\ell-1)(\ell+2)}{r^2}\mathscr{J}. \tag{5.42}$$

where the $a_i$ are to be evaluated on the background. Expressions for $\{\mathscr{F}, \mathscr{G}, \mathscr{H}, \mathscr{J}\}$ are given by

$$\mathscr{F} = 2\left(G_4 - \frac{q^2}{B}G_{4X}\right),$$

$$\mathscr{G} = 2\left(G_4 + \left(\frac{q^2}{B} - 2X\right)G_{4X}\right),$$

$$\mathscr{H} = 2(G_4 - 2XG_{4X}),$$

$$\mathscr{J} = 2q\psi'G_{4X}, \tag{5.43}$$

where in our case the $G$-functions take the form given by (5.7) but the formula applies to more general theories [574, 577]. Note that $\mathscr{J} \neq 0$ only due to the presence of the scalar hair $q \neq 0$. The quadratic action (5.41) contains two fields $(h_0, h_1)$, but describes only one dynamical degree of freedom. As shown in [574], the action can be rewritten to make this manifest.

$$S^{(2)} = \frac{\ell(\ell+1)}{4(\ell-1)(\ell+2)} \int dt dr \Big[b_1\dot{Q}^2 - b_2Q'^2 + b_3\dot{Q}Q'$$
$$- (\ell(\ell+1)b_4 + V_{\text{eff}}(r))Q^2\Big], \tag{5.44}$$

where the new variable $Q$ can be written in terms of the old ones as

$$Q = \dot{h}_1 - h_0' + \frac{2}{r}h_0 \tag{5.45}$$

and the $b$-coefficients are given by

$$b_1 = \frac{\mathscr{F}\mathscr{H}^2r^2}{B\mathscr{F}\mathscr{G} + B\mathscr{J}^2},$$

$$b_2 = \frac{B\mathscr{G}\mathscr{H}^2r^2}{\mathscr{F}\mathscr{G} + \mathscr{J}^2},$$

$$b_3 = \frac{2\mathscr{H}^2\mathscr{J}r^2}{\mathscr{F}\mathscr{G} + \mathscr{J}^2},$$

$$b_4 = \mathscr{H}. \tag{5.46}$$

Note the presence of the cross term $b_3$ which includes one time and one radial derivative. By performing a time redefinition as described in [577, 706]

$$\tilde{t} = t + \int \frac{b_3}{2b_2}dr, \tag{5.47}$$

we can diagonalise the kinetic part of the Lagrangian and rewrite the quadratic action in the standard form

$$S^{(2)} = \frac{\ell(\ell+1)}{4(\ell-1)(\ell+2)} \int dt dr \Big[ \tilde{b_1}(\partial_t Q)^2 - b_2 Q'^2$$
$$- (\ell(\ell+1)b_4 + V_{\text{eff}}(r))Q^2 \Big], \qquad (5.48)$$

with

$$\tilde{b_1} = b_1 - \frac{b_3^2}{4b_2} = \frac{\mathscr{H}^2 r^2}{B\mathscr{G}}. \qquad (5.49)$$

The potential is given by

$$V_{\text{eff}}(r) = r^2 \mathscr{H} \left( b_2 \left( \frac{1}{r^2 \mathscr{H}} \right)' \right)' - 2\mathscr{H}. \qquad (5.50)$$

$$= -2 \left( r^2 \left( \frac{b_2}{r^3} \right)' + \mathscr{H} \right). \qquad (5.51)$$

where the second equation is obtained using the fact that $\mathscr{H} = 2$ in this theory and thus $\mathscr{H}' = 0$.

### 5.3.5 Stability conditions

Having written the quadratic action in the form of (5.44), we can easily identify the following conditions in order for perturbations to be well-behaved and not grow over the background

$$\tilde{b_1} > 0, \qquad b_2 \geq 0, \qquad b_4 \geq 0. \qquad (5.52)$$

Since $b_4 = \mathscr{H} = 2$, the last inequality, which ensures the avoidance of tachyonic instabilities, is always satisfied. Before analysing the first two conditions, it is important to show that they indeed are legitimate measures of stability. As was shown in [575], the Hamiltonian density in the $(t, r)$ coordinates, i.e. as obtained from (5.44), might be unbounded from below. However, Hamiltonian densities are coordinate-dependent quantities, and by performing a time coordinate transformation of the type (5.47) and showing the boundedness from below of the Hamiltonian in the new $(\tilde{t}, r)$ coordinates, which is also encapsulated by conditions (5.52), it suffices to guarantee stability [575]. In our background of interest, $\tilde{b_1}$ and $b_2$ take the form

$$\tilde{b_1} = 2r^2 \frac{1}{\frac{|q|}{\sqrt{2}} \sqrt{\lambda^2 + \eta \lambda r^2} + B(r)} > 0,$$

$$b_2 = 2r^2 \frac{\frac{|q|}{\sqrt{2}} \sqrt{\lambda^2 + \eta \lambda r^2} + B(r)}{1 + \frac{|q|}{\sqrt{2}} \frac{\lambda^2}{\sqrt{\lambda^2 + \eta \lambda r^2}}} \geq 0. \qquad (5.53)$$

Note that $B(r)$ is a positive definite function of $r$, so to satisfy both inequalities and ensure the coefficients are real we require

$$\lambda^2 + \eta \lambda r^2 \geq 0. \qquad (5.54)$$

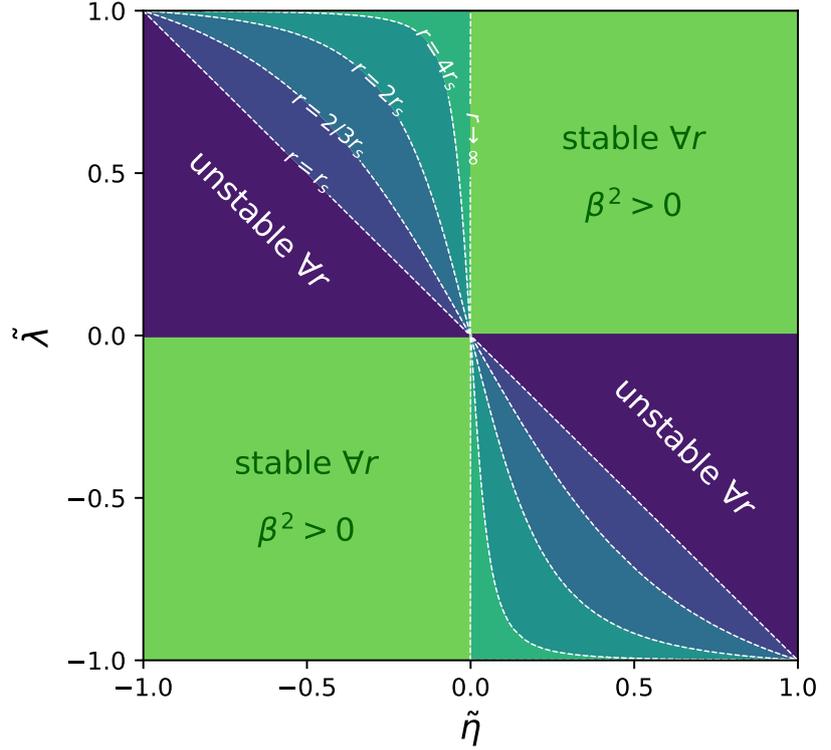

**Figure 5.2:** Stability plot for black holes with a time-dependent scalar (5.2) in cuscuton-like theory (5.8). In the bright green region, stability is ensured for all $r$. At $r = r_s \equiv 2M$, the stable region is extended to cover up to the diagonal lines (5.55). In between the two extremes, there is a smooth $r$-dependent transition as shown by the dashed contours. Finally, the dark blue region corresponds to parameter values violating the stability conditions for all $r$. The plot we show here is a 2D representation of what really is a 3D space, where the extra dimension is given by $r$. In [4] we include an interactive 3D version of this plot which might help in its understanding.

Investigating this condition in the short and long distance limits and for a fiducial mass of $M = 1/2$, one finds

$$\begin{cases} \lambda^2 + \eta\lambda \geq 0 & \text{at } r = 2M, \\ \eta\lambda \geq 0 & \text{at } r \to \infty. \end{cases} \tag{5.55}$$

Note that the second condition is stronger than the first, meaning that $\eta\lambda \geq 0$ is a sufficient condition to guarantee stability for all $r$. It is instructive to point out that we have glossed over the difference between Schwarzschild and SdS metric backgrounds in deriving (5.55). In SdS it does not make sense to take $r \to \infty$, but instead there the cosmological horizon $r_c$ serves as an appropriate long distance limit for $r$. However, $r_c \gg r_s$ by $\sim 20$ orders of magnitude, so corrections to (5.55) arising from this finite long-distance limit are very strongly suppressed and we can therefore work with (5.55) to very high accuracy even for SdS.

To visualise the stability conditions in the full parameter space, we can compactify the infinite range of $r$ as well as $\eta$ and $\lambda$, i.e. $-\infty \leq (\gamma, \kappa) \leq \infty$ into a finite range. This can be done with the choices[17]

17: Note that inverting this one gets

$$\eta \equiv \frac{\tilde\eta}{\sqrt{1 - \tilde\eta^2}}, \qquad \lambda \equiv \frac{\tilde\lambda}{\sqrt{1 - \tilde\lambda^2}}. \tag{5.56}$$

$$\tilde\eta = \frac{\eta}{\sqrt{1 + \eta^2}}, \qquad\qquad \tilde\lambda = \frac{\lambda}{\sqrt{1 + \lambda^2}}. \tag{5.57}$$

Now, the full range of $\tilde\eta$ and $\tilde\kappa$ is given by $-1 \leq (\tilde\eta, \tilde\lambda) \leq 1$. Figure 5.2 summarises the results of our stability analysis using these variables.

As can be easily appreciated from the figure, stability conditions share certain symmetries in the $\eta$ and $\lambda$ plane. This is indeed not surprising, since (as dscussed at the start of this section) one can rewrite the theory we consider in terms of only one free parameter, namely the ratio between $\lambda$ and $\eta$. More specifically, we can redefine the scalar field $\phi \to \phi/\eta$, which maps the action (5.8) to the following

$$S = \int d^4x \sqrt{-g} \Big[ M_{\text{Pl}}^2 R - 2\Lambda + 2M_{\text{Pl}}^2 \sqrt{X} + 2\beta^2 R\sqrt{X}$$
$$+ \frac{\beta^2}{\sqrt{X}} \left( (\Box\phi)^2 - (\phi_{\mu\nu})^2 \right) \Big], \tag{5.58}$$

where we have temporarily suspended geometric units to make powers of $M_{\text{Pl}}$ explicit and have implicitly defined the parameter $\beta^2$ as

$$\beta^2 \equiv \frac{\lambda}{2\eta} \geq 0, \tag{5.59}$$

where the final inequality is mandated by the stability conditions (5.55). Note that defining the parameter as $\beta^2$ is precisely motivated by those stability conditions, but the square root structure of the theory will mean several predictions in the following Sections are controlled by $\beta \equiv \sqrt{\beta^2}$.

## 5.4 Quasinormal modes

Having set up the relevant perturbation theory and discussed stability properties in the previous section, we are now in a position to derive the corresponding quasinormal mode frequencies in the odd sector. As we will see, deviations from standard GR predictions will (as one may expect) be controlled by the $\beta$ parameter introduced above.

### 5.4.1 Modified Regge-Wheeler equation

In order to obtain the analogue of the Regge-Wheeler equation we will follow the procedures described in [706]. We start by obtaining the equation of motion for the master variable $Q$ of the odd parity perturbations by applying the variational principle to (5.44).

$$-\partial_t^2 Q + \frac{b_2}{\tilde{b}_1} Q'' + \frac{b_2'}{\tilde{b}_1} Q' - \frac{\ell(\ell+1)b_4 + V_{\text{eff}}}{\tilde{b}_1} Q = 0. \tag{5.60}$$

To remove the single radial derivative term, we express the equation above in a generalisation of the tortoise coordinate given by[18]

$$r_* = \int \sqrt{\frac{\tilde{b}_1}{b_2}} dr \tag{5.61}$$

and redefine $Q$ as

$$\Psi = FQ \tag{5.62}$$

where

$$F = \left( \tilde{b}_1 b_2 \right)^{1/4}. \tag{5.63}$$

Doing this, we obtain an expression in the form of the Regge-Wheeler equation

$$\left( \partial_*^2 - \partial_t^2 - V \right) \Psi = 0, \tag{5.64}$$

[706]: Nakashi et al. (2023), "Black hole ringdown from physically sensible initial value problem in higher-order scalar-tensor theories"

18: Note that for $\lambda = 0$ we recover the tortoise coordinate in GR $dr_* = \frac{1}{B}dr$.

where we have expressed the tortoise derivative as $\frac{\partial}{\partial r_*} \equiv \partial_*$. The potential is given by

$$V = \frac{\ell(\ell+1)b_4 + V_{\text{eff}}}{\tilde{b}_1} + \frac{1}{F}\partial_*^2 F.$$

(5.65)

and $V_{\text{eff}}$ is given by (5.51). The full analytical expression for $V$ can be written as

$$
\begin{aligned}
V = &V_{RW}\left(1 + \frac{q\beta\sqrt{2\beta^2+r^2}}{B}\right) + \frac{q\beta(B + q\beta\sqrt{2\beta^2+r^2})}{4r^2(2\beta^2+r^2)(2q\beta^3 + \sqrt{2\beta^2+r^2})^3} \times \\
&\times \Bigg[ q^2\beta^6\left(192\beta^4 + 104\beta^2 r^2 + 3r^4\right) + 2(2\beta^2+r^2)\left(24\beta^4 + 17\beta^2 r^2 + 2r^4\right) \\
&\quad + 6q\beta^3\sqrt{2\beta^2+r^2}\left(384\beta^4 + 416\beta^2 r^2 + 117r^4\right) - \beta^2 B\left(2\left(48q^2\beta^6(2\beta^2+r^2) + 48\beta^4 + 56\beta^2 r^2 + 19r^4\right)\right) \\
&\quad + \frac{q\beta^3}{\sqrt{2\beta^2+r^2}}\left(384\beta^4 + 416\beta^2 r^2 + 117r^4\right) \Bigg].
\end{aligned}
$$

(5.66)

where $V_{\text{RW}}$ is the well-known Regge-Wheeler potential in GR (2.58)

$$V_{\text{RW}} = B\left(\frac{\ell(\ell+1)}{r^2} - \frac{6M}{r^3}\right),$$

(5.67)

and, as discussed above, square roots in (5.66) denote principal square roots (this is also the origin of odd powers of $\beta$ seen in the potential).

In the next Section we will investigate the quasinormal frequencies and damping times of this potential. There, we will find that, at leading order, deviations from GR are in fact controlled by a single effective parameter, namely

$$\hat{\beta} \equiv \beta\frac{Mq}{M_{\text{Pl}}^3}.$$

(5.68)

Let us now look at how the potential $V$ looks analytically when $\hat{\beta} \ll 1$, so as to better understand its parametric form. To this end it is instructive to write the potential as

$$V = V_{\text{RW}} + \sum_{i=1}\delta V_i \times \hat{\beta}^i.$$

(5.69)

with the first orders of $\delta V_i$ given by

$$
\begin{aligned}
\delta V_1 &= \frac{1}{Mr}\left(M_{\text{Pl}}^2(\ell(\ell+1)+1) - \frac{8M}{r}\right), \\
\delta V_2 &= \frac{M_{\text{Pl}}^4}{M^2}, \\
\delta V_3 &= \frac{M_{\text{Pl}}^2}{q^2 M^3 r^3}\left(M_{\text{Pl}}^4(\ell(\ell+1)-4) + \frac{21MM_{\text{Pl}}^2}{r} - \frac{38M^2}{r^2}\right), \\
\delta V_4 &= \frac{M_{\text{Pl}}^6}{q^2 M^4 r^2}\left(\frac{56M}{r} - 15M_{\text{Pl}}^2\right),
\end{aligned}
$$

(5.70)

where we have again temporarily suspended geometric units to make powers of $M_{\text{Pl}}$ (and hence mass dimensions) explicit. Higher orders can be found in the companion `Mathematica` notebook [4]. The effect of $\delta V$ on the un-perturbed $V_{RW}$ is shown in Figure 5.3, where one can appreciate almost a constant shift in the amplitude of the potential throughout the range of the radial coordinate.

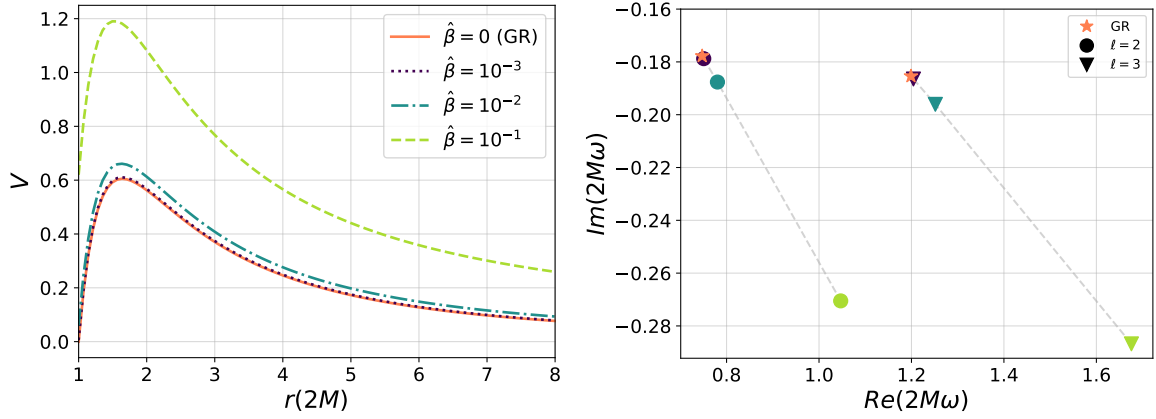

**Figure 5.3: Left panel**: Modified Regge-Wheeler potential for different magnitudes of the deviation parameter $\hat{\beta}$ for $\ell = 2$. We show $r$ in units of $r_s = 2M$, so that $r = 1$ corresponds to $r = r_s$. One can observe that the main effect is an enhancement in the overall amplitude. Also note that the maximum of the potential is slightly shifted to lower values of $r$. **Right panel**: Quasinormal modes for $\ell = 2$ (corresponding to Table **??**) and $\ell = 3$ for increasing magnitudes of $\hat{\beta}$. Both stars correspond to the GR values and colours correspond to $\hat{\beta}$ values on the left panel.

### 5.4.2 Quasinormal frequencies

Having obtained the effective potential for the modified Regge-Wheeler equation, we can now study its quasinormal mode solutions.[19] Looking for quasinormal mode solutions where $\Psi$ has a time dependence of $e^{-i\omega t}$, we can write the Regge-Wheeler equation as

$$\left( \partial_*^2 + \omega^2 - V \right) \Psi = 0. \tag{5.71}$$

Recall from Section 2.4.2 that upon imposing the relevant dissipative boundary conditions (corresponding to outgoing waves at spatial infinity and ingoing waves at the black hole horizon), the $\omega$ solutions to (5.71) become complex and can therefore be written as

$$\omega_{\ell m} = 2\pi f_{\ell m} + \frac{i}{\tau_{\ell m}}. \tag{5.72}$$

Here, the real part corresponds to the physical oscillation frequency of a mode, while the imaginary part corresponds to its damping time. Due to the axisymmetric nature of our background modes with different $(\ell, m)$ indices do not mix.[20] Here we will mainly focus on the dominant mode $(2, 2)$.

There are a number of techniques one can use to obtain the QNM themselves (see e.g. [231, 232, 243, 253, 652–655]) but we refer to [194] for an extensive review of those. Here, we will use the WKB method, first applied to black holes in [243] and subsequently extended to higher orders in [244–247].[21] We do this by adapting the `Mathematica` package which can be found in [247, 252].

The WKB method provides a straightforward and precise technique of obtaining quasinormal frequencies from an effective potential. However, in

|  | Re($2M\omega$) | Im($2M\omega$) |
|---|---|---|
| $\hat{\beta} = 0$ (GR) | 0.7472 | $-0.1778$ |
| $\hat{\beta} = 10^{-3}$ | 0.7506 | $-0.1788$ |
| $\hat{\beta} = 10^{-2}$ | 0.7803 | $-0.1876$ |
| $\hat{\beta} = 10^{-1}$ | 1.0469 | $-0.2705$ |

22: Note that, as is also the case for the GR potential in 5.3, the second turning point for the modified potentials is hidden at $r < 1$, where $V$ approaches $+\infty$ as $r \to 0$.

order for it to be justifiably applied, the potential needs to satisfy some criteria laid out in [247]. Most notably, the potential must have one local maximum, be asymptotically constant, and contain two turning points. As can be observed form Figure 5.3, these are all satisfied by our modified effective potential.[22]

Let us now briefly introduce how the method works practically. The rationale is to match the asymptotic solutions given by (2.133) respectively with a Taylor expansion around the maximum of the potential located at $r_0$. For a differential equation written in the form of (5.71), the matching of solutions at the different regions imposes

$$\omega^2 = V_0 - i\left(n + \frac{1}{2}\right)\sqrt{2V_0^{(2)}} - i\sqrt{2V_0^{(2)}} \sum_{i=2}^{N} \Lambda_j. \qquad (5.73)$$

Here, $V_0$ denotes the potential evaluated at the maximum $r_0$, and we use $V_0^{(2)}$ to denote the second tortoise derivative of $V$ evaluated at the maximum. $n$, being the overtone number, is set to zero when focusing on the fundamental mode. Finally, $\Lambda_j$ are functions of higher order derivatives of the potential, where $j$ denotes the order to which the WKB expansion is carried out. The first application of this method to black holes was carried out in [243] and included only the first order. [244] then extended the WKB formula to 3rd order by computing $\Lambda_2$ and $\Lambda_3$. This was then extended in [245] to 6th order and in [246] to 13th order. However, going up in WKB order does not necessarily mean improving accuracy. For instance, the fundamental mode in GR is best approximated by 6th order WKB [252], and we find that this is also true when including our corrections.[23] Therefore, we perform calculations to 6th order WKB in what follows.

23: We show in [4] that this is the case by computing how error estimation increases with WKB order. As said, we find that this is minimised for 6th order.

Note that obtaining QNMs via the WKB method involves taking derivatives at the maximum of the potential. The location of the maximum, however, changes as a function of $\hat{\beta}$. The WKB package [252] allows one to obtain numerical values for the QNMs while taking this into account automatically, and we show some examples in Table **??** and Figure 5.3 calculated this way.

24: Note that the maximum is approximately located around the light ring, which in Schwarzschild GR is at $r = 3M$. In Figure 5.3, because $r_s = 1$ has been chosen, the three maxima appear around $r(2M) \approx 1.5$.

Nonetheless, we also want 'semi-analytical' expressions for $\omega$ as a function of $\hat{\beta}$ that we can then use in our forecast analysis. For this, we employ the *light ring expansion* [637] to find the location of the maximum as a function of $\hat{\beta}$. The expansion works under the assumption that the geometry is "quasi-Schwarzschild" or, in other words, that the location of the potential maximum is only a small deviation away from the Schwarzschild value $r_\star^{\max} = r_\star^{\max,\,\text{GR}} + \delta r_\star$. This is certainly true in our case of study with small $\hat{\beta}$, as can be seen from Figure 5.3.[24] Upon employing the light ring expansion we can expand the defining property of the maximum of the potential

$\delta_r V \big|_{r=r_*^{max,GR}+\delta r_*} = 0$ to approximate to first order [637]

$$\delta r_* = -\frac{\partial_r V}{\partial_r^2 V}\bigg|_{r=r_*^{max,GR}}. \tag{5.74}$$

With this, we then have expressions for the maximum as a function of $\hat{\beta}$ that we can then substitute in all the potential derivatives in (5.73). While the full expression of $r_*^{\max}(\hat{\beta})$ is quite lengthy – see [4] for full details – this simplifies considerably for small $\hat{\beta}$, i.e. the case we are focusing on here. But rather than immediately truncating to the lowest order correction in $\hat{\beta}$, it is instructive to examine the few lowest order terms. We find

$$\frac{M_{\mathrm{Pl}}^2}{M} r_*^{\max} = 3.2808 - 3.0306 \cdot \hat{\beta} + 0.8316 \cdot \hat{\beta}^2$$
$$- \left(0.22819 + 1.8502 \frac{M_{\mathrm{Pl}}^8}{M^4 q^2}\right) \cdot \hat{\beta}^3 + \mathcal{O}(\hat{\beta}^4) \tag{5.75}$$

where we have set $\ell = 2$ in deriving this expression, and the first term represents the GR value $r_*^{\max, GR} = 3.2808 \cdot M/M_{\mathrm{Pl}}^2$. This expression highlights three important points:

1. First, the regime where it makes sense to truncate the above expansion is $\hat{\beta} \ll 1$. This corresponds to $\beta M q \ll M_{\mathrm{Pl}}^3$, which puts an implicit bound on the (until now unrestricted) parameter $q$ if we are to demand working in the small $\hat{\beta}$ regime, i.e. in the regime where potential deviations and the shift of $r_*^{\max}$ are small. As an example, taking $\beta \sim \mathcal{O}(1)$ and considering black holes with $M \sim \mathcal{O}(10)M_{\odot}$, where $M_{\odot} \sim 10^{30}$kg while $M_{\mathrm{Pl}} \sim 10^{-8}$kg, this requirement becomes $\sqrt{q} \ll 10^{-28}$kg or, equivalently, $\sqrt{q} \ll 10^8$ eV.

2. At cubic order in $\hat{\beta}$ we effectively see that the theory ultimately is controlled by two parameters (in addition to $M$ and $M_{\mathrm{Pl}}$): $\beta$ and $q$. While at lower (leading) orders these only enter in the form of the single effective parameter $\hat{\beta}$, from cubic order onwards we can see $q$ entering independently. Note that this is expected in light of the potential corrections (5.70), which share this feature. If we require these qualitatively different terms (e.g. the second term in the second line of (5.75)) to be suppressed with respect to the solely $\hat{\beta}$-dependent terms, we in addition require $M_{\mathrm{Pl}}^8 \ll M^4 q^2$. Taking the same example masses as above, this is akin to requiring $\sqrt{q} \gg 10^{-47}$ kg or $\sqrt{q} \gg 10^{-11}$ eV, when this bound is satisfied, these additional terms can be safely dropped for black hole mass ranges close to the example chosen here.

3. For small $\hat{\beta}$, the linear $\hat{\beta}$ term in (5.75) is the leading order correction. An immediate consequence of this is that the maximum of the potential always decreases in such cases, getting closer to the light ring at $r = 3M$ compared to GR.

Equipped with (5.75), we can substitute this into the WKB formula (5.73) to obtain an expression for the QNM frequencies in terms of $\hat{\beta}$, i.e. $\omega(\hat{\beta})$. Since the full expression is quite cumbersome, we do not show it here (but leave it available in [4] for the interested reader). We have checked that the $\omega$ values found with this analytical expression agree with those in Table **??**. Working in a small $\hat{\beta}$ expansion, one then finds

$$\frac{2M\omega}{M_{\mathrm{Pl}}^2} = \frac{2M\omega_0}{M_{\mathrm{Pl}}^2} + (3.36 - 1.00i)\hat{\beta} - (4.78 - 0.32i)\hat{\beta}^2 + \mathcal{O}(\hat{\beta}^3), \tag{5.76}$$

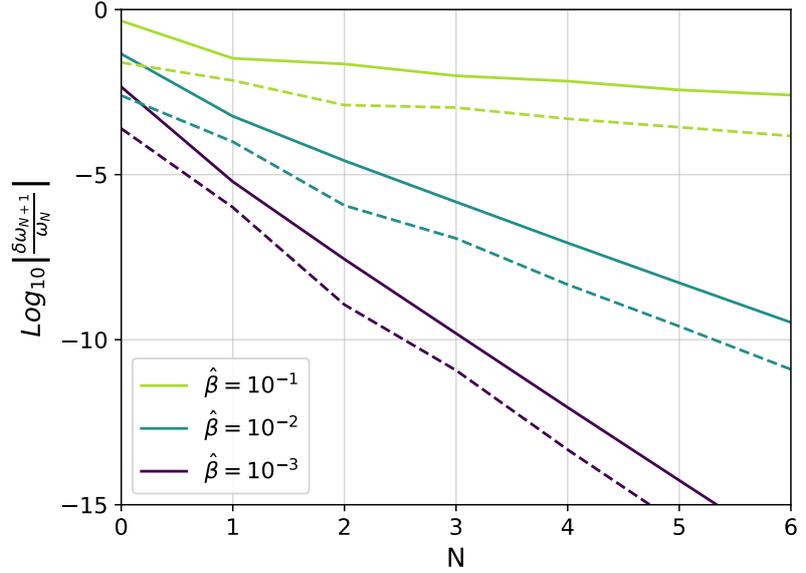

**Figure 5.4:** This plot displays the achievable accuracy in the real (solid) and imaginary (dashed) parts of quasinormal modes for $\ell = 2$ for increasing orders of $N$ in (5.77). The y-axis represents the fractional contribution to the quasinormal frequencies gained by going to the next $N+1$ order as compared to $\omega_N$. As can be seen, these fall of quickly, resulting e.g. in $O(10^{-10})$ corrections for $N \geq 6$ for $\hat{\beta} \sim 0.01$.

where $\omega_0$ is the GR prediction, now satisfying $2M\omega_0/M_{\text{Pl}}^2 = (0.74-0.18i)$, and the other terms correspond to entries in Table **??**. More precise and extensive coefficients for this expansion are given in this table, with corrections to the QNM frequencies parametrised via

$$\omega_N = \omega_0 + \sum_{i=1}^{i=N} \delta\omega_i \hat{\beta}^i. \tag{5.77}$$

The numerical precision of this expansion is shown in Figure 5.4 for increasing $N$.

**Table 5.3:** Real and imaginary components of the corrections to the $\ell = 2$ quasinormal frequencies $\omega$.

| $i$ | $2M\delta\omega_i$ | | $2M\omega_i(\hat{\beta} = 10^{-1})$ | |
|---|---|---|---|---|
| | Re | Im | Re | Im |
| 0 | / | / | 0.7472 | −0.1778 |
| 1 | 3.3501 | −0.9935 | 1.0822 | −0.2771 |
| 2 | −4.7710 | 0.3282 | 1.0345 | −0.2738 |
| 3 | 20.8065 | −4.0438 | 1.0553 | −0.2779 |
| 4 | −118.2672 | 18.8014 | 1.0435 | −0.2760 |
| 5 | 674.1128 | −123.6702 | 1.0503 | −0.2772 |
| 6 | −4153.5112 | 789.1050 | 1.0461 | −0.2765 |
| 7 | 26 259.0482 | −5274.1042 | 1.0487 | −0.2770 |

The methodology for deriving such semi-analytical expressions for the QNM frequencies from the WKB method is described in more detail in Appendix A, where the method is exemplified and validated by deriving $\Lambda$-dependent QNM expressions.

## 5.5 Forecasted constraints

In the previous Section we obtained the QNM spectrum for a hairy black hole with a time-dependent scalar, where deviations from GR are controlled by the parameter $\hat{\beta}$, which encodes information about the underlying ST theory.

Our aim now is to forecast how well current and future GW experiments will be able to constrain $\hat{\beta}$ using solely the ringdown. Table 5.4 collects our main results.

We employ a Fisher forecast analysis to estimate the precision with which $\hat{\beta}$ will be measurable. Our analysis ubiquitously uses techniques developed in [216] - for a detailed summary of the forecasting setup specialised to our present analysis see Appendix 3. We will here focus on a simplified scenario where all the standard waveform parameters are already known ($A, \phi^+, ...$), leaving $\hat{\beta}$ as the only free parameter. Given this idealised setup, the constraints forecasted here should be interpreted as optimistic/optimal estimates for expected achievable precision. Forecasts with full joint constraints for all waveform parameters will be left for future work.

For a setup as considered here, where the only waveform parameters we want to constrain are those appearing inside the quasinormal frequencies $\omega$, general expressions for the achievable precision can be derived analytically. These only depend on the number of parameters one wants to constrain. Here, GR deviations are solely controlled by the parameter $\hat{\beta}$ so we employ the expression for single-parameter constraints

$$\sigma_{\hat{\beta}}^2 \rho^2 = \frac{1}{2} \left( \frac{f}{Qf'} \right)^2 , \qquad (5.78)$$

where the prime denotes a derivative with respect to $\hat{\beta}$ – we refer to Appendix 3 for details on the derivation of (5.78). In (5.78), $\rho$ refers to the standard signal-to-noise-ratio (SNR) and the quality factor $Q$ is defined as

$$Q_{\ell m} = \pi f_{\ell m} \tau_{\ell m}. \qquad (5.79)$$

In the context of astrophysical binary compact objects, the $(\ell, m) = (2, 2)$ mode is generically the one with largest amplitude and dominates the ringdown signal [216–221]. It is therefore usually referred to as the dominant mode. For non-rotating black holes, which are our focus here, the equations of motion do not depend on $m$ [212]. Consequently, even though $m = 0$ is typically set for simplicity, as we have done here, the results are valid for any $m$. For discussions on the amplitude and detectability of subdominant modes we refer to [217, 219, 222, 223], but here highlight that not only does the $\ell = 2$ mode dominate in typical scenarios, but also note that for binary systems which have orbited each other sufficiently long, the orbits will have circularised, additionally enhancing the $\ell = 2$ mode relative to other modes [224–226]. That being said, our analysis here can straightforwardly be repeated for higher $\ell$ modes,[25] while we leave conducting a multi-mode analysis and extending this study to slowly rotating black holes (see e.g. [656–660]) for future work. Note that, in the context of constraints purely derived from ringdown, in order to use black hole spectroscopy to both constrain the standard GR black hole parameters (mass and angular momentum), as well as additional beyond-GR parameters (such as $\hat{\beta}$ here), the measurement of at least two QNMs is required. This means an additional (subdominant) mode is required in addition to the $(2, 2)$ mode discussed above, which typically entails an order of magnitude reduction in the constraining power, see e.g. [707]. So when using a single mode analysis here, this is based on the idealised assumption discussed at the start of this section, that all parameters except for $\hat{\beta}$ are (already) known to sufficient accuracy (e.g. from the inspiral), and hence only one mode is needed to constrain the single remaining

25: Note that the dipole mode $\ell = 1$ requires special treatment, as the Regge-Wheeler gauge does not completely fix the gauge in this case. For detailed information see [649], where it is also shown that $\ell = 1$ contributions are negligible in set ups as the one considered here.

| Detector(s) | Ringdown SNR ($\rho$) | Error on $\hat\beta$ |
|---|---|---|
| LVK | 10 [183, 259, 260] | $10^{-2}$ |
| ET / CE | $10^2$ [182–185] | $10^{-3}$ |
| pre-DECIGO | $10^2$ [261] | $10^{-3}$ |
| DECIGO / AEDGE | $10^3$ [186, 262]$^*$ | $10^{-4}$ |
| LISA | $10^3$ [263–266] | $10^{-4}$ |
| TianQin | $10^3$ [265, 266] | $10^{-4}$ |
| AMIGO | $10^5$ [267] | $10^{-6}$ |

free parameter.

We can now use the semi-analytical expressions found for the $\hat\beta$-dependent QNMs together with the error expression (5.78) to place estimated order-of-magnitude constraints on $\hat\beta$ and hence on the gravity model (5.1). Reading off $f$ and $Q$ from (5.77), as defined in (5.72) and (5.79), and substituting them into the single-parameter error expression (5.78), we obtain an estimate on its detectability in the same fashion as [1, 455].[26] This gives us[27]

$$\sigma_{\hat\beta}\rho \approx 0.08. \tag{5.80}$$

26: Note that, in evaluating the final expression, we set $\hat\beta$ to zero. This should simply be understood as capturing the leading order contributions to the error – depending on the actual value of $\hat\beta$ the precise error can differ by $\leq \mathcal{O}(1\%)$ for $\hat\beta \lesssim 10^{-1}$.

27: A value with more significant figures is provided in [4]. Ultimately, we will approximate $\sigma_{\hat\beta}\rho \approx \mathcal{O}(10^{-1})$ in Table 5.4, as we are purely interested in the robust order-of-magnitude constraints here.

The precision with which $\hat\beta$ can be constrained therefore inversely depends on the achievable SNR, which varies for different existing and upcoming detectors. Table 5.4 collects updated estimates on the optimistic obtainable ringdown SNRs together with the corresponding order-of-magnitude constraint on $\hat\beta$ for several ground and space-based detectors. $3^{rd}$ generation space based detectors such as LISA are predicted to be able to achieve a ringdown SNR as high as $10^3$, which would entail a constraint on $\hat\beta$ of

$$\sigma_{\hat\beta}^{\text{LISA/TianQin}} \sim 10^{-4}. \tag{5.81}$$

Let us stress here that the primary bounds discussed in this analysis are projected from a single ringdown observation with the SNR as described in Table 5.4. In fact, as the number of equivalent events $N$ increases, such bounds are predicted to improve as $N^{1/2}$ [671, 672]. For the LISA band, this could entail an improvement in the constraint on $\hat\beta$ of up to two orders of magnitude, i.e. $\sigma_{\hat\beta}^{\text{LISA/TianQin}} \sim 10^{-6}$, as the estimated rates of SMBH mergers, despite somewhat uncertain, lie in the $\mathcal{O}(10-100)$ per year range – see e.g. [190–197].[28] Having pointed out how constraints may improve going forward, let us however also re-emphasise that the bounds quoted are from an idealised analysis using only the dominant ringdown mode to constrain $\hat\beta$, with other (GR) parameters inferred to high accuracy from the inspiral and merger phases. As discussed above, a pure ringdown constraint would require the measurement of another subdominant ringdown mode, typically entailing an order of magnitude reduction in the constraining power [707]. Finally, let us point out that shifts in the QNM frequencies and damping times are detectable with approximately the same precision as constraints on $\hat\beta$, e.g. $\delta\omega \sim 10^{-4}$ will be detectable with LISA/TianQin. In our context this can e.g. be seen by noticing that the precision with which $\hat\beta$ is constrainable is inversely proportional to SNR (5.80), while the shift in QNM frequencies and damping times scales linearly with $\hat\beta$ (in the small $\hat\beta$ regime we are investigating) at leading order with order one coefficient(s) $\delta\omega_1 \sim 1.68 - 0.50i$.

28: This however assumes an optimistic scenario of $N$ events with identical SNR, while many events will have lower SNR values, e.g. mergers at higher redshifts.

## 5.6 Conclusions

In this Chapter, which summarises the results of [2], we have investigated the stability and quasinormal modes of odd parity perturbations of hairy black hole solutions with a linearly time-dependent scalar. In particular, we have focused on the specific ST theory identified in [679] and given by (5.1). This theory possesses an exact SdS solution for the metric ((5.4) and (5.2)) and the scalar field is indeed linearly time-dependent for this solution, but (unlike for other currently known exact solutions in ST theories with these characteristics) the canonical scalar kinetic term $X \equiv -\frac{1}{2}\phi_\mu \dot{\phi}^\mu$ is not constant. This solution is of particular interest since known constant $X$ solutions are generically plagued by instabilities. Our key findings are:

▸ The solution we investigate can support stable odd parity perturbations, where a non-trivial bound on the theory's coupling constant $\hat{\beta}^2 > 0$ is placed by requiring odd sector stability. This parameter is effectively a dimensionless measure of the strength of interactions in the theory (and suggestively written as $\hat{\beta}^2$, given this bound - note, however, that the square root structure of the theory means several theory predictions are controlled by $\hat{\beta} \equiv \sqrt{\hat{\beta}^2}$).

▸ We have derived the modified Regge Wheeler equation governing odd parity perturbations in this setup, and provided its quasinormal mode solutions. QNMs in this setting are controlled by the same single parameter $\hat{\beta}$ highlighted above. We have computed the QNM spectrum and quantified deviations from the GR values as a function of $\hat{\beta}$, as illustrated in Figure 5.3, finding the following qualitative features: 1) The maximum of the modified Regge-Wheeler potential shifts to smaller radii, lying closer to the light ring than in GR, 2) The overall amplitude of the modified Regge-Wheeler potential is enhanced (i.e. experiences a positive-definite shift, as above). The size of all three effects is controlled by $\hat{\beta}$.

▸ We derive forecasted constraints on $\hat{\beta}$ and the resulting shifts in the QNM spectrum from upcoming ringdown observations. A single LISA or TianQin event can constrain these shifts down to $\mathcal{O}(10^{-4})$, with potential improvements of up to two orders of magnitude by stacking events. In the small-shift/small-$\hat{\beta}$ regime considered, the shifts scale linearly with $\hat{\beta}$, making it similarly constrainable at $\mathcal{O}(10^{-4})$ (or better with stacking). Loud LVK events could provide analogous constraints at the $\mathcal{O}(10^{-2})$ level. Table 5.4 summarizes the achievable constraints across various current and future detectors.

Several avenues therefore suggest themselves for future work extending the above: Most importantly, it will be interesting to see what complementary bounds the even sector will place and whether a fully stable corner of parameter space remains for such theories when considering both odd and even parity perturbations. Considering the dynamics of the theory (5.1) in detail in other regimes, e.g. its cosmological dynamics/limit as briefly discussed here, promises to further constrain the relevant parameter space. More generally speaking, we hope the work presented in this Chapter provides a stepping stone towards identifying the landscape of which (if any) black hole solutions with time-dependent scalar hair are fully stable and hence are of particular interest for both theoretical and observational follow-up investigations.

# Inverting no-hair theorems | 6

自然は曲線を創り人間は直線を創る

<div style="text-align: right">Hideki Yukawa</div>

---

**Chapter summary**

Black hole solutions in general scalar-tensor theories are known to permit hair. Imposing that some such solutions should exist, the space of scalar-tensor theories is strongly restricted. In this Chapter, we investigate precisely what these restrictions are within general quadratic/cubic higher-order scalar-tensor theories for stealth solutions, whose metric is given by that in GR, supporting time-dependent scalar hair with a constant kinetic term. In Section (§6.2), we derive, in a fully covariant approach, the Euler-Lagrange equations and the conditions under which they admit all (or a specific set of) exact GR solutions. Focusing on static and spherically symmetric black hole spacetimes, we study in Section (§6.3) the dynamics of linear odd-parity perturbations. We construct a second-order covariant Lagrangian collecting all possible contributions from HOST functions, and show which of them contribute in our setup. In order to diagnose potential departures from GR, we inspect the component form of the quadratic Lagrangian on Schwarzschild(-de Sitter) black holes. Importantly, we find that requiring the existence of all stealth solutions prevents any deviations from GR in the odd-parity sector. In less restrictive scenarios, in particular for theories only requiring the existence of Schwarzschild(-de Sitter) black holes, we identify allowed deviations from GR, derive the stability conditions for the odd modes in Section (§6.4), and investigate the generic deviation of a non-trivial speed of gravitational waves. In Section (§6.5) we discuss an issue in deriving the master equation for odd-parity perturbations in non-shift-symmetric theories. Finally, we summarise the results and discuss several future directions in Section (§6.6).

This Chapter is based on the findings in [3], and all calculations shown here are reproducible in the companion `Mathematica` notebook [4].

## 6.1 Introduction

The study of black hole solutions has been pivotal in advancing our understanding of gravity, offering a fertile testing ground for General Relativity (GR) and alternative theories of gravity. Among these alternatives, scalar-tensor (ST) theories have attracted significant attention due to their capacity to modify GR (potentially addressing some of the challenges in describing e.g. dark energy, dark matter, or inflation) while maintaining a predictive and well-posed theoretical structure. In particular, the degenerate higher-order scalar-tensor (DHOST) theories [348–350], which generalise the Horndeski framework [355, 356, 391], provide one of the richest known and well-defined landscapes for exploring deviations from GR.[1] The theories explored in this Chapter therefore represent some of the most comprehensive and general ST frameworks developed to date.[2]

In the context of such ST theories, a number of no-hair theorems exist, showing that (for large classes of ST theories) the presence of the extra scalar degree of freedom does not alter the background black hole solutions in the theory with respect to GR—see e.g. [114, 558, 680, 713] and Sections 15 for a more detailed exposition. Black holes therefore retain their remarkable simplicity in these ST extensions of GR and (at the background level) are still fully specified by their mass and spin (and, if present, charge). However, in general ST theories so-called 'hairy' black hole solutions exist, where the scalar has a non-trivial profile and, importantly, black hole solutions for the metric can also be affected by the presence of the scalar field—see Section 3.4 for more details. As such, general ST theories may also contain (some or all) no-hair black hole solutions as present in GR, as well as other solutions with hair. It is then interesting to turn the spirit of no-hair theorems on its head and ask the reverse question: If we require the presence of all, or some given specific, black hole solutions we are familiar with from GR, then we can ask:

1. To what extent does this restrict the theories in question?
2. How do the resulting restrictions affect the black hole dynamics, especially also the behaviour of perturbations around black holes as e.g. observable via black hole ringdown?

In this Chapter we will address these questions within the framework of cubic (and quadratic) higher-order scalar-tensor (HOST) theories. The motivation for focusing on theories that admit stealth solutions is threefold. First, from a phenomenological perspective, stealth solutions provide a compelling mechanism to evade current observational constraints on deviations from GR in strong-gravity regimes. By construction, these solutions allow for non-trivial scalar field configurations that leave the metric unchanged, making them compatible with existing observations while still allowing for novel scalar dynamics. Second, from a theoretical standpoint, the requirement that a theory admits stealth solutions acts as a non-trivial structural constraint on its operator content. This makes it a useful tool for carving out physically interesting subspaces within the larger theory space. In particular, some of the ST theories studied for cosmology can also admit stealth configurations, potentially linking late-time scalar field dynamics to the strong-field regime in a coherent framework. Furthermore, as shall be explained below, one needs to go beyond DHOST by introducing so-called scordatura terms in order to avoid strong coupling of perturbations. This promotes a stealth solution to an approximately stealth one that behaves as stealth for any practical purposes at the level of the background and that is free from the strong coupling issue for perturbations. Since scordatura terms are of order unity (and not necessarily small) in the unit of the cutoff of the theory, it is expected that the space of scalar-tensor theories admitting approximately stealth solutions is significantly broader than that admitting exact stealth ones. The present Chapter should be considered as the first step towards the important problem of finding such a broader theory space.

Black hole solutions in theories beyond GR that describe spacetimes indistinguishable from their GR counterparts (at the level of the background metric, but may be accompanied by a non-trivial scalar field profile, i.e. they may still have 'scalar hair') are typically referred to as 'stealth' black hole solutions. For example, a shift-symmetric $k$-essence described by the action $\int d^4x \sqrt{-g}\, P(X)$, coupled to GR, admits stealth solutions if $P'(X) \equiv dP(X)/dX$ has a non-trivial root $X = X_0 \neq 0$, where $X \equiv -g^{\mu\nu}\partial_\mu\phi\partial_\nu\phi/2$ and $g$ is the determinant of the metric. Indeed, for $X = X_0$, the stress-energy tensor $T_{\mu\nu} = P'(X)\partial_\mu\phi\partial_\nu\phi + P(X)g_{\mu\nu} = P(X_0)g_{\mu\nu}$ is equivalent to that of an effective cosmological constant $\Lambda_{\rm eff} = -P(X_0)/M_{\rm Pl}^2$ (with $M_{\rm Pl}$ denoting the reduced Planck mass) which can be adjusted to any desired value by adding a bare cosmological constant to the $k$-essence action, meaning that any GR solution with or without the cosmological constant can be promoted to a stealth solution as far as the spacetime geometry admits a scalar profile with $X = X_0$. Actually, any regular spacetime locally admits a scalar profile with $X = X_0$: for $X_0 > 0$ (or $X_0 < 0$) one only needs to specify a spacelike (or timelike) hypersurface, to construct a congruence of geodesics orthogonal to the hypersurface and then identify $\phi = \sqrt{2|X_0|}\,\tau$ up to a constant shift, where $\tau$ is the proper time (or distance) along each geodesic.[3] The stealth Schwarzschild solution of this type based on the Lemaître coordinates was first found in [714] and then later generalised to the stealth Schwarzschild-de Sitter (SdS) solution in Horndeski theory [565, 566] and in DHOST theory [571–573, 577, 715]. It is even possible to obtain a stealth Kerr solution in DHOST theories [572, 696]. While perturbations around those stealth solutions suffer from strong coupling in the scalar sector [578], one can easily solve this problem and render those perturbations weakly coupled by taking into account higher derivative terms as in the effective field theory of ghost condensate [716], if and only if the scalar profile is timelike. This mechanism dubbed scordatura [701] was already taken into account in [714] and the stealth Schwarzschild solution in $k$-essence was promoted to the approximately stealth solution in ghost condensate that behaves as stealth for any practical purposes in astrophysical scales at the level of the background [717] and is free from the strong coupling issue for perturbations. For approximately stealth solutions in more general quadratic HOST with scordatura terms, see [718]. In this way, once a stealth solution with a timelike scalar profile is found, one can easily promote it to an approximately stealth solution without the strong coupling issue. The timelike nature of such scalar profiles also offers the intriguing possibility that the associated time-dependence connects with cosmological dynamics driven by the scalar in the long distance limit, although it is of course non-trivial to have the same scalar yield leading order effects at long and short distance scales, see e.g. [667, 719].

Cast in the above language, we are here investigating the effect of requiring the presence of stealth solutions in general ST theories. More concretely, we consider solutions in ST theories that satisfy the Einstein equation in GR: $G_{\mu\nu} = M_{\rm Pl}^{-2} T_{\mu\nu} - \Lambda g_{\mu\nu}$. Here, recall that $G_{\mu\nu}$ is the Einstein tensor, $T_{\mu\nu}$ is the stress-energy tensor of matter field(s), and $\Lambda$ is the (effective) cosmological constant. Specifically, we will consider the effects of requiring the following stealth solutions:

1. General stealth GR with matter: *Any* metric satisfying $G_{\mu\nu} = M_{\rm Pl}^{-2} T_{\mu\nu} - \Lambda g_{\mu\nu}$ is a solution.
2. General stealth GR in vacuum: *Any* metric satisfying $G_{\mu\nu} = -\Lambda g_{\mu\nu}$ is a solution.
3. Stealth SdS: The SdS metric $g_{\mu\nu}^{\rm SdS}$ (6.25) is a solution for any $\Lambda$.
4. Stealth Schwarzschild: The Schwarzschild metric $g_{\mu\nu}^{\rm Schw}$ (6.24) is a so-

**Table 6.1:** This table organises the main results of this Chapter as well as previous literature. We classify different combinations of theories and stealth metric solutions. The third column collects the works where the conditions ensuring the existence of the corresponding stealth solutions were first derived. The fourth column indicates whether GR deviations are present in each case in linear odd-parity perturbations on stealth S(dS) solution, and the subscript in ✓ counts the number of independent combinations of beyond-GR parameters present. (See Table 6.6 for a summary of the beyond-GR parameters for each case.) The final column refers to the stability conditions of each case, with a ✓ denoting that stability conditions are automatically satisfied under the existence conditions. In the main body of the Chapter we use a simplified nomenclature for the different cases, namely $^{\text{symmetry}}\textbf{Theory}_{\text{solution required}}$, so e.g., $^{\text{SS}}\textbf{Cubic}_{\text{SdS}}$ for shift-symmetric cubic HOST theories where we impose the conditions requiring the existence of stealth SdS solutions.

| HOST theory | Stealth metric | Existence conditions | GR-deviations in odd modes on S(dS) | Stability conditions |
|---|---|---|---|---|
| Cubic | General GR with matter | this Chapter [3] (6.16) | ✗ (this Chapter [3]) (6.56) | ✓ |
| | General GR vacuum | this Chapter [3] (6.17) | ✓$_1$ (this Chapter [3]) (6.57) | (6.67) |
| | SdS | this Chapter [3] (6.18) | ✓$_2$ (this Chapter [3]) (6.58) | (6.69) |
| | Schwarzschild | this Chapter [3] (6.19) | ✓$_3$ (this Chapter [3]) (6.60) | (6.71) |
| Shift-sym cubic | General GR with matter | this Chapter [3] (B.1) | ✗ [720, 721] (6.56) | ✓ |
| | General GR vacuum | this Chapter [3] (B.4) | ✗ [720, 721] (C.1) | (6.68) |
| | SdS | [715] (B.7) | ✓$_1$ [720, 721] (C.2) | (6.70) |
| | Schwarzschild | [715] (B.10) | ✓$_2$ [720, 721] (C.3) | (6.72) |
| Quadratic | General GR with matter | [572] (B.2) | ✗ (this Chapter [3]) (6.56) | ✓ |
| | General GR vacuum | [572] (B.5) | ✓$_1$ (this Chapter [3]) (C.1) | (6.68) |
| | SdS | this Chapter [3] (B.8) | ✓$_2$ (this Chapter [3]) (C.2) | (6.70) |
| | Schwarzschild | this Chapter [3] (B.11) | ✓$_2$ (this Chapter [3]) (C.2) | (6.70) |
| Shift-sym quadratic | General GR with matter | [572] (B.3) | ✗ [577, 579] (6.56) | ✓ |
| | General GR vacuum | [572] (B.6) | ✗ [577, 579] (C.1) | (6.68) |
| | SdS | [571] (B.9) | ✓$_1$ [577, 579] (C.2) | (6.70) |
| | Schwarzschild | [571] (B.12) | ✓$_1$ [577, 579] (C.2) | (6.70) |

lution.

Solutions listed earlier include all of the later conditions listed—e.g. requiring SdS solutions clearly includes requiring the presence of Schwarzschild solutions—so demanding the presence of the first set of solutions is a stronger requirement than for the second, which is stronger than for the third.

This work then focuses on investigating the above stealth solutions in the context of HOST theories up to cubic order in double scalar derivatives (henceforth referred to as cubic HOST) with a linearly time dependent scalar field whose kinetic term is constant. We summarise the main findings in Table 6.1. All calculations performed in this Chapter are reproducible via 2 companion `Mathematica` notebooks [4], with calculations of Section 6.2 appearing in 'Inverting-no-hair-theorems-I.nb', and those of Sections 6.3 and 6.4 in 'Inverting-no-hair-theorems-II.nb'. These notebooks construct an adaptable general formalism for the study of cubic HOST theories at the background and perturbative levels, which can be tuned to specific models.

## 6.2 Field equations and existence conditions for stealth GR solutions

### 6.2.1 Quadratic/cubic HOST theories

We examine the following action composed of the metric $g_{\mu\nu}$ and the scalar field $\phi$ [348–350]:

$$S_{\text{grav}} = \int d^4x \sqrt{-g}\Bigg[ F_0(\phi, X) + F_1(\phi, X)\Box\phi + F_2(\phi, X)R + \sum_{I=1}^{5} A_I(\phi, X)L_I^{(2)}$$

$$+ F_3(\phi, X)G_{\mu\nu}\phi^{\mu\nu} + \sum_{J=1}^{10} B_J(\phi, X)L_J^{(3)} \Bigg] + \int d^4x \sqrt{-g}\, L_{\text{m}} \ , \quad (6.1)$$

where $L_{\text{m}}$ is the matter Lagrangian (assumed to be minimally coupled to gravity)[4], and recall that we have used the shorthands $X \equiv -\phi_\mu\phi^\mu/2$, $\phi_\mu \equiv \nabla_\mu\phi$, and $\phi_{\mu\nu} \equiv \nabla_\nu\nabla_\mu\phi$. $L_I^{(2)}$ and $L_J^{(3)}$ comprise all possible terms built from $\phi_\mu$ and $\phi_{\mu\nu}$ which are quadratic and cubic in $\phi_{\mu\nu}$, respectively, and are written explicitly as

$$L_1^{(2)} = \phi_{\mu\nu}\phi^{\mu\nu} \ , \qquad L_1^{(3)} = (\Box\phi)^3 \ , \qquad L_2^{(3)} = (\Box\phi)\phi_{\mu\nu}\phi^{\mu\nu} \ ,$$

$$L_2^{(2)} = (\Box\phi)^2 \ , \qquad L_3^{(3)} = \phi_{\mu\nu}\phi^{\nu\rho}\phi_\rho^\mu \ , \qquad L_4^{(3)} = (\Box\phi)^2\phi_\mu\phi^{\mu\nu}\phi_\nu \ ,$$

$$L_3^{(2)} = (\Box\phi)\phi^\mu\phi_{\mu\nu}\phi^\nu \ , \quad L_5^{(3)} = \Box\phi\phi_\mu\phi^{\mu\nu}\phi_{\nu\rho}\phi^\rho \ , \quad L_6^{(3)} = \phi_{\mu\nu}\phi^{\mu\nu}\phi_\rho\phi^{\rho\sigma}\phi_\sigma \ ,$$

$$L_4^{(2)} = \phi^\mu\phi_{\mu\rho}\phi^{\rho\nu}\phi_\nu \ , \quad L_7^{(3)} = \phi_\mu\phi^{\mu\nu}\phi_{\nu\rho}\phi^{\rho\sigma}\phi_\sigma \ , \quad L_8^{(3)} = \phi_\mu\phi^{\mu\nu}\phi_{\nu\rho}\phi^\rho\phi_\sigma\phi^{\sigma\lambda}\phi_\lambda \ ,$$

$$L_5^{(2)} = (\phi^\mu\phi_{\mu\nu}\phi^\nu)^2, \qquad L_9^{(3)} = \Box\phi(\phi_\mu\phi^{\mu\nu}\phi_\nu)^3 \ , \qquad L_{10}^{(3)} = (\phi_\mu\phi^{\mu\nu}\phi_\nu)^3 \ .$$

$$(6.2)$$

The action introduced above encompasses both standard Horndeski [355, 356, 391] and beyond-Horndeski/DHOST [348–350, 400, 412, 413] theories as particular limits. For instance, the Horndeski action in the standard form with the Galileon functions can be recovered with the choices:

$$F_0 = G_2 \ , \quad F_1 = G_3 \ , \quad F_2 = G_4 \ , \quad F_3 = G_5 \ , \quad A_1 = -A_2 = -G_{4X} \ ,$$

$$6B_1 = -2B_2 = 3B_3 = -G_{5X} \ , \qquad (6.3)$$

and $A_I = B_J = 0$ for $I = 3, 4, 5$ and $J = 4, \cdots, 10$. The full classification of quadratic/cubic DHOST theories can be found in [350], and there are a large number of subclasses distinguished by different sets of degeneracy conditions. Among these subclasses, there is one that can be obtained from the Horndeski class via invertible conformal/disformal transformation, which was called "disformal Horndeski" class in [415].[5] (Note that an invertible conformal/disformal transformation preserves the number of physical degrees of freedom [417, 418].) In the terminology of [350], this class corresponds to a sum of the quadratic DHOST of class $^2$N-I and the cubic DHOST of class $^3$N-I, see Table 3.1 for reference. For the quadratic part (characterised by $F_2$ and $A_I$'s), the degeneracy conditions are given by[6]

$$A_2 = -A_1 \ , \qquad (6.4)$$

$$A_4 = \frac{1}{2(F_2 + 2XA_1)^2}\Big[ (3F_2 + 8XA_1 + 2X^2A_3)(A_1 + XA_3 + F_{2X})^2$$

$$- 2A_3(3XA_1 + X^2A_3 + F_2 + XF_{2X})^2 \Big] \ , \qquad (6.5)$$

$$A_5 = \frac{(A_1 + XA_3 + F_{2X})\big[ A_1(A_1 - 3XA_3 + F_{2X}) - 2A_3F_2 \big]}{2(F_2 + 2XA_1)^2} \ , \qquad (6.6)$$

where $F_2\,(\neq 0)$, $A_1$, and $A_3$ are free functions and the condition $F_2 + 2XA_1 \neq 0$ is assumed. For the cubic HOST characterised by $F_3$ and $B_J$'s, we have

$$-\frac{B_2}{3} = \frac{B_3}{2} = B_1\,, \quad B_5 = -B_7 = \frac{4XB_4F_{3X} - (6B_1 + F_{3X})^2}{12XB_1}\,,$$

$$B_8 = \frac{B_5(4XB_4 - 6B_1 - F_{3X})}{12XB_1}\,, \quad B_6 = -B_4\,,$$

$$B_8 = \frac{B_5(4XB_4 - 6B_1 - F_{3X})}{12XB_1}\,, \quad B_{10} = \frac{B_4(4XB_4 - 6B_1 - F_{3X})^2}{24X^2B_1^2}\,, \quad (6.7)$$

where $F_3$, $B_1\,(\neq 0)$, and $B_4$ are arbitrary functions. Moreover, when both the quadratic and cubic parts are present, one has to impose the following conditions on top of Eqs. (6.6) and (6.7):

$$A_3 = \frac{A_1(4XF_{2X} - 3F_2)}{XF_2} + \frac{XF_{2X} - F_2}{X^2} - \frac{F_{3X}(F_2 + 2XA_1)^2}{6X^2F_2B_1}\,,$$

$$B_4 = \frac{6B_1(F_2 - XF_{2X}) + F_{3X}(F_2 + XA_1)}{2XF_2}\,. \quad (6.8)$$

Having said that, in what follows, we do not necessarily impose the degeneracy conditions for the following reason. When we consider perturbations about stealth solutions (i.e., those with the metric of GR solutions) with a timelike scalar profile, the perturbations would be strongly coupled in the asymptotic Minkowski/de Sitter region [701]. Therefore, in order to render those perturbations weakly coupled, one has to take into account deviation from the degeneracy conditions, which is known as the scordatura mechanism [701]. Of course, any deviation from the degeneracy conditions leads to the appearance of an Ostrogradsky ghost in general, and therefore we assume the deviation is tiny so that the mass of the Ostrogradsky ghost is heavy enough. (An exception is U-DHOST, where the scordatura mechanism is implemented by default while the Ostrogradsky ghost is intrinsically absent [702].) Note that we will mainly focus on odd-parity perturbations about static and spherically symmetric background, where the problems of the strong coupling and Ostrogradsky ghost are irrelevant.[7] The above comment on the breaking of the degeneracy conditions and the scordatura mechanism applies when we investigate even-parity perturbations, which we leave for future study.

## 6.2.2  Background field equations

Now we assume a stealth solution with $X = X_0 = \text{const.}$, which captures aspects of a broad range of known solutions. For instance, a de Sitter attractor with constant $X$ was found in [374]. Also, there have been extensive studies on stealth S(dS) solutions [565, 566, 571–573, 577, 714, 715] as well as stealth Kerr solutions [572, 696] in the context of Horndeski and (D)HOST theories. As such, constant-$X$ solutions are motivated due to their link to exact stealth solutions while still having non-trivial scalar dynamics, therefore offering a controlled and simplified way to explore departures from GR. However, exact stealth solutions have also been found with non-constant $X$ in [679] and their stability and effect on quasinormal modes have been quantified in [2, 577]. Hence, looking forward, non-constant-$X$ solutions offer yet a richer and promising avenue to explore further viable GR departures. Under the condition that $X$ is a constant, one can express higher-derivative terms of $\phi$

as[8]

$$\phi^\mu \phi_{\mu\nu} = 0 \,, \qquad \phi^\lambda \nabla_\mu \phi_{\nu\lambda} = -\phi^\lambda_\mu \phi_{\nu\lambda} \,, \qquad \phi^\lambda \Box \phi_\lambda = -\phi^\beta_\alpha \phi^\alpha_\beta \,,$$

$$\phi^\lambda \nabla_\lambda \phi_{\mu\nu} = -R_{\mu\lambda\nu\sigma} \phi^\lambda \phi^\sigma - \phi^\lambda_\mu \phi_{\lambda\nu} \,, \qquad \phi^\lambda \nabla_\lambda \Box \phi = -R_{\lambda\sigma} \phi^\lambda \phi^\sigma - \phi^\beta_\alpha \phi^\alpha_\beta \,. \tag{6.9}$$

Then, the Euler-Lagrange (EL) equation for the metric is given by $\mathscr{E}_{\mu\nu} = 0$, with

$$
\begin{aligned}
\mathscr{E}_{\mu\nu} =& (F_2 - X_0 F_{3\phi}) G_{\mu\nu} \\
& - \frac{1}{2} \Big\{ F_0 + 2X_0 (F_{1\phi} + 2F_{2\phi\phi}) + (A_1 + A_2 + 2X_0 B_{2\phi}) \phi^\beta_\alpha \phi^\alpha_\beta - 2 \left[ F_{2\phi} + X_0 (F_{3\phi\phi} - 2A_{2\phi}) \right] \Box\phi \\
& + (F_{3\phi} - A_2) \left[ (\Box\phi)^2 - \phi^\beta_\alpha \phi^\alpha_\beta - 2\phi^\sigma \phi^\beta R_{\alpha\beta} \right] - 2B_1 \Box\phi \left[ (\Box\phi)^2 - 3\phi^\beta_\alpha \phi^\alpha_\beta - 3\phi^\sigma \phi^\beta R_{\alpha\beta} \right] \\
& + 6X_0 B_{1\phi} (\Box\phi)^2 + (2B_2 + B_3) \phi^\beta_\alpha \phi^\gamma_\beta \phi^\alpha_\gamma + 2B_2 \phi^\alpha \phi^\sigma \phi^{\alpha\beta} R_{\lambda\alpha\sigma\beta} \Big\} g_{\mu\nu} \\
& - \frac{1}{2} \Big\{ F_{0X} + 2(F_{1\phi} + F_{2\phi\phi}) + (F_{2X} - F_{3\phi}) R + \left( F_{1X} - F_{3\phi\phi} - \frac{1}{2} R F_{3X} + 4A_{2\phi} - 2X_0 A_{3\phi} \right) \Box\phi \\
& + \left[ A_{1X} + 2(B_{2\phi} - X_0 B_{6\phi}) \right] \phi^\beta_\alpha \phi^\alpha_\beta + \left[ A_{2X} + 2(3B_{1\phi} - X_0 B_{4\phi}) \right] (\Box\phi)^2 + (B_{1X} + B_4)(\Box\phi)^3 \\
& + F_{3X} \phi^{\alpha\beta} R_{\alpha\beta} + A_3 \left[ (\Box\phi)^2 - \phi^\beta_\alpha \phi^\alpha_\beta - \phi^\sigma \phi^\beta R_{\alpha\beta} \right] + (B_{3X} - 2B_6) \phi^\beta_\alpha \phi^\gamma_\beta \phi^\alpha_\gamma \\
& + (B_{2X} - 2B_4 + B_6) \Box\phi \phi^\beta_\alpha \phi^\alpha_\beta - 2B_4 \Box\phi \phi^\alpha \phi^\beta R_{\alpha\beta} - 2B_6 \phi^\lambda \phi^\sigma \phi^{\alpha\beta} R_{\lambda\alpha\sigma\beta} \Big\} \phi_\mu \phi_\nu \\
& - \Big\{ F_{2\phi} + X_0 (F_{3\phi\phi} + 2A_{1\phi}) - (F_{3\phi} + A_1 - 2X_0 B_{2\phi}) \Box\phi - B_2 \left[ (\Box\phi)^2 - \phi^\beta_\alpha \phi^\alpha_\beta - R_{\alpha\beta} \phi^\alpha \phi^\beta \right] \Big\} \phi_{\mu\nu} \\
& - \frac{1}{2} \left[ 2(F_{3\phi} + A_1 + 3X_0 B_{3\phi}) + (2B_2 - 3B_3) \Box\phi \right] \phi_{\mu\lambda} \phi^\lambda_\nu - (2B_2 + 3B_3) \Box\phi \phi_{\lambda}{}_{(\mu} \phi^\lambda_{\nu)} \\
& - (F_{3\phi} + A_1 + B_2 \Box\phi) \phi^\lambda \phi^\sigma R_{\mu\lambda\nu\sigma} + 2B_2 \phi^\lambda R_{\lambda\sigma} \phi_{(\mu} \phi^\sigma_{\nu)} - 2(F_{3\phi} - A_2 - 3B_1 \Box\phi) \phi^\lambda \phi_{(\mu} R_{\nu)\lambda} \\
& - 2 \left[ A_1 + A_2 + (3B_1 + B_2) \Box\phi \right] \phi_{(\mu} \Box\phi_{\nu)} - 3B_3 \phi_{\lambda\sigma} \phi^\lambda_{(\mu} \phi^\sigma_{\nu)} - (2B_2 + 3B_3) \phi^{\lambda\sigma} \phi_{(\mu} \phi_{\nu)\lambda\sigma} \\
& + 3B_3 \phi^\lambda \phi^\sigma \phi^\rho_{(\mu} R_{\nu)\lambda\sigma\rho} + 2B_2 \phi^\lambda \phi^{\sigma\rho} \phi_{(\mu} R_{\nu)\sigma\lambda\rho} - T_{\mu\nu} \,, \tag{6.10}
\end{aligned}
$$

where the coupling functions and their derivatives are evaluated at the background solution with $\phi = \phi_0$ (not necessarily a constant) and $X = X_0$. Here, subscripts $\phi$ and $X$ denote derivatives with respect to $\phi$ and $X$ respectively. The stress-energy tensor for the matter sector is given by $T_{\mu\nu} \equiv -\frac{2}{\sqrt{-g}} \frac{\delta(\sqrt{-g} L_m)}{\delta g^{\mu\nu}}$. It should be noted that, unless $\phi = $ const., the EL equation for the scalar field, which we denote by $\mathscr{E}_\phi = 0$, is automatically satisfied for any configuration $(g_{\mu\nu}, \phi)$ that satisfies $\mathscr{E}_{\mu\nu} = 0$ and the equations of motion for the matter field thanks to the Noether identity associated with general covariance, i.e. $\nabla^\mu \mathscr{E}_{\mu\nu} \propto \phi_\nu \mathscr{E}_\phi$ (see [723] for related discussions). In other words, $\mathscr{E}_\phi = 0$ can be reproduced from other EL equations and hence is a redundant equation. Note also that the terms with $A_4$, $A_5$, $B_5$, $B_7$, $B_8$, $B_9$, and $B_{10}$ do not contribute to the EL equations under the condition $X = $ const.. Because nothing about the metric background structure is assumed in Eq. (6.10), this is our main equation that we will use to derive the existence conditions for general GR metrics, either with minimally coupled matter field(s) or in vacuum.

Equation (6.10) is the master field equation derived in this work. From it, by taking different limits, one can recover the relevant field equations for the different cases collected in Table 6.1. More concretely, one can easily restrict

Eq. (6.10) to shift-symmetric cubic HOST by rewriting all the functions as $F(\phi, X) = F(X)$, and therefore setting their $\phi$ derivatives to zero, i.e. $F_\phi = F_{\phi\phi} = 0$. It is also straightforward to restrict to quadratic HOST theories by setting the cubic functions to zero, i.e. $B_J = F_3 = 0$. In the latter case, Eq. (6.10) recovers the expressions derived in [572].

Let us see now how the equation of motion gets simplified as we progressively weaken the requirement on background solutions and only require the presence of subsets of the previous solutions. Assuming that $G_{\mu\nu} = -\Lambda g_{\mu\nu}$, Eq. (6.10) reduces to

$$
\begin{aligned}
\mathscr{E}_{\mu\nu} = & -\frac{1}{2}\Big\{ F_0 + 2\Lambda F_2 + 2X_0 \Big[ F_{1\phi} + 2F_{2\phi\phi} + 2\Lambda\Big(\frac{1}{2}F_{3\phi} - A_1 - 2A_2 + 6X_0 B_{1\phi}\Big)\Big] \\
& - 2\Big[F_{2\phi\phi} + X_0(F_{3\phi\phi} - 2A_{2\phi} + 2\Lambda(5B_1 + 2B_2))\Big]\,\Box\phi + B_3\phi_\alpha^\beta\phi_\beta^\gamma\phi_\gamma^\alpha \\
& + \Big[A_1 + A_2 + 2X_0(3B_1 + B_2) + 2(2B_1 + B_2)\Box\phi\Big]\,\phi_\alpha^\beta\phi_\beta^\alpha \Big\}g_{\mu\nu} \\
& -\frac{1}{2}\Big\{ F_{0X} + 2F_{1\phi} + 2F_{2\phi\phi} + 2\Lambda\big[2F_{2X} - F_{3\phi} - A_1 - 2A_2 + X_0(2A_{2X} + 3A_3 + 4(3B_{1\phi} - X_0 B_{4\phi}))\big]\\
& + \Big[F_{1X} - F_{3\phi\phi} + 4A_{2\phi} - 2X_0 A_{3\phi} - \Lambda(F_{3X} + 12B_1 + 2B_2 - 4X_0(B_{1X} + 2B_4 + B_6))\Big]\,\Box\phi \\
& + \Big[A_{1X} + A_{2X} + 2(3B_{1\phi} + B_{2\phi} - X_0(B_{4\phi} + B_{6\phi})) + (B_{1X} + B_{2X} - B_4 - B_6)\Box\phi\Big]\phi_\alpha^\beta\phi_\beta^\alpha \\
& + B_{3X}\phi_\alpha^\beta\phi_\beta^\gamma\phi_\gamma^\alpha \Big\}\phi_\mu\phi_\nu \\
& -\Big[F_{2\phi} + X_0(F_{3\phi\phi} + 2A_{1\phi} - 2\Lambda B_2) + B_2\phi_\alpha^\beta\phi_\beta^\alpha + 2X_0 B_{2\phi}\Box\phi\Big]\phi_{\mu\nu} - 3\Big(X_0 B_{3\phi} - \frac{1}{2}B_3\Box\phi\Big)\phi_{\mu\lambda}\phi_\nu^\lambda \\
& - 2\Big[A_1 + A_2 + (3B_1 + B_2)\Box\phi\Big]\phi_{(\mu}\Box\phi_{\nu)} - 3B_3\Big(\phi_{\lambda\sigma}\phi_\mu^\lambda\phi_\nu^\sigma + \phi_{(\mu}\phi_{\nu)\lambda\sigma}\phi^{\lambda\sigma} - \phi^\lambda\phi^\sigma\phi^\sigma_{(\mu}R_{\nu)\lambda\sigma\rho}\Big) \\
& - B_2\Big(\phi_{(\mu}\phi_{\nu)\lambda\sigma}\phi^{\lambda\sigma} - \phi^\lambda\phi^{\sigma\rho}\phi_{(\mu}R_{\nu)\sigma\lambda\rho}\Big) - (2B_2 + 3B_3)\,\Box\phi\phi_\lambda\phi_{(\mu}\phi_{\nu)}^\lambda \,,
\end{aligned}
$$
(6.11)

where we have used the identities

$$
\begin{aligned}
(\Box\phi)\phi_{\mu\nu} - \phi_\mu^\lambda\phi_{\lambda\nu} - R_{\mu\lambda\nu\sigma}\phi^\lambda\phi^\sigma &= \Lambda\big(2X_0 g_{\mu\nu} + \phi_\mu\phi_\nu\big)\,, \\
(\Box\phi)^2 - \phi_\alpha^\beta\phi_\beta^\alpha &= 4\Lambda X_0 \,.
\end{aligned}
$$
(6.12)

Notice that the above identities hold true in the case of the S(dS) spacetime with a linearly time-dependent scalar field $\phi = qt + \psi(r)$,[9] so that $X_0 = q^2/2$. Here $q$ has a mass dimension two for a scalar with mass dimension one.[10] Having more simplified equations of motion then leads to less restrictive existence conditions, and hence the case **Cubic**$_{\text{SdS}}$, where only SdS solutions are required to exist, will generally allow a bigger region of theory space than in case **Cubic**$_{\text{GR-mat}}$. In the limit of including only shift-symmetric quadratic terms (i.e. $^{\text{SS}}$**Quadratic**$_{\text{SdS}}$), Eq. (6.11) above reduces to Eq. (2.9) in [572].

We can further weaken the requirement on the background geometry to be that of a Schwarzschild black hole, i.e. if we are interested in vacuum solutions where $G_{\mu\nu} = 0$ (hence with $\Lambda = 0$), then Eq. (6.11) is further simplified to

$$\mathscr{E}_{\mu\nu} = -\frac{1}{2}\Big\{F_0 + 2X_0\left(F_{1\phi} + 2F_{2\phi\phi}\right) - 2\left[F_{2\phi} + X_0(F_{3\phi\phi} - 2A_{2\phi})\right]\Box\phi$$
$$+ \left[A_1 + A_2 + 2X_0(3B_{1\phi} + B_{2\phi}) + 2\left(2B_1 + B_2 + \frac{11}{18}B_3\right)\Box\phi\right]\phi_\alpha^\beta\phi_\beta^\alpha\Big\}g_{\mu\nu}$$
$$-\frac{1}{2}\Big\{F_{0X} + 2F_{1\phi} + 2F_{2\phi\phi} + \left(F_{1X} - F_{3\phi\phi} + 4A_{2\phi} - 2X_0A_{3\phi}\right)\Box\phi$$
$$+ \Big[A_{1X} + A_{2X} + 2(3B_{1\phi} + B_{2\phi} - X_0(B_{4\phi} + B_{6\phi}))$$
$$+ \left(B_{1X} + B_{2X} + \frac{5}{9}B_{3X} - \frac{1}{3}B_3 - B_4 - B_6\right)\Box\phi\Big]\phi_\alpha^\beta\phi_\beta^\alpha\Big\}\phi_\mu\phi_\nu$$
$$- \left[F_{2\phi} + X_0(F_{3\phi\phi} + 2A_{1\phi}) + \frac{1}{2}(2B_2 + B_3)\phi_\alpha^\beta\phi_\beta^\alpha + 2X_0B_{2\phi}\Box\phi\right]\phi_{\mu\nu} - 3X_0B_{3\phi}\Box\phi\phi_{\mu\lambda}\phi_\nu^\lambda$$
$$- 2\left[A_1 + A_2 + \left(3B_1 + 2B_2 + \frac{7}{6}B_3\right)\Box\phi\right]\phi_{(\mu}\Box\phi_{\nu)} - (2B_2 + B_3)\Box\phi_\lambda\phi_{(\mu}\phi_{\nu)}^\lambda \,, \tag{6.13}$$

where we have used the following identities:

$$\phi_\alpha^\beta\phi_\beta^\gamma\phi_\gamma^\alpha = \frac{5}{9}\phi_\alpha^\beta\phi_\beta^\alpha\Box\phi \,, \qquad X_0\phi^{\sigma\lambda}\phi_{\mu\sigma\lambda} = \left[X_0\Box\phi_\mu - \frac{2}{9}(\Box\phi)^2\phi_\mu\right]\Box\phi \,,$$
$$X_0R_{\mu\sigma\lambda\rho}\phi^{\sigma\rho}\phi^\lambda = -\frac{2}{5}\phi_\alpha^\beta\phi_\beta^\gamma\phi_\gamma^\alpha\phi_\mu \,, \qquad R_{\mu\sigma\lambda\rho}\phi^\sigma\phi^\lambda\phi_\nu^\rho = \phi_\mu^\sigma(-\Box\phi\phi_{\nu\sigma} + \phi_\nu^\rho\phi_{\sigma\rho}) \,, \tag{6.14}$$
$$X_0\left[9\Box\phi\phi_\mu^\sigma\phi_{\nu\sigma} + 4\phi_{(\mu}\left(\Box\phi\Box\phi_{\nu)} + 3\phi_{\nu)}^\sigma\Box\phi_\sigma\right)\right] = (\Box\phi)^2\left[3X_0\phi_{\mu\nu} + \Box\phi\left(2X_0g_{\mu\nu} + 3\phi_\mu\phi_\nu\right)\right] \,,$$

which are valid for the stealth Schwarzschild background. Indeed, Eq. (6.13) can be used to derive the existence conditions for Schwarzschild black holes. We will see that weakening of the nature of the background allows more freedom in the remaining valid theory space, and hence also in the potential number of GR deviations, as can be seen in Table 6.1 and Figure 6.1c.

Finally, for completeness, let us also comment on the trivial case where $\phi = $ const., and hence all derivatives of $\phi$ vanish. The equations of motion for the metric and scalar are given by

$$\mathscr{E}_{\mu\nu} = F_2G_{\mu\nu} - \frac{1}{2}F_0g_{\mu\nu} - T_{\mu\nu} = 0 \,,$$
$$\mathscr{E}_\phi = F_{0\phi} + F_{2\phi}R = 0 \,, \tag{6.15}$$

which coincide with the expressions in [572] for quadratic HOST. Note that the equation of motion of $\phi$ cannot be derived from that of the metric in this case. In what follows, we only consider the case where the scalar field has a non-trivial gradient and therefore $\mathscr{E}_\phi = 0$ automatically follows from $\mathscr{E}_{\mu\nu} = 0$.

### 6.2.3  Existence conditions for stealth GR solutions

We can now obtain the conditions that need to be satisfied for each class of stealth solutions to exist. We will show here the conditions for general cubic theories (i.e. cases **Cubic**$_{\text{GR-mat/GR-vac/SdS/Schw}}$), while an exhaustive list of existence conditions for shift-symmetric and/or quadratic theories can be found in Table 6.3.

Note that we obtain the existence conditions for stealth solutions in such a way that the covariant equations of motion are trivially satisfied when the Einstein equation in GR (i.e. $G_{\mu\nu} = M_{\rm Pl}^{-2}T_{\mu\nu} - \Lambda g_{\mu\nu}$) is imposed. Once we assume such existence conditions, the metric is determined by solving the Einstein equation under the spacetime symmetry of interest, and then the scalar field profile is fixed via $X_0 = -g^{\mu\nu}\partial_\mu\phi_0\partial_\nu\phi_0/2$ [572, 696]. It should also be noted that the existence conditions are written in terms of the functions of HOST theories evaluated at the background solution with $\phi = \phi_0$ and $X = X_0$. Since we are focusing on solutions with $X_0 = $ const., even if the existence conditions impose, e.g. $A_1(\phi_0, X_0) = 0$, this does not necessarily mean $A_{1X}(\phi_0, X_0) = 0$. However, the condition $A_1(\phi_0, X_0) = 0$ implies $A_{1\phi}(\phi_0, X_0) = 0$ when $\phi_0$ is a non-trivial function of spacetime. (Recall that we do not impose the shift symmetry from the outset, and therefore the functions can depend on $\phi$ explicitly.) This property has been used to simplify the conditions.

### General stealth GR with minimally coupled matter

Let us begin by investigating the case **Cubic**$_{\rm GR\text{-}mat}$, in which Eq. (6.10) is required to allow general GR solutions (i.e. which satisfy $G_{\mu\nu} = M_{\rm Pl}^{-2}T_{\mu\nu} - \Lambda g_{\mu\nu}$). If the following conditions,

$$F_0 + 2\Lambda M_{\rm Pl}^2 = -2X_0(F_{1\phi} + 2F_{2\phi\phi}) = X_0(F_{0X} - 2F_{2\phi\phi}) \,,$$
$$X_0 F_{1X} = -3F_{2\phi} = -3X_0 F_{3\phi\phi} \,, \quad X_0^{-1}(F_2 - M_{\rm Pl}^2) = F_{2X} = F_{3\phi} = -A_1 = A_2 \,,$$
$$F_{3X} = A_{1X} = A_{2X} = A_3 = 0 \,,$$
$$B_1 = B_{1X} = B_2 = B_{2X} = B_3 = B_{3X} = B_4 = B_6 = 0 \,,$$

$$(6.16)$$

are satisfied at $X = X_0$,[11] this sufficiently ensures that Eq. (6.10) is satisfied for any stealth GR solution with arbitrary time-dependent scalar background with constant $X_0$, and any minimally coupled matter field—this is visualised in Figure 6.1a.[12] In that sense, they are the most restrictive set of existence conditions presented in this Chapter. The conditions (B.1) for $^{\rm SS}$**Cubic**$_{\rm GR\text{-}mat}$ can be obtained from Eq. (6.16) by imposing shift symmetry on all the functions. Additionally, the conditions (B.2) for **Quadratic**$_{\rm GR\text{-}mat}$ can be obtained by removing all cubic-order interactions, and then similarly the conditions (B.3) for $^{\rm SS}$**Quadratic**$_{\rm GR\text{-}mat}$ can be obtained by then taking the shift-symmetric limit, recovering the expressions in [572].

11: In writing the conditions in (6.16), we have also employed the relation $R = 4\Lambda - M_{\rm Pl}^{-2}T$, where $T \equiv T_\mu^\mu$ is the trace of the stress-energy tensor.

12: Note that the specific colour chosen for the boxes in Figure 6.1a is not important. Instead, the number of different colours indicates the number of independent combinations of HOST functions after the imposition of existence conditions. In the present case, we see that after the imposition of the conditions (6.16), all non-zero HOST functions can be rewritten in terms of four independent functions, e.g. $\{F_0, F_{1\phi}, F_2, F_{2\phi}\}$.

### General stealth GR in vacuum

In the case of stealth GR metric solutions in vacuum, i.e. if we focus on **Cubic**$_{\rm GR\text{-}vac}$, the existence conditions for cubic HOST theories now read

$$F_0 + 2\Lambda(F_2 - X_0 F_{3\phi}) = -2X_0(F_{1\phi} + 2F_{2\phi\phi}) \,,$$
$$F_{0X} = -2[F_{1\phi} + F_{2\phi\phi} + \Lambda(2F_{2X} - 2F_{3\phi} + X_0 A_{1X})] \,,$$
$$3F_{2\phi} + X_0(F_{1X} - \Lambda F_{3X}) = 2X_0^2(A_{3\phi} - 2\Lambda B_4) \,, \quad F_{2\phi} = X_0 F_{3\phi\phi} \,,$$
$$F_{3\phi} = -A_1 = A_2 \,, \quad A_3 = A_{1X} = -A_{2X} + 2X_0 B_{4\phi} \,,$$
$$2B_4 = -2B_{1X} = B_{2X} \,, \quad B_1 = B_2 = B_3 = B_{3X} = B_6 = 0 \,. \quad (6.17)$$

From the above conditions, we see that less functions are required to be set to zero, therefore allowing a bigger region in the theory parameter space. Similarly as before, the conditions for $^{SS}$**Cubic**$_{GR\text{-}vac}$ (B.4), **Quadratic**$_{GR\text{-}vac}$ (B.5), and $^{SS}$**Quadratic**$_{GR\text{-}vac}$ (B.6) can be obtained from Eq. (6.17) in the appropriate limits. The results we have obtained for **Quadratic**$_{GR\text{-}vac}$ and $^{SS}$**Quadratic**$_{GR\text{-}vac}$ coincide with those found in [572].

**Schwarzschild-de Sitter**

Here we focus on extracting the existence conditions for SdS background solutions in cubic HOST theories, i.e. **Cubic**$_{SdS}$. From Eq. (6.11) we obtain

$$F_0 + 2\Lambda F_2 = -2X_0(F_{1\phi} + 2F_{2\phi\phi} + \Lambda F_{3\phi} + 2\Lambda A_1) \,, \quad F_{2\phi} = -X_0(F_{3\phi\phi} - 2A_{2\phi}) \,,$$
$$F_{0X} + 4\Lambda F_{2X} = -2\{F_{1\phi} + F_{2\phi\phi} - \Lambda[F_{3\phi} - A_1 - X_0(2A_{2X} + 3A_3 - 4X_0 B_{4\phi})]\} \,,$$
$$A_1 = -A_2 \,, \quad A_{1X} + A_{2X} = 2X_0(B_{4\phi} + B_{6\phi}) \,,$$
$$F_{1X} - \Lambda F_{3X} = F_{3\phi\phi} - 4A_{2\phi} + 2X_0 A_{3\phi} - 4\Lambda X_0(2B_4 + B_6 + B_{1X}) \,,$$
$$B_1 = B_2 = B_3 = B_{3X} = 0 \,, \quad B_4 + B_6 = B_{1X} + B_{2X} \,. \tag{6.18}$$

These conditions are displayed in Figure 6.1b. When shift-symmetry is imposed, i.e. $^{SS}$**Cubic**$_{SdS}$, the above conditions reduce to those derived in [715]. We can also obtain the conditions up to quadratic interactions, i.e. **Quadratic**$_{SdS}$, from the equation above by setting all $B_J = 0$ as well as $F_3 = 0$ [see (B.8)]. Finally, in the case of shift-symmetric quadratic theories, i.e. $^{SS}$**Quadratic**$_{SdS}$, Eq. (6.18) recovers Eq. (42) of [571] [see (B.9)].

**Schwarzschild**

Here we consider the conditions that HOST functions are required to satisfy in order to admit the existence of stealth Schwarzschild solutions (**Cubic**$_{Schw}$). From Eq. (6.13) we obtain

$$F_0 = -2X_0(F_{1\phi} + 2F_{2\phi\phi}) \,, \quad F_{2\phi} = -X_0(F_{3\phi\phi} - 2A_{2\phi}) \,,$$
$$F_{0X} = -2(F_{1\phi} + F_{2\phi\phi}) \,, \quad A_1 = -A_2 \,,$$
$$A_{1X} + A_{2X} = 2X_0(B_{4\phi} + B_{6\phi}) \,, \quad F_{1X} = F_{3\phi\phi} - 4A_{2\phi} + 2X_0 A_{3\phi} \,,$$
$$18B_1 = 2B_2 = -B_3 \,, \quad 3B_3 = X_0[9(B_{1X} + B_{2X} - B_4 - B_6) + 5B_{3X}] \,,$$
$$B_{1\phi} = B_{2\phi} = B_{3\phi} = 0 \,. \tag{6.19}$$

These conditions are shown in Figure 6.1c. In this case, it is important to clarify why certain functions—such as $F_2$—appear uncoloured and unshaded in Figure 6.1c, in contrast to their coloured or shaded appearance in Figures 6.1a and 6.1b. Taking $F_2$ as an example, we note that in a Ricci-flat geometry like Schwarzschild, its contribution becomes redundant, since $F_2 G_{\mu\nu} = 0$ automatically holds in Eq. (6.10). This does not imply that $F_2$ is inadmissible in this case; rather, it means that $F_2$ is unconstrained by the imposition of the existence of a Schwarzschild background and is therefore allowed complete freedom. Similarly to the previous case, in the shift-symmetric limit, corresponding to $^{SS}$**Cubic**$_{Schw}$, these conditions exactly reproduce the results in [715] and given in (B.10). For quadratic theories, i.e. **Quadratic**$_{Schw}$, the existence conditions can also be obtained from the equation above by setting all $B_J = 0$ as well as $F_3 = 0$ [see (B.11)]. Finally, in the case of shift-symmetric quadratic theories, i.e. $^{SS}$**Quadratic**$_{Schw}$, Eq. (6.18) recovers the results of Eq. (23) in [571] and given in (B.12).

At this point, it is interesting to inspect the existence conditions obtained so far, specifically in regards to the degeneracy conditions summarised in section 6.2.1. One notices that the existence conditions (6.16), (6.17), and (6.18) impose $B_1 = 0$ when evaluated at the background. Also, the existence condition (6.19) for the stealth Schwarzschild solution imposes $18B_1 = 2B_2 = -B_3$, which implies $B_1 = 0$ when the degeneracy condition (6.7) for cubic DHOST of class $^3$N-I is assumed. Recall that for cubic DHOST of class $^3$N-I, one requires that $B_1 \neq 0$ [see (6.7) and (6.8), where $B_1$ appears in the denominator]. This means that, if we work within DHOST theories of class N-I and require the existence of stealth solutions with constant $X$, the cubic DHOST part is prohibited, and we are left with quadratic DHOST of class $^2$N-I (see also [715]).[13] Regarding the degeneracy condition (6.6) for quadratic DHOST, one easily sees the compatibility with the existence conditions. Indeed, all the existence conditions (6.16)–(6.19) satisfy $A_2 = -A_1$, and the functions $A_4$ and $A_5$ are irrelevant to the existence conditions. In the remaining of the Chapter, however, and as motivated in the introduction, we will not be requiring the satisfaction of degeneracy conditions. So far, we have explicitly shown and discussed the existence conditions for the following cases: **Cubic**$_{GR\text{-}mat}$ (6.16), **Cubic**$_{GR\text{-}vac}$ (6.17), **Cubic**$_{SdS}$ (6.18), and **Cubic**$_{Schw}$ (6.19). As the most general cases, these are the most interesting expressions to show, also because the conditions for simpler theories can be directly obtained from them in the appropriate limits. With the aim of providing a comprehensive review, we list the conditions for the remaining cases in Table 6.3.

Some expressions in the Table have already been obtained in previous work. The conditions for $^{(SS)}$**Quadratic**$_{GR\text{-}mat}$ and $^{(SS)}$**Quadratic**$_{GR\text{-}vac}$ can be obtained from Eq. (2.6) in [572]. In addition, the conditions for $^{SS}$**Cubic**$_{SdS}$ and $^{SS}$**Cubic**$_{Schw}$ correspond to Eq. (21) and Eq. (18) respectively in [715]. Similarly, the conditions for $^{SS}$**Quadratic**$_{SdS}$ and $^{SS}$**Quadratic**$_{Schw}$ correspond respectively to Eqs. (42) and (23) in [571]. Note that when comparing these expressions with the corresponding ones in the literature one needs to take into account different conventions for $X$, which is often defined without the factor of $-1/2$.

13: Here, we have assumed that the theory is valid all the way from the black hole scale to the cosmological scale. Recall that, as mentioned in section 6.2.1, the class N-I (i.e., a sum of quadratic DHOST of class $^2$N-I and cubic DHOST of class $^3$N-I) is the only subclass of DHOST that allows for viable cosmology.

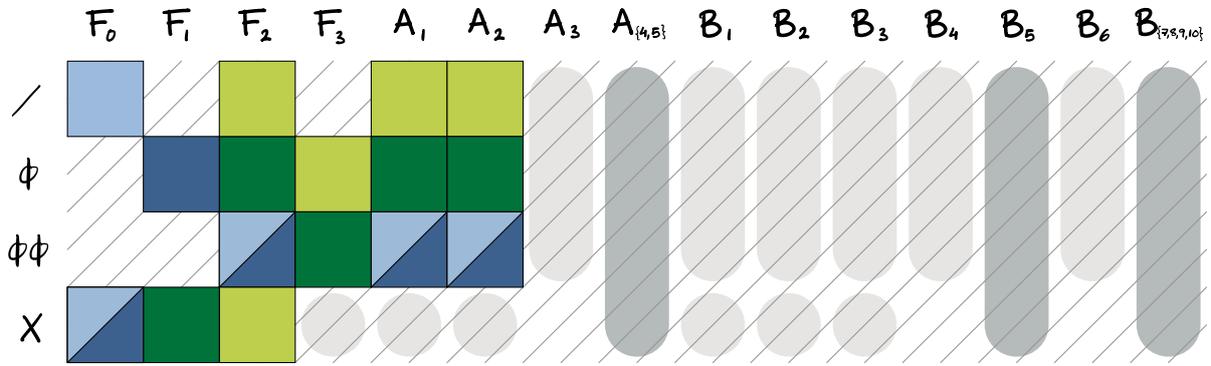

**(a)** Existence conditions for general GR stealth solutions with matter (**Cubic**$_{\text{GR-mat}}$) (6.16).

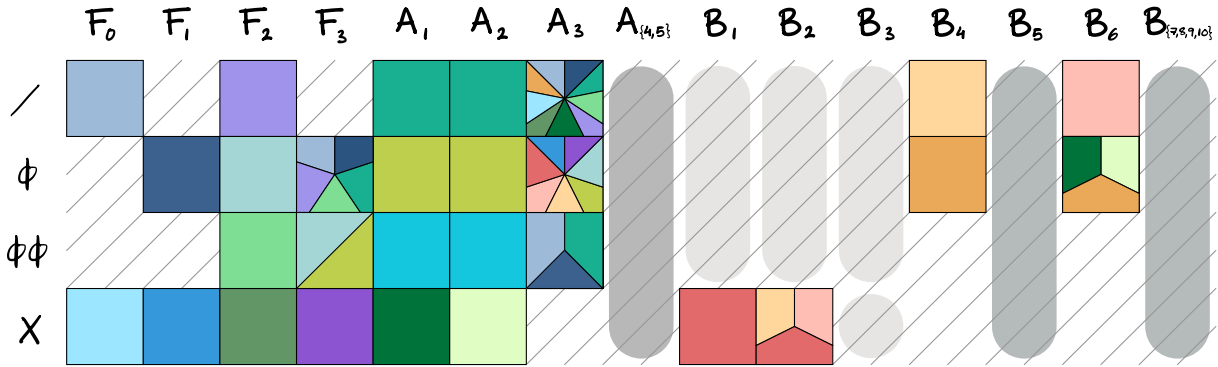

**(b)** Existence conditions for stealth SdS solutions (**Cubic**$_{\text{SdS}}$) (6.18).

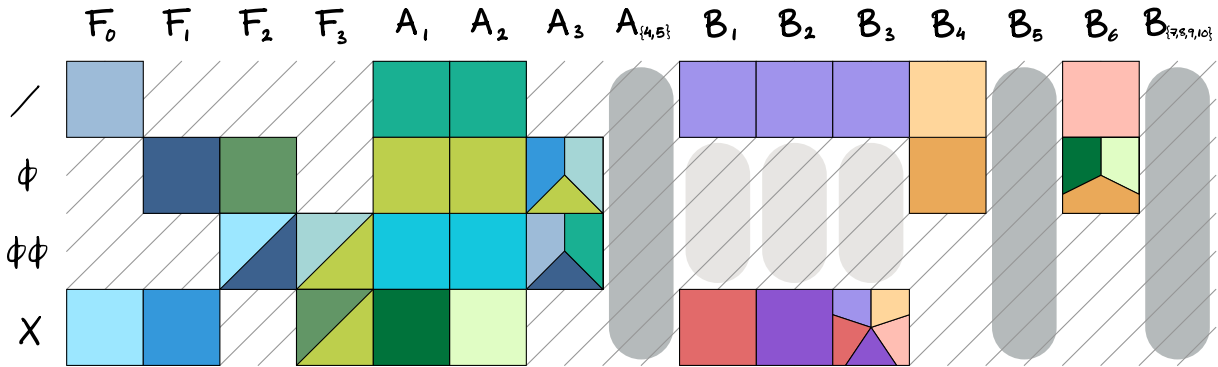

**(c)** Existence conditions for stealth Schwarzschild solutions (**Cubic**$_{\text{Schw.}}$) (6.19).

**Figure 6.1:** Here we display the theory space of cubic HOST theories which allow the existence of different stealth solutions. (Rows) columns represent (derivatives of) cubic HOST functions, e.g. $\phi\phi$-row and $F_0$-column refers to $F_{0\phi\phi}$. In coloured boxes, we show (combinations of) free functions that remain independent after the imposition of the corresponding existence conditions. When a single box is split into regions of different colours, this indicates that such term can be rewritten as a distinct combination of functions corresponding to such colours. The shaded region is either not allowed due to the constant-$X$ nature of our background (dark grey), or due to the existence conditions (light grey). Finally, uncoloured and unshaded regions simply do not contribute to the equations of motion. Note that the allowed theory-space region increases as the assumptions on the background geometry are weakened.

**Table 6.3:** Here we collect an exhaustive list of existence conditions for all models not shown in the main text.

| Model | Existence conditions | |
|---|---|---|
| $^{SS}$**Cubic**$_{GR-mat}$ | $F_0 + 2\Lambda M_{Pl}^2 = 0, \quad F_2 - M_{Pl}^2 = 0, \quad F_{0X} = F_{1X} = F_{2X} = F_{3X} = 0$ $A_1 = A_2 = A_{1X} = A_{2X} = A_3 = 0,$ $B_1 = B_{1X} = B_2 = B_{2X} = B_3 = B_{3X} = B_4 = B_6 = 0.$ | (B.1) |
| **Quadratic**$_{GR-mat}$ | $F_0 + 2\Lambda M_{Pl}^2 = -2X_0(F_{1\phi} + 2F_{2\phi\phi}) = X_0(F_{0X} - 2F_{2\phi\phi})$ $F_2 - M_{Pl}^2 = 0, \quad F_{1X} = F_{2\phi} = F_{2X} = A_1 = A_2 = A_{1X} = A_{2X} = A_3 = 0.$ | (B.2) |
| $^{SS}$**Quadratic**$_{GR-mat}$ | $F_0 + 2\Lambda M_{Pl}^2 = 0, \quad F_2 - M_{Pl}^2 = 0,$ $F_{0X} = F_{1X} = F_{2X} = A_1 = A_2 = A_{1X} = A_{2X} = A_3 = 0.$ | (B.3) |
| $^{SS}$**Cubic**$_{GR-vac}$ | $F_0 + 2\Lambda F_2 = 0, \quad F_{0X} = -2\Lambda(2F_{2X} + X_0 A_{1X}), \quad F_{1X} - \Lambda F_{3X} = -4\Lambda X_0 B_4,$ $A_3 = A_{1X} = -A_{2X}, \quad 2B_4 = -2B_{1X} = B_{2X},$ $A_1 = A_2 = B_1 = B_2 = B_3 = B_{3X} = B_6 = 0.$ | (B.4) |
| **Quadratic**$_{GR-vac}$ | $F_0 + 2\Lambda F_2 = -2X_0(F_{1\phi} + 2F_{2\phi\phi}),$ $F_{0X} = -2[F_{1X} + F_{2\phi\phi} + \Lambda(2F_{2X} + X_0 A_{1X})],$ $3F_{2\phi} + X_0 F_{1X} = 2X_0^2 A_{3\phi}, \quad F_{2\phi} = A_1 = A_2 = 0, \quad A_3 = A_{1X} = -A_{2X}.$ | (B.5) |
| $^{SS}$**Quadratic**$_{GR-vac}$ | $F_0 + 2\Lambda F_2 = 0, \quad F_{0X} = -2\Lambda(2F_{2X} + X_0 A_{1X}),$ $F_{1X} = 0, \quad A_3 = A_{1X} = -A_{2X}, \quad A_1 = A_2 = 0.$ | (B.6) |
| $^{SS}$**Cubic**$_{SdS}$ | $F_0 + 2\Lambda F_2 = -4\Lambda X_0 A_1, \quad F_{0X} + 4\Lambda F_{2X} = -2\Lambda[A_1 + X_0(2A_{2X} + 3A_3)],$ $A_1 = -A_2, \quad A_{1X} = -A_{2X}, \quad B_1 = B_2 = B_3 = B_{3X} = 0,$ $F_{1X} - \Lambda F_{3X} = -4\Lambda X_0(2B_4 + B_6 + B_{1X}), \quad B_4 + B_6 = B_{1X} + B_{2X}.$ | (B.7) |
| **Quadratic**$_{SdS}$ | $F_0 + 2\Lambda F_2 = -2X_0(F_{1\phi} + 2F_{2\phi\phi} + 2\Lambda A_1), \quad F_{2\phi} = 2X_0 A_{2\phi},$ $F_{0X} + 4\Lambda F_{2X} = -2[F_{1\phi} + F_{2\phi\phi} - A_1 - X_0(2A_{2X} + 3A_3)],$ $A_1 = -A_2, \quad A_{1X} = -A_{2X}, \quad F_{1X} = -4A_{2\phi} + 2X_0 A_{3\phi}.$ | (B.8) |
| $^{SS}$**Quadratic**$_{SdS}$ | $F_0 + 2\Lambda F_2 = -4\Lambda X_0 A_1,$ $F_{0X} + 4\Lambda F_{2X} = -2\Lambda[A_1 + X_0(2A_{2X} + 3A_3)],$ $A_1 = -A_2, \quad A_{1X} = -A_{2X}, \quad F_{1X} = 0.$ | (B.9) |
| $^{SS}$**Cubic**$_{Schw}$ | $F_0 = F_{0X} = F_{1X} = 0, \quad A_1 = -A_2, \quad A_{1X} = -A_{2X},$ $18B_1 = 2B_2 = -B_3, \quad 3B_3 = X_0[9(B_{1X} + B_{2X} - B_4 - B_6) + 5B_{3X}].$ | (B.10) |
| **Quadratic**$_{Schw}$ | $F_0 = -2X_0(F_{1\phi} + 2F_{2\phi\phi}), \quad F_{2\phi} = 2X_0 A_{2\phi},$ $F_{0X} = -2(F_{1\phi} + F_{2\phi\phi}),$ $A_1 = -A_2, \quad A_{1X} = -A_{2X}, \quad F_{1X} = -4A_{2\phi} + 2X_0 A_{3\phi}.$ | (B.11) |
| $^{SS}$**Quadratic**$_{Schw}$ | $F_0 = F_{0X} = F_{1X} = 0, \quad A_1 = -A_2, \quad A_{1X} = -A_{2X}.$ | (B.12) |

## 6.3  Quadratic Lagrangian in the odd sector

In the previous Section we have investigated the background evolution in cubic HOST theories as well as in a number of specific subcases, with a particular focus on understanding the constraints imposed by requiring stealth GR solutions. Having done so, we arrived at a reduced set of cubic HOST theories and in this Section we now consider perturbations about the stealth solutions to ultimately understand how restrictions imposed by requiring GR stealth solutions affect the behaviour of perturbations (and ultimately observable quasinormal modes). Specifically, we here investigate the dynamics of linear odd-parity perturbations about a static and spherically symmetric background given by

$$ds^2 = -A(r)dt^2 + \frac{dr^2}{B(r)} + r^2\left(d\theta^2 + \sin^2\theta\, d\varphi^2\right),\qquad (6.20)$$

accompanied by a spherical and (linearly) time-dependent scalar field $\phi = qt + \psi(r)$. In this section, we will focus on S(dS) solutions, i.e. the unique static and spherically symmetric vacuum solutions of GR, therefore satisfying $A = B$. We, however, choose to keep $A$ and $B$ independent for now, as the equations we derive can also be applied to study the dynamics of per-turbations for non-stealth metrics with $A \neq B$, e.g. those corresponding to hairy black holes [565, 566, 569]. We begin by studying the dynamics of odd-parity perturbations as an initial step toward identifying stable models within the broad spectrum of theories explored in this work, reserving a more comprehensive analysis of the more complex even-parity modes for future investigation.[14]

In the so-called Lemaître coordinates, the line element (6.20) can be written as:

$$ds^2 = -d\tau^2 + (1-A)d\rho^2 + r^2 \left( d\theta^2 + \sin^2\theta \, d\varphi^2 \right) \,, \tag{6.21}$$

where $\tau$ and $\rho$ are defined so that

$$d\tau = dt + \sqrt{\frac{1-A}{AB}} \, dr \,, \qquad d\rho = dt + \frac{dr}{\sqrt{AB(1-A)}} \,. \tag{6.22}$$

We then see that the coordinate $r$ is a function of $\rho - \tau$, satisfying

$$\partial_\rho r = -\dot{r} = \sqrt{\frac{B(1-A)}{A}} \,, \tag{6.23}$$

where a dot denotes the derivative with respect to $\tau$. As a reminder, recall that in the previous sections we have distinguished four different cases in relation to the nature of the background solution: 1) general GR solutions in the presence of matter, 2) general GR solutions in vacuum, 3) SdS, and 4) Schwarzschild black holes. As a result, for cases 1) and 2) one can in principle study the dynamics of perturbations for metrics other than S(dS), e.g. Kerr for cases 1) and 2) and Reissner-Nordström for case 1). However, in this section we will focus on static and spherically symmetric stealth GR solutions in vacuum, i.e., stealth S(dS) solutions. Written explicitly, the metric functions for these solutions are as follows:

$$\text{Schwarzschild:} \qquad A = B = 1 - \frac{r_s}{r} \,, \tag{6.24}$$

$$\text{SdS:} \qquad A = B = 1 - \frac{r_s}{r} - \frac{1}{3}\Lambda r^2 \,, \tag{6.25}$$

where $r_s (> 0)$ is a constant of length dimension (corresponding to the horizon radius for the Schwarzschild metric). Note that for the S(dS) metric, the following relation applies:

$$A' - \frac{1-A}{r} = -\Lambda r \,. \tag{6.26}$$

We now introduce metric perturbations $h_{\mu\nu}$ as

$$h_{\mu\nu} \equiv g_{\mu\nu} - \bar{g}_{\mu\nu} \,, \tag{6.27}$$

where $\bar{g}_{\mu\nu}$ is the background metric given by Eq. (6.21). As we are considering the odd-parity sector and hence no scalar perturbations will be present in our analysis, we will (in an abuse of notation) use the same symbol for the scalar field $\phi$ and its background value, i.e. $\phi_0 = \phi$. Note that, if we assume that the Lemaître coordinate $\tau$ is compatible with the time coordinate in the unitary gauge (i.e., $\phi \propto \tau$), then $X_0$ is a constant.

As introduced in detail in Section 2.4.1, we know that odd-parity metric perturbations are usually decomposed in terms of the spherical harmon-

14: The additional complexity arises from the higher number of even-parity functions as well as their coupling to the scalar mode.

ics $Y_{\ell m}(\theta, \varphi)$. Note that one can set $m = 0$ without loss of generality thanks to the spherical symmetry of the background,[15] and hence one can employ the Legendre polynomials $P_\ell(\cos\theta)$ instead. In the Regge-Wheeler gauge, the odd-parity metric perturbations look like

$$h_{\mu\nu}^{\text{odd}} = \begin{pmatrix} 0 & 0 & 0 & h_0 \\ 0 & 0 & 0 & h_1 \\ 0 & 0 & 0 & 0 \\ h_0 & h_1 & 0 & 0 \end{pmatrix} r^2 \sin\theta\, \partial_\theta P_\ell(\cos\theta)\,, \qquad (6.28)$$

where the $\ell$-dependence of $h_0$ and $h_1$ has been suppressed as modes with different $\ell$ evolve independently. Note in passing that the Regge-Wheeler gauge can be achieved by a complete gauge fixing, and therefore one can impose it at the level of Lagrangian [723]. Here, $h_0$ and $h_1$ are functions of $(\tau, \rho)$, and it should be pointed out that, as can be seen above, they differ from definitions used in other works (e.g. [1, 577, 644, 720]) by an overall factor of $r^2$ which has been included for later convenience. It should also be noted that we focus on generic higher multipoles with $\ell \geq 2$, where one expects to have one propagating degree of freedom in the odd sector. For $\ell = 1$, the odd-parity perturbations are non-dynamical and correspond to a slow rotation of the black hole.

We write the perturbed quadratic covariant action as:

$$S_{\text{grav}}^{(2)} = \frac{1}{4} \int d^4x \sqrt{-g} \sum_a \left[ \sum_{K=0}^{3} \delta\mathscr{L}_{F_{Ka}} F_{Ka} + \sum_{I=1}^{5} \delta\mathscr{L}_{A_{Ia}} A_{Ia} + \sum_{J=1}^{10} \delta\mathscr{L}_{B_{Ja}} B_{Ja} \right], \qquad (6.29)$$

where we have introduced the new notation $a = \{\varnothing, \phi, X, \phi\phi, XX, \phi X\}$ denoting the different $\phi$- and $X$-derivatives, with $\varnothing$ referring to a lack of derivatives. This enables one to easily identify which terms contribute at perturbative level and also directly confirm which features of the background solution cause the other terms to vanish. For our setup with linear odd-parity perturbations with $X = \text{const.}$, out of the potential 114 contributions to (6.29), only 23 are non-zero, which are diagrammatically collected in Fig. 6.2 and whose expressions we also show below, where the indices of the perturbed metric are raised/lowered by the background metric, and $R$, $R_{\mu\nu}$, $G_{\mu\nu}$, and the covariant derivatives are evaluated on the background. [16] Precisely speaking, the coefficients presented here are those that we see just after expanding (the gravitational part of) the Lagrangian for cubic HOST theories up to the quadratic order in odd-parity perturbations. The explicit form of the Lagrangian changes when one performs integration by parts. These expressions can be found and directly used in the provided `Mathematica` files [4].

All other contributions are shown to be zero for $X = \text{const.}$ [i.e. by employing the relations in Eq. (6.9)] and/or restricting to odd-parity modes, as described in Fig. 6.2. The simplifying relations valid for odd-parity modes on a spherically symmetric background with the component form of Eq. (6.28) are given below. These relations (as well as derivatives of some of them) have been used to simplify the expressions in Eqs. (6.31)–(6.53).

$$h_\mu^\mu = \phi^\mu \phi^\nu h_{\mu\nu} = \phi^{\mu\nu} h_{\mu\nu} = \phi^\mu \nabla_\nu h_\mu^\nu = \phi^\mu \phi^\nu \nabla_\sigma h_\mu^\sigma \nabla_\rho h_\nu^\rho =$$
$$= \phi^\mu \phi^\nu \phi^\sigma \nabla_\sigma h_{\mu\nu} = \phi^{\mu\nu} \phi^\sigma \nabla_\nu h_{\sigma\mu} = 0\,. \qquad (6.30)$$

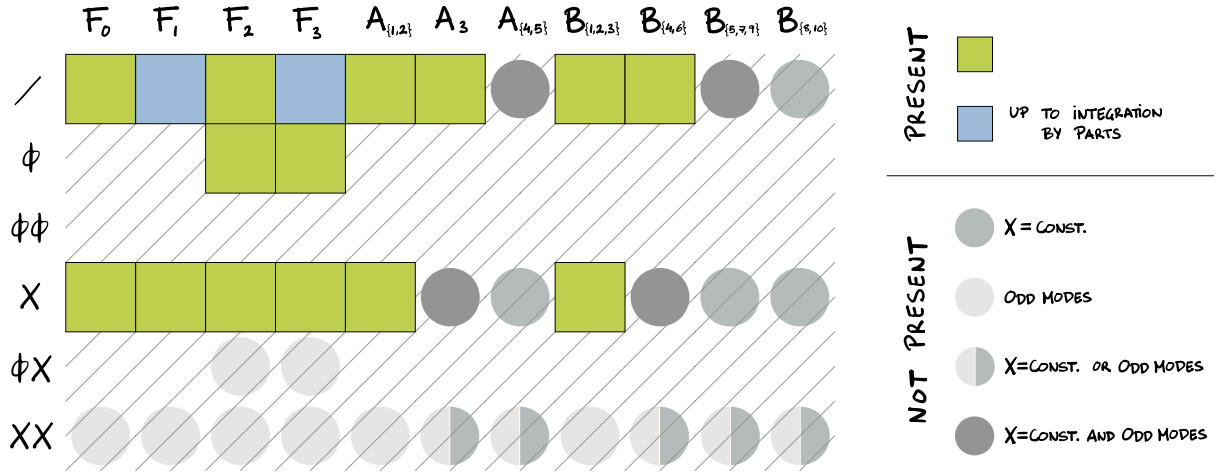

**Figure 6.2:** Coefficients in the quadratic Lagrangian in covariant form (6.29). In green we show the terms that provide non-zero contributions to the covariant quadratic Lagrangian for odd-parity perturbations, Eq. (6.29). In blue we show terms that contribute to the quadratic action only if the corresponding functions are not constants, i.e. $F_1, F_3 \neq$ const. Otherwise, they vanish up to total derivatives. The remaining coefficients are not present due to the reasons shown in the legend.

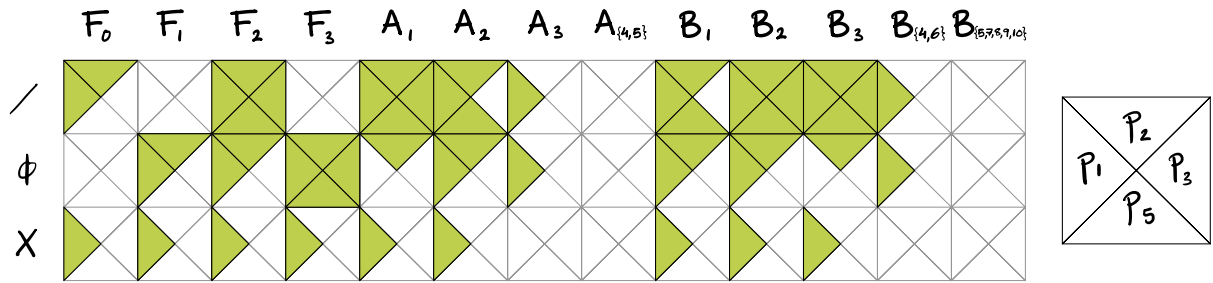

**Figure 6.3:** Contribution from all cubic HOST functions to the different $p$-coefficients. We show in green when a specific HOST function appears in a given $p_i$ and we leave in blank the cases where they do not, e.g. $F_0$ appears in the full expression for $p_1$ and $p_2$, but not in the one for $p_3$ and $p_4$. The full expressions can be found in the companion repository in [4].

$$\delta\mathscr{L}_{F_0} = -h^\nu_\mu h^\mu_\nu \,, \tag{6.31}$$

$$\delta\mathscr{L}_{F_{0X}} = -2\phi^\mu\phi^\nu h^\sigma_\mu h_{\nu\sigma} \,, \tag{6.32}$$

$$\delta\mathscr{L}_{F_1} = h_{\nu\sigma}(4\phi^\mu h^\sigma_\mu - \Box\phi h^{\nu\sigma}) + 2\phi^\mu[2h^\nu_\mu\nabla_\sigma h^\sigma_\nu - h^{\nu\sigma}(\nabla_\mu h_{\nu\sigma} - 2\nabla_\sigma h_{\mu\nu})] \,, \tag{6.33}$$

$$\delta\mathscr{L}_{F_{1X}} = -2\phi^\mu\phi^\nu\Box\phi h^\sigma_\mu h_{\nu\sigma} \,, \tag{6.34}$$

$$\delta\mathscr{L}_{F_2} = 4R_{\mu\nu}h^{\mu\sigma}h^\nu_\sigma - Rh^\nu_\mu h^\mu_\nu + \nabla^\sigma h^{\mu\nu}(2\nabla_\nu h_{\mu\sigma} - \nabla_\sigma h_{\mu\nu}) \,, \tag{6.35}$$

$$\delta\mathscr{L}_{F_{2\phi}} = 4\phi^\mu[h^{\nu\sigma}(\nabla_\sigma h_{\mu\nu} - \nabla_\mu h_{\nu\sigma}) + h^\nu_\mu\nabla_\sigma h^\sigma_\nu] \,, \tag{6.36}$$

$$\delta\mathscr{L}_{F_{2X}} = -2\phi^\mu[R\phi^\sigma h^\nu_\mu h_{\nu\sigma} + (\phi^\nu\nabla^\sigma h_{\mu\nu} + 2h_{\mu\nu}\phi^{\nu\sigma})\nabla_\rho h^\rho_\sigma] \,, \tag{6.37}$$

$$\begin{aligned}
\delta\mathscr{L}_{F_3} = {}& \phi^\mu\{2G^{\nu\sigma}[h^\rho_\mu(2\nabla_\sigma h_{\nu\rho} - \nabla_\rho h_{\nu\sigma}) + 2h^\rho_\nu(\nabla_\sigma h_{\mu\rho} + \nabla_\rho h_{\mu\sigma} - \nabla_\mu h_{\sigma\rho})] - 4\nabla^\sigma h^\nu_\mu\nabla_\rho\nabla_{(\nu}h^\rho_{\sigma)} \\
& + \nabla_\mu h^{\nu\sigma}(2\nabla_\rho\nabla_\sigma h^\rho_\nu - \Box h_{\nu\sigma}) + 2\nabla_\nu h^\nu_\mu\nabla_\rho\nabla_\sigma h^{\sigma\rho} + 2\nabla^\sigma h^\nu_\mu\Box h_{\nu\sigma} - h^{\nu\sigma}[R(\nabla_\mu h_{\nu\sigma} - 2\nabla_\sigma h_{\mu\nu}) + 2R_{\nu\sigma}\nabla_\rho h^\rho_\mu]\} \\
& + \phi^{\mu\nu}[4G^{\sigma\rho}h_{\mu\sigma}h_{\nu\rho} + 8G^\sigma_\nu h^\rho_\mu h_{\sigma\rho} - G_{\mu\nu}h^\sigma_\rho h^\rho_\sigma - 2R_{\sigma\rho\mu}h_\nu h^{\sigma\rho} + 4Rh^\sigma_\mu h_{\nu\sigma} \\
& + \nabla_\mu h^{\sigma\rho}(4\nabla_\rho h_{\nu\sigma} - \nabla_\nu h_{\sigma\rho}) - 2(\nabla^\sigma h_{\mu\nu}\nabla_\rho h^\rho_\sigma + \nabla_\rho h_{\nu\sigma}\nabla^\rho h^\sigma_\mu - \nabla_\sigma h_{\nu\rho}\nabla^\rho h^\rho_\mu)] \\
& + \Box\phi\left[\left(\frac{1}{2}\nabla_\rho h_{\nu\sigma} - \nabla_\rho h_{\nu\rho}\right)\nabla^\rho h^{\nu\sigma} - 2R_{\sigma\rho}h^\rho_\nu h^{\nu\sigma}\right] + 4\phi^\mu_{[\mu\nu]}h^{\sigma\rho}\nabla^\nu h_{\sigma\rho} - 2\phi_\mu{}^{\mu\nu}(h^{\sigma\rho}\nabla_\rho h_{\nu\sigma} + h^\sigma_\nu\nabla_\rho h^\rho_\sigma) \\
& + 2\phi^{\mu\nu\sigma}[h^\rho_\sigma(2\nabla_\mu h_{\nu\rho} - \nabla_\rho h_{\mu\nu}) + 2h^\rho_\mu(\nabla_\nu h_{\sigma\rho} - \nabla_\sigma h_{\nu\rho} + \nabla_\rho h_{\nu\sigma}) - h_{\mu\nu}\nabla_\rho h^\rho_\sigma]
\end{aligned} \tag{6.38}$$

$$\begin{aligned}
\delta\mathscr{L}_{F_{3\phi}} = {}& 2\phi^\mu\{\phi^{\nu\rho}[2h^\sigma_\nu(\nabla_\sigma h_{\mu\rho} + \nabla_\rho h_{\mu\sigma} - \nabla_\mu h_{\sigma\rho}) + h^\sigma_\mu(2\nabla_\nu h_{\sigma\rho} - \nabla_\sigma h_{\nu\rho})] \\
& + \Box\phi[h^{\sigma\rho}(\nabla_\mu h_{\sigma\rho} - \nabla_\sigma h_{\mu\rho}) - h^\rho_\mu\nabla_\rho h^\sigma_\sigma]\} \,, \tag{6.39}
\end{aligned}$$

$$\begin{aligned}
\delta\mathscr{L}_{F_{3X}} = {}& \phi^\mu\{2h_{\mu\nu}[\phi^\lambda\phi^{\sigma\rho}G_{\sigma\rho}h^\nu_\lambda + \phi^{\nu\sigma}\phi^{\rho\lambda}(\nabla_\sigma h_{\rho\lambda} - 2\nabla_\lambda h_{\sigma\rho}) + \Box\phi\phi^{\nu\sigma}\nabla_\rho h^\rho_\sigma] \\
& + \phi^\nu\nabla^\sigma h_{\mu\nu}[\phi^{\rho\lambda}(\nabla_\sigma h_{\rho\lambda} - 2\nabla_\lambda h_{\rho\sigma}) + \Box\phi\nabla_\rho h^\rho_\sigma]\} \,, \tag{6.40}
\end{aligned}$$

$$\begin{aligned}
\delta\mathscr{L}_{A_1} = {}& \phi^\mu\phi^\nu[\nabla_\mu h^{\sigma\rho}(\nabla_\nu h_{\sigma\rho} - 4\nabla_\sigma h_{\nu\rho}) + 4\nabla^\sigma h^\rho_\mu\nabla_\sigma(h_{\rho\nu)\nu}] + \phi^{\mu\nu}[h_{\sigma\rho}(8h^\rho_\mu\phi^\sigma_\nu - h^{\sigma\rho}\phi_{\mu\nu}) + 4h_{\mu\sigma}h_{\nu\rho}\phi^{\sigma\rho}] \\
& + 4\phi^\mu\phi^{\nu\sigma}[h^\rho_\mu(\nabla_\sigma h_{\nu\rho} + \nabla_\sigma h_{\nu\rho} - \nabla_\rho h_{\nu\sigma}) + 2h^\rho_\sigma(\nabla_\rho h_{\mu\nu} - \nabla_\mu h_{\nu\rho} + \nabla_\nu h_{\mu\rho})] \,, \tag{6.41}
\end{aligned}$$

$$\delta\mathscr{L}_{A_{1X}} = -2\phi^\mu\phi^\nu h^\sigma_\mu h_{\nu\sigma}\phi^\alpha_\beta\phi^\beta_\alpha \,, \tag{6.42}$$

$$\delta\mathscr{L}_{A_2} = \Box\phi\{h_{\mu\nu}[-\Box\phi h^{\mu\nu} + 8h^\sigma_\sigma\phi^{\mu\nu} - 4\phi^\sigma(\nabla_\sigma h^{\mu\nu} - 2\nabla^\nu h^\mu_\sigma)] + 8\phi^\sigma h^\mu_\sigma\nabla_\nu h^\nu_\mu\} \,, \tag{6.43}$$

$$\delta\mathscr{L}_{A_{2X}} = -2\phi^\mu\phi^\nu(\Box\phi)^2 h^\sigma_\mu h_{\nu\sigma} \,, \tag{6.44}$$

$$\delta\mathscr{L}_{A_3} = 2\phi^\mu\phi^\nu\Box\phi[\phi^\sigma h^\rho_\mu(2\nabla_\rho h_{\nu\sigma} + \nabla_\sigma h_{\nu\rho}) + 2h_{\mu\sigma}h_{\nu\rho}\phi^{\sigma\rho}] \,, \tag{6.45}$$

$$\delta\mathscr{L}_{B_1} = (\Box\phi)^2 h^{\mu\nu}[-\Box\phi h_{\mu\nu} + 12\nabla_\sigma(\phi_\mu h^\sigma_\nu) - 6\phi^\sigma(\nabla_\sigma h_{\mu\nu} - 2\nabla_\nu h_{\sigma\mu})] \,, \tag{6.46}$$

$$\delta\mathscr{L}_{B_{1X}} = -2\phi^\mu\phi^\nu(\Box\phi)^3 h^\sigma_\mu h_{\nu\sigma} \,, \tag{6.47}$$

$$\begin{aligned}
\delta\mathscr{L}_{B_2} = {}& \Box\phi\phi^\mu\phi^\nu[\nabla_\mu h^{\sigma\rho}\nabla_\nu h_{\sigma\rho} + 4\nabla^\sigma h^\rho_\mu(\nabla_\sigma(h_{\rho)\nu} - \nabla_\nu h_{\sigma\rho})] + 4\Box\phi\phi^{\mu\nu}\phi^\sigma[h^\rho_\mu(4\nabla_{[\nu}h_{\sigma]\rho} + \nabla_\rho h_{\sigma\nu}) + 2h^\rho_\sigma\nabla_{[\mu}h_{\rho]\nu}] \\
& + 2\phi^\alpha_\beta\phi^\beta_\alpha\phi^\sigma(2h^\nu_\sigma\nabla_\nu h^\mu_\mu + 2h^{\mu\nu}\nabla_\nu h_{\sigma\mu} - h^{\mu\nu}\nabla_\sigma h_{\mu\nu}) \\
& + h_{\mu\nu}[-\Box\phi\phi^\alpha_\beta\phi^\beta_\alpha h^{\mu\nu} + 4h^\sigma_\nu(2\Box\phi\phi_\rho\phi^{\rho\sigma} + \phi^\alpha_\beta\phi^\beta_\alpha\phi^{\sigma\mu}) + 4\Box\phi\phi^{\mu\sigma}\phi^{\nu\rho}h_{\sigma\rho}] \,, \tag{6.48}
\end{aligned}$$

$$\delta\mathscr{L}_{B_{2X}} = -2\phi^\mu\phi^\nu\Box\phi\phi^\alpha_\beta\phi^\beta_\alpha h^\sigma_\mu h_{\nu\sigma} \,, \tag{6.49}$$

$$\begin{aligned}
\delta\mathscr{L}_{B_3} = {}& 3\phi^\mu\{\frac{1}{3}\phi^\sigma_\nu h_{\rho\lambda}(12\phi^\lambda_\sigma h^\nu_\mu - \phi_{\sigma\mu}h^{\rho\lambda}) + 4\phi^\sigma_\nu\phi^{\rho\lambda}h_{\mu\lambda}h_{\sigma\rho} + 2\phi^\sigma\phi^{\rho\lambda}h_{\mu\lambda}(\nabla_\sigma h_{\nu\rho} + \nabla_\nu h_{\sigma\rho} + \nabla_\rho h_{\sigma\nu}) \\
& + 2\phi^\sigma\phi^\rho_\nu[h^\lambda_\sigma(\nabla_\mu h_{\rho\lambda} + \nabla_\rho h_{\mu\lambda} - \nabla_\lambda h_{\mu\rho}) + 2h^\lambda_\mu(-\nabla_\sigma h_{\rho\lambda} + \nabla_\rho h_{\sigma\lambda} + \nabla_\lambda h_{\sigma\rho})] \\
& + \phi^\sigma\phi^\rho[2\nabla_{[\nu}h_{\rho]\lambda}(\nabla_\mu h^\lambda_\sigma + 2\nabla^\lambda h_{\mu\sigma}) + 2\nabla_\rho h_{\nu\lambda}\nabla_{[\sigma}h^\lambda_{\mu]} + \nabla_\lambda h_{\rho\nu}\nabla^\lambda h_{\sigma\mu}]\} \,, \tag{6.50}
\end{aligned}$$

$$\delta\mathscr{L}_{B_{3X}} = -2\phi^\mu\phi^\nu\phi^\alpha_\beta\phi^\beta_\gamma\phi^\gamma_\alpha h^\sigma_\mu h_{\nu\sigma} \,, \tag{6.51}$$

$$\delta\mathscr{L}_{B_4} = 2\phi^\mu\phi^\nu(\Box\phi)^2[\phi^\rho h^\sigma_\mu(2\nabla_\rho h_{\nu\sigma} + \nabla_\sigma h_{\nu\rho}) + 2h_{\mu\sigma}h_{\nu\rho}\phi^{\sigma\rho}] \,, \tag{6.52}$$

$$\delta\mathscr{L}_{B_6} = 2\phi^\mu\phi^\nu\phi^\alpha_\beta\phi^\beta_\alpha[\phi^\rho h^\sigma_\mu(2\nabla_\rho h_{\nu\sigma} + \nabla_\sigma h_{\nu\rho}) + 2h_{\mu\sigma}h_{\nu\rho}\phi^{\sigma\rho}] \,, \tag{6.53}$$

The quadratic Lagrangian for odd-parity perturbations (6.29) can be written in component form by substituting in the expressions for the background metric (6.21), the background scalar, and metric perturbations (6.28). Written explicitly in terms of $h_0$ and $h_1$, we have

$$\frac{2\ell + 1}{2\pi j^2}\mathcal{L}_2 = p_1 h_0^2 + p_2 h_1^2 + p_3[(\dot{h}_1 - \partial_\rho h_0)^2 + 2p_4 h_1 \partial_\rho h_0] + p_5 h_0 h_1 \,, \quad (6.54)$$

where $j^2 \equiv \ell(\ell + 1)$. Note that the background equations of motion have not been used at this stage, and we have adopted the notation in [699]. Analytic expressions for the $p$-coefficients in terms of cubic HOST functions are very extensive and are therefore not included here. They can be found and used, however, in the corresponding `Mathematica` notebook in [4]. Figure 6.3 summarises the cubic HOST functions (in green) that contribute to the $p_i$ coefficients. The fact that the green coefficients appear differently in Figures 6.2 and 6.3 is due to the use of integration by parts to write the quadratic Lagrangian in component form as in Eq. (6.54), which necessarily rearranges the function content.

It is interesting to highlight that in our general setup without assuming shift and reflection symmetries and even without using the background equations of motion we find that

$$p_4 = 0 \,. \quad (6.55)$$

In other words, for a static and spherically symmetric metric background (not necessarily stealth) with a linearly time-dependent scalar field, the specific combination of $h_0(\tau, \rho)$ and $h_1(\tau, \rho)$ associated with $p_4$ does not appear in cubic HOST theories even without imposing the degeneracy conditions. This happens presumably due to the specific structure of the cubic HOST Lagrangian. Indeed, the $p_4$ term is present in general in the context of EFT of BH perturbations with a timelike scalar profile [699], which encompasses HOST theories in principle. It was also shown in [699] that the presence of $p_4$ forbids the existence of slowly rotating BH solutions (or otherwise leads to a diverging sound speed). In this sense, the vanishing of $p_4$ in cubic HOST is phenomenologically desirable.

In the following Subsections, we will explicitly show expressions for the $p_i$ coefficients under several existence conditions, and identify the presence of any potential deviations from GR. We will show here the expressions for the $p_i$ coefficients for general cubic theories (i.e. **Cubic**$_{\text{GR-mat/GR-vac/SdS/Schw}}$), and an exhaustive list for shift-symmetric and/or quadratic theories can be found in Table 6.5. Moreover, the beyond-GR parameter(s) for all models will be summarised in Table 6.6.

### 6.3.1 General stealth GR with minimally coupled matter

Let us first consider the case where all GR solutions in the presence of matter are required to exist. Upon employing the conditions (6.16) and specifying a SdS background, the coefficients $p_i$'s are given by

$$p_1 = r^2\sqrt{1 - A}(j^2 - 2)M_{\text{Pl}}^2 \,, \qquad p_2 = -\frac{r^2}{\sqrt{1 - A}}(j^2 - 2)M_{\text{Pl}}^2 \,,$$

$$p_3 = \frac{M_{\text{Pl}}^2 r^4}{\sqrt{1 - A}} \,, \qquad\qquad p_5 = 0 \,. \quad (6.56)$$

These expressions for the $p_i$ coefficients are precisely the ones obtained in GR. Therefore, in this case the dynamics of the odd-parity perturbations are the same as in GR, implying that ringdown observables associated with such perturbations, like the quasinormal mode frequencies, will be indistinguishable between the two cases.

### 6.3.2   General stealth GR in vacuum

Upon employing the conditions for the existence of stealth solutions (6.17) and specifying an SdS background, the $p$-coefficients for the case **Cubic**$_{\text{GR-vac}}$ are simplified to

$$p_1 = r^2\sqrt{1-A}(j^2-2)(F_2 - X_0 F_{3\phi}) , \quad p_2 = -\frac{r^2}{\sqrt{1-A}}(j^2-2)(F_2 - X_0 F_{3\phi}) ,$$

$$p_3 = \frac{r^4}{\sqrt{1-A}}(F_2 - X_0 F_{3\phi}) , \qquad\qquad p_5 = 0 . \tag{6.57}$$

From Eq. (6.57) we see that the presence of $F_{3\phi}$ shifts the value of $F_2$ in the same way for all non-zero coefficients $p_1$, $p_2$ and $p_3$. In this sense, we here conclude that there is one independent combination of functions for the case **Cubic**$_{\text{GR-vac}}$ corresponding to $F_2 - X_0 F_{3\phi}$ controlling the behaviour of odd-parity perturbations. We can now distinguish two cases. First, for non-shift-symmetric models, these functions (and hence also their one independent combination) can be non-trivial functions of spacetime and, as a result, deviations from GR can appear. Because of this, we use the symbol $\checkmark_1$ in Table 6.1 to denote that there is one independent combination of HOST functions which governs the dynamics of odd modes in a way that might differ from GR. As will be discussed in Section 6.5, this actually complicates the definition of quasinormal mode frequencies. Second, if $F_2 - X_0 F_{3\phi}$ is a constant when evaluated on the background (note that this does not entail shift-symmetry, as it is possible that both $F_2$ and $F_{3\phi}$ are non-trivial functions of spacetime but $F_2 - X_0 F_{3\phi}$ is a constant), then all ringdown phenomenology in the odd sector will be indistinguishable from GR, albeit from a constant shift in the effective Planck mass [c.f. (6.57) with (6.56)]. In this case, for instance, quasinormal mode frequencies will have the same numerical values as in standard GR. In the shift-symmetric and/or quadratic limit, i.e. for the cases $^{\text{SS}}$**Cubic**$_{\text{GR-vac}}$, $^{\text{SS}}$**Quadratic**$_{\text{GR-vac}}$, and **Quadratic**$_{\text{GR-vac}}$, the contribution from the coefficient $F_{3\phi}$ disappears and $F_2$ is the only contributing function [see (C.1)]. In these cases, if $F_2$ is shift-symmetric we also recover GR predictions (therefore represented with a $\times$ in Table 6.1). However, for the case **Quadratic**$_{\text{GR-vac}}$, $F_2$ can generally be a non-trivial function of spacetime, hence potentially sourcing deviations from GR. Because of this, this case is represented with the symbol $\checkmark_1$ in Table 6.1.

### 6.3.3   Schwarzschild-de Sitter

As we have seen, requiring the existence of SdS solutions as opposed to requiring general stealth GR solutions leads to a broader allowed region of theory space. Imposing the existence conditions (6.18) for SdS spacetimes on cubic theories leads to the following $p_i$ coefficients:

$$p_1 = r^2\sqrt{1-A}(j^2-2)\Big[F_2 + X_0(F_{3\phi} + 2A_1)\Big] \,,$$

$$p_2 = -\frac{r^2}{\sqrt{1-A}}(j^2-2)(F_2 - X_0F_{3\phi}) \,,$$

$$p_3 = \frac{r^4}{\sqrt{1-A}}\Big[F_2 + X_0(F_{3\phi} + 2A_1)\Big] \,,$$

$$p_5 = 0 \,. \tag{6.58}$$

Inspecting Eq. (6.58) we see that 2 independent combinations of HOST functions fully characterise the all $p_i$ coefficients, corresponding to

$$F_2 - X_0F_{3\phi} \,, \qquad\qquad A_1 + F_{3\phi} \,. \tag{6.59}$$

Note that $p_1$ and $p_3$ are proportional to a linear combination of these two quantities.

In the shift-symmetric limit, all contributions from cubic functions disappear and hence for $^{SS}\textbf{Cubic}_{SdS}$ and $^{SS}\textbf{Quadratic}_{SdS}$, and actually also the model $\textbf{Quadratic}_{SdS}$, we obtain the coefficients given by (C.2), which is equivalent to Eq. (3.6) in [579] for shift-symmetric quadratic theories. In those cases, only $A_1$ survives as an additional contribution on top of $F_2$. When $F_2$ and $A_1$ are shift-symmetric (and therefore time-independent when evaluated on the background), the corresponding odd-parity quasinormal mode frequencies can be obtained by simple rescaling of those in GR [690, 706]. Written explicitly, for stealth Schwarzschild solutions, the relation is given by $\omega = \omega_{\text{GR}}[F_2/(F_2 + 2X_0A_1)]^{3/2}$.

As mentioned before, when the shift-symmetry is not imposed, recall that HOST functions can in principle contain explicit time dependences, which, as stated before, and discussed in more detail in Section 6.5, makes deriving the master equation in the odd-parity sector more challenging.

### 6.3.4 Schwarzschild

In this Subsection we express the $p_i$ coefficients after imposing the existence conditions for Schwarzschild black hole solutions [Eq. (6.19)]. They are given by

$$p_1 = r\sqrt{1-A}(j^2-2)\Big\{r[F_2 + X_0(F_{3\phi} + 2A_1)] - 81\sqrt{2X_0^3(1-A)}\,B_1\Big\} \,,$$

$$p_2 = -\frac{r^2}{\sqrt{1-A}}(j^2-2)(F_2 - X_0F_{3\phi}) \,,$$

$$p_3 = \frac{r^4}{\sqrt{1-A}}\Big[F_2 + X_0(F_{3\phi} + 2A_1)\Big] \,,$$

$$p_5 = 0 \,. \tag{6.60}$$

In this case, $\textbf{Cubic}_{Schw}$, we find the following 3 independent combinations of HOST functions as fully characterising the $p_i$ coefficients:

$$F_2 - X_0F_{3\phi} \,, \qquad A_1 + F_{3\phi} \,, \qquad B_1 \,. \tag{6.61}$$

Interestingly, one cubic function, $B_1$, survives in the shift-symmetric limit. Hence, $^{SS}\textbf{Cubic}_{Schw}$ contains 2 potential beyond-GR parameters (note that, as explained before, in shift-symmetric models $F_2$ is a constant and therefore not regarded as a beyond-GR parameter) in the evolution of odd-parity perturbations corresponding to $A_1$ and $B_1$. The presence of $B_1$ here leads to a non-trivial $r$-dependent radial speed for GWs, something which will be discussed in Section 6.4 together with the corresponding stability conditions. In [720] odd-parity perturbations for Schwarzschild black holes in shift-symmetric cubic HOST theories were studied and, in particular, the contribution from $B_1$ to the fundamental quasinormal mode was calculated (while setting $A_1 = 0$). In Section 6.4 we show how a non-zero $B_1$ affects the radial speed of GWs.

**Table 6.5:** Here we collect the list of $p$-coefficients in the quadratic Lagrangian (6.54) for odd-parity perturbations. In the main text we have explicitly shown the expressions for the following cases: **Cubic**$_{\text{GR-mat}}$ (6.56), **Cubic**$_{\text{GR-vac}}$ (6.57), **Cubic**$_{\text{SdS}}$ (6.58), and **Cubic**$_{\text{Schw}}$ (6.60). For the cases $^{\text{SS}}$**Cubic**$_{\text{GR-mat}}$ and $^{\text{(SS)}}$**Quadratic**$_{\text{GR-mat}}$, it is trivial that the $p$-coefficients are given by (6.56). Here we exhaustively collect the expressions for all the remaining models in Table 6.1. Recall that we found $p_4 = 0$ to be generically true in cubic HOST theories (as one would expect for theories admitting slowly-rotating black hole solutions [699]).

| Model | $p$-coefficients | |
|---|---|---|
| $^{\text{SS}}$**Cubic**$_{\text{GR-vac}}$, $^{\text{(SS)}}$**Quadratic**$_{\text{GR-vac}}$. | $p_1 = r^2\sqrt{1-A}(j^2-2)F_2,$ <br> $p_2 = -\frac{r^2}{\sqrt{1-A}}(j^2-2)F_2,$ <br> $p_3 = \frac{r^4}{\sqrt{1-A}}F_2,$ <br> $p_5 = 0.$ | (C.1) |
| $^{\text{SS}}$**Cubic**$_{\text{SdS}}$, $^{\text{(SS)}}$**Quadratic**$_{\text{SdS}}$, $^{\text{(SS)}}$**Quadratic**$_{\text{Schw}}$. | $p_1 = r^2\sqrt{1-A}(j^2-2)(F_2+2X_0A_1),$ <br> $p_2 = -\frac{r^2}{\sqrt{1-A}}(j^2-2)F_2,$ <br> $p_3 = \frac{r^4}{\sqrt{1-A}}(F_2+2X_0A_1),$ <br> $p_5 = 0.$ | (C.2) |
| $^{\text{SS}}$**Cubic**$_{\text{Schw}}$ | $p_1 = r\sqrt{1-A}(j^2-2)\Big[r(F_2+2X_0A_1)-81\sqrt{2X_0^3(1-A)}\,B_1\Big],$ <br> $p_2 = -\frac{r^2}{\sqrt{1-A}}(j^2-2)F_2,$ <br> $p_3 = \frac{r^4}{\sqrt{1-A}}(F_2+2X_0A_1),$ <br> $p_5 = 0.$ | (C.3) |

Both $^{\text{(SS)}}$**Quadratic**$_{\text{Schw}}$ contain only $A_1$ as an additional function and do in fact also fall into the same category where $p_i$ coefficients are given by (C.2) (i.e. Eq. (3.6) in [579], which was found for shift-symmetric quadratic HOST). As such, the same discussion below Eq. (6.58) related to the relation between quasinormal mode frequencies in this case and the ones for Schwarzschild black holes in GR also applies here. As mentioned in the previous cases, when HOST functions above contain an implicit time dependence, the master equation for odd-parity perturbations cannot be converted to an ODE, and as a PDE, it makes the definition of quasinormal modes ambiguous. This will be discussed in more detail in Section 6.5.

To conclude this Section, we have obtained the $p_i$ coefficients in the quadratic Lagrangian (6.54) for odd-parity perturbations in general cubic theories under the existence conditions for stealth solutions, i.e. for the models denoted by **Cubic**$_{\text{GR-mat/GR-vac/SdS/Schw}}$. An exhaustive list of the $p_i$ coefficients for the remaining cases can be found in Table 6.5 and the beyond-GR parameter(s) for each case are summarised in Table 6.6.

**Table 6.6:** Here we collect the number of independent combinations of beyond-GR parameters for all different models (i.e. for each set of {*symmetry*, *theory*, *stealth solution*}). The last column shows the symbol we use to condense this information in Table 6.1. Note how the interpretation for $F_2$ differs for shift- vs. non-shift-symmetric models, where in the former, as a constant, does not count as a potential deviation from GR while in the latter, as potentially depending non-trivially on spacetime coordinates, is included as a beyond-GR parameter.

| Model | Beyond-GR parameter(s) | Symbol |
|---|---|---|
| $^{(SS)}$**Cubic**$_{GR-mat}$, $^{(SS)}$**Quadratic**$_{GR-mat}$, $^{SS}$**Cubic**$_{GR-vac}$, $^{SS}$**Quadratic**$_{GR-vac}$ | None | ✗ |
| **Cubic**$_{GR-vac}$ | $F_2 - X_0 F_{3\phi}$ at $(\phi, X) = (\phi_0, X_0)$ | ✓$_1$ |
| **Cubic**$_{SdS}$ | $F_2 - X_0 F_{3\phi}$, $A_1 + F_{3\phi}$ at $(\phi, X) = (\phi_0, X_0)$ | ✓$_2$ |
| **Cubic**$_{Schw}$ | $F_2 - X_0 F_{3\phi}$, $A_1 + F_{3\phi}$, $B_1$ at $(\phi, X) = (\phi_0, X_0)$ | ✓$_3$ |
| **Quadratic**$_{GR-vac}$ | $F_2$ at $(\phi, X) = (\phi_0, X_0)$ | ✓$_1$ |
| **Quadratic**$_{SdS}$, **Quadratic**$_{Schw}$ | $F_2$, $A_1$ at $(\phi, X) = (\phi_0, X_0)$ | ✓$_2$ |
| $^{SS}$**Cubic**$_{SdS}$, $^{SS}$**Quadratic**$_{SdS}$, $^{SS}$**Quadratic**$_{Schw}$ | $A_1$ at $X = X_0$ | ✓$_1$ |
| $^{SS}$**Cubic**$_{Schw}$ | $A_1$, $B_1$ at $X = X_0$ | ✓$_2$ |

# 6.4 Stability and speeds of odd-parity perturbations

In the previous sections we have derived the conditions under which different stealth black hole solutions with time-dependent scalar hair exist for cubic HOST theories (Section 6.2) and obtained the quadratic Lagrangian for odd-parity perturbations (Section 6.3). These kinds of solutions are known to exist for several large classes of ST theories (see e.g. [562, 565, 566, 568–573, 679, 693, 696]). Note in passing that stealth solutions in DHOST theories have been shown to generically suffer from instability or strong coupling issues when the even sector is taken into account [578, 579, 639] (see Table I in [2] for a more comprehensive summary). In this section, as a first step towards evaluating the overall stability of the different models considered, we derive the conditions that HOST functions need to satisfy in order for linear odd-perturbations to remain stable. In the EFT context, as an initial step towards deriving generalised master equations for the even sector without strong coupling issues, the dynamics of monopole ($\ell = 0$) even perturbations have been studied with the inclusion of the scordatura term, whose presence allows for the avoidance of strong coupling issues [648].

## 6.4.1 Stability conditions

In order to rewrite the quadratic Lagrangian (6.54) in terms of one variable, we introduce an auxiliary field $\chi$ and integrate out the variables $h_0$ and $h_1$. For a detailed description of the procedure, see, e.g. [579, 699]. We have seen in the previous section that $p_4 = p_5 = 0$ in cubic HOST under the existence conditions for stealth solutions, and therefore the corresponding terms in Eq. (6.54) will be ignored in what follows. After some manipulations, we obtain the quadratic Lagrangian for $\chi$ as[17]

17: The field $\chi$ is an effective combination of $h_0$ and $h_1$, capturing the one-degree-of-freedom nature of perturbations in the odd-parity sector. More concretely, for $p_4 = 0$, we have $\chi = \dot{h}_1 - \partial_\rho h_0$.

18: Precisely speaking, the quantities $c_\rho^2$ and $c_\theta^2$ correspond to the sound speed of GWs in unit of the speed of photons which are assumed to be minimally coupled to gravity.

$$\frac{(j^2-2)(2\ell+1)}{2\pi j^2}\mathcal{L}_2 = s_1\dot{\chi}^2 - s_2(\partial_\rho\chi)^2 - s_3\chi^2 \,, \tag{6.62}$$

where the parameters $s_i$'s are defined as

$$s_1 \equiv -\frac{(j^2-2)p_3^2}{p_2} \,, \qquad s_2 \equiv \frac{(j^2-2)p_3^2}{p_1} \,,$$
$$s_3 \equiv (j^2-2)p_3\left[1-\left(\frac{\dot{p}_3}{p_2}\right)^{\cdot} - \partial_\rho\left(\frac{\partial_\rho p_3}{p_1}\right)\right]. \tag{6.63}$$

Note that these expressions apply for the case when the $p_i$-coefficients are generic functions of $\tau$ and $\rho$, and hence generalise the expressions in [579, 639, 699] which were derived under the condition that $p_i$'s are functions only of $r = r(\rho - \tau)$.

The variable $\chi$ now represents the propagating degree of freedom in the odd-parity sector. As usual, we define the sound speed squared of GWs (i.e. the speed of GWs) in the radial and angular directions as[18]

$$c_\rho^2 = \frac{\bar{g}_{\rho\rho}}{|\bar{g}_{\tau\tau}|}\frac{s_2}{s_1} \,, \qquad c_\theta^2 = \lim_{\ell\to\infty}\frac{r^2}{|\bar{g}_{\tau\tau}|}\frac{s_3}{j^2 s_1} \,. \tag{6.64}$$

The absence of ghost and gradient instabilities requires that $s_1$, $c_\rho^2$, and $c_\theta^2$ are positive definite:

$$s_1 > 0 \,, \qquad c_\rho^2 > 0 \,, \qquad c_\theta^2 > 0 \,. \tag{6.65}$$

Using the following parametrisation to characterise deviations from unity (i.e. the GR prediction) in the propagation speeds,

$$c_\rho^2 = 1 + \alpha_T^{(\rho)} \,, \qquad c_\theta^2 = 1 + \alpha_T^{(\theta)} \,, \tag{6.66}$$

the stability requirements $c_{\rho/\theta}^2 > 0$ are equivalent to $\alpha_T^{(\rho/\theta)} > -1$ and the GR result is obtained when $\alpha_T^{(\rho/\theta)} = 0$.

We can now assess what these stability conditions imply for all the different cases in Table 6.1. As shown, cases $^{(SS)}$**Cubic**$_{GR\text{-}mat}$ and $^{(SS)}$**Quadratic**$_{GR\text{-}mat}$ recover GR results at the level of linear odd-parity perturbations and hence stability conditions are automatically satisfied.

For general GR solutions in vacuum, more concretely for **Cubic**$_{GR\text{-}vac}$, we obtain the following stability criterion:

$$F_2 - X_0 F_{3\phi} > 0 \,. \tag{6.67}$$

From the above we see that in the cases $^{SS}$**Cubic**$_{GR\text{-}vac}$ and $^{(SS)}$**Quadratic**$_{GR\text{-}vac}$ the stability condition becomes

$$F_2 > 0 \,. \tag{6.68}$$

For the case **Cubic**$_{SdS}$, the stability of perturbations requires that

$$F_2 - X_0 F_{3\phi} > 0 \,, \qquad F_2 + X_0(F_{3\phi} + 2A_1) > 0 \,. \tag{6.69}$$

Similarly, the stability conditions for cases $^{SS}$**Cubic**$_{SdS}$ and $^{(SS)}$**Quadratic**$_{SdS}$ can be straightforwardly obtained from the condition above. In fact, the same conditions also apply for cases $^{(SS)}$**Quadratic**$_{Schw}$, and these are given by

$$F_2 > 0 \, , \qquad F_2 + 2X_0 A_1 > 0 \, . \tag{6.70}$$

This leaves us with two remaining cases. First, we have **Cubic**$_{Schw}$, from which we obtain the following stability conditions:

$$F_2 - X_0 F_{3\phi} > 0 \, , \qquad F_2 + X_0(F_{3\phi} + 2A_1) > 0 \, ,$$
$$r[F_2 + X_0(F_{3\phi} + 2A_1)] - 81\sqrt{2X_0^3(1-A)}\, B_1 > 0 \, , \tag{6.71}$$

where the last condition was derived from requiring $c_\rho^2 > 0$. Second, the stability conditions for $^{SS}$**Cubic**$_{Schw}$ can be straightforwardly obtained by imposing shift symmetry to the condition (6.71),

$$F_2 > 0 \, , \qquad F_2 + 2X_0 A_1 > 0 \, ,$$
$$r(F_2 + 2X_0 A_1) - 81\sqrt{2X_0^3(1-A)}\, B_1 > 0 \, . \tag{6.72}$$

From the above expressions, one sees that setting the cubic HOST function to zero one recovers the condition (6.70) for cases $^{(SS)}$**Quadratic**$_{Schw}$.

### 6.4.2 Speed of gravity

In this Subsection, for illustrative purposes, we analyse the speed of GWs in the odd-parity sector. Let us focus on the cases $^{(SS)}$**Cubic**$_{Schw}$ since they contain non-trivial deviations from GR in the $p_i$ coefficients. Using Eqs. (6.60) and (6.63) in Eq. (6.64), the parameters $\alpha_T^{(\rho/\theta)}$ defined in Eq. (6.66) are given by

$$\alpha_T^{(\rho)} = \frac{-2rX_0(A_1 + F_{3\phi}) + 81\sqrt{2X_0^3(1-A)}\, B_1}{r[F_2 + X_0(2A_1 + F_{3\phi})] - 81\sqrt{2X_0^3(1-A)}\, B_1} \, ,$$
$$\alpha_T^{(\theta)} = -\frac{2X_0(A_1 + F_{3\phi})}{F_2 + X_0(2A_1 + F_{3\phi})} \, . \tag{6.73}$$

From the expressions for the $\alpha_T^{(\rho/\theta)}$ parameters above, it is interesting to understand how they are related to current observational constraints, most notably from the GW event GW170817 and the gamma-ray burst 170817A emitted from a binary neutron star merger, which constrained $|c_{GW} - 1| \lesssim 10^{-15}$ at the frequency scales probed by LIGO-Virgo-KAGRA (LVK) [538, 581–584].

The implication of the bound on GW speed on HOST theories in the context of cosmology is as follows. On a homogeneous and isotropic cosmological background described by the Friedmann-Lemaître-Robertson-Walker metric, the deviation of $c_{GW}^2$ from unity is given by [416]

$$\alpha_T^{cosm} \equiv c_{GW}^2 - 1 = -\frac{2X[A_1 + F_{3\phi} - 6HX(B_2 + B_3)]}{F_2 + X[2A_1 + F_{3\phi} - 12HX(B_2 + B_3)]} \, , \tag{6.74}$$

where $H$ denotes the Hubble parameter. In order for $\alpha_T^{cosm}$ to vanish irrespective of the matter content of the Universe (i.e. irrespective of how $H$

19: Recall that we consider here non-shift-symmetric theories in general, and therefore time dependence can show up through the explicit $\phi$-dependence. However, even if the HOST functions are not strictly constant, assuming that the scalar field is responsible for dark energy for instance, one would expect their timescale of evolution to be much longer than the timescale associated with ringdown observables here, making the HOST functions effectively constant in such an environment.

20: Note that this coincides with the first condition in Eq. (6.75). If we adopt the second condition $B_2 + B_3 = 0$ in our context, this can be understood as $B_1 = 0$ as the existence condition (6.19) for $^{(SS)}\mathbf{Cubic}_{Schw}$ imposes $18B_1 = 2B_2 = -B_3$.

21: As discussed in [720], for $\mathscr{B} < 0$, Eq. (6.78) has only one positive solution for $r_g$, which satisfies $r_g > r_s$. For $0 < \mathscr{B} < \mathscr{B}_c$ with $\mathscr{B}_c \equiv \sqrt{108/3125}$, there exist two positive solutions to (6.78), of which the larger one is identified as $r_g$ and in this case we have $r_g < r_s$. Finally, for $\mathscr{B} > \mathscr{B}_c$, the solution for (6.78) ceases to exist.

22: The event GW170817 was observed at a distance of $40^{+8}_{-14}$ Mpc and the total mass of the binary neutron stars was approximately $2.8M_\odot$ [538], with $M_\odot$ being the mass of the Sun, i.e. $1.99 \times 10^{30}$ kg. The distance in units of $r_s$ (evaluated with the total mass of the system) is given by $r \approx 1.5 \times 10^{20} r_s$.

evolves), one requires that

$$A_1 + F_{3\phi} = 0 , \qquad B_2 + B_3 = 0 . \tag{6.75}$$

Having said that, note that non-trivial $\alpha_T^{cosm}$ is not ruled out for theories whose cutoff scale is lower than (or close to) the frequencies probed by LVK observations. More specifically, higher-order operators suppressed by the scale $\Lambda_3 \equiv (M_{pl}H_0^2)^{1/3}$ (chosen so that these operators give $\mathscr{O}(1)$ contributions to cosmological dynamics), with $H_0$ being the present Hubble parameter, lead to a cutoff close to or below the LVK frequency band (see [545] for theoretical background and [1, 601, 628, 724, 725] for related GW phenomenology and constraints).

We can in principle also apply the bound on GW speed in our context, provided that the spacetime is described by the stealth Schwarzschild solution throughout the propagation of GWs. As previously discussed in, e.g. [690, 719], while applying the GW170817 constraint suggests that $\alpha_T^{(\rho)}$ must approach zero at large distances, it may nonetheless have non-trivial configurations at short distances (i.e. in the black hole environment). In order to provide tangible results, we now assume that the relevant HOST functions are constant when evaluated at the background, i.e. $F_2$, $F_{3\phi}$, $A_1$, $B_1$ are all constant.[19] Then, requiring $\alpha_T^{(\rho)}$ to vanish at large $r$ imposes $A_1 + F_{3\phi} = 0$.[20] In that case, we can rewrite the $\alpha_T^{(\rho)}$ parameter in (6.73) in the following form:

$$\alpha_T^{(\rho)} = \frac{\left(\frac{r_s}{r}\right)^{3/2} \mathscr{B}}{1 - \left(\frac{r_s}{r}\right)^{3/2} \mathscr{B}} , \qquad \mathscr{B} \equiv \frac{81\sqrt{2X_0^3}\, B_1}{r_s(F_2 - X_0 F_{3\phi})} , \tag{6.76}$$

where, in the absence of $F_{3\phi}$, the dimensionless quantity $\mathscr{B}$ exactly matches that introduced in Eq. (59) of [720], from which we adopt the nomenclature. We plot $\alpha_T^{(\rho)}$ as a function of $r/r_g$ for some (negative) values of $\mathscr{B}$ in Figure 6.4, where $r_g$ corresponds to the graviton horizon (i.e. the horizon for the odd modes). Note that, for $\alpha_T^{(\rho)} \neq 0$, $r_g$ is different than the radius of the photon horizon $r_s$ (i.e. the horizon for particles travelling at the speed of light).[21] Following the same discussion as in [690, 719], one can show that $r_g$ satisfies

$$A(r_g) + \alpha_T^{(\rho)}(r_g) = 0 . \tag{6.77}$$

Specifically, on a Schwarzschild background where $A(r) = 1 - r_s/r$, the two horizons are related via

$$r_s = [1 + \alpha_T^{(\rho)}(r_g)]r_g . \tag{6.78}$$

In Figure 6.4 we see that $\alpha_T^{(\rho)}$ transitions smoothly from $-1$ as $r \to 0$ and to zero as $r \to \infty$. In other words, as we approach the singularity ($r \to 0$) the odd modes 'freeze', i.e. $c_\rho = 0$, while at spatial infinity we recover the GR prediction, i.e. $c_\rho = c_{light} (\equiv 1)$. As a first approximation, assuming the constraint on GW speed from the event GW170817 applies to black holes, and taking them to imply that $|\alpha_T^{(\rho)}| \lesssim 10^{-15}$ at $r \approx 10^{20} r_s$,[22] then we can conclude that roughly $|\mathscr{B}| \lesssim 10^{15}$, suggesting that such a measurement is not really effective for constraining this theoretical setup.

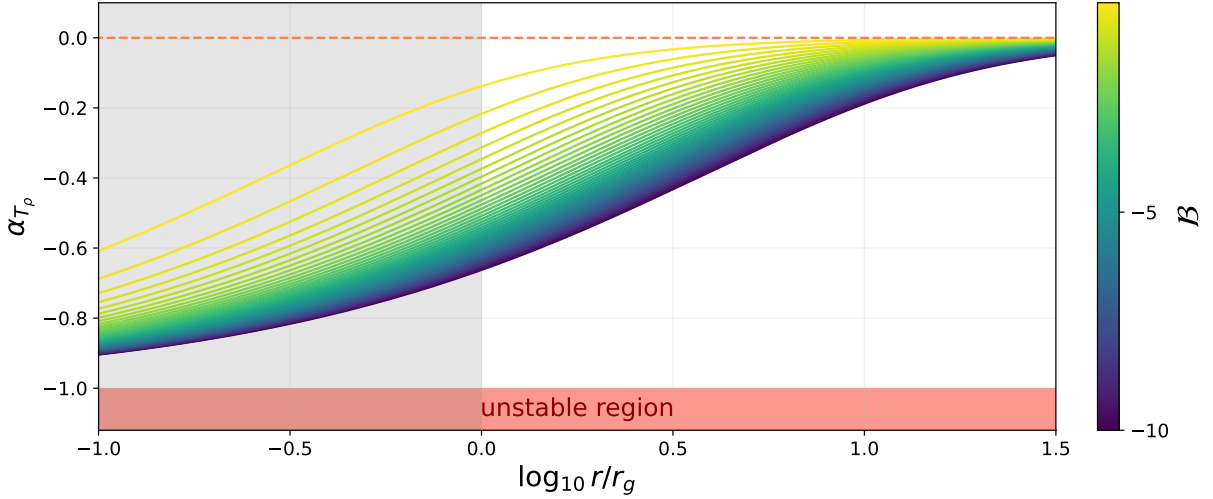

**Figure 6.4:** Speed of GWs in the radial direction as a function of distance for different values of the beyond-GR parameter $\mathscr{B}$ (6.76). The distance is plotted relative to the graviton horizon radius $r_g$, with the region inside $r_g$ shaded in grey (note that $r_g > r_s$ for $\mathscr{B} < 0$). Shaded in red is the region where stability conditions are violated. In dashed orange we show the GR prediction (i.e. $\alpha_T^{(\rho)} = 0$).

## 6.5 Master equation for odd-parity perturbations

In order to further connect the models investigated here with actual ringdown GW observables (e.g. quasinormal mode frequencies), one first needs to derive a master equation for odd modes from the quadratic Lagrangian (6.62). In this Section, we discuss how this derivation becomes more subtle once HOST functions are allowed to carry explicit time dependencies, as is the case for non-shift-symmetric models.

In order to clarify the subtlety, let us follow the derivation of such a master equation in e.g. [579, 690, 699] until some difficulty arises. First, let us rewrite the quadratic Lagrangian (6.62) in terms of the Schwarzschild coordinates $\{t, r, \theta, \phi\}$. After some manipulations, we obtain

$$\frac{(j^2-2)(2\ell+1)}{2\pi j^2}\tilde{\mathscr{L}}_2 = a_1(\partial_t\chi)^2 - a_2(\partial_r\chi)^2 + 2a_3(\partial_t\chi)(\partial_r\chi) - a_4\chi^2 \,, \quad (6.79)$$

with

$$a_1 = \frac{s_1 - (1-A)^2 s_2}{\sqrt{A^3 B(1-A)}} \,, \qquad a_2 = \sqrt{\frac{B(1-A)}{A}}(s_2 - s_1) \,,$$
$$a_3 = \frac{(1-A)s_2 - s_1}{A} \,, \qquad a_4 = \sqrt{\frac{A}{B(1-A)}}s_3 \,. \quad (6.80)$$

As mentioned earlier, though we mainly focus on the S(dS) background in the present Chapter, here we have kept the metric functions $A$ and $B$ independent for generality.[23]

Notice that the Lagrangian (6.79) contains the cross term $(\partial_t\chi)(\partial_r\chi)$. In the case where all the $a_i$ coefficients defined above are static with respect to the Killing time $t$ for the background metric, one can remove the cross term by performing a transformation of the time coordinate, and it is straightforward to define a tortoise coordinate associated with the effective metric for odd modes. Note that this is the case for all shift-symmetric theories, but

23: Note that we have taken into account the Jacobian determinant associated with the coordinate transformation:

$$\tilde{\mathscr{L}}_2 \equiv \left|\frac{\partial(\tau,\rho)}{\partial(t,r)}\right|\mathscr{L}_2. \quad (6.81)$$

can also be applied in non-shift-symmetric cases as long as all HOST functions appearing in these coefficients do not carry an implicit $t$-dependence. In this case, the master equation in the form of a wave equation (i.e., the generalised Regge-Wheeler equation) was obtained in [690, 699] in the context of EFT of BH perturbations.

However, the situation changes for non-shift-symmetric models, where the $a_i$ coefficients in (6.80) are generic functions of $\tau$ and $\rho$. In these cases, these coefficients in Schwarzschild coordinates are no longer independent of $t$, and therefore a redefinition of $t$ that removes the cross term does not exist in general. Moreover, the position of the graviton horizon would also be $t$-dependent, which would make the definition of the tortoise coordinate subtle. As a result, it is not possible to write the master equation in the standard form of the generalised Regge-Wheeler equation as in [690, 699]. In other words, the master equation cannot be converted from a PDE to a simple ODE for a mode with a fixed frequency when the coefficients are allowed to depend on $t$. Consequently, we lose a clear definition for quasinormal mode frequencies. This implies that in order to solve the PDE one would need to resort to a numerical approach. Let us also point out, nonetheless, that when the timescale of the change of the coefficients is long enough compared to that of perturbations, those coefficients in Eq. (6.79) can be effectively regarded as constant and the effect of the time dependence may be treated perturbatively. This is a reasonable assumption as the time dependence of the coefficients appears only through the explicit dependence on $\phi$, which evolves in a cosmological time scale if, e.g., the scalar field is responsible for dark energy. In this case, we expect that the effect of the time dependence can be treated in a perturbative manner. A more in-depth exploration in this direction is left for future work.

## 6.6　Conclusions

In this Chapter, which collects the results in [3], we have explored the landscape of stealth solutions—i.e., solutions that remain identical to those in General Relativity (GR) despite the presence of a non-trivial scalar field—in the framework of general quadratic/cubic higher-order scalar-tensor (HOST) theories. We have considered configurations where the scalar field exhibits time-dependent hair while maintaining a constant background kinetic term. Our analysis has focused on deriving the precise conditions required for either all, or some given specific (Schwarzschild and Schwarzschild-de Sitter), stealth solutions to exist. Furthermore, we have examined the behaviour of odd-parity perturbations and derived the conditions that ensure such perturbations remain stable. Table 6.1 collects the existence conditions, the nature of odd-parity perturbations, and the stability conditions for all combinations of theory setups with different classes of stealth solutions. Our key findings are:

▶ Requiring all GR solutions to exist in the presence of generic matter leads to odd modes that display the same behaviour as in GR. Within the context of general cubic HOST theories, in order to encounter departures from GR one therefore requires one of the following: a) non-stealth metric solutions (see e.g. [562, 565, 566, 569]), b) scalars with a non-constant kinetic term (see e.g. [2, 580, 679]), c) study the even sector (see [579]), or d) relax the requirement of the existence of all GR solutions in the presence of matter and employing less restrictive conditions (this leads to the conclusions in the following 2 bullet points). We leave detailed further explorations on these directions as future work.

▸ Requiring all GR solutions to exist in vacuum also leads to similar results. In the generic case of cubic HOST, we have found one potential beyond-GR parameter in the odd sector. However, imposing shift-symmetry and/or restricting to quadratic interactions recovers the standard GR form for the evolution of odd-parity perturbations. In these cases, departures from GR can also only appear if one takes any of the options a)–c) spelled out in the previous bullet point.

▸ When requiring specific (Schwarzschild and Schwarzschild-de Sitter) stealth solutions to exist, we have found that a large plethora of the (theory + stealth solution) combinations considered in this Chapter results in a reduced set of potential deviations from GR in the odd modes, as described in Table 6.1. In most cases, odd-parity quasinormal mode frequencies can be obtained from those in GR via a simple rescaling, in the same fashion as in [690, 706].

▸ We have identified a unique deviation from GR that does not fall in the previous category for Schwarzschild black holes in cubic HOST (shift-symmetric or otherwise). We have shown how this deviation, denoted by the parameter $\mathscr{B}$, modifies the propagation speed of odd modes in a non-trivial way, in particular with an $r$-dependent $\alpha_T$ parameter. We have shown that the speed of gravity is modified in the black hole environment (while still satisfying stability conditions) and approaches the speed of light at cosmological distances, hence making this an interesting healthy model in light of the constraint on the propagation speed of GWs from the event GW170817. The fundamental quasinormal mode frequency for a shift-symmetric version of this model was investigated in [720]. Given the uniqueness of the $\mathscr{B}$ signature in the large class of models we studied, a further investigation of the quasinormal mode spectrum with contributions from $\mathscr{B}$, as well as an assessment of the observability of $\mathscr{B}$ by current and future GW detectors constitute interesting directions for future research.

▸ We have established that $p_4 = 0$ in the quadratic Lagrangian (6.54) for odd-parity perturbations about static and spherically symmetric background in general cubic HOST theories. In general, non-zero $p_4$ appears in the context of EFT of BH perturbations with a timelike scalar profile [690]. We have shown here that in covariant cubic HOST theories, contributions to $p_4$ from individual terms cancel out in a non-trivial manner. The relevance of this result is that the presence of $p_4 \neq 0$ is associated with the exclusion of slowly rotating black hole solutions (or otherwise the divergence of the radial sound speed at spatial infinity) [690], and so we show that cubic HOST theories do not suffer from this problem.

▸ In the context of non-shift-symmetric theories, we have shown how the master equation for odd-parity perturbations cannot be written in the standard ODE form of the generalised Regge-Wheeler equation due to the time-dependence of its coefficients. We have argued that solving the corresponding PDE would require numerical methods and/or further approximations.

▸ When we require the degeneracy conditions (6.6), (6.7), and (6.8) for quadratic/cubic DHOST theories of class N-I [350] as well as the ex-istence conditions derived in this work, we have concluded that the cubic DHOST part of the theory is not allowed. (The authors of [715] already pointed out that the existence conditions for stealth S(dS) solutions are not compatible with cubic DHOST.) In this case, we are left with the quadratic DHOST of class $^2$N-I, for which the compatibility between the degeneracy and existence conditions is guaranteed.

In this Chapter, we have shown how restrictive it is to require the existence of exact stealth solutions with timelike scalar profile in the space of scalar-tensor theories. On the other hand, as briefly reviewed in the introduction, one needs to introduce so-called scordatura terms in order to avoid strongly coupled perturbations, promoting the background solution to an approximately stealth one that behaves as stealth in practice at the level of the background and that is free from the strong coupling issue for perturbations. Since scordatura terms are of order unity (and not necessarily small) in the unit of the cutoff of the theory, it is expected that relaxing the existence condition of exact stealth solutions to that of approximately stealth ones should significantly broaden the space of scalar-tensor theories. We leave detailed investigations of this important problem for future work.

# Conclusions and Outlook

This thesis has explored the theoretical and phenomenological implications of scalar-tensor (ST) theories of gravity through the lens of black hole solutions and gravitational wave observations, with a focus on the ringdown regime.

In the *Preliminaries* part of this thesis, we have laid the foundational groundwork for the later exposition of three different *Investigations*. This has included an overview of our understanding of gravity throughout history in Chapter 1, where we have highlighted some remarkable similarities between concepts in cutting-edge research and some millennia-old ideas. We have also discussed views on the progress of science, with the goal of better understanding the current state of gravitational physics, and concluding that there are signs of an approaching paradigm shift in cosmology and gravitation.

Chapters 2 and 3 have respectively reviewed the theoretical structure and observational tests of General Relativity (GR) and scalar-tensor (ST) theories, focusing on black hole solutions and ringdown probes. As a recap, we have exposed how the motivation for considering alternatives theories of gravity, in particular ST, arises from limitations of GR when applied to cosmological and astrophysical phenomena. While GR has been remarkably successful in describing gravitational interactions in a wide range of settings, it requires the introduction of dark energy and dark matter to account for the observed accelerated expansion of the universe and galaxy rotation curves. These components, though effective phenomenologically, remain unexplained within the standard framework. Additionally, while the mathematical structure of GR results in very unique settings describing the nature of black holes and their gravitational wave emission, these still remain largely unconstrained. Therefore, testing the robustness of these predictions in strong-field regimes is essential. ST theories, as investigated in this thesis, provide a well-motivated class of extensions to GR, introducing a scalar field that modifies the gravitational dynamics. These theories enable the construction of black hole solutions with richer dynamics—i.e. hairy black holes—and serve as testbeds for probing deviations from GR. We have shown how the rise of gravitational wave (GW) science offers a powerful new observational window into the strong-field regime of gravity, allowing us to test the fundamental predictions of GR and its alternatives with unprecedented precision. Among the various phases of a GW signal from a compact binary merger, in this thesis we have focused on the ringdown phase, dominated by the black hole's quasinormal modes (QNMs), which is particularly well-suited for probing the nature of gravity in the vicinity of black holes. This stage is modelled with the framework of black hole perturbation theory. QNMs are determined entirely by the spacetime geometry and the underlying theory, and are independent of any other idiosyncrasies of the particular source. As such, they provide a theoretically clean and observationally accessible means of testing the validity of GR in the strong-field regime. Deviations in the QNM spectrum from GR predictions can signal the presence of new gravitational degrees of freedom or modifications to the theory, making black hole spectroscopy a direct and robust probe of beyond-GR physics.

The *Investigations* part of this thesis collects three Chapters in which the new findings of the presented research presented have been discussed. Some details for each of the projects can be seen in Table 6.7 and here we briefly summarise the main conclusions.

**Table 6.7:** Summary of the main aspects of the investigations presented in Chapters 4–6. Note that Chapter 4 contains no explicit reference to the stability of the solution. This is because such solution, instead of being exact, is constructed by adding small parametrised deviations to a Schwarzschild black hole with a constant scalar field, a configuration which is stable by nature, and therefore the stability of this solutions is guaranteed by the hairy perturbative hierarchy. The asterisk $*$ in the observational constraints column for Chapter 6 is there to denote that, rather than an actual forecasted constraint, such huge number should be understood as the insensitivity of GWs to this beyond-GR parameter. This number comes from an interpretation of the GW170817 result, while the other two results in this column refer to forecasted constraints in the LISA band from one event.

| Theory | Background | | Stability | QNMs | Observational constraints |
|---|---|---|---|---|---|
| | $g_{\mu\nu}$ | $\phi$ | | | |
| **Ch 4** [1] | Horndeski w/: $G_{4\phi} = 0$ | Schwarzschild | $\phi(r)$ static (paramterised) | Satisfied | Parametrised ringdown formalism | $|\alpha_T| \leq 10^{-2}$ |
| **Ch 5** [2] | Horndeski w/: $G_2 = -2\Lambda - 2\eta\sqrt{X}$, $G_4 = 1 + \lambda\sqrt{X}$, $G_3 = 0 = G_5$ | Schwarzschild -de Sitter | $\phi(\tau)$ time-dependent (exact) | Derived conditions | Semi-analytical WKB | $\hat{\beta} \leq 10^{-4}$ |
| **Ch 6** [3] | cubic HOST | stealth | $\phi(\tau)$ time-dependent (exact) | Derived conditions | Connected to other literature | $\mathcal{B} \leq 10^{15*}$ |

## Speed of gravity—based on [1]

In Chapter 4, we investigated the speed of gravitational waves as a phenomenological probe of beyond-GR physics in the ringdown phase. Focusing on hairy black hole solutions with static scalar hair, we demonstrated that upcoming space-based detectors such as LISA and TianQin will be capable of constraining deviations from the speed of light—quantified by the parameter $\alpha_T$—at the percent level from a single supermassive black hole merger. While several studies have explored constraints on $\alpha_T$, these have predominantly been conducted in the context of a cosmological background, see e.g. [541–543, 585]. In contrast, our work provides the first analysis of $\alpha_T$ from a black hole background. The ringdown constraints we forecast are particularly relevant in the LISA frequency band, where ST theories can remain within their regime of validity—i.e., below the cutoff scale associated with their interpretation as effective dark energy theories [545]. Moreover, the ambiguity of electromagnetic counterparts in this band enhances the value of purely gravitational constraints, such as the one presented here. We further showed that, within ST theories, odd parity QNMs exhibit $\alpha_T$ dependence only in the presence of scalar hair. Consequently, a non-detection of $\alpha_T$ deviations in the ringdown signal can also be interpreted as placing constraints on the presence and magnitude of black hole scalar hair.

## Time-dependent scalar—based on [2]

Chapter 5 expanded the scope of investigation to include black holes endowed with time-dependent scalar hair. We focused on a recently discovered exact solution in which the scalar field exhibits non-trivial kinetic structure, characterised by a non-constant background kinetic term $X$ [679]. Through stability analysis and QNM calculations of odd parity perturbations, we found that deviations from GR in the ringdown spectrum are potentially stable and observable with current and future detectors. This chapter provided concrete constraints on the parameter controlling the deviation, $\hat{\beta}$, with potential bounds at the level from LVK and space-based interferometers, respectively. Time-dependent scalar field configurations are of particular interest due to their possible connections with cosmologically relevant limits. However, previously known solutions of this kind have generally been plagued by instabilities, linked to their assumption of a constant background kinetic term $X = \text{const}$ [578]. In contrast, the theory studied here circumvents this limitation by allowing for a non-constant $X$. Our findings therefore represent a significant step toward identifying viable families of stealth black hole solutions with time-dependent scalar hair.

## Inverting no-hair theorems—based on [3]

Finally, in Chapter 6, we adopted a reverse-engineering approach to determine the precise restrictions on ST theories—in the broad context of cubic higher-order scalar-tensor (HOST) theories—that allow for exact stealth black hole solutions, i.e. those with a GR-like metric but non-trivial scalar structure, in this case also with linear time-dependent hair. Focusing on theories that admit stealth solutions is motivated by their ability to reconcile non-trivial scalar dynamics with current observational constraints, their role as a structural filter within ST theory space, and their potential to connect cosmological and strong-field physics. We derived covariant equations of motion for such theories and showed that requiring the existence of particular sets of stealth solutions imposes strong theoretical constraints on the allowed theory space. We derived these precise conditions both generically and for specific cases such as Schwarzschild and Schwarzschild-de Sitter backgrounds. Our analysis of odd-parity perturbations revealed that, in most cases, the QNM spectrum matches that of GR up to simple rescalings, as in [690, 706]. However, we identified a unique deviation parameter, $\mathscr{B}$, in cubic HOST theories that induces a non-trivial, radially dependent modification to the gravitational wave speed $\alpha_T$, offering an intriguing phenomenology consistent with current observational bounds. Furthermore, we demonstrated that covariant cubic HOST theories generically avoid the issue of diverging radial sound speed often associated with EFT formulations as in [690], and therefore allow in principle for slowly-rotating black hole solutions. We also clarified the limitations of obtaining well-posed perturbation equations in non-shift-symmetric cases. Finally, we showed that imposing degeneracy and existence conditions simultaneously restricts the viable DHOST theories, in particular excluding cubic DHOST models, and therefore restricting to quadratic DHOST of class $^2$N-I. These findings underscore how demanding the requirement of exact stealth solutions is, and suggest that relaxing this requirement in favour of approximately stealth configurations—made viable by introducing so-called scordatura terms [701]—may signifi-cantly enlarge the space of phenomenologically viable scalar-tensor theories.

## Future work

Building on the findings of this thesis, several promising avenues for future research emerge:

- ▶ **Rotating black holes.** All astrophysical black holes are expected to rotate to some degree, and extending beyond static solutions is essential for realistic modelling. While some QNM results for rotating black holes in modified gravity exist, a broader, systematic investigation remains to be done.
- ▶ **Even parity sector.** This thesis has focused on odd-parity perturbations, which are often more tractable analytically. However, even-parity perturbations—where scalar degrees of freedom typically couple more directly—offer rich phenomenological signatures and are essential for a complete understanding of deviations from GR. Extending the current methods to this sector is a natural and important next step. In particular, an even-parity study of the configuration in Chapter 5 will reveal whether a fully stable corner of parameter space remains for such theories when considering both odd and even parity perturbations
- ▶ **Higher-dimensional parameter constraints.** The application of black hole spectroscopy in this thesis has been idealised in the sense that we have assumed the 'usual' binary black hole merger parameters (masses, amplitudes, phases) to be known—e.g. from the inspiral phase—and focused on the effect of novel beyond-GR parameters. A more comprehensive analysis would extend such analysis by investigating degeneracies and constraints in higher-dimensional parameter spaces.
- ▶ **Using GW data.** While this thesis has provided forecasted constraints for several beyond-GR parameters, it would be interesting to employ the already in hand data from LVK GW observations to set similar constraints. Data analysis tools for such analyses already exist, and have been mostly implemented under model-agnostic parametrisations. The work presented in this thesis has set some ground in order to enable theory-specific constraints on ST parameters from actual GW ringdown observations. This effort, which relies on partially developing the previous bullet points, is already underway.
- ▶ **Non-stealth solutions.** This thesis focused on exact stealth black hole solutions. However, as shown in Chapter 6, the existence of exact stealth configurations impose strong theoretical constraints. An alternative approach involves relaxing these conditions and considering non-stealth solutions. One way is to consider black hole solutions with primary hair, where the spacetime geometry is already modified at the background level. Multiple such solutions have been found and it would therefore be interesting to investigate their QNM phenomenology further. Another way is to consider approximate stealth solutions, made viable through the inclusion of so-called scordatura terms. These terms resolve strong-coupling issues without spoiling stealth-like behaviour at the background level. Investigating the landscape of approximately stealth solutions would likely broaden the viable theory space significantly and offer richer phenomenology in the GWs, and therefore potentially stronger constraints.

In summary, this thesis has explored the interplay between black hole physics, gravitational wave observations, and scalar-tensor modifications of gravity. From theoretical foundations to phenomenological applications, the results presented here contribute to the growing effort to test gravity in its strong regimes. As new observational data continues to arrive with increasing precision, further theoretical work in the lines of this thesis will be required in order to translate such exceptional experimental milestones into theory-informed statements about fundamental gravity. With it, the study of black holes and their associated observational signatures will remain a powerful probe of fundamental physics.

# Appendix

# A

## `ringdown-calculations`: a repository with `Mathematica` notebooks

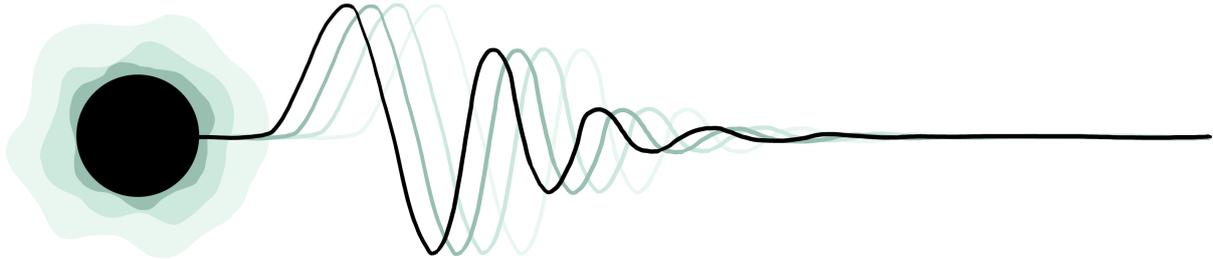

**Figure A.1:** Logo of `ringdown-calculations`.

All new calculations in this thesis are made openly available and reproducible in the `Git-Hub` repository `ringdown-calculations` [4]. Here we briefly describe the contents of such repository, which mainly includes different `Mathematica` notebooks employing the `xAct` tensor algebra [208]. As the notebooks are fully documented themselves, we will not include here any descriptions of how calculations are actually carried out, but will instead direct the interested reader to the notebooks.

[4]: Sirera (2023), *ringdown-calculations*

[208]: "Martín-García (), *xAct*

▸ **Basic examples.** This folder contains two notebooks with standard examples for the application of black hole perturbation theory in General Relativity (GR) and scalar-tensor (ST) theories respectively. These serve as more approachable introductions to the techniques used in the companion notebooks for the papers.

   · `Schwarzschild-GR.nb`. This includes the derivation of Regge-Wheeler and Zerilli equations for a Schwarzschild black hole in GR by applying the perturbations at the level of the action. The calculations here correspond to the contents of Section 2.4.

   · `Schwarzschild-Horndeski.nb`. In this notebook we derive the three master equations (gravitational odd, gravitational even, scalar even) for a Schwarzschild black hole in Horndeski gravity. The calculations here correspond to the contents of Section 3.5.

▸ `Testing-speed-gravity.nb`. This is the companion notebook to the paper [1], or equivalently, the one containing the calculations in Chapter 4, where we investigate how the speed of gravitational waves can be tested with ringdown observations and focusing on odd parity quasinormal modes of hairy black holes in Horndeski theories. The notebook contains all calculations starting from the covariant action to the final Fisher parameter estimation and all details in between including: quadratic action, modified Regge-Wheeler equation, scalar hair and other parametrisations, quasinormal mode calculation, construction of waveform with relevant parameters, calculation of SNR and errors, plots.

[1]: Sirera et al. (2023), "Testing the speed of gravity with black hole ringdowns"

▶ `Stability-QNMs-SdS-t-dep.nb`. This is the companion notebook to the paper [2], or equivalently, the one containing the calculations in Chapter 5, where we investigate the stability and quasinormal modes of black holes with time-dependent scalar hair. The notebook contains all calculations starting from the covariant action to the final Fisher parameter estimation and all details in between including: background equations of motion, quadratic action, stability conditions, modified Regge-Wheeler equation, WKB quasinormal mode calculation, construction of waveform with relevant parameters, calculation of SNR and errors. This notebooks make use of the WKB package in [252].

  • `plots-Stability-QNMs-SdS-t-dep.ipynb`. Here we make available the python code for all plots in [2].

▶ **Inverting-no-hair-theorems**. In this folder we collect the companion notebooks to the paper [3]—i.e. Chpater 6—where we explore black hole solutions in cubic/quadratic higher-order scalar-tensor (HOST) theories. These notebooks construct an adaptable general formalism for the study of cubic HOST theories at the background and perturbative levels, which can be tuned to specific models.

  • `Inverting-no-hair-theorems-I.nb`. In this notebook we define cubic HOST theories, derive the covariant equations of motion and obtain the conditions for the existence of stealth GR solutions. This corresponds to the calculations in Section 6.2.
  • `Inverting-no-hair-theorems-II.nb`: In this notebook we derive the quadratic Lagrangian for black hole perturbations in cubic HOST theories both in covariant and component form, where the latter is associated to a static and spherically symmetric background. Focusing on odd parity modes, we also derive their stability conditions. This corresponds to the calculations in Sections 6.3 and 6.4.
  • **Precomputed expressions**: The notebooks above contain calculations that can take several minutes to complete. This can be avoided by loading the precomputed expressions contained in this folder. To do so, one needs to place these files in a path accessible by `Mathematica`. You can check the list of accessible paths by `Mathematica` by typing `$Path` in a `Mathematica` notebook. You can append a new path with `AppendTo[$Path, "your_path"]`. The precomputed expressions in this folder are:

| | | |
|---|---|---|
| ∗ `metricEOMs.mx` | ∗ `deltaLF2comp.mx` | ∗ `deltaLA3comp.mx` |
| ∗ `scalarEOM.mx` | ∗ `deltaLF2phicomp.mx` | ∗ `deltaLB1comp.mx` |
| ∗ `deltaL-cubic-HOST-odd.mx` | ∗ `deltaLF2Xcomp.mx` | ∗ `deltaLB1Xcomp.mx` |
| ∗ `deltaLcomp2.mx` | ∗ `deltaLF3comp.mx` | ∗ `deltaLB2comp.mx` |
| ∗ `deltaLcomp3.mx` | ∗ `deltaLF3phicomp.mx` | ∗ `deltaLB2Xcomp.mx` |
| ∗ `deltaLF0comp.mx` | ∗ `deltaLF3Xcomp.mx` | ∗ `deltaLB3comp.mx` |
| ∗ `deltaLF0Xcomp.mx` | ∗ `deltaLA1comp.mx` | ∗ `deltaLB3Xcomp.mx` |
| ∗ `deltaLF1comp.mx` | ∗ `deltaLA1Xcomp.mx` | ∗ `deltaLB4comp.mx` |
| ∗ `deltaLF1Xcomp.mx` | ∗ `deltaLA2comp.mx` | ∗ `deltaLB6comp.mx` |
| | ∗ `deltaLA2Xcomp.mx` | |

# B

# Semi-analytic WKB method: $\Lambda$ effects on QNMs

In this Appendix, we explore the effect that a cosmological constant term $\Lambda$ has on the emission of quasinormal modes, and use this as a complementary check of the semi-analytical WKB method used in Section 5.4.2 of Chapter 5. One would expect that such an effect is strongly suppressed for two reasons: 1) the small value of the cosmological constant, and 2) the fact that quasinormal mode emission should be governed by the local dynamics around the source, and $\Lambda$ contributes as a large-$r$ term to the spacetime metric

$$B = 1 - \frac{2M}{r} - \frac{1}{3}\Lambda r^2. \tag{B.1}$$

Whether this really impacts at large $r$ or not is of course dependent on the value of $\Lambda$. As was shown in [290], for cosmological values of $\Lambda$ (i.e. $\Lambda/M_{\text{pl}}^2 = 10^{-52}\text{m}^{-2}$ in natural units) its effect on the QNMs is negligible. For example, for a black hole of $10^6 M_\odot$, the correction to the quasinormal modes appears at the $10^{-34}$ level.

The SdS solution also allows for values of $\Lambda$ larger than the one for the cosmological constant. In particular, the range of values allowed for $\Lambda$ which preserve the 2 horizon nature of SdS is $0 \leq \Lambda \leq \frac{1}{9M^2}$,[1] QNM values spanning this full range of $\Lambda$ were calculated in [726]. In the cases considered there, where the black hole and cosmological horizon are of comparable size, it is not surprising that deviations in the frequencies become quite significant.

We employ here the WKB method to obtain quasinormal frequencies in the two same ways as has been done in Section 5.4.2. First, we apply the package [252] directly and obtain numerical solutions for specific values of $\Lambda$. By doing so, we find that our results agree well with [726]. Secondly, we use the $6^{th}$ order WKB formula (5.73) and take derivatives of the potential while keeping $\Lambda$ unspecified. Hence, we are able to obtain an expression for $\omega(\Lambda)$. As was the case with $\omega(\beta)$ in Section 5.4.2, the expression is cumbersome and therefore unfeasible to write here. One can find such expression in [4]. Here, we instead Taylor expand around the zero value for $\Lambda$ and obtain[2]

$$\begin{aligned} M\omega = M\omega_0 &- [1.67631 - 0.33463i] \cdot M^2\Lambda \\ &- [3.85847 - 0.89970i] \cdot M^4\Lambda^2 \\ &- [17.4684 - 6.04082i] \cdot M^6\Lambda^3 + \mathcal{O}(\Lambda^4). \end{aligned} \tag{B.2}$$

where we recover $M\omega_0$ as the Schwarzschild quasinormal frequencies. Using this semi-analytical expressions, we are also able to recover the same results as in [726], gaining precision for smaller $\Lambda$. This provides a non-trivial validation test of the semi-analytical prescription, which we developed in Chapter 5 to constrain $\hat{\beta}$. Note that the maximum of the potential is shifted due to $\Lambda$. Employing the *light-ring expansion* [637], for small values of $\Lambda$, this

is given by

$$r_*^{\mathrm{max}} = 3.28 \cdot M - 2.85 \cdot M^3\Lambda + \mathcal{O}(\Lambda^2).$$ (B.3)

3: Note the the mass of the Sun in geometric units is around $1.5\mathrm{km} \simeq 10^3\mathrm{m}$.

Finally, for a black hole of $10^6 M_\odot \simeq 10^9\mathrm{m}^3$ and a cosmological constant of $\Lambda = 10^{-52}\mathrm{m}^{-2}$, we can easily use equation (B.2) to confirm that the first correction to the quasinormal frequencies appears at the $10^{-34}$ level, in agreement with [290].

# C

# Fisher forecast formalism

In this Appendix, we spell out the details in the construction of the Fisher forecast formalism, which is used in Chapters 4 (Section 4.4) and 5 (Section 5.5) to derive forecasted constraints on beyond-GR parameters. Full details on the implementation of this formalism can be found in the corresponding companion `Mathematica` notebooks in [4].

We begin by modelling the waveform as[1] [216]

$$h = h_+ F_+ + h_\times F_\times,$$ (C.1)

where $h_{+,\times}$ represent the strain in the two polarisations of the gravitational wave. These are in principle functions of all coordinates, i.e. $h_{+,\times}(t, r, \theta, \phi)$. However, to distinguish between time and frequency domains, we will only make $t$ (or $\nu$) explicit and take the dependence on $r, \theta, \phi$ as understood. $F_{+,\times}$ are functions encoding the geometry of the problem (i.e. they depend on the angles specifying the orientation of the source with respect to the detector). The strain functions for the ringdown are given by

$$h_+(t) = \sum_{\ell m} A_{\ell m}^+ e^{-\frac{\pi t f_{\ell m}}{Q_{\ell m}}} S_{\ell m} \cos(\phi_{\ell m}^+ + 2\pi t f_{\ell m}),$$

$$h_\times(t) = \sum_{\ell m} A_{\ell m}^+ N_\times e^{-\frac{\pi t f_{\ell m}}{Q_{\ell m}}} S_{\ell m} \sin(\phi_{\ell m}^\times + 2\pi t f_{\ell m}),$$ (C.2)

where these are the strain functions as emitted by the source and we will implicitly assume that these trivially propagate to the detector.[2] In (C.2) we have absorbed any overall constant normalization factors into the amplitude parameters $A_{\ell m}^+$,[3] and where $f_{\ell m}$ and $\tau_{\ell m}$ characterise the real and imaginary parts of $\omega_{\ell m}$ in the following way

$$\omega_{\ell m} = 2\pi f_{\ell m} + \frac{i}{\tau_{\ell m}}, \qquad\qquad Q_{\ell m} = \pi f_{\ell m} \tau_{\ell m},$$ (C.3)

where $Q_{\ell m}$ is the quality factor. $\{A_{\ell m}^+, A_{\ell m}^\times = A_{\ell m}^+ N_\times, \phi_{\ell m}^+, \phi_{\ell m}^\times\}$ are the amplitudes and phases for the two polarisations. Finally, $S_{\ell m}$ are spheroidal functions carrying angular dependencies. Because modes with different $(\ell, m)$ do not mix at linear level due to the nature of our background, the $\ell m$ indices in $\{\omega_{\ell m}, f_{\ell m}, \tau_{\ell m}, Q_{\ell m}, S_{\ell m}, A_{\ell m}^+, \phi_{\ell m}^+, \phi_{\ell m}^\times\}$ will play no role for the time being, so we will obviate them to simplify our notation.[4]

Using the above strain functions, we compute the signal-to-noise-ratio (SNR) with the usual

$$\rho^2 \equiv (h|h) = 4 \int_0^\infty d\nu \, \frac{\tilde{h}(\nu)^* \tilde{h}(\nu)}{S_h(\nu)},$$ (C.4)

where $S_h(\nu)$ is the noise spectral density characteristic of the detector and $\tilde{h}(\nu)$ is the Fourier transform of $h(t)$.[5]

1: Note that, throughout this section, we ubiquitously use the techniques developed in [216] for our analysis.

2: The propagation of GWs in GR is exposed in Section 2.3.3, and effects altering these waveforms throughout their cosmological evolution in ST theories are discussed in detail in Section 2.3.3.

3: The strain functions $h_{+/\times}$ appear with different normalisation factors in the literature depending on the setup in question, e.g. with a factor of $1/2\sqrt{10\pi}$, an extra geometrical $\sqrt{3/4}$ for LISA or a $\frac{1}{r}$ factor [216, 217, 219, 651]. We choose to remain general and absorb all such factors into the amplitudes $A^+$. This does not affect the calculations presented in this section, as these factors only enter trough the signal-to-noise-ratio $\rho$, for which the appropriate detector-specific values will be discussed and used in the next section.

4: Which $\ell m$ modes are of interest in physically sensible scenarios is discussed where relevant in the main text.

5: Note that in (C.4) we use $\nu$ rather than $f$ for the frequency domain representation (or Fourier transform) of the time coordinate. This is to distinguish it from the real component of the quasinormal modes $f_{\ell m}$ as defined in (C.3), especially since we will be omitting the $\ell m$ indices.

We now make use of the following set of simplifying assumptions: $\langle F_+ \rangle = \langle F_\times \rangle = 1/5$, $\langle F_+ F_\times \rangle = 0$, $\langle |S|^2 \rangle = 1/4\pi$, $A^+ = A$. We also make use of the fact that we can approximate $S_h(v)$ to be constant. For details on (and explicit checks of) these assumptions see [216]. We hence obtain

$$\rho^2 = \frac{QA^2((1+N_\times^2)(1+4Q^2) + \cos 2\phi^+ - N_\times^2 \cos 2\phi^\times)}{(1+N_\times^2)(1+4Q^2)\pi f S_h}. \qquad \text{(C.5)}$$

Now, we further assume that $\phi^+ = \phi^\times$ and $N_\times = 1$ (i.e. $A^\times = A^+ = A$) [216]. However, we stress that this is not a necessary assumption to recover the expression for the single-parameter error (C.11). We do find it necessary to recover the double-parameter error expressions (C.12). Using these assumptions, (C.4) can be re-expressed as

$$\rho^2 = \frac{QA^2}{\pi f S_h}. \qquad \text{(C.6)}$$

To derive error estimates we make use of the Fisher Information Matrix, defined by

$$\Gamma_{ab} \equiv \left( \frac{\delta h}{\delta \theta^a} \Big| \frac{\delta h}{\delta \theta^b} \right), \qquad \text{(C.7)}$$

where $\theta^a$ is the set of parameters for our theory and the noise-weighted product $(\cdot|\cdot)$ is defined as

$$(h_1|h_2) \equiv 2 \int_0^\infty dv \frac{\tilde{h_1}^* \tilde{h_2} + \tilde{h_2}^* \tilde{h_1}}{S_h(v)}. \qquad \text{(C.8)}$$

Then, we can calculate the parameter errors by inverting the Fisher matrix (which gives the covariance matrix $\Sigma$). The error for a parameter $a$ is given by

$$\sigma_a = \sqrt{\Sigma_{aa}} = \sqrt{\Gamma_{aa}^{-1}}. \qquad \text{(C.9)}$$

In the work described in this Thesis, we study the simplified case where all the usual parameters of the waveform are known ($A, \phi^+, \dots$) and our only free parameters are the beyond-GR parameters, here denoted as ($\alpha, \beta, \gamma, \dots$). We leave forecasting full joint constraints to future work. This simplified setup effectively corresponds to an initial estimate providing upper bounds on the precision one can expect for such parameters.

6: Note that we need two variables to represent the real and imaginary parts of $\omega$ and, similarly to previous literature [455] we choose to work with the pair $\{f, Q\}$ rather than $\{f, \tau\}$. We note that, if working with the latter, the SNR equation (C.6) would be further simplified to

$$\rho^2 = \frac{\tau A^2}{S_h}. \qquad \text{(C.10)}$$

7: This simple expression also implicitly makes use of the large-$Q$ limit (or equivalently large damping times $\tau$). More terms appear at the $Q^{-4}$ order. For details on the validity of this approximation we again refer to [216] and the full 'unapproximated' expressions are available in the companion notebook [4].

Hence, for a setup as considered here, where the only waveform parameters we want to constrain are those appearing inside the quasinormal frequencies $\omega$ (i.e. inside $f$ and $Q$)[6], general expressions for the errors can be analytically derived. These only depend on the number of parameters one wants to constrain. In this Thesis we constrain up to two beyond-R parameters together so we provide here the expression for single-parameter constraints[7]

$$\sigma_\alpha^2 \rho^2 = \frac{1}{2} \left( \frac{f}{Qf'} \right)^2, \qquad \text{(C.11)}$$

where the prime denotes a derivative with respect to $\alpha$, and for double-

parameter constraints

$$\sigma_\alpha^2 \rho^2 = \frac{\dot{f}^2}{2} \frac{(2Q)^2 + (1 - \frac{f\dot{Q}}{\dot{f}Q})^2}{(\dot{Q}f' - \dot{f}Q')^2},$$

$$\sigma_\beta^2 \rho^2 = \frac{f'^2}{2} \frac{(2Q)^2 + (1 - \frac{fQ'}{f'Q})^2}{(\dot{Q}f' - \dot{f}Q')^2}, \tag{C.12}$$

where again a prime denotes a derivative with respect to $\alpha$, and a dot represents a derivative with respect to $\beta$. This matches analogous expressions in [455].

# Bibliography

Here are the references in citation order.